\documentclass[11pt]{book}

\usepackage{graphicx}
\usepackage[paperwidth=22cm,paperheight=27.5cm]{geometry}
\usepackage{balance}
\usepackage{amsmath}
\usepackage{amsfonts}
\usepackage[applemac]{inputenc}
\usepackage{braket}
\usepackage{cite}

\usepackage{amsthm}
\usepackage{MnSymbol}
\usepackage{mdframed}
\newtheorem*{result}{Result}
\newenvironment{fresult}
  {\hspace*{0,5cm}\begin{mdframed}\begin{result}}
  {\end{result}\end{mdframed}}

\usepackage{afterpage}

\usepackage{fancyhdr}
\usepackage{wrapfig}
\usepackage{dsfont}

\usepackage{titlesec}

\titleformat*{\section}{\LARGE\bfseries}
\titleformat*{\subsection}{\Large\bfseries}
\titleformat*{\subsubsection}{\large\bfseries}

\usepackage{array}
\newcolumntype{M}[1]{>{\centering\arraybackslash}m{#1}}

\usepackage[urlcolor={red},citecolor=false]{hyperref}

\begin{document}


\begin{titlepage} 
\vspace*{2cm}
\begin{center}
\hspace{0.6cm}
\includegraphics[width=0.65\textwidth]{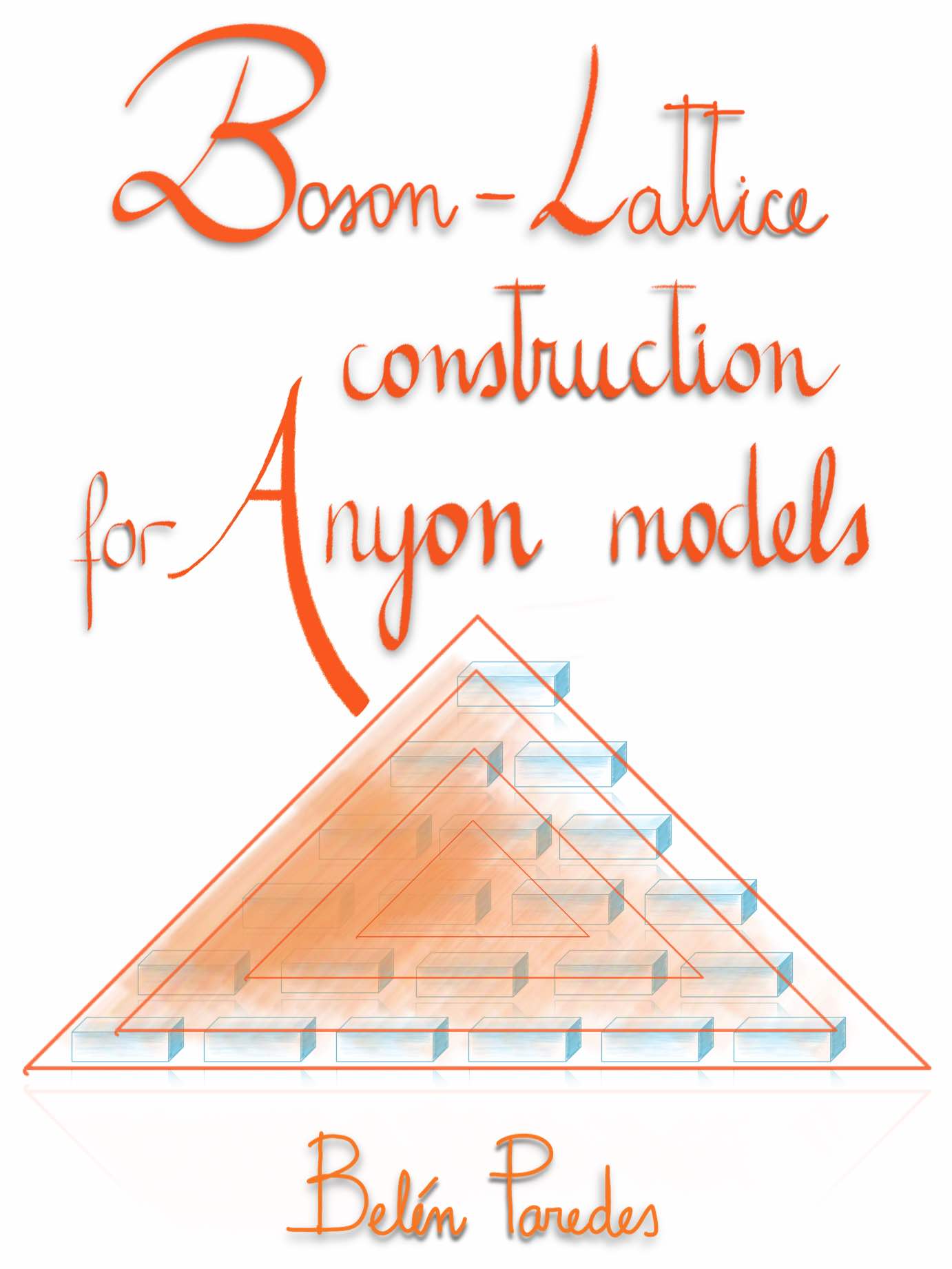}
\end{center}
\end{titlepage}

\newpage
\thispagestyle{empty}
\vspace*{10cm}
\hspace*{7cm}
\includegraphics[width=0.4\textwidth]{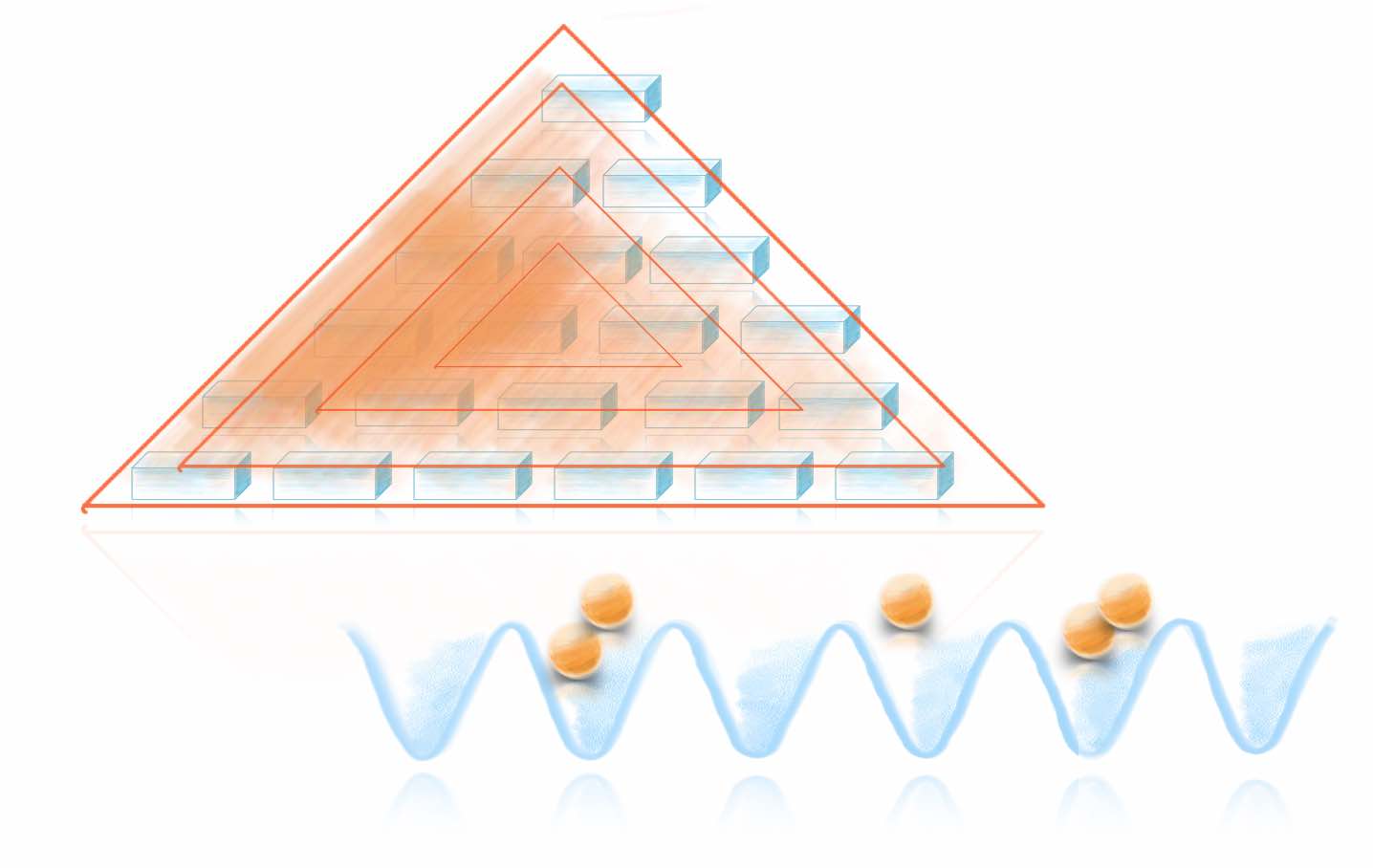}

\newpage
\thispagestyle{empty}
\vspace*{1cm}
\begin{center}
\hspace*{0.7cm}
\includegraphics[width=0.8\textwidth]{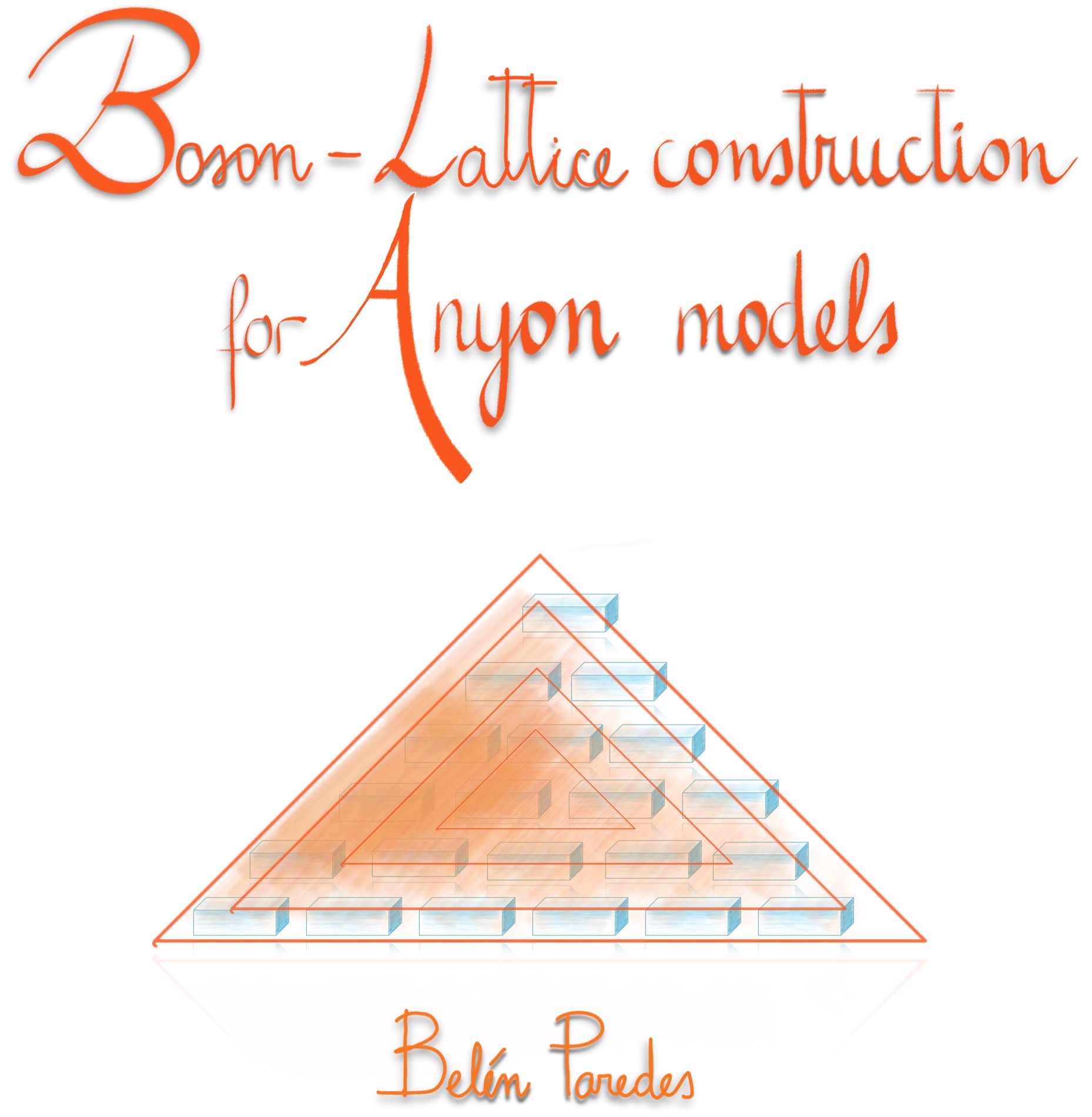}
\end{center}
\begin{center}
\hspace*{1cm}
\Large Arnold Sommerfeld Center for Theoretical Physics\\
\hspace*{1cm}
Ludwig Maximilian University\\
\vspace*{1cm}
\hspace*{1cm}
Munich, January 2018
\end{center}


\newpage
\thispagestyle{plain}
\vspace*{2cm}
\begin{center}
\hspace*{1.5cm}
\parbox{13cm}
{\parskip=5pt
\bf This work is an attempt to unveil the skeleton of anyon models. 

I present a construction to systematically generate anyon models.
The construction uses a set of elementary pieces or fundamental anyon models, which constitute the building blocks to construct other, more complex, anyon models. A principle of assembly is established that dictates how to articulate the building blocks, setting out the global blueprint for the whole structure. Remarkably, the construction generates essentially all tabulated anyon models. Moreover, novel anyon models (non-tabulated, to my knowledge) arise. 

To embody the construction I develop a very physical, visual and intuitive lexicon. An anyon model corresponds to a system of bosons in a lattice. By varying the number of bosons and the number of lattice sites, towers of more and more complex anyon models are built up. It is a Boson-Lattice construction. A self-similar anatomy is revealed: an anyon model is a graph that is filled with bosons to engender a new graph that is again filled with bosons. 

And further, bosons curve the graph that the anyon model is: I disclose a geography in the space of anyon models, where one is born from another by deforming the geometry of space. I advance an alluring duality between anyon models and gravity. }
\end{center}

\newpage
\thispagestyle{plain}
\vspace*{2cm}
\begin{center}
\hspace*{1.5cm}
\parbox{13cm}
{\parskip=15pt
In this manuscript I present the full theory of the Boson-Lattice construction, carrying out a complete analysis with thorough proofs, all combined with enlightening illustrations and discussions. 

For a {\bf condensed manuscript} containing the key essential ideas and results of the Boson-Lattice construction, I direct the reader to my manuscript\\
\begin{wrapfigure}{l}{0.5\textwidth}
\vspace*{-1cm}
\includegraphics[width=0.5\textwidth]{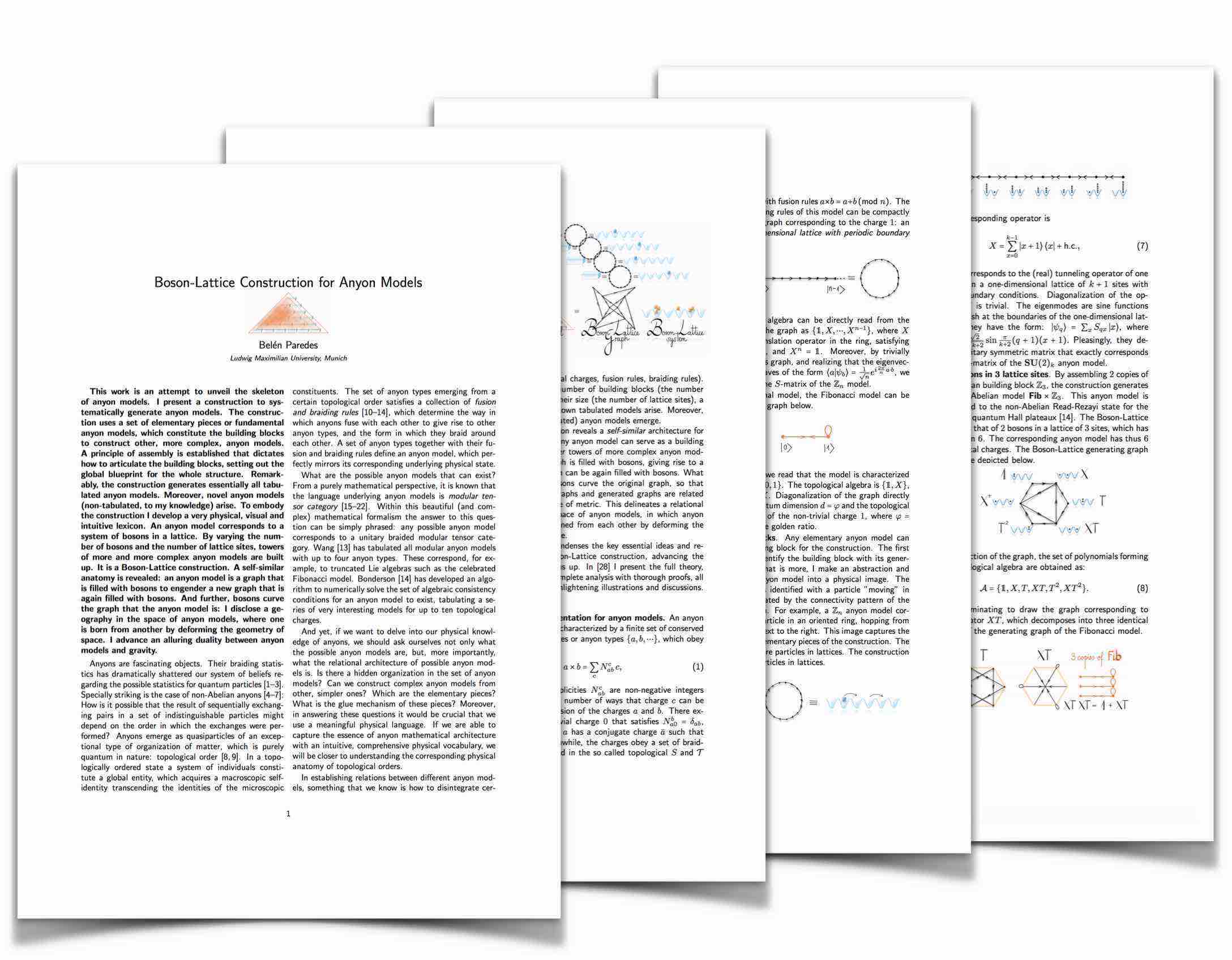}
\end{wrapfigure}
\url{http://www.theorie.physik.uni-muenchen.de/agparedes/publications/bl-construction-corto.pdf}\\

In this article the reader can also find further generalizations and implications of the Boson-Lattice formalism.}
\end{center}

\newpage
\thispagestyle{plain}
\section*{}
\addcontentsline{toc}{section}{Introduction}
\normalsize

Anyons \cite{A1,A2,A3,A4,A5} are probably the most interesting objects I have encountered in physics.

Anyons captivate because of their mathematical beauty, because studying them means to enter into the heart of topology, of knot and braid theories, of tensor categories.

Meanwhile anyons are fascinating because of their deep physical meaning. They emerge as quasiparticles of an exceptional type of organization of matter, which is purely quantum in nature:  topological order \cite{TO1, TO2}.
In contrast to all other types of orders (crystals, ferromagnets, or even superfluids or superconductors) topological order cannot be described by a classical local field. It is not readable from the individual components (atoms, electrons, photons), it is hidden in the pattern of many-particle entanglement established among them.

In a topologically ordered state a system of individuals constitute a global entity, which acquires a macroscopic self-identity that transcends the identities of the microscopic constituents. The laws that govern the emergent collectivity are topological laws: they are invariant under local deformations of the system. This is in radical opposition to the nature of the original microscopical laws (electromagnetic or gravitational forces), which are strongly dependent on geometrical details such as distances or angles.

Anyons are profoundly counterintuitive. Their braiding statistics has dramatically shattered our system of beliefs regarding the possible statistics for quantum particles \cite{A1,A2,A3,A4,A5}. Specially striking is the case of non-Abelian anyons \cite{NA1,NA2,NA3,NA4,NA5,NA6,NA7,NA8,NA9,NA10}: How is it possible that the result of sequentially exchanging pairs in a set of indistinguishable particles might depend on the order in which the exchanges were performed?

We do have mathematical languages to describe anyons and represent their bizarre properties. We could say, for example, that an anyon can be identified with an irreducible representation of the group of braids, and that its non-Abelian character is a natural consequence of the composition of braids being non-commutative. 
But that we are able to name or represent anyons with appropriate mathematical tools does not mean that we thoroughly understand what they are, nor, specially, does it mean that we can explain under which conditions they emerge from a physical system. 

There is still a large gap between the topological mathematical rules governing anyons and the physical laws dictating the behavior of the underlying physical system. 
What is the correspondence between a certain type of anyon statistics and the pattern of many-particle entanglement that gave it birth? Which particular combination of microscopic degrees of freedom, interactions among them and external fields, made such pattern emerge? We can still not give precise answers to these questions.

\newpage
\thispagestyle{plain}
\vspace*{1cm}
It appears to me that finding physically meaningful languages to describe anyons can help us sharpen our knowledge about them. I believe that if we are able to capture the essence of anyons using an intuitive, comprehensive physical vocabulary, we will be closer to fill the lacunas between the mathematical and the physical, between the global and the local faces of topological orders.

\vspace*{1cm}
\begin{center}
\includegraphics[width=0.9\textwidth]{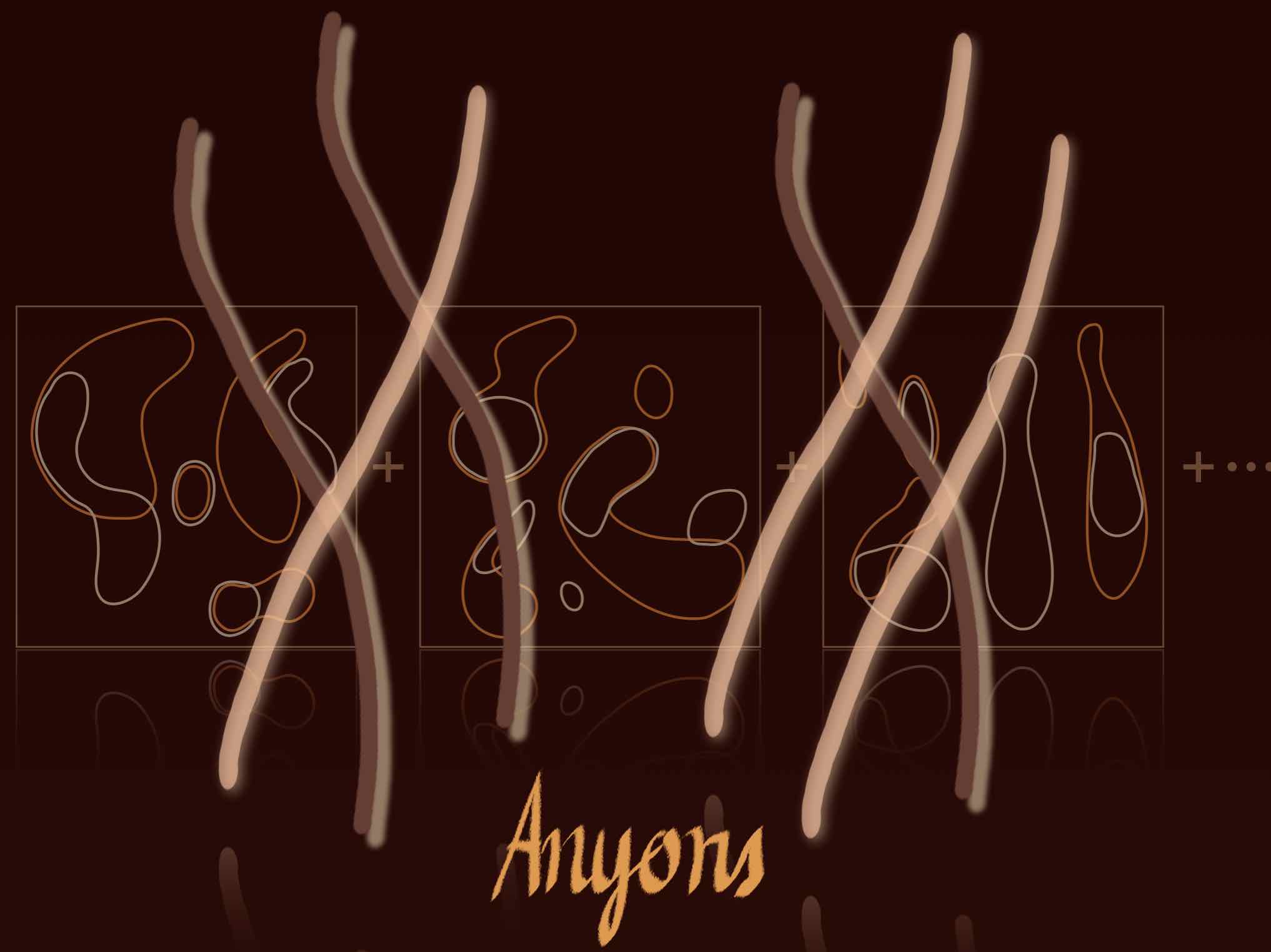}
\end{center}

\newpage
\thispagestyle{plain}
\section*{}
\addcontentsline{toc}{section}{Introduction}

The set of anyon types emerging from a certain topological order satisfies a collection of {\em fusion and braiding rules} \cite{Kitaev1,Kitaev2,Preskill,Wang,Bonderson}. These rules determine the way in which anyons fuse with each other to give rise to other anyon types, and the form in which they braid around each other. A set of anyon types together with their fusion and braiding rules define an anyon model. 
Different topological orders give rise to distinct anyon models, which perfectly mirror their corresponding underlying physical states.

A first natural question we can ask  is: What are the possible anyon models that can exist? The answer to this question can be easily expressed in a formal manner.
For an anyon model to exist its fusion and braiding rules cannot be arbitrary. They have to fulfill a collection of consistency conditions which can be written as a set of equations, known as the Pentagon and Hexagon equations \cite{Preskill, Bonderson}. An anyon model is therefore a solution of these equations. 
Wang \cite{Wang} has tabulated all {\em modular} anyon models with up to four anyon types. These correspond, for example, to truncated Lie algebras such as $\mathbf{SU}(2)_2$, or $\mathbf{SO}(3)_3$, the celebrated Fibonacci model.
Bonderson \cite{Bonderson} has developed an algorithm to numerically solve the pentagon and hexagon equations under certain conditions, tabulating a series of very interesting anyon models for up to ten topological charges. 

From a purely mathematical perspective, it is known that the language underlying anyon models is {\em modular tensor category}. Firstly, the structures of anyon models originated from conformal field theory \cite{CFT1, CFT2} and Chern-Simons theory \cite{CHS1}. They were further developed in terms of algebraic quantum field theory \cite{AT1, AT2} and then made mathematically rigorous in the language of braided tensor categories \cite{TC1,TC2,TC3}. Within this beautiful (and complex) mathematical formalism the answer to the question above can be also simply phrased:
any possible anyon model corresponds to a unitary braided modular tensor category. 

But if we want to delve into our physical knowledge of anyons, we should refine the question above.  We should ask ourselves not only what the possible anyon models are, but, more importantly, what the relational architecture of possible anyon models is. We should be able to find answers to questions such as: Is there a hidden organization in the set of anyon models? Can we construct complex anyon models from other simpler ones? Which are the elementary pieces? What is the glue mechanism of these pieces?
In clarifying these questions, it seems unclear whether generating solutions to the Pentagon and Hexagon equations or enumerating possible unitary braided theories can be by themselves illuminating enough.

\newpage
\thispagestyle{plain}
\vspace*{1cm}

In establishing relations between different anyon models, something that we know is how to disintegrate certain complex anyon models into other simpler ones.  This procedure, called {\em anyon condensation} \cite{AC1,AC2,AC3,AC4,AC5,AC6,AC7,AC8,AC9,AC10}, works by making two or more different anyon types become the same. 
Though no fully general description is known, for the special case in which the condensing anyons have trivial statistics, it is possible to systematically obtain a condensed anyon model from a more complex (uncondensed) one.
Anyon condensation does not tell us, however, about the reverse process, that is, about how to build up more complex anyon models by putting simpler ones together. To go in this "up" direction, we have only straightforward operations at our disposal, such as, for instance, making the tensor product of two or more given anyon models. 

Here, I believe it is crucial to develop pathways to orderly construct anyon models.  This can help us enormously to apprehend the subjacent texture of anyon models and thereby the anatomy of topological orders.

\newpage
\thispagestyle{plain}
\vspace*{0cm}
\begin{center}
\includegraphics[width=0.7\textwidth]{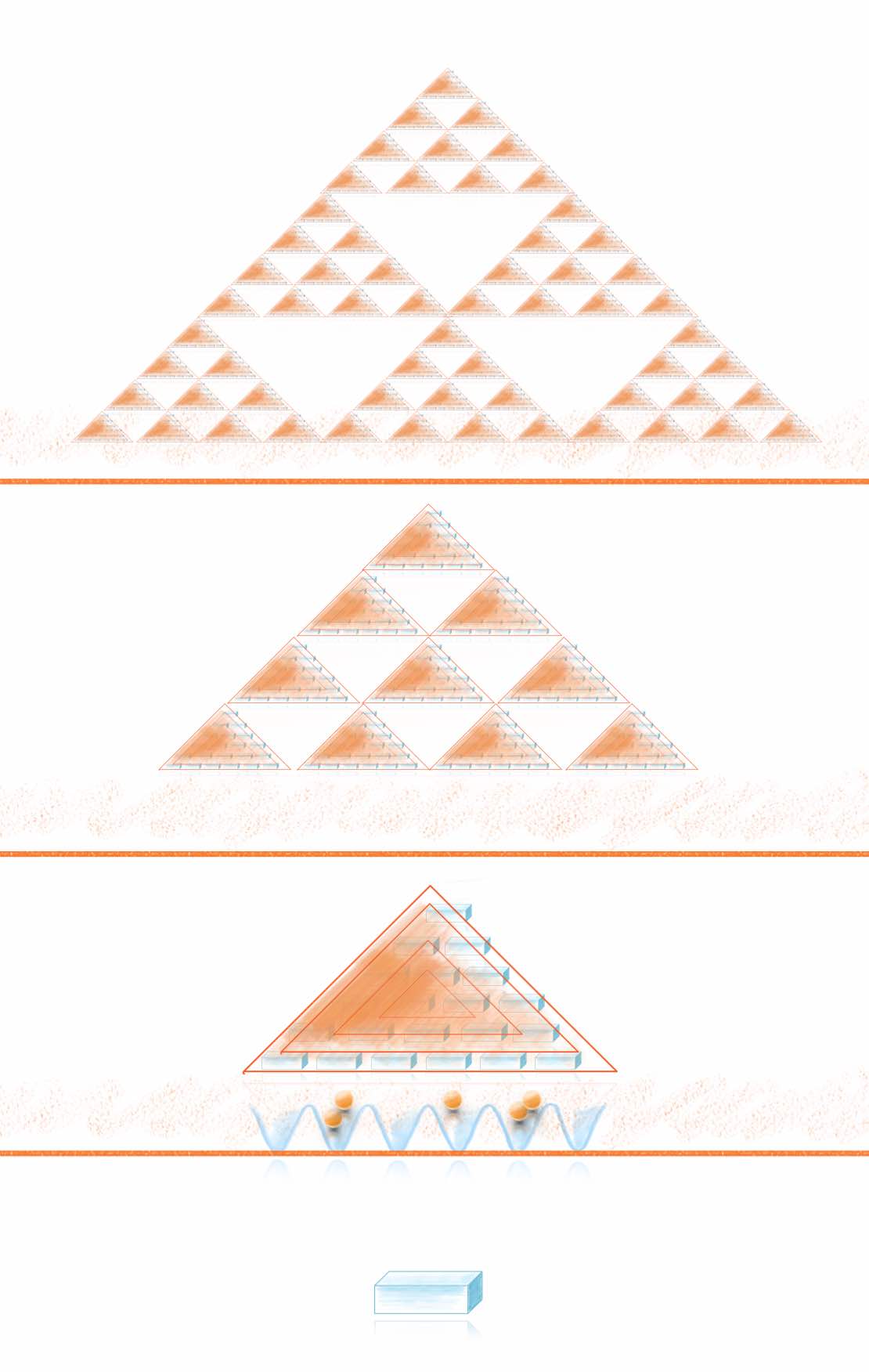}
\end{center}

\newpage
\thispagestyle{plain}
\section*{}
\addcontentsline{toc}{section}{Introduction}

This work is an attempt to find out the skeleton of anyon models. 

I present a construction to systematically generate anyon models.
The construction uses a set of elementary pieces or fundamental anyon models, which constitute the building blocks to construct other, more complex, anyon models.
A principle of assembly is established that dictates how to articulate the building blocks, setting out the global blueprint for the whole structure. 
Remarkably, the construction generates essentially all tabulated anyon models \cite{Wang, Bonderson}. Moreover, novel anyon models (non-tabulated, to my knowledge) arise.

The construction is formulated in a very physical, visual and intuitive manner. An anyon model corresponds to a system of bosons in a lattice.
By varying the number of bosons and the number of lattice sites, towers of more and more complex anyon models are built up. It is a Boson-Lattice construction.
Importantly, the Boson-Lattice systems used in the construction are not real physical systems from which anyon models emerge. Here, Boson-Lattice systems {\em are} themselves  anyon models.

To formulate the construction I develop a language for anyon models. In this language an anyon model is represented by a graph or a collection of graphs, which encode the properties of the anyon model. Topological charges (anyon types) are represented by graph vertices. Fusion rules can be read from the connectivity pattern of the graphs.
Braiding rules are obtained through diagonalization of the graphs.
This graph language is the first contribution of this work. It provides an enlightening way to encode anyon models,  allowing to both easily visualize and extract their properties.

The elementary pieces of the Boson-Lattice construction are the Abelian $\mathbb{Z}_n$ anyon models. In the language of graphs these models are represented by periodic one-dimensional lattices, in which each lattice site is connected to its next (to the right) neighbour. 
Triggered by this graph representation, the first key idea to develop the construction arises: I make a conceptual leap by identifying a $\mathbb{Z}_n$ model with a single particle in a periodic lattice with $n$ sites.
This visual image condenses the essence of the anyon model into a particle in a lattice. It inspires the next crucial step: the conception of the principle of assembly. The assembly of the building blocks is defined as a {\em bosonization procedure}, in which particles corresponding to different building blocks are made indistinguishable. The resulting Boson-Lattice system  characterizes the constructed anyon model.

\newpage
\thispagestyle{plain}
\vspace*{2.3cm}
\begin{center}
\includegraphics[width=\textwidth]{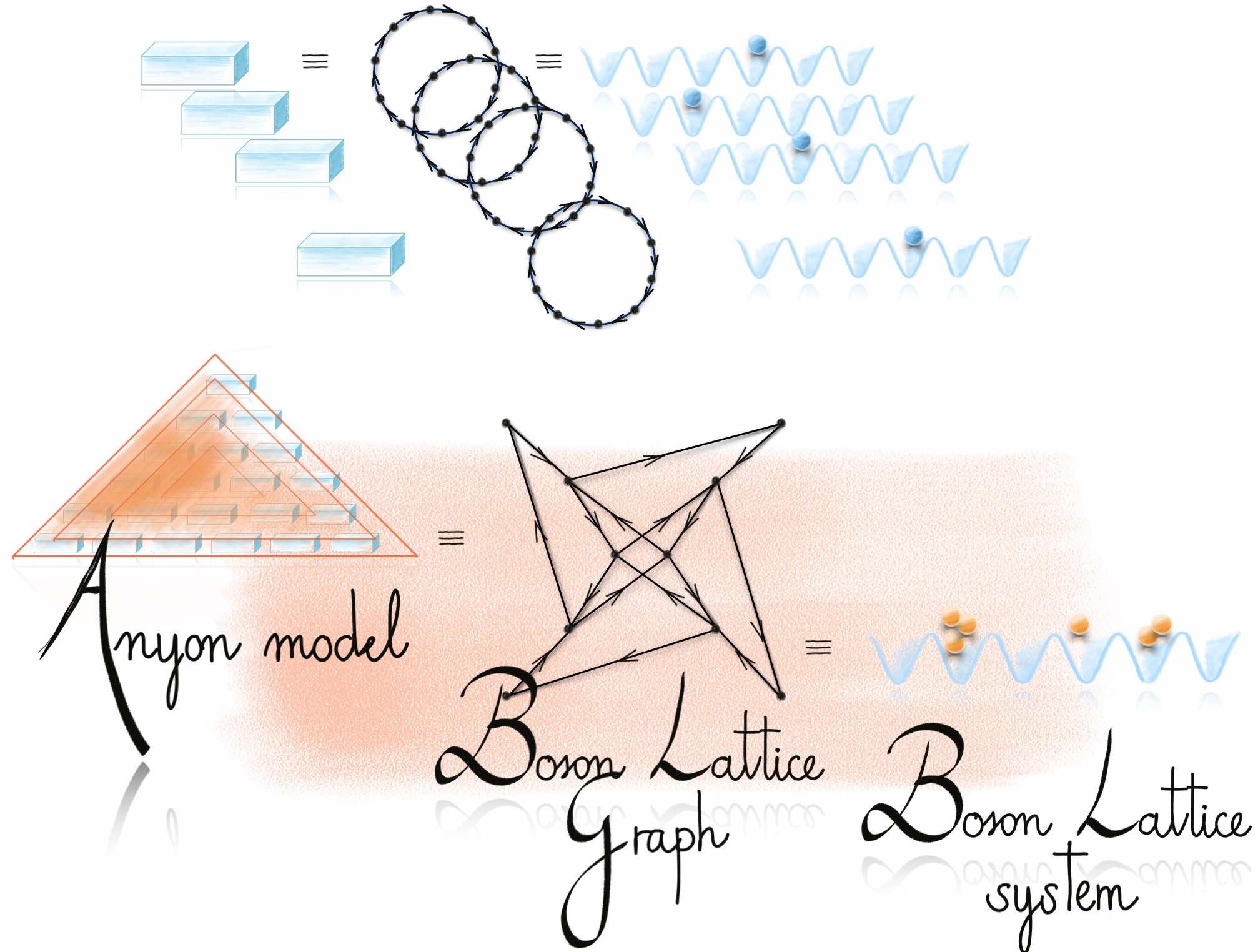}
\end{center}

\newpage
\thispagestyle{plain}
\vspace*{0.5cm}

I give a prescription to assign a graph to the Boson-Lattice system. The graph is defined as the {\em connectivity graph} of the corresponding Fock states.
The central result of this work states that, with this definition, the Boson-Lattice graph always embodies a modular anyon model. A dictionary is established between the elements of the Boson-Lattice graph (Fock states, connectivity pattern, eigenvalues and eigenstates), and the properties of the anyon model (topological charges, fusion rules, braiding rules).
The special features of the Boson-Lattice graph assure that the extracted properties are well defined and satisfy the required consistency conditions.

To illustrate the construction I consider several examples of anyon models arising within the Boson-Lattice formalism. It is pleasing to see how series of well known anyon models are generated by varying the number of bosons and the number of lattice sites. $\mathbf{SU}(2)_k$ anyon models are constructed as Boson-Lattice systems of $k$ bosons in a lattice of $2$ sites. The Fibonacci anyon model corresponds to a Boson-Lattice system of $2$ bosons in a lattice of $3$ sites. The series of $\mathbf{SO}(3)_k$ models are built up as systems of $2$ bosons in lattices of $k$ sites. Furthermore, other non-tabulated anyon models emerge, as, for instance, those corresponding to Boson-Lattice systems of $2$ bosons in $4$ lattice sites, or $3$ bosons in $3$ lattice sites.

The Boson-Lattice construction is a fractal construction. Anyon models created by assembling the building blocks $\mathbb{Z}_n$ can be used themselves as elementary pieces to generate new models at a second level of the construction. Nicely, the principle of assembly is the same at any level of the construction, giving rise to a self-similar pattern that replicates itself at any scale. Might this fractal architecture be the one behind anyon models? 

The formalism can be generalized by adding internal degrees of freedom to the bosons participating, by using multidimensional lattices, or, additionally, by considering fermions instead of bosons. It is very interesting to see how further series of anyon models, such as, for instance, {\em quantum double models} \cite{Kitaev1}, can arise from such generalizations.

The construction reveals an anatomy for anyon models. 
I have focused here on building up {\em modular} anyon models, for which corresponding topological field theories and conformal field theories exist. It would be revealing  to
investigate how known structures and concepts in topological quantum field theory and conformal field theory can be interpreted within the language of the Boson-Lattice construction. And, conversely, to see how the construction might shed light onto the skeleton of topological field theories and conformal field theories themselves.

\newpage
\thispagestyle{plain}
\vspace*{3cm}

\begin{center}
\includegraphics[width=0.85\textwidth]{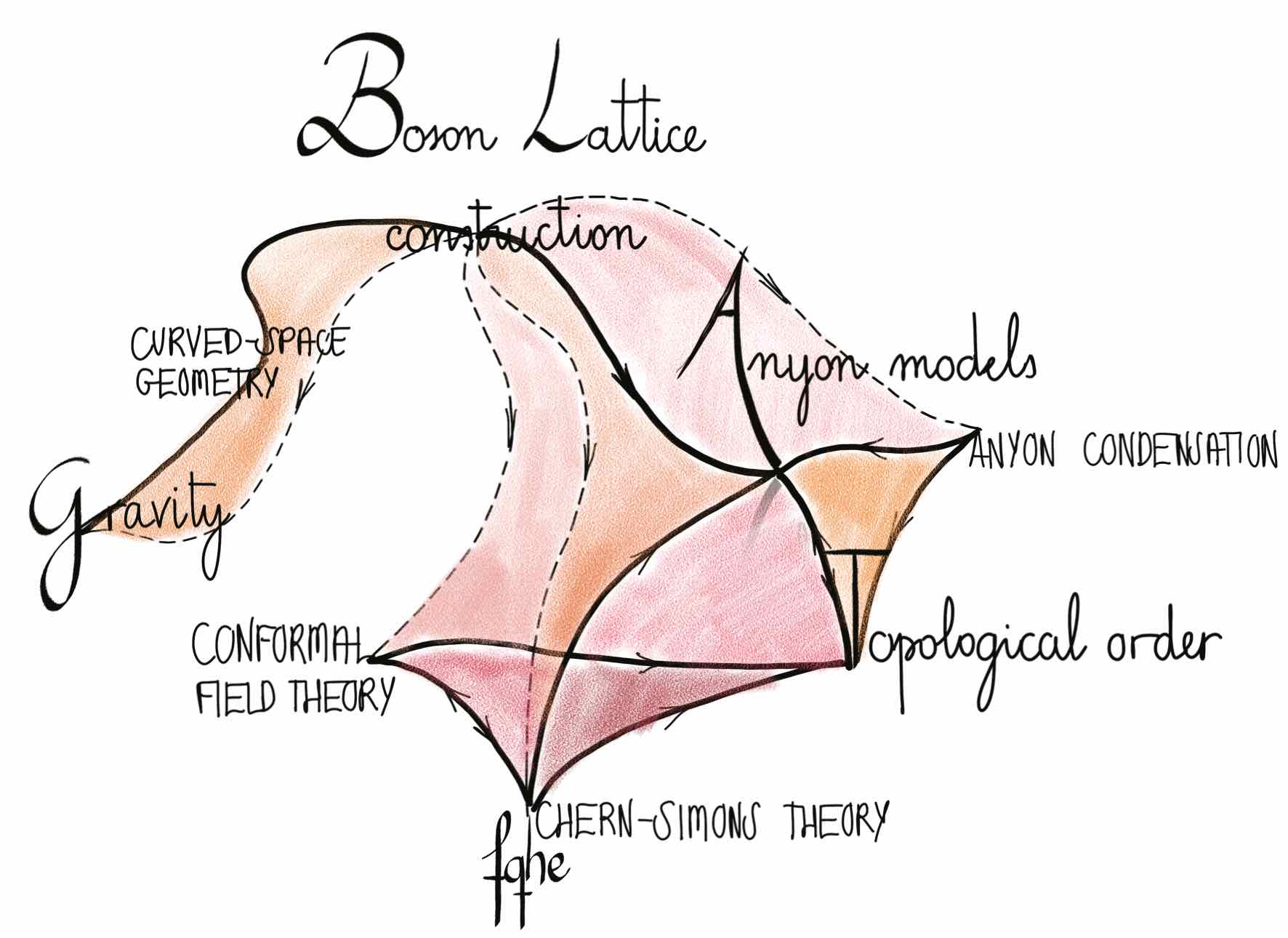}
\end{center}

\newpage
\thispagestyle{plain}
\vspace*{2cm}

The Boson-Lattice approach can guide us to develop a construction for topological states and Hamiltonians at the microscopic level. The Boson-Lattice blueprint can serve as a dual blueprint to design the corresponding many-particle wave functions.
Moreover, the actual Boson-Lattice system that abstractly represents the anyon model, can help us to design the microscopic degrees of freedom and interactions composing the topological model from which the anyon model emerges. I believe the Boson-Lattice system can be regarded as a dual physical entity, able to encode at the same time the global and the local, the mathematical and the physical ingredients of a topological order.

I find extremely interesting to draw a map of connections among the many-body wave functions and Hamiltonians generated by the Boson-Lattice construction and those arising in seminal topological systems and models, such as {\em fractional quantum Hall systems} \cite{NA1,NA2,NA3,NA4,NA5,NA6,NA7}, {\em quantum loop models} \cite{Kitaev2, QLM1, QLM2, QLM3} and {\em string-net models} \cite{SNC1,SNC2,SNC3,SNC4}.

As an essential outcome, this work reveals that the mathematical language describing anyon models can be identical to the one describing bosonic lattice systems. It states that the fusion rules and braiding rules characterizing anyon models can be represented by simple, intuitive physical objects, such as Fock states or tunneling Hamiltonians.

It is indeed remarkable that the connectivity graph of Fock states of a bosonic lattice system can encode the non-trivial consistency conditions required for an anyon model to exist.
While developing the Boson-Lattice formalism I was thrilled to see how the construction succeeded in always generating anyon models with the correct properties, for any number of bosons and lattice sites, at any level of the construction.
I was urged to understand the reason behind this extraordinary coincidence. 
Nicely, in trying to find intuitive grounds for it, an unexpected connection emerged: a correspondence between Boson-Lattice graphs and {\em curved space geometries}. 

This connection anticipates an intriguing duality between anyon models and curved space-time geometries, between anyon models and {\em gravity}. Understanding and developing this duality is a challenge I feel compelled to achieve.

\newpage
\thispagestyle{empty}
\vspace*{10cm}
\hspace*{7cm}
\includegraphics[width=0.4\textwidth]{ConstructionTitleFigure.jpg}

\newpage
\thispagestyle{plain}

\section*{}
\large
\includegraphics[width=0.5\textwidth]{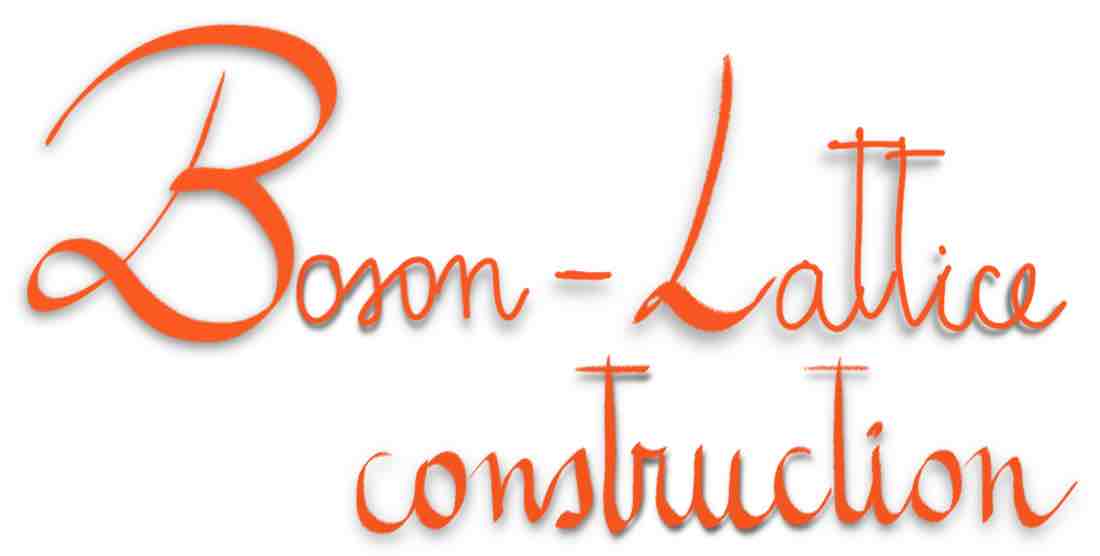}

\hspace*{1cm}
\begin{tabular}{m{2.5cm}m{10cm}m{1cm}}
\includegraphics[width=0.1\textwidth]{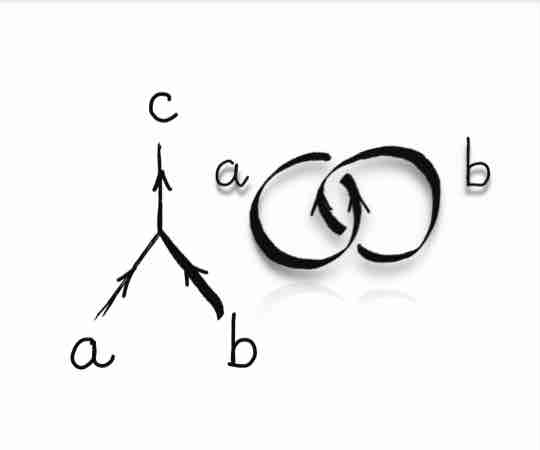}&\large Preliminaries & \normalsize17\\
\includegraphics[width=0.1\textwidth]{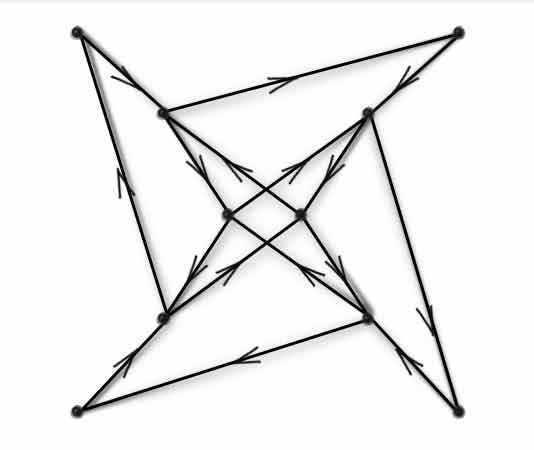}&\large Topological graphs. The language of the construction & \normalsize 21\\
\includegraphics[width=0.1\textwidth]{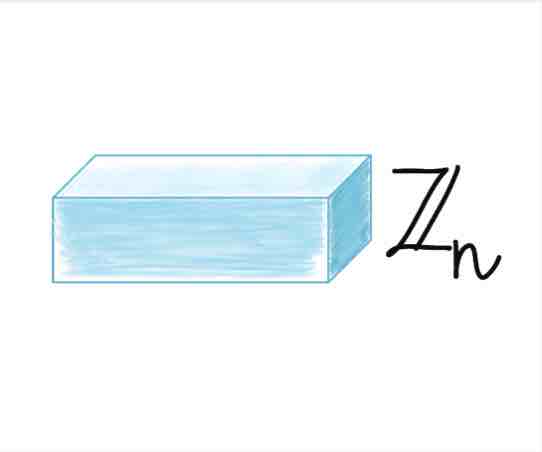}&\large The building blocks &  \normalsize 41\\
\includegraphics[width=0.1\textwidth]{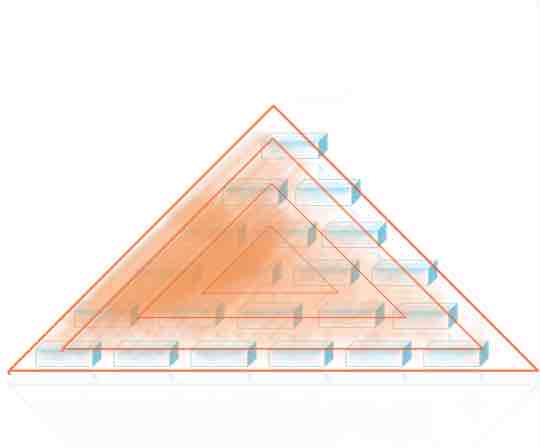}&\large The Boson-Lattice construction & \normalsize 49\\
\includegraphics[width=0.1\textwidth]{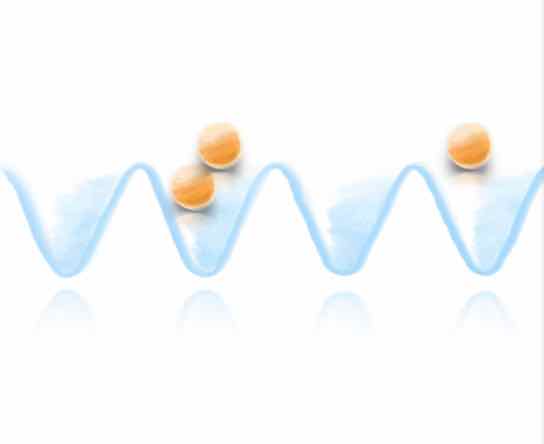}&\large Boson-Lattice examples & \normalsize 73\\
\includegraphics[width=0.1\textwidth]{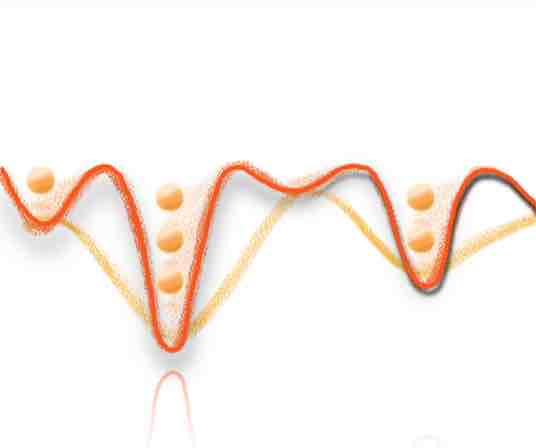}&\large Why does the construction work? & \normalsize 123\\
\includegraphics[width=0.1\textwidth]{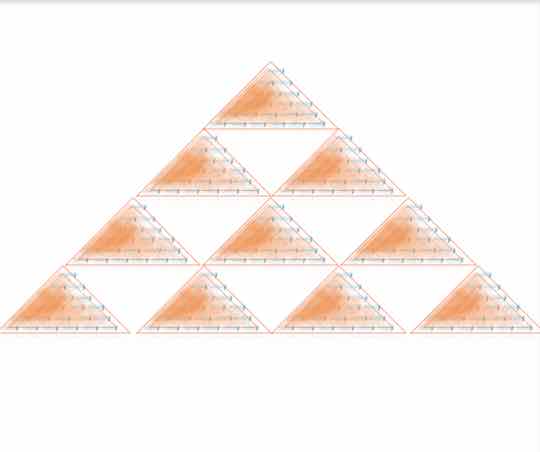}&\large Higher levels of the construction &\normalsize 129 \\
\includegraphics[width=0.1\textwidth]{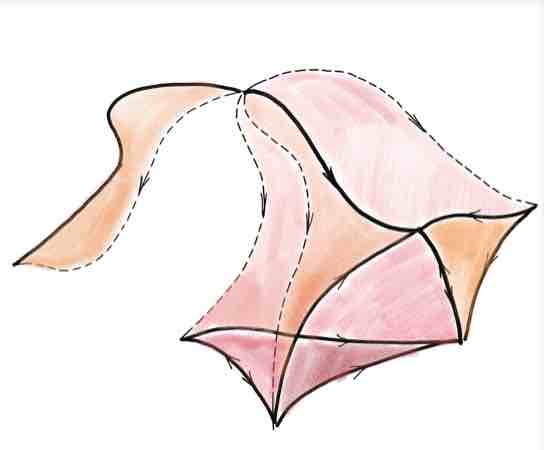}&\large Closures and openings & \normalsize 133\\
\end{tabular}


\newpage
\pagestyle{fancy}
\fancyhf{}
\lhead{Preliminaries}
\vspace*{1cm}
\begin{center}
\includegraphics[width=\textwidth]{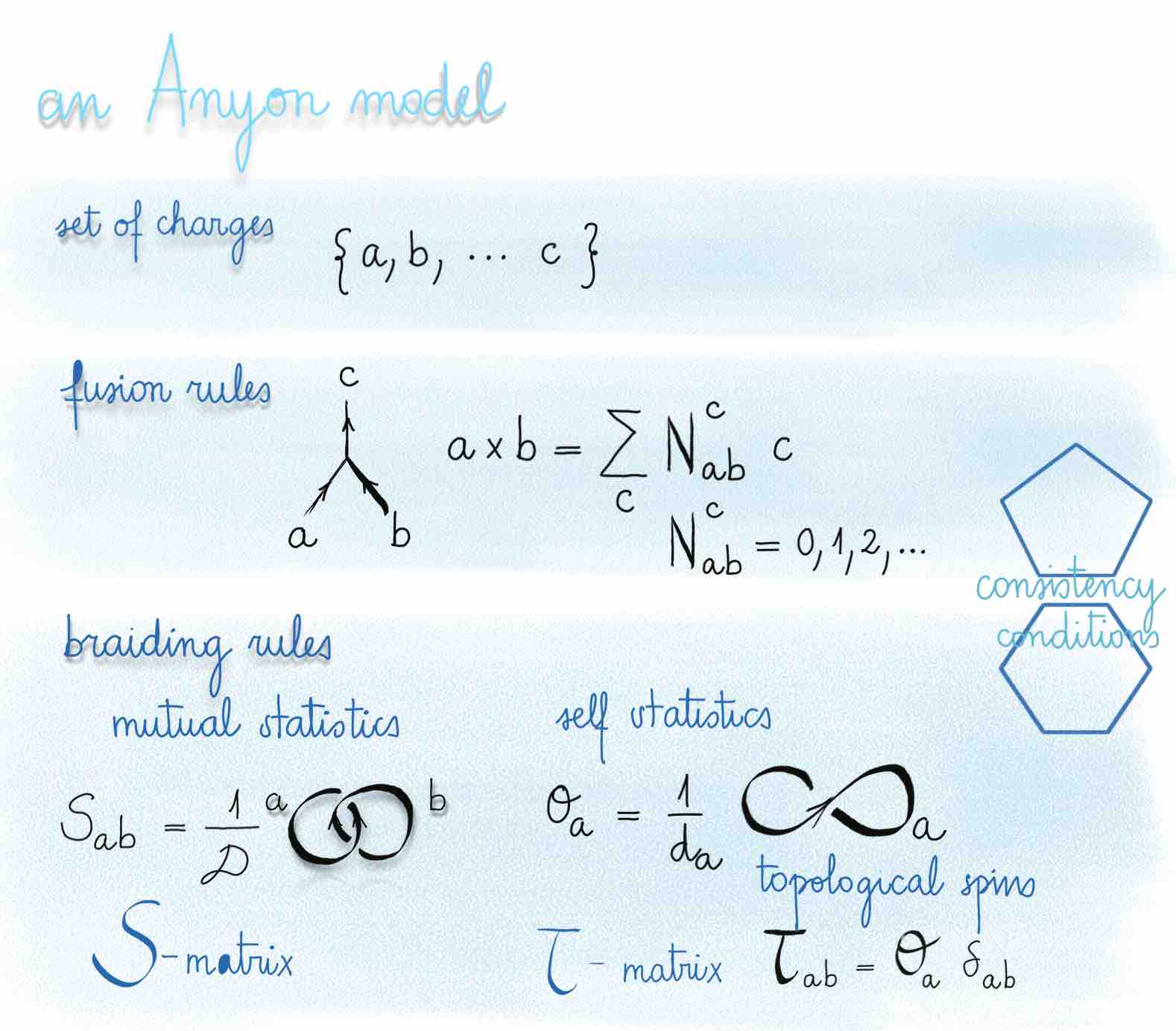}
\end{center}

\newpage
\pagestyle{fancy}
\fancyhf{}
\rhead{An Anyon model}
\rfoot{\thepage}
\section*{An anyon model}
\addcontentsline{toc}{section}{An anyon model}
\normalsize

An anyon model \cite{Kitaev1, Kitaev2, Preskill, Wang, Bonderson} is characterized by a finite set of conserved topological charges or anyon types:
\begin{eqnarray}
\{a,b,\cdots,c\}.
\end{eqnarray}
These charges obey the fusion algebra:
\begin{eqnarray}
a \times b =\sum_c N_{ab}^c \, c,
\end{eqnarray}
where the multiplicities $N_{ab}^c$ are non-negative integers that indicate the number of ways that charge $c$ can be obtained from fusion of the charges $a$ and $b$.
The fusion algebra is commutative and associative. 

There exists a unique trivial charge $0$ that satisfies $N_{a0}^b=\delta_{ab}$.

Each charge $a$ has a conjugate charge $\bar a$ such that $N_{a b}^0=\delta_{b \bar a}$.

The fusion multiplicities obey the relations:
\begin{eqnarray}
&N_{ab}^c=N_{ba}^c=N_{b\bar c}^{\bar a}=N_{\bar a\bar b}^{\bar c}&\\
&\sum_e N_{ab}^eN_{ec}^d=\sum_f N_{af}^dN_{bc}^f.&
\end{eqnarray}

Meanwhile, the charges obey a set of braiding rules that determine the way in which they braid around each other. Self-braiding of charge $a$ with itself is encoded in the topological spin $\theta_a$, which is a root of unity. The diagonal matrix of topological spins is called the topological $T$-matrix of the anyon model:
\begin{eqnarray}
T_{ab}=\theta_a \delta_{ab}.
\end{eqnarray}
The mutual braiding of charges $a$ and $b$ is given by the elements of the topological $S$-matrix, which is a symmetric matrix defined as:
\begin{eqnarray}
S_{ab}=\sum_c N_{a\bar b}^c \frac{\theta_c}{\theta_a\theta_b} d_c,
\end{eqnarray}
where $d_c$ is the {\em quantum dimension} of charge $c$, determined through the fusion multiplicities.

When the topological $S$-matrix is {\em unitary}, the anyon model is called {\em modular}. A modular anyon model corresponds to a {\em topological quantum field theory}.


\newpage
\pagestyle{fancy}
\fancyhf{}
\lhead{Preliminaries}
\includegraphics[width=0.7\textwidth]{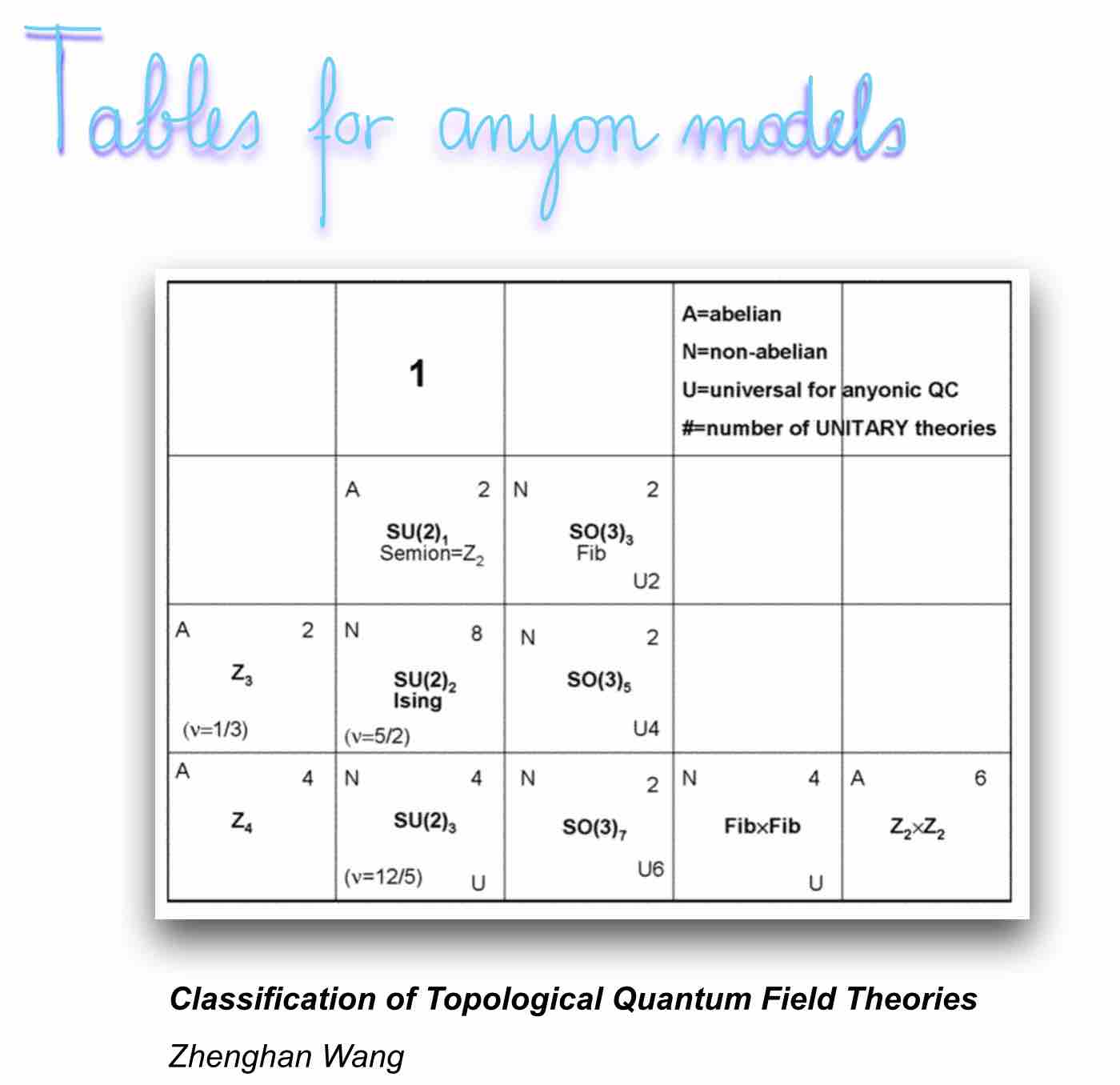}

\begin{center}
\includegraphics[width=0.75\textwidth]{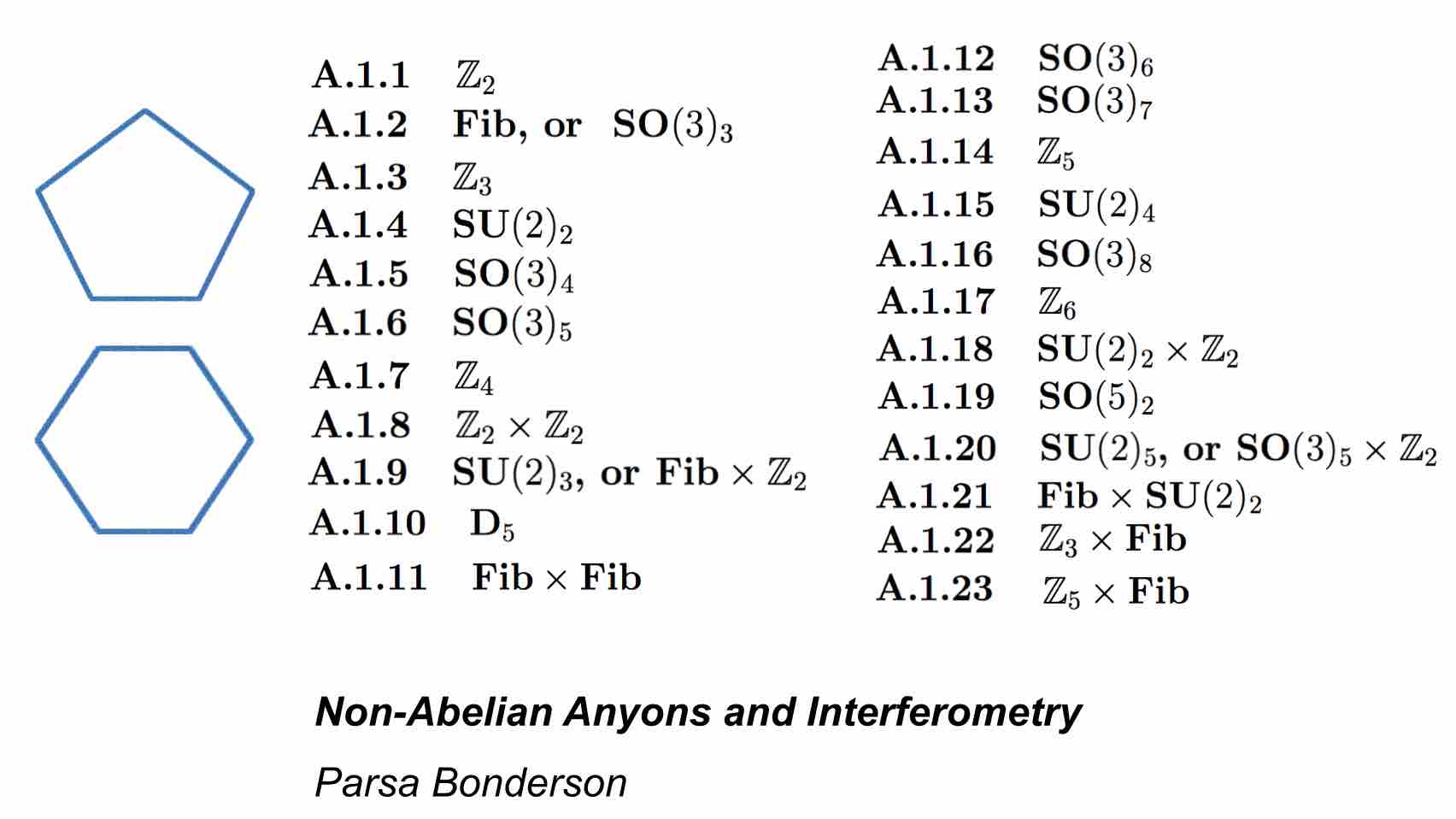}
\end{center}

\newpage
\pagestyle{fancy}
\fancyhf{}
\rhead{Tabulating Anyon models}
\rfoot{\thepage}
\vspace*{2cm}
\section*{Tabulating anyon models}
\addcontentsline{toc}{section}{Tabulating anyon models}

The fusion rules and braiding rules of an anyon model fulfill a set of multivariate polynomial equations known as the {\em Pentagon and Hexagon equations}.
Finding all possible anyon models seems then easy. We just need to find all possible solutions to these equations.
However, the number of variables and equations involved grows rapidly with the number of charges, which makes difficult to systematically solve them.

By classifying all topological quantum field theories up to four topological charges, Wang \cite{Wang} has tabulated all possible modular anyon models with up to four particle types.

By using a numerical program, Bonderson \cite{Bonderson} has been able to solve the Pentagon and Hexagon equations for many interesting fusion rules. This has allowed to tabulate a list of anyon models with up to 6 particles restricted to multiplicity-free fusion rules ($N_{ab}^c=0,1$). Several additional models relevant for non-Abelian quantum Hall states \cite{NA1,NA4} have been listed for $10$ and $12$ particles.

\newpage
\thispagestyle{empty}
\vspace*{3,5cm}
\hspace*{1cm}
\includegraphics[width=0.9\textwidth]{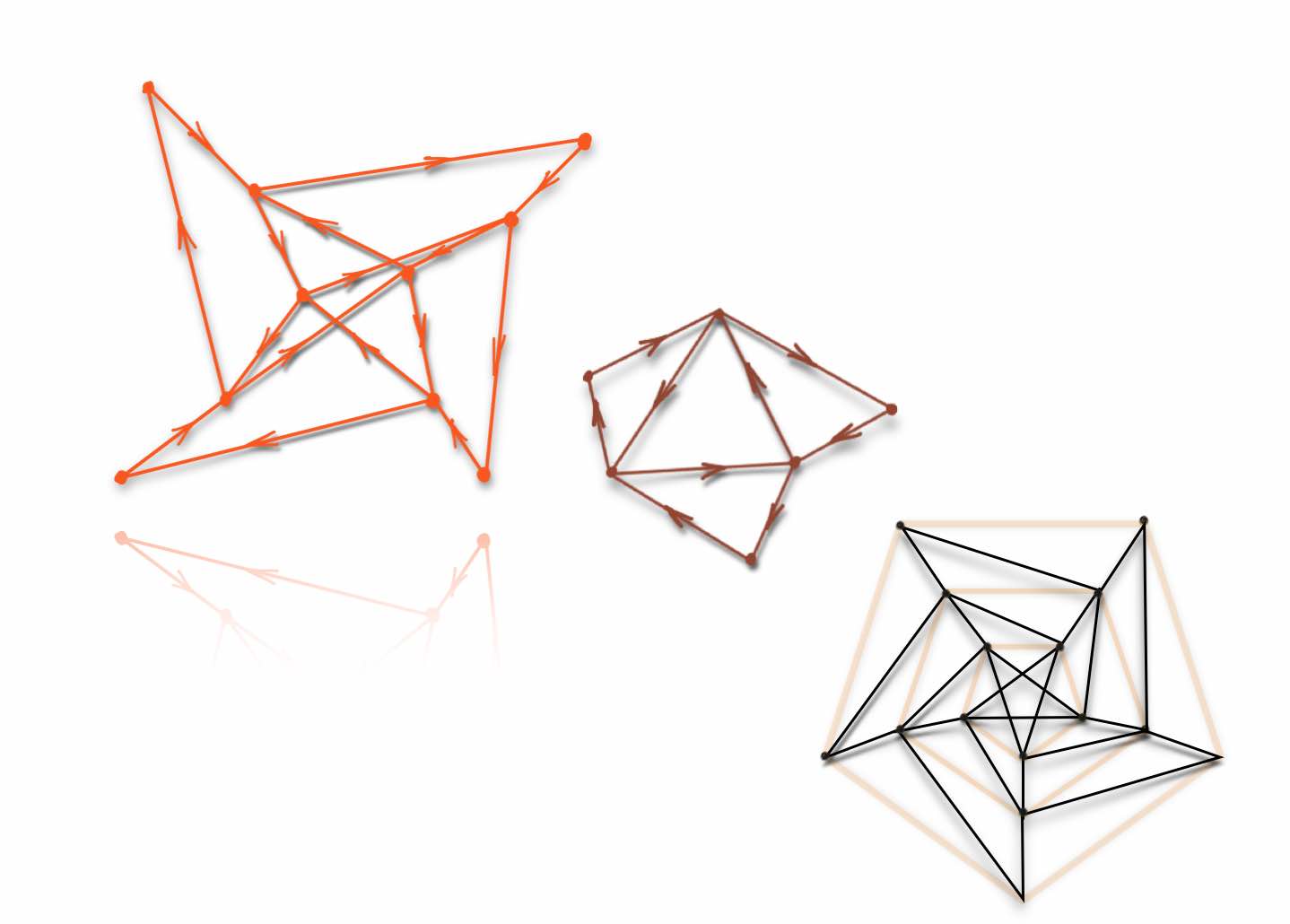}

\newpage
\thispagestyle{empty}
\section*{}
\addcontentsline{toc}{section}{The language of the construction}
\vspace*{-3cm}
\hspace*{1cm}
\includegraphics[width=10cm]{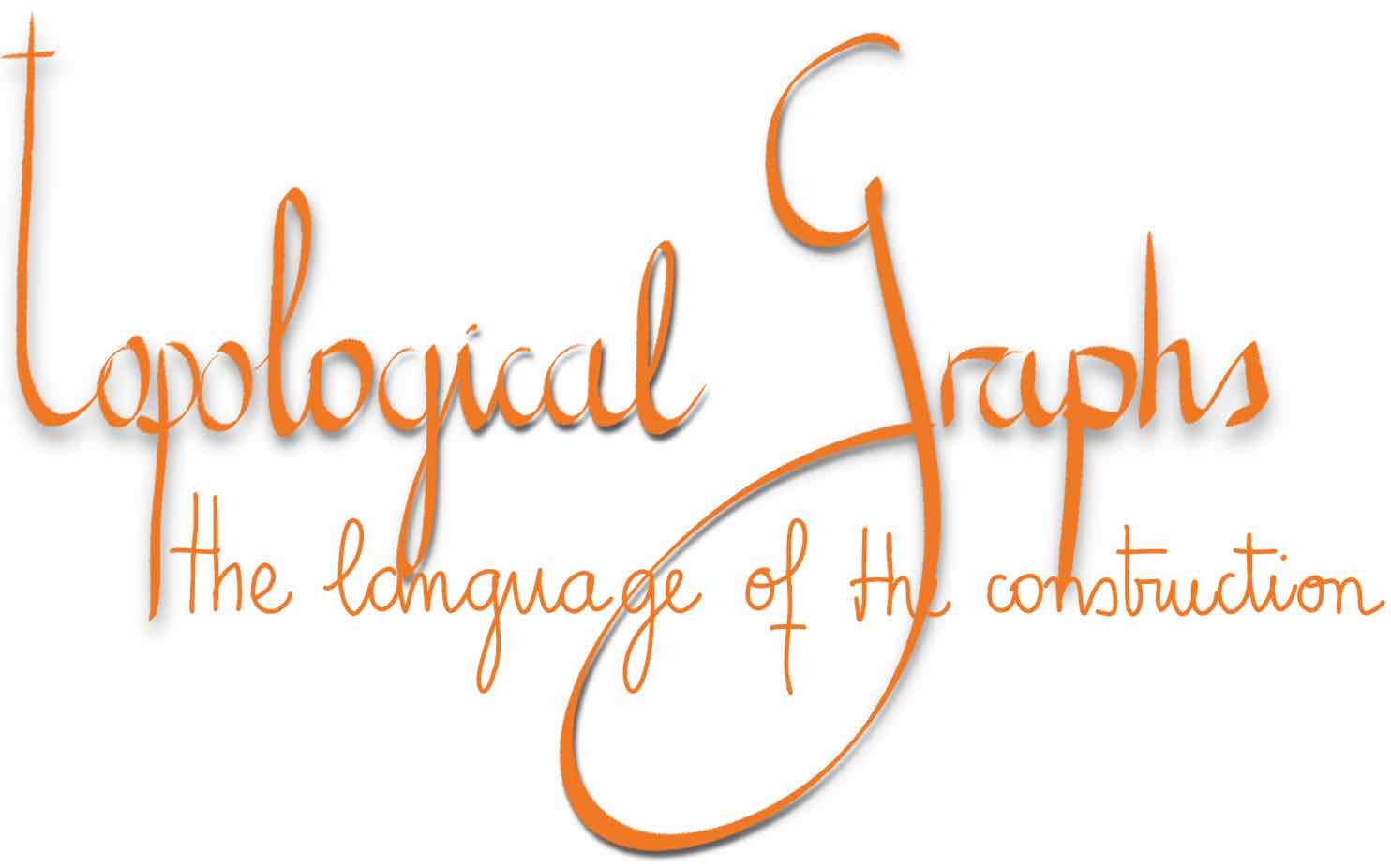}

\hspace*{2cm}
\parbox{13cm}
{I introduce a language to describe anyon models. 
\parskip=5pt
This language encodes the properties of an anyon model in a collection of graphs, which I call {\bf topological graphs}.
Graph encoding provides a visual and enlightening way of representing anyon models.

First, I introduce the concept of a {\bf topological algebra}, an algebra of operators able to encode the fusion rules of an anyon model. I analyze in depth its properties, as well as the conditions for an algebra to be topological. I give special focus to the fact that a topological algebra can also encode valuable information about the braiding properties of an anyon model. Everyone familiar with anyon models has learnt as a mantra the beautiful result by Verlinde \cite{Verlinde}: {\em the topological $S$-matrix of a modular anyon model diagonalizes the fusion rules}. Yet I think that the meaning and the consequences or extensions of this idea have been neither realized nor exploited enough.
The results and connections I present  are not the review from texts I have read. They have followed from the genuine need to give an orderly structure to the concepts that were naturally emerging in my way to conform (from the pure definition of fusion and braiding rules) an appropriate language to express the Boson-Lattice construction.

Finally, I represent a topological algebra with a collection of graphs. This representation allows to easily visualize the properties of the anyon model. Topological charges are represented by graph vertices. Fusion rules can be read from the connectivity patterns of the graphs. Braiding rules are obtained through diagonalization of the graphs.

Graphs have been extensively used in physics and mathematics to represent matrices. Here, I reveal that topological graphs compose a useful language to embody anyon models. They constitute a befitting language to formulate the Boson-Lattice construction I will develop in the following sections.}

\newpage
\pagestyle{fancy}
\fancyhf{}
\lhead{The language}
\lfoot{\thepage}
\newgeometry{bottom=0.1cm}
\vspace*{0,5cm}
\begin{center}
\includegraphics[width=0.85\textwidth]{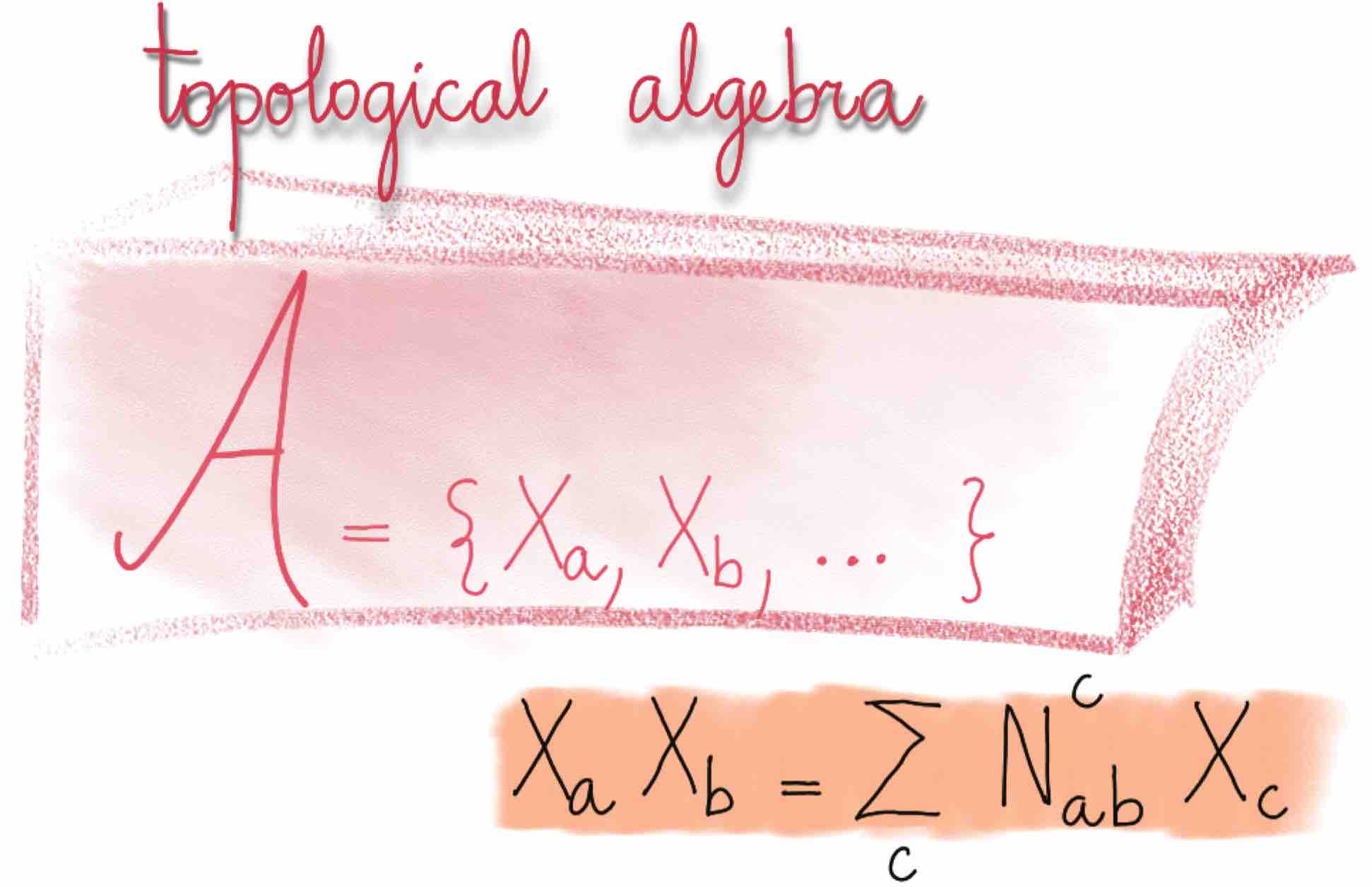}

\hspace*{0.5cm}
\includegraphics[width=0.8\textwidth]{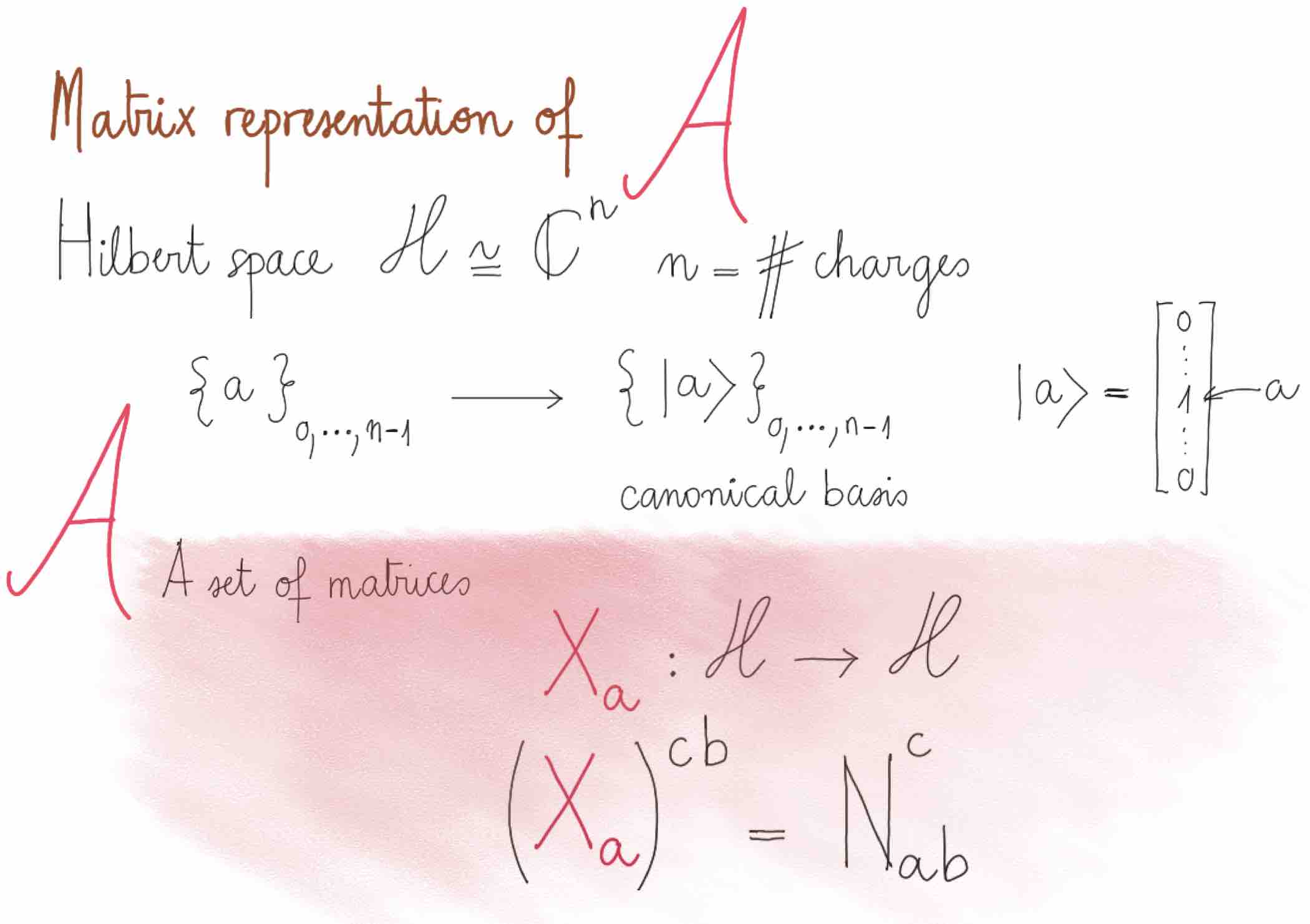}
\end{center}

\newpage
\restoregeometry
\pagestyle{fancy}
\fancyhf{}
\rhead{Topological Algebra}
\rfoot{\thepage}
\vspace*{0cm}
\section*{Topological algebras}
\addcontentsline{toc}{section}{Topological algebras}
\normalsize

I introduce the concept of topological algebra, as a useful way to encode, firstly, the fusion rules of an anyon model.

{\em \bf Definition.} An algebra of operators 
\begin{eqnarray}
\mathcal{A}=\{X_a,X_b,\ldots\},
\end{eqnarray}
is a topological algebra if it fulfills:
\begin{eqnarray}
X_aX_b=\sum_c N_{ab}^c\,X_c,
\label{AlgebraCondition1}
\end{eqnarray}
with the tensors $N_{ab}^c$ defining a set of well defined fusion rules as described above. 
The operators in the algebra are in one to one correspondence with the topological charges of the anyon model.

{\em \bf The topological algebra of an anyon model: matrix representation.}

Given an anyon model with topological charges $\{a,b,\ldots\}$ and fusion rules $a \times b= \sum_c N_{ab}^c \,c$, 
we can always find a topological algebra associated with the anyon model in the following way.

Let me consider a Hilbert space of dimension $n$ equal to the number of topological charges in the anyon model.
Let me denote the canonical basis in this Hilbert space by
\begin{eqnarray}
\{\ket{a},\ket{b},\ldots\},
\end{eqnarray}
where each state is associated with a charge in the anyon model.
I define a set of $n\times n$ matrices with matrix elements given by:
\begin{eqnarray}
\braket{c|X_a|b}=\ N_{ab}^c.
\label{AlgebraDef}
\end{eqnarray}

\begin{result}
With the definition (\ref{AlgebraDef}) the set of matrices $\{X_a\}$ satisfies the condition (\ref{AlgebraCondition1}) and is thus a topological algebra associated with the anyon model.
\end{result}
\begin{proof}
Using the associative property of the fusion rules we have:
\begin{eqnarray}
\braket{i|X_aX_b|j}&=&\sum_k\braket{i|X_a|k}\braket{k|X_b|j}=\nonumber\\
&=&\sum_kN_{ak}^iN_{bj}^k=\sum_cN_{ab}^cN_{cj}^i=\sum_cN_{ab}^c\braket{i|X_c|j}.\qedhere
\end{eqnarray}
\end{proof}



\newpage
\pagestyle{fancy}
\fancyhf{}
\lhead{The language}
\lfoot{\thepage}
\includegraphics[width=\textwidth]{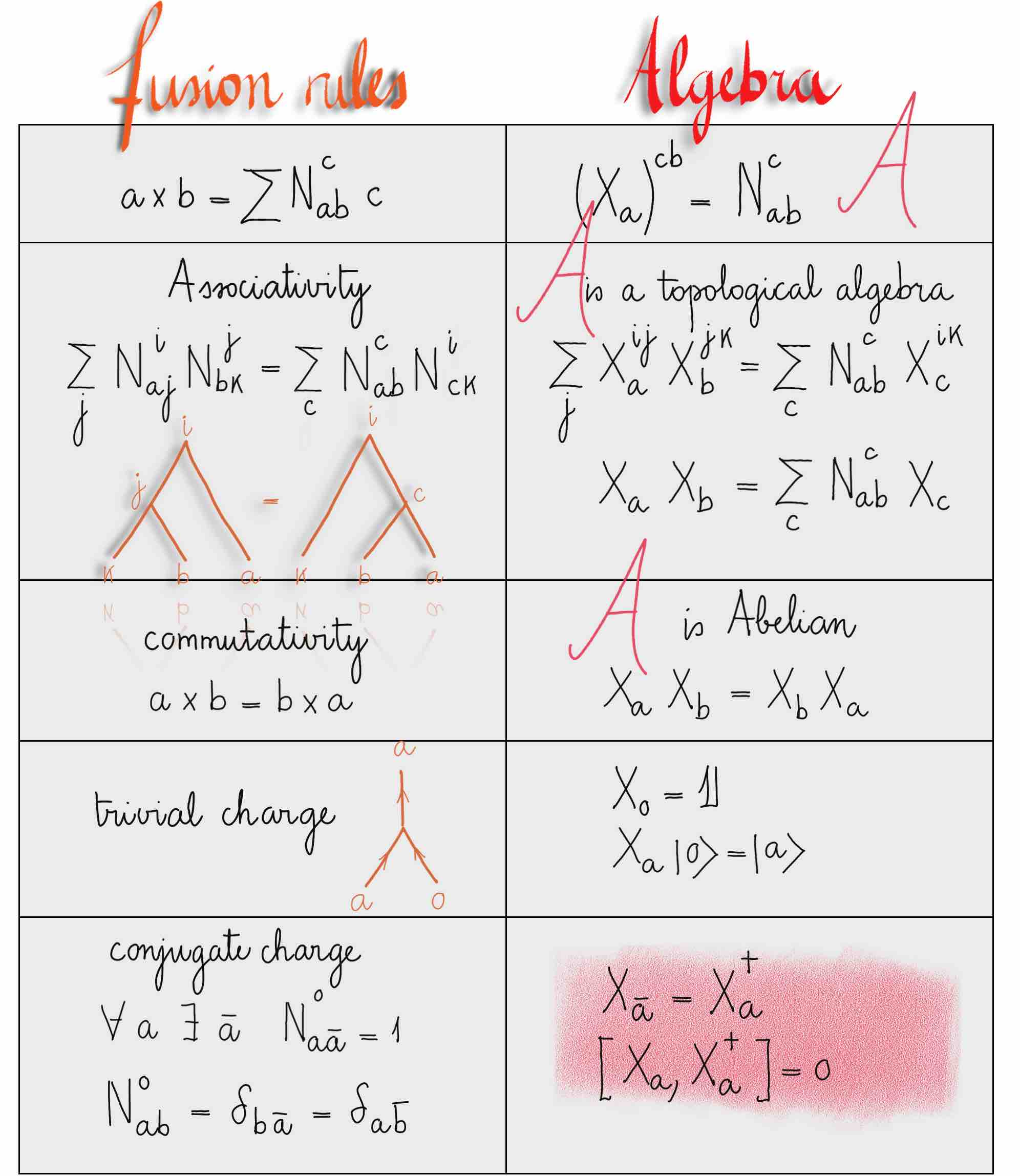}

\newpage
\pagestyle{fancy}
\fancyhf{}
\rhead{Topological Algebra}
\rfoot{\thepage}
\subsection*{Properties of a topological algebra}
\addcontentsline{toc}{subsection}{Properties of a topological algebra}

A topological algebra exhibits the following properties, which are inherited from the properties of fusion rules:
\begin{itemize}
\item {\bf the algebra is Abelian}. From the commutativity property of the fusion rules encoded by the algebra, it follows that:
\begin{eqnarray}
X_aX_b=\sum_c N_{ab}^c\,X_c=\sum_c N_{ba}^c\,X_c=X_bX_a.
\end{eqnarray}

\item {\bf the identity operator} is the operator corresponding to the trivial charge $0$:
\begin{eqnarray}
\braket{a|X_0|b}=N_{0b}^a=\delta_{ab}.
\end{eqnarray}
\item {\bf acting on the trivial state}, the operator $X_a$ gives 
\begin{eqnarray}
X_a\ket{0}=\ket{a},
\end{eqnarray}
since we have:
\begin{eqnarray}
\braket{b|X_a|0}=N_{a0}^b=\delta_{ab}.
\end{eqnarray}
\item {\bf the operator associated with the conjugate charge is the adjoint operator}. From the properties of the fusion rules it follows that:
\begin{eqnarray}
\braket{c|X_{\bar a}^{}|b}=N_{\bar ab}^c=N_{a\bar b}^{\bar c}=N_{ac}^b=\braket{b|X_{a}^{}|c}=\braket{c|X^\dagger_a|b},
\end{eqnarray}
and therefore
\begin{eqnarray}
X_{\bar a}=X_a^\dagger.
\end{eqnarray}
\item {\bf the operators in the algebra are normal}. Combining the properties above we have:
\begin{eqnarray}
[X_a^{},X_a^\dagger]=[X_a^{},X_{\bar a}]=0.
\end{eqnarray}
\end{itemize}

It is interesting to see how the properties of the fusion rules are translated into a set of illuminating properties of the algebra: the operators are {\bf normal and mutually commuting}, which allows for their simultaneous diagonalization.

\newpage
\thispagestyle{empty}

\newpage
\pagestyle{fancy}
\fancyhf{}
\lhead{The language}
\lfoot{\thepage}
\vspace*{0cm}
\subsection*{When is an algebra a topological algebra?}
\addcontentsline{toc}{subsection}{When is an algebra a topological algebra?}

We have seen above how to define an algebra encoding the fusion rules of an anyon model.

But if we are given a certain algebra, how do we know that this algebra is topological?, that is, how do we know that it defines a set of well defined fusion rules?

In the following result I give the {\bf necessary and sufficient conditions} that an algebra needs to fulfill in order to be topological.
This result will be very useful when building up topological algebras in the Boson-Lattice construction I develop in the next sections.

\indent
\hangindent=0,6cm
\begin{fresult}

Let us consider the following set of $n$ linear operators on a Hilbert space of dimension $n$:
\begin{eqnarray}
\mathcal{A}=\{X_0=\mathds{1},X_1,\cdots,X_{n-1}\}. 
\end{eqnarray}
This set defines a topological algebra if and only if the following conditions are satisfied:
\begin{enumerate}
\item The operators commute with each other: $\left[X_a,X_b\right]=0$, $\forall a,b$.
\item For each $X_a\in \mathcal{A}$ there exists $X_{\bar a}\in \mathcal{A}$ such that $X_{\bar a}=X_a^\dagger$.
\item There exists a state $\ket{0}$ for which the set of states $\{\ket{a}=X_a\ket{0}\}$ defines an orthonormal basis.
\item In such basis the operators have natural entries: $\braket{c|X_a|b}=0,1,2,\cdots.$
\end{enumerate}

With these conditions the set of operators $\mathcal{A}$ is an algebra satisfying:
\begin{eqnarray}
\label{AlgebraCondition}
X_aX_b=\sum_cN_{ab}^cX_c,
\end{eqnarray}
with $N_{ab}^c=\braket{c|X_a|b}$ defining a set of well defined {\em fusion rules}.
\end{fresult}

\begin{proof}
\small
It is clear that a topological algebra fulfills the set of conditions listed above.

To see that an algebra fulfilling conditions 1-4 is topological, we proceed as follows.

\newpage
\pagestyle{fancy}
\fancyhf{}
\rhead{Topological Algebra}
\rfoot{\thepage}
\vspace*{0cm}

First, we prove that $\mathcal{A}$ fulfills Eq.(\ref{AlgebraCondition}). From properties 1. and 3. we have:
\begin{eqnarray}
X_a\ket{b}=X_aX_b\ket{0}=X_b\ket{a},
\end{eqnarray}
and therefore
\begin{eqnarray}
X_aX_b\ket{d}=X_aX_d\ket{b}=\sum_cX_d\ket{c}\braket{c|X_a|b}=\sum_c\braket{c|X_a|b}X_c\ket{d}=\sum_c N_{ab}^c X_c\ket{d}.
\end{eqnarray}

Second, we prove that $N_{ab}^c=\braket{c|X_a|b}$ are well defined fusion rules, since they fulfill:
\begin{itemize}
\item $N_{a b}^c=N_{b a}^c$,\\
since we have $N_{a b}^c=\braket{c\,|X_a|\,b}=\braket{c\,|X_b|\,a}=N_{b a}^c,$.
\item $N_{a 0}^b=\delta_{ab}$,\\
since we have  $N_{a 0}^b=\braket{b\,|X_a|\,0}=\delta_{ab}$.
\item $N_{ab\phantom{\bar b}}^c=N_{a \bar c\phantom{\bar b}}^{\bar b}=N_{\bar a \bar b}^{\bar c}$,\\
since we have
$\braket{c\,|X_a|\,b}=\braket{\bar b\,|X_a|\,\bar c}=\braket{\bar c\,|X_a^\dagger|\,\bar b}$\\
$\leftrightarrow \braket{0\,|X^\dagger_cX_aX_b|\,0}=\braket{0\,|X_bX_aX_c^\dagger|\,0}=\braket{0\,|X_cX^\dagger_aX^\dagger_b|\,0}$.\qedhere

\end{itemize}
\end{proof}

\begin{center}
\includegraphics[width=0.75\textwidth]{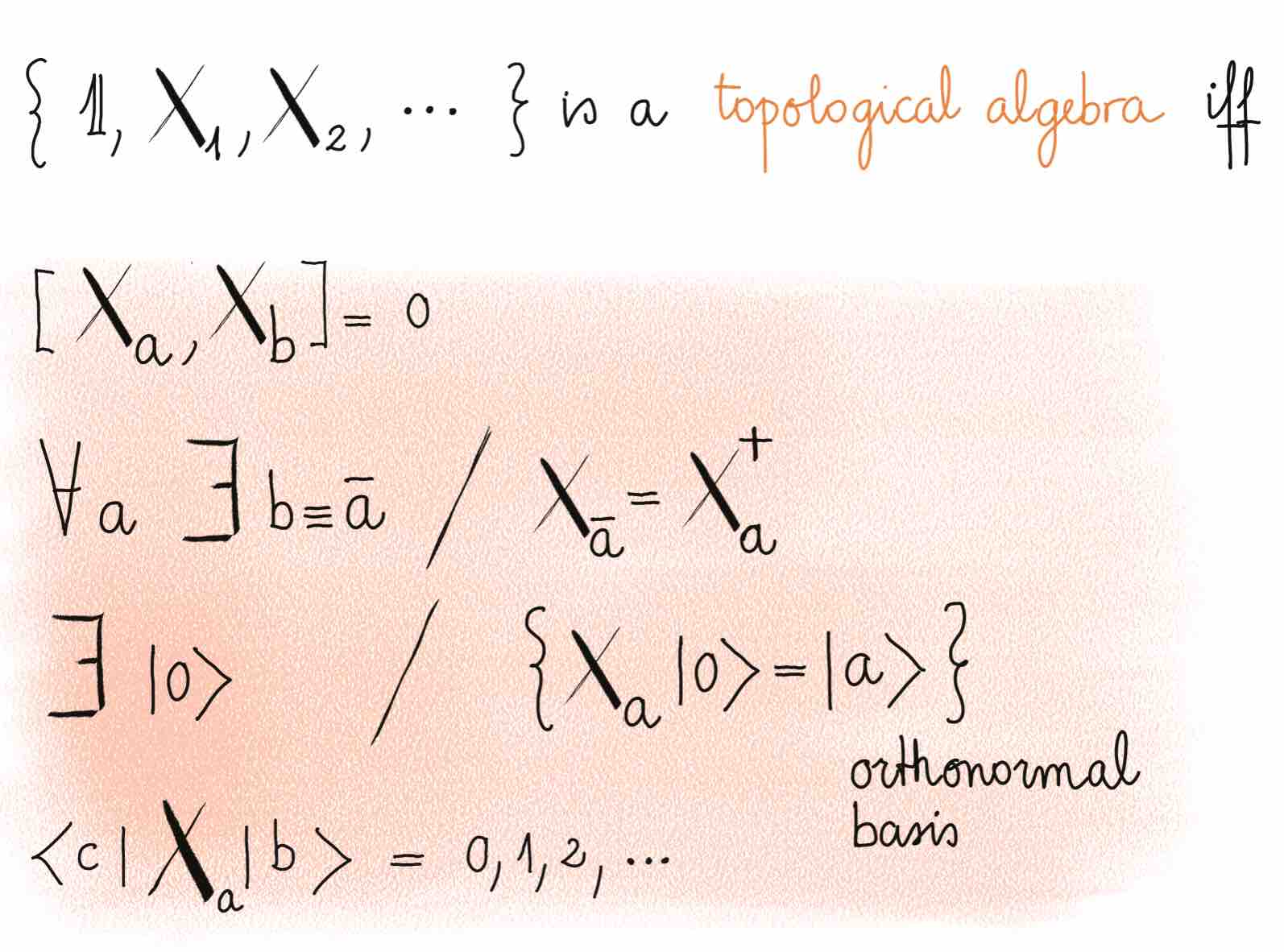}
\end{center}


\normalsize
\newpage
\pagestyle{fancy}
\fancyhf{}
\lhead{The language}
\lfoot{\thepage}
\vspace*{1.5cm}
\begin{center}
\includegraphics[width=0.9\textwidth]{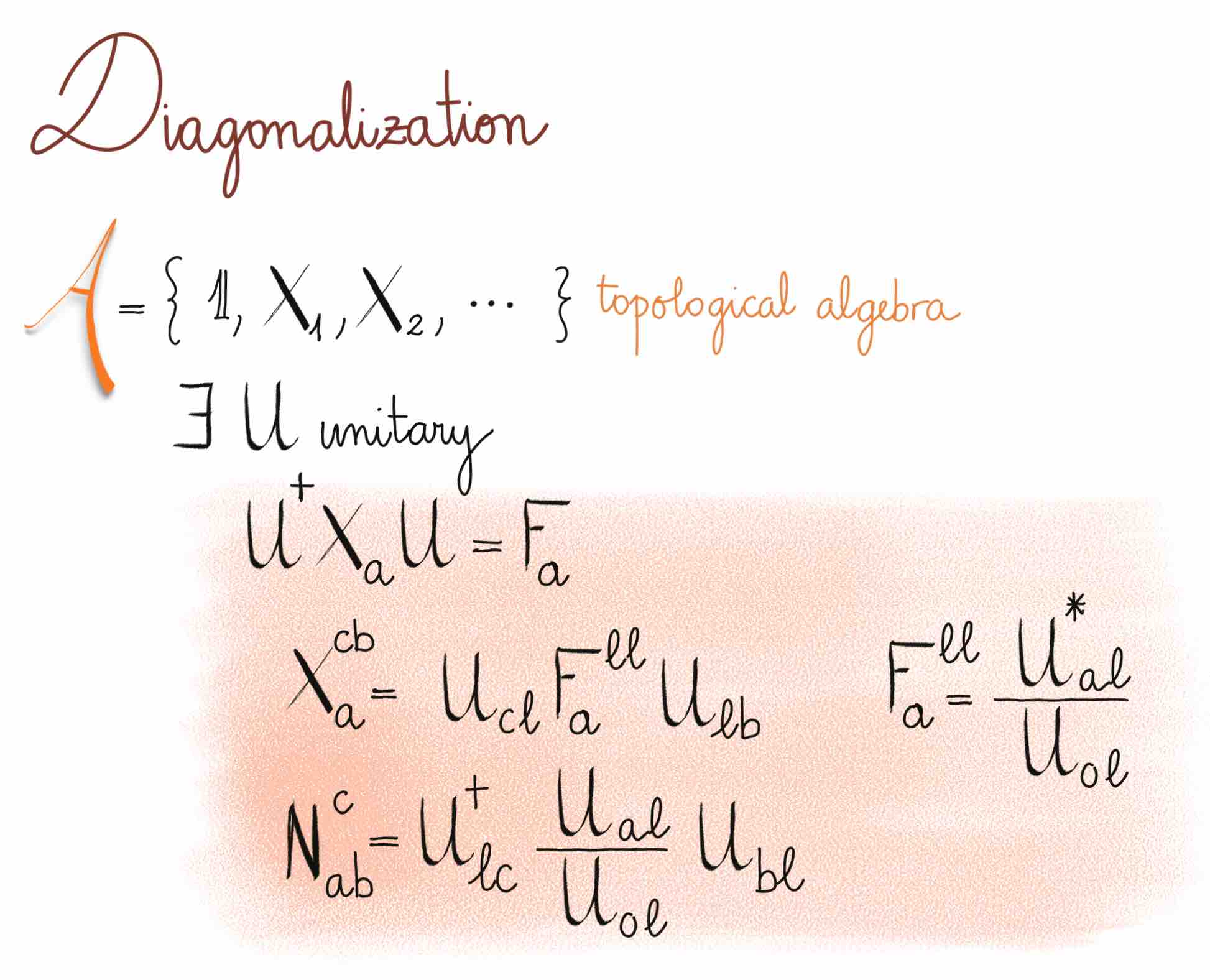}
\end{center}

\newpage
\thispagestyle{empty}
\subsection*{Diagonalization of a topological algebra}
\addcontentsline{toc}{subsection}{Diagonalization of a topological algebra}

A topological algebra $\mathcal{A}$ is a collection of {\em mutually commuting normal operators}. Therefore there exists an {\em orthonormal basis of eigenstates common to all operators} in the algebra. 
For an anyon model, these eigenstates encode valuable information about the {\bf braiding properties} of the model.

The following result holds for any topological algebra.

\begin{fresult}
There exists a unitary matrix $U$ that simultaneously diagonalizes the operators $\{X_a\}$ of the topological algebra, so that:
\begin{eqnarray}
U^\dagger X_aU=F_a,
\end{eqnarray}
with $\{F_a\}$ a collection of diagonal matrices.
The unitary matrix $S$ satisfies the equation:
\begin{eqnarray}
N_{ab}^{c}=\sum_\ell U_{\ell c}^{\dagger} \,\frac{U_{a\ell}^{}}{U_{0\ell}^{}}\,U_{b\ell}^{},
\label{VerlindeEquation}
\end{eqnarray}
where $U_{\ell \ell^\prime}=\braket{\ell|U|\ell^\prime}$ are the matrix elements of $U$.
\end{fresult}

Equation (\ref{VerlindeEquation}) reminds us of {\bf Verlinde's equation} \cite{Verlinde, Witten, VerlindeEquation1, VerlindeEquation2} for a modular anyon model, which relates the fusion rules $N_{ab}^{c}$ with the topological $S$-matrix.
Indeed, as we will see later, for a modular anyon model, the topological $S$-matrix exactly corresponds to a symmetric choice of the unitary matrix $U$. Here, it is important to note that the result above is valid for any topological algebra, independently of whether it corresponds to a modular anyon model or not. 

\begin{proof}
Let me denote the orthonormal basis of common eigenstates of the algebra $\{X_a\}$ by $\{\ket{\psi_a}\}$, with
\begin{eqnarray}
X_b\ket{\psi_a}=\lambda_b^{(a)}\ket{\psi_a},
\end{eqnarray}
and $\lambda_b^{(a)}$ the eigenvalue corresponding to the operator $X_b$.
Encoding the eigenvectors in the unitary matrix $U$ and the set of eigenvalues in the set of diagonal matrices $F_b$:
\begin{eqnarray}
\braket{a^\prime|U_{}^{}|a}&=&\braket{a^\prime|\psi_a}\nonumber\\
\braket{a^\prime|F_b^{}|a}&=&\lambda_b^{(a)}\delta_{aa^\prime},\nonumber
\end{eqnarray}
we have:
\begin{eqnarray}
U^\dagger X_aU=F_a.
\end{eqnarray}

\newpage
\pagestyle{fancy}
\fancyhf{}
\lhead{The language}
\lfoot{\thepage}

To prove equation (\ref{VerlindeEquation}) we solve $X_a$ from the above expression and take matrix elements to obtain:
\begin{eqnarray}
\braket{c|X_a|b}&=&\sum_\ell\braket{c|U|\ell}\braket{\ell|F_a|\ell}\braket{\ell|U^\dagger|b}\nonumber\\
N_{ab}^c&=&\sum_\ell U^\dagger_{\ell c}\braket{\ell |F^*_a|\ell}U^{}_{b\ell}\label{VerlindePrevious}.
\end{eqnarray}

Comparing (\ref{VerlindePrevious}) with (\ref{VerlindeEquation}) it suffices to prove that:
\begin{eqnarray}
\frac{U_{a\ell}}{U_{0\ell}}=\braket{\ell |F^*_a|\ell}.
\label{1DRepresentation_1}
\end{eqnarray}

To prove (\ref{1DRepresentation_1}) let me consider a set of states of the form:
\begin{eqnarray}
\ket{\varphi_\ell}=\sum_a\braket{\ell|F_a^*|\ell}\ket{c}.
\end{eqnarray}
Since we have that
\begin{eqnarray}
\braket{c|X_a|\varphi_\ell}&=&\sum_b\braket{c|X_a|b}\braket{b|\varphi_\ell}=\sum_b\braket{\ell|N_{ab}^cF_b^*|\ell}=\sum_b\braket{\ell|N_{a\bar c}^{\bar b}F_{\bar b}|\ell}\nonumber\\
&=&\braket{\ell|F_aF_{\bar c}|\ell}=\braket{\ell|F_a|\ell}\braket{c|\varphi_\ell},
\end{eqnarray}
it follows that 
\begin{eqnarray}
X_a\ket{\varphi_\ell}=\lambda_a^{(\ell)}\ket{\varphi_\ell},
\end{eqnarray}
and therefore the state $\ket{\varphi_\ell}$ is proportional to the eigenstate $\ket{\psi_\ell}$. 
We thus have
\begin{eqnarray}
\frac{U_{a\ell}}{U_{0\ell}}=\frac{\braket{a|\psi_\ell}}{\braket{0|\psi_\ell}}
=\frac{\braket{\ell|F_a^*|\ell}}{\braket{\ell|F_0^*|\ell}}=\braket{\ell|F_a^*|\ell}.\qedhere
\end{eqnarray}
\end{proof}
\normalsize
\newpage
\pagestyle{fancy}
\fancyhf{}
\rhead{Topological Algebra}
\subsubsection*{One dimensional representations of the topological algebra}
\addcontentsline{toc}{subsubsection}{One dimensional representations of the topological algebra}

The result above tells us that the common eigenstates of a topological algebra are in one to one correspondence with the \textit{one-dimensional representations of the algebra}.

To see this we note that the set of eigenvalues 
\begin{eqnarray}
\{\lambda_a^{(\ell)}=\braket{\ell|F_a|\ell}\,\,\}_{a=0,\cdots,n-1}
\end{eqnarray}
is (for each $\ell=0,\cdots,n-1$) a one dimensional representation of the topological algebra $\mathcal{A}$, since we have 
\begin{eqnarray}
F_aF_b=\sum_cN_{ab}^cF_c,
\end{eqnarray}
and therefore
\begin{eqnarray}
\braket{\ell|F_a|\ell}\braket{\ell|F_b|\ell}=\sum_cN_{ab}^c\braket{\ell|F_c|\ell}.
\end{eqnarray}
Since equation (\ref{1DRepresentation_1}), proven above, states that the common eigenstates of the algebra have components proportional to the eigenvalues:
\begin{eqnarray}
\braket{\psi_\ell|a}\propto\braket{\ell|F_a|\ell}=\lambda_a^{(\ell)},
\end{eqnarray}
it follows that these eigenstates are in one to one correspondence with the one-dimensional representations of the algebra.

\subsubsection*{A common eigenvector with all positive components}
\addcontentsline{toc}{subsubsection}{A common eigenvector with all positive components}

A beautiful property of a topological algebra is the existence of a common eigenvector, whose components are all positive. Without loss of generality, this eigenvector can be written as
\begin{eqnarray}
\ket{\psi_0}=\frac{1}{\mathcal{D}}\sum_a d_a \ket{a},
\end{eqnarray}
where $d_a>0$, $d_0=1$, and $\mathcal{D}=\sqrt{\sum_a d_a^2}$.
The positive numbers $d_a$ correspond to the largest eigenvalues of the operators $X_a$.

This property follows from Perron-Frobenius theorem \cite{Perron, Frobenius}, which applies to non-negative irreducible matrices. 
In the language of graphs that I introduce later, it becomes transparent that the operators of a topological algebra are direct sums of irreducible operators, so that the theorem applies.


\newpage
\pagestyle{fancy}
\fancyhf{}
\lhead{The language}
\lfoot{\thepage}
\subsection*{Diagonalization and the S-matrix}
\addcontentsline{toc}{subsection}{Diagonalization and the S-matrix}

In light of the results above and taking into account the definition of the topological $S$-matrix, we can state the following useful connections between the $S$-matrix of an anyon model and the eigenvectors of the corresponding topological algebra \footnote{The proof of the results above is straightforward by combining the definition of the topological $S$-matrix with the information given before regarding diagonalization of a topological algebra. I will present the details of this proof elsewhere.}

\begin{itemize}

\item
The S-matrix of an anyon model is a symmetric matrix of eigenvectors of the topological algebra. 
We have:
\begin{eqnarray}
S_{ab}=\braket{a|\psi_b},
\end{eqnarray}
where $\{\ket{\psi_b}\}$ is a set of eigenvectors fulfilling
$\braket{a|\psi_b}=\braket{b|\psi_a}$.

The {\em quantum dimensions} of the anyon model correspond to the components of the common eigenvector $\ket{\psi_0}$, whose components are all positive:
\begin{eqnarray}
d_c=\frac{\braket{c|\psi_0}}{\braket{0|\psi_0}}.
\end{eqnarray}

These results hold for any anyon model, modular or not.

\item For a modular anyon model, the $S$-matrix is a unitary symmetric matrix. It therefore corresponds to a symmetric orthonormal basis of eigenstates of the topological algebra.
As any orthonormal basis of the topological algebra, it fulfills:
\begin{eqnarray}
N_{ab}^{c}=\sum_\ell S_{\ell c}^{\dagger} \,\frac{S_{a\ell}^{}}{S_{0\ell}^{}}\,S_{b\ell}^{}.
\end{eqnarray}
This equation is Verlinde's equation \cite{Verlinde}.

\item For a non-modular anyon model the $S$-matrix is not unitary. It corresponds to a non-orthonormal set of eigenvectors of the topological algebra. 
It fulfills the equation:
\begin{eqnarray}
X_aS=SE_a,
\end{eqnarray}
where $E_a$ is a diagonal matrix with elements $\braket{b |E_a| c}=\delta_{bc} \braket{\psi_b | X_a |\psi_b}$, corresponding to a set of eigenvalues of $X_a$.
Element by element  we have:
\begin{eqnarray}
\sum_b N_{ab}^c S_{b\ell}^{}=S_{c\ell}\frac{S_{a\ell}^*}{S_{0\ell}}.
\end{eqnarray}

This is a generalization of Verlinde's equation. It is valid for any anyon model, modular or not.
\end{itemize}


\newpage
\pagestyle{fancy}
\fancyhf{}
\rhead{Topological Algebra}
\rfoot{\thepage}
\subsection*{Topological algebras and anyon models}
\addcontentsline{toc}{subsection}{Topological algebras and anyon models}

Given an anyon model there is always a topological algebra associated with it, which encodes its fusion rules.
However, given a topological algebra (a set of well defined fusion rules) there is not necessarily an anyon model satisfying the corresponding fusion rules.

For example, it is clear that if the topological algebra does not admit a symmetric set of eigenvectors, there will be no anyon model corresponding to it.
Also, if the algebra does not admit a symmetric eigenbasis, we can be sure that there will be no modular anyon model with such fusion rules.

Remarkably, even if there is a symmetric set of eigenvectors, it is not guaranteed that an anyon model exists \footnote{It is illuminating to construct examples of topological algebras which do not correspond to anyon models, even when a symmetric eigenbasis exists. In a forthcoming work I will give explanatory examples of different interesting situations.}.
The following result summarizes the conditions under which an anyon model can exist with the fusion rules of a given topological algebra.

\begin{mdframed}
Given a topological algebra, an anyon model associated with it corresponds to a symmetric choice $S_{ab}$ of one dimensional representations of the algebra satisfying the equation:
\begin{eqnarray}
S_{ab}=\sum_c N_{a\bar b}^c \frac{\theta_c}{\theta_a\theta_b} d_c, \label{S-T-Relation}
\end{eqnarray}
where $N_{ab}^c$ are the fusion multiplicities, $\theta_a$ are roots of unity, with $\theta_0=1$, and $d_c$ are the components of the algebra eigenvector with all positive components.
\end{mdframed}

This result is indeed an alternative formulation of the Pentagon and Hexagon equations. 
It can provide us with a guided route to obtain the possible anyon models associated with a given set of fusion rules. First, we search for the possible symmetric sets of eigenvectors of the algebra. This step highly reduces the possible candidates for topological $S$-matrices. Then, we check whether these matrices fulfill equation (\ref{S-T-Relation}) for a certain choice of the $\theta_a$.

The phrasing above can be enlightening. For example, it becomes clear that anyon models with the same fusion rules have $S$-matrices corresponding to different symmetric choices of a set of eigenvectors (for example, they can correspond to reorderings of the same set of eigenvectors). 

This formulation will be very useful to prove the existence of modular anyon models corresponding to topological algebras in the Boson-Lattice construction I describe in the next sections.


\newpage
\pagestyle{fancy}
\fancyhf{}
\lhead{The language}
\lfoot{\thepage}
\vspace*{1.5cm}
\begin{center}
\includegraphics[width=\textwidth]{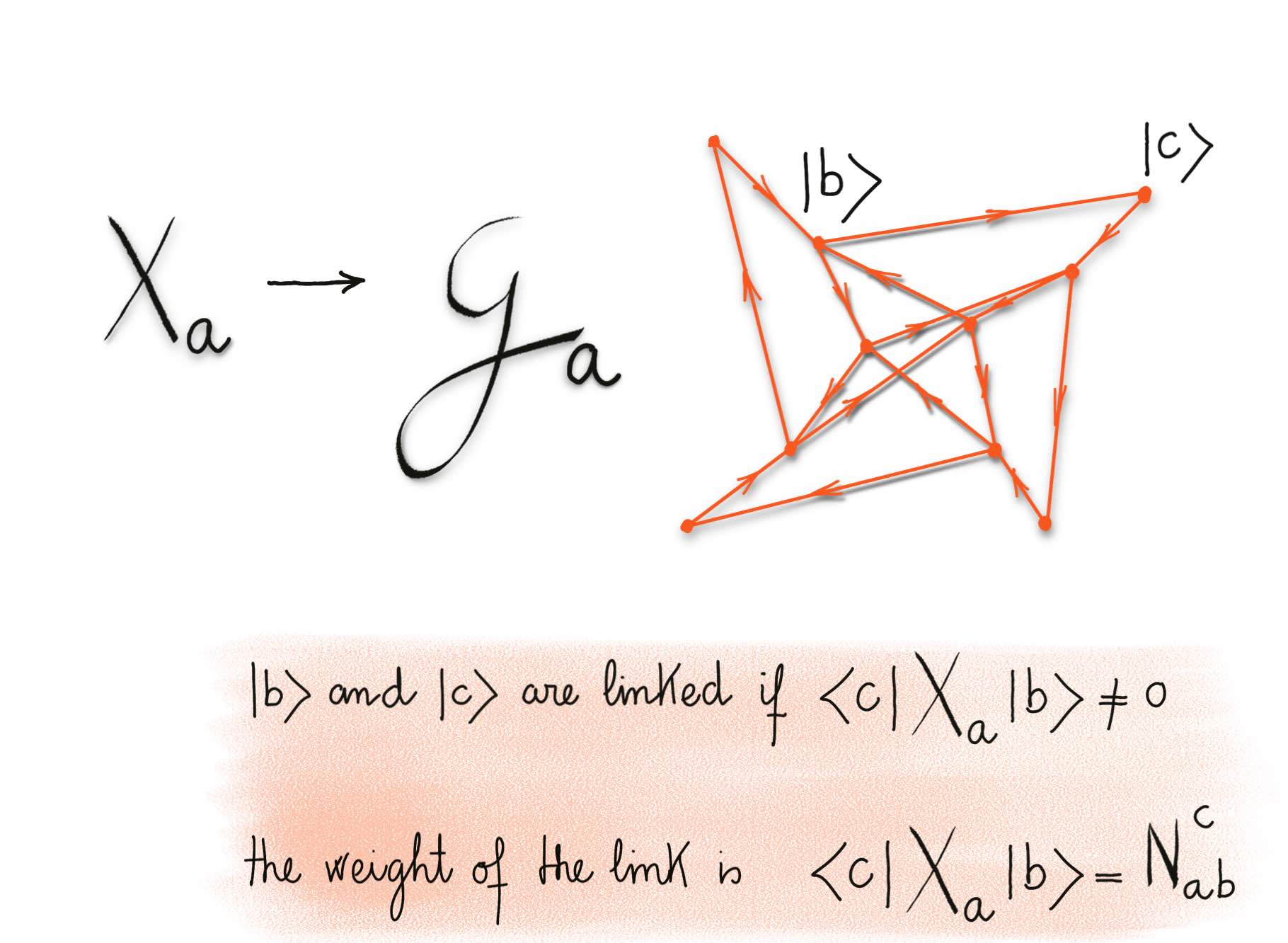}
\end{center}

\newpage
\pagestyle{fancy}
\fancyhf{}
\rhead{Topological Graphs}
\rfoot{\thepage}
\vspace*{0cm}
\normalsize
\section*{Topological graphs}
\addcontentsline{toc}{section}{Topological graphs}

I represent a topological algebra with a collection of graphs. 

Each operator $X_a$ in the topological algebra is represented by a {\bf weighted directed graph} $\mathcal{G}_a$ defined as follows:
\hspace*{0,5cm}
\begin{mdframed}
{\bf Vertices}. The vertices of the graph are in one to one correspondence to the states of the canonical basis, each of them corresponding to a charge of the anyon model.

{\bf Connectivity}. Two vertices $\ket{b}$ and $\ket{c}$ are connected if the matrix element 
$\braket{c|X_a|b}$ is different from zero. The link is oriented from $\ket{b}$ to $\ket{c}$.

{\bf Links-weight}. The link connecting vertex $\ket{b}$ to vertex $\ket{c }$ has weight $\braket{c|X_a|b}=N_{ab}^c$.
A link with weight $n=0,1,2,\cdots$ is represented by a $n$-multiple line.

\end{mdframed}

The table below shows an example (for an anyon model of five charges) of graph encoding of a set of fusion rules.

\begin{center}
\includegraphics[width=0.75\textwidth]{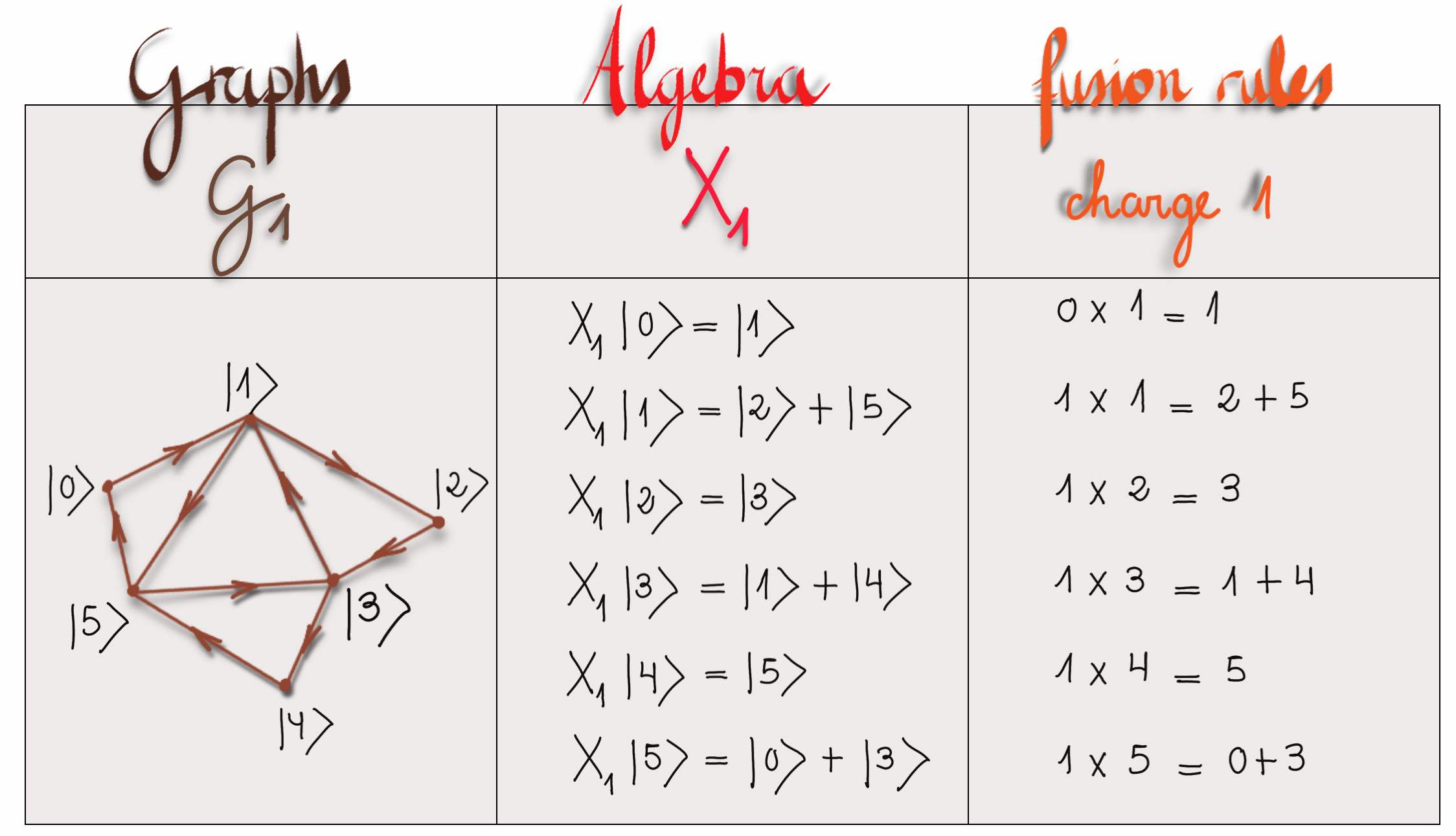}
\end{center}

\newpage
\newgeometry{bottom=0.1cm}
\pagestyle{fancy}
\fancyhf{}
\lhead{The language}
\lfoot{\thepage}
\begin{center}
\includegraphics[width=0.9\textwidth]{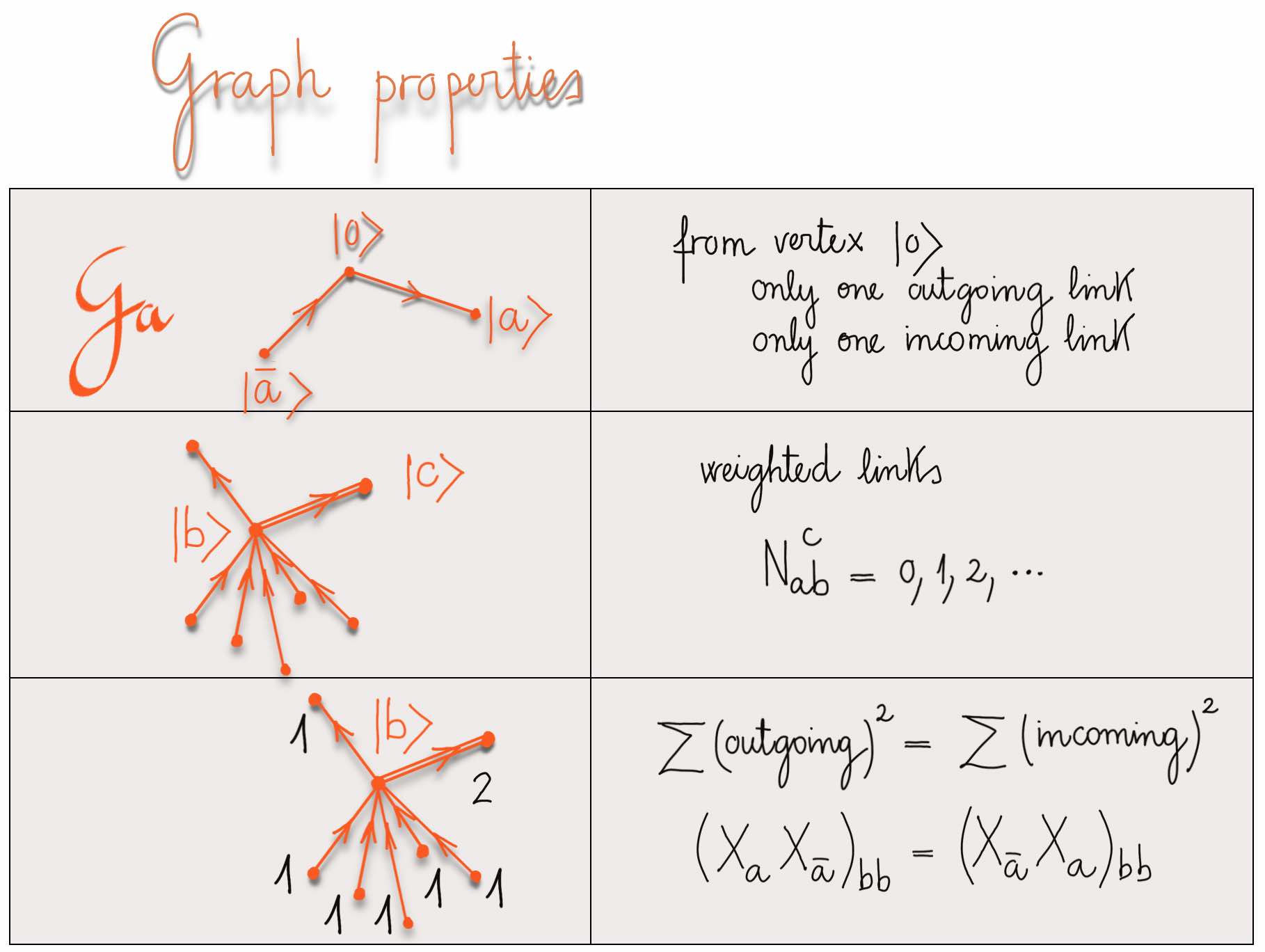}

\includegraphics[width=0.895\textwidth]{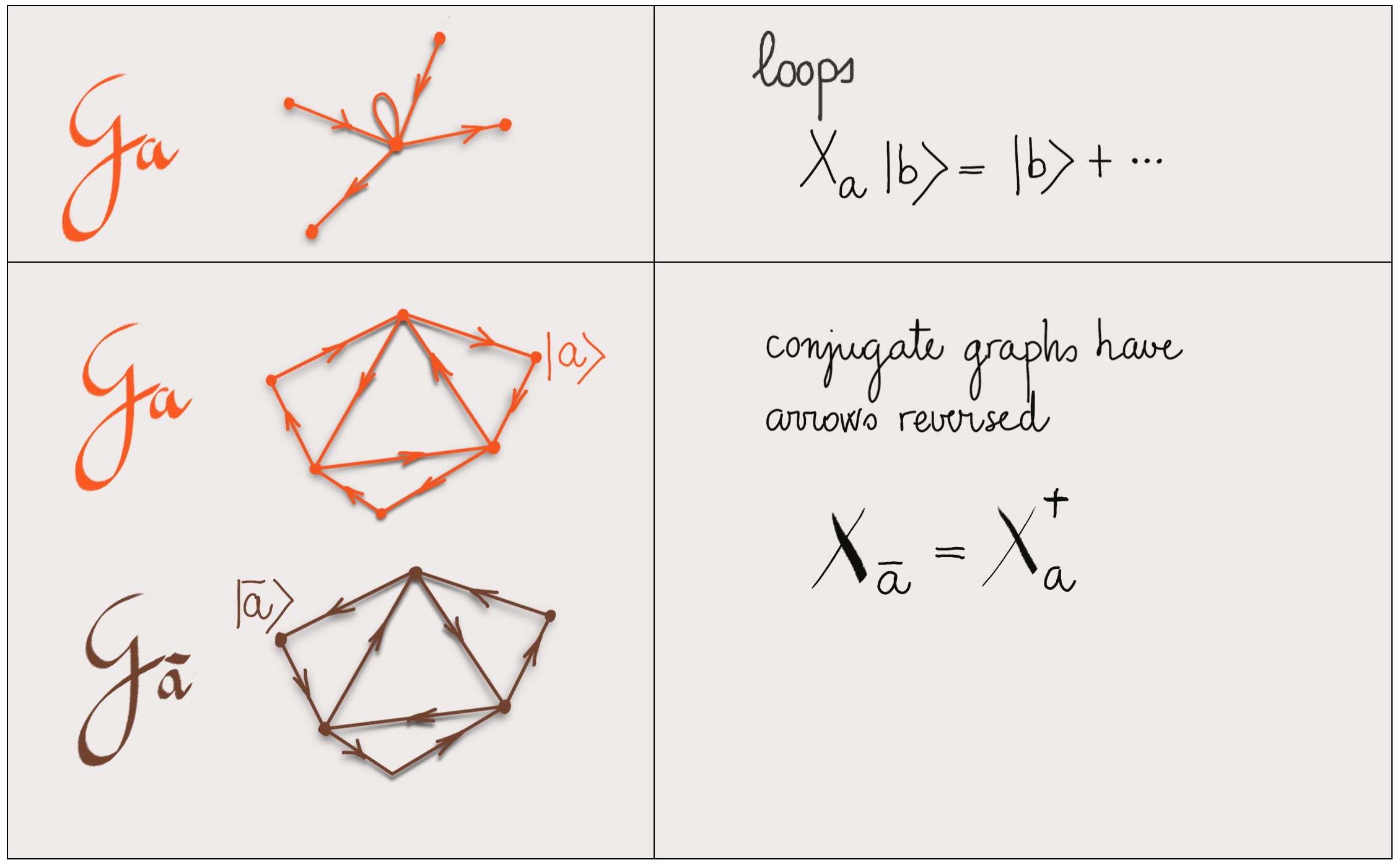}
\end{center}
\newpage
\restoregeometry
\pagestyle{fancy}
\fancyhf{}
\rhead{Topological Graphs}
\rfoot{\thepage}
\vspace*{0cm}
\subsection*{Properties of topological graphs}
\addcontentsline{toc}{subsection}{Properties of topological graphs}
\normalsize
Topological graphs exhibit the following properties, which are inherited from the properties of the topological algebra.

\begin{itemize}
\setlength{\itemindent}{-0.1in}
\item {\bf From the vertex $\ket{0}$}  there is only one outgoing link (to the vertex $\ket{a}$ in graph $\mathcal{G}_a$) and one incoming link (from the vertex $\ket{\bar a}$).
\item {\bf Current conservation law}. The sum of the squares of the multiplicities of links entering a vertex is equal to the one of links going out from it. For a graph $\mathcal{G}_a$ and a vertex $\ket{b}$ we have:
\begin{eqnarray}
\sum_cN_{ab}^cN_{ab}^c=\braket{b|X_a^\dagger X_a|b}=\braket{b|X_aX_a^\dagger|b}=\sum_dN_{ad}^bN_{ad}^b.
\end{eqnarray}
For a graph with weights either $0$ or $1$, the number of links is conserved at each vertex:
\begin{eqnarray}
\sum_cN_{ab}^c=\sum_d N_{ad}^b.
\end{eqnarray}
\item {\bf Loops.} A vertex can be connected to itself, forming a loop. This occurs for non-vanishing diagonal matrix elements 
\begin{eqnarray}
\braket{b|X_a|b}\ne0.
\end{eqnarray}
\item {\bf Conjugate graphs.} Graphs corresponding to conjugate charges have the same links, with arrows reversed:
\begin{eqnarray}
\braket{c|X_a|b}=\braket{b|X_{\bar a}|c}.
\end{eqnarray}
Similarly, conjugate vertices share the same links with arrows reversed:
\begin{eqnarray}
\braket{\bar c|X_a| \bar b}=\braket{b|X_{a}|c}.
\end{eqnarray}
\item {\bf Connectivity.} A topological graph is always a disjoint union of {\em connected} graphs.

\end{itemize}


\newpage
\pagestyle{fancy}
\fancyhf{}
\lhead{The language}
\lfoot{\thepage}
\vspace*{2cm}
\includegraphics[width=\textwidth]{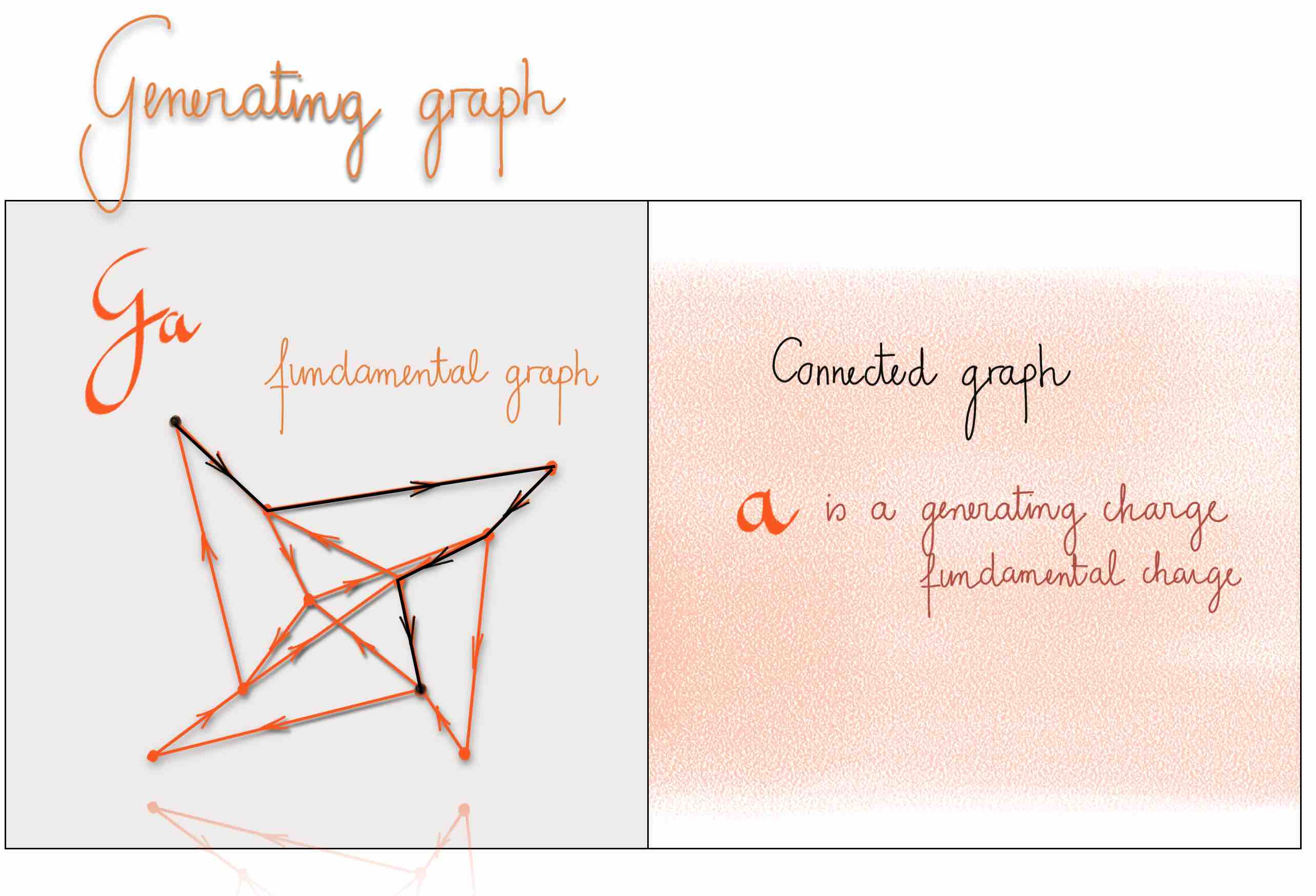}

\newpage
\vspace*{3cm}
\subsection*{Generating topological graph}
\pagestyle{fancy}
\fancyhf{}
\rhead{Topological Graphs}
\rfoot{\thepage}
\addcontentsline{toc}{subsection}{Generating topological graph}

A graph is called {\bf connected} if there is a path that connects any pair of vertices.

The operator $X$ corresponding to a connected graph is {\em irreducible}. It fulfills that for every pair $i,j$ there exists a positive integer $m$ such that:
\begin{eqnarray}
\braket{i|X^m| j}\ne0.
\end{eqnarray}

A connected topological graph defines a very interesting kind of topological graph. All other graphs in the topological algebra can be derived from it. They are indeed {\em polynomials} of this graph.
I will call it generating or fundamental graph, since it encodes the complete topological algebra.

In the Boson-Lattice construction I develop here, an anyon model will be encoded in a generating graph, from which all properties of the model can be read.

A general topological graph is always the disjoint union of connected graphs 
\footnote{The graph language provides an enlightening way to prove this property. The details of this proof will be presented elsewhere.}. Therefore the corresponding operator is the direct sum of irreducible operators.
Thanks to this property, the Perron-Frobenius theorem applies, and the existence of an eigenvector with all positive components is guaranteed.

\newpage
\thispagestyle{empty}
\vspace*{6cm}
\includegraphics[width=0.95\textwidth]{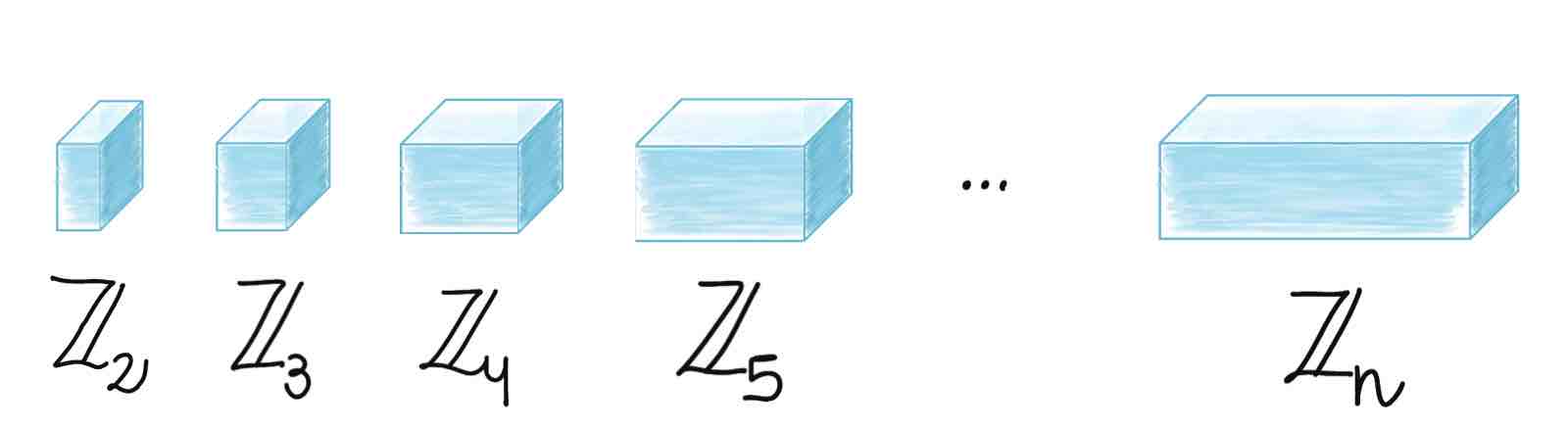}

\newpage
\section*{}
\thispagestyle{empty}
\addcontentsline{toc}{section}{Building blocks}
\vspace*{2.5cm}
\hspace*{1cm}
\includegraphics[width=0.75\textwidth]{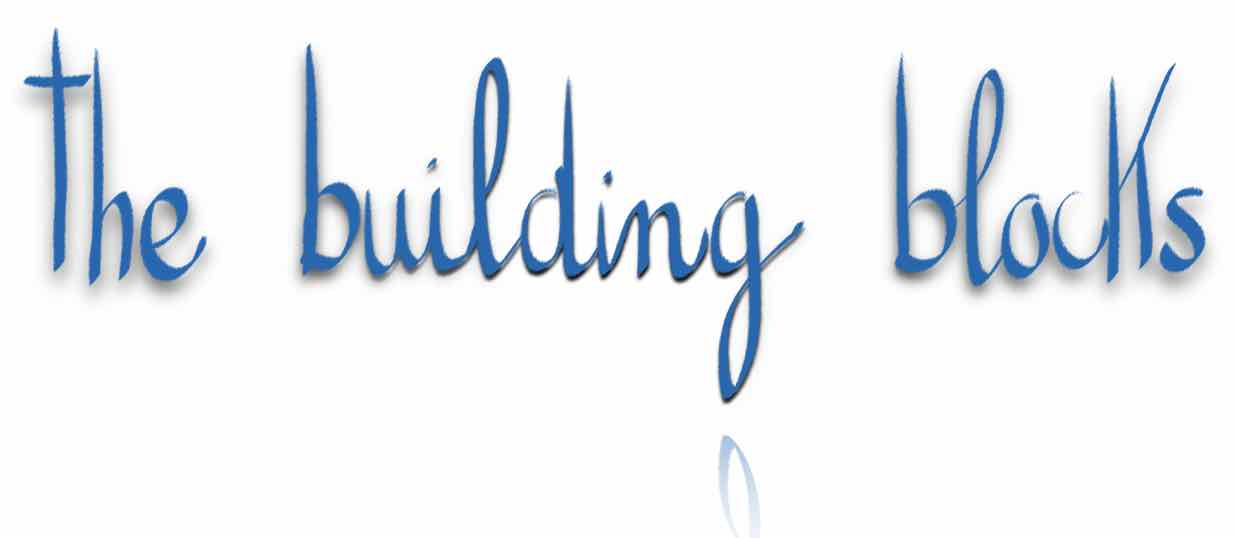}
\vspace*{1cm}

\large
\hspace*{2.5cm}
\parbox{11cm}{I introduce the building blocks of the Boson-Lattice construction.}

\hspace*{2.5cm}
\parbox{11cm}{These are the Abelian anyon models $\mathbb{Z}_n$. I describe them using the language of topological graphs introduced in the previous section.}

\hspace*{2.5cm}
\parbox{11cm}{A conceptual leap is made by identifying a $\mathbb{Z}_n$ model with {\em a particle in a one-dimensional periodic lattice} of $n$ sites. With this identification, the elementary pieces of the Boson-Lattice construction are defined as {\em particles in lattices}.}


\newpage
\pagestyle{fancy}
\fancyhf{}
\lhead{The building blocks}
\lfoot{\thepage}
\vspace*{2.5cm}
\begin{center}
\includegraphics[width=\textwidth]{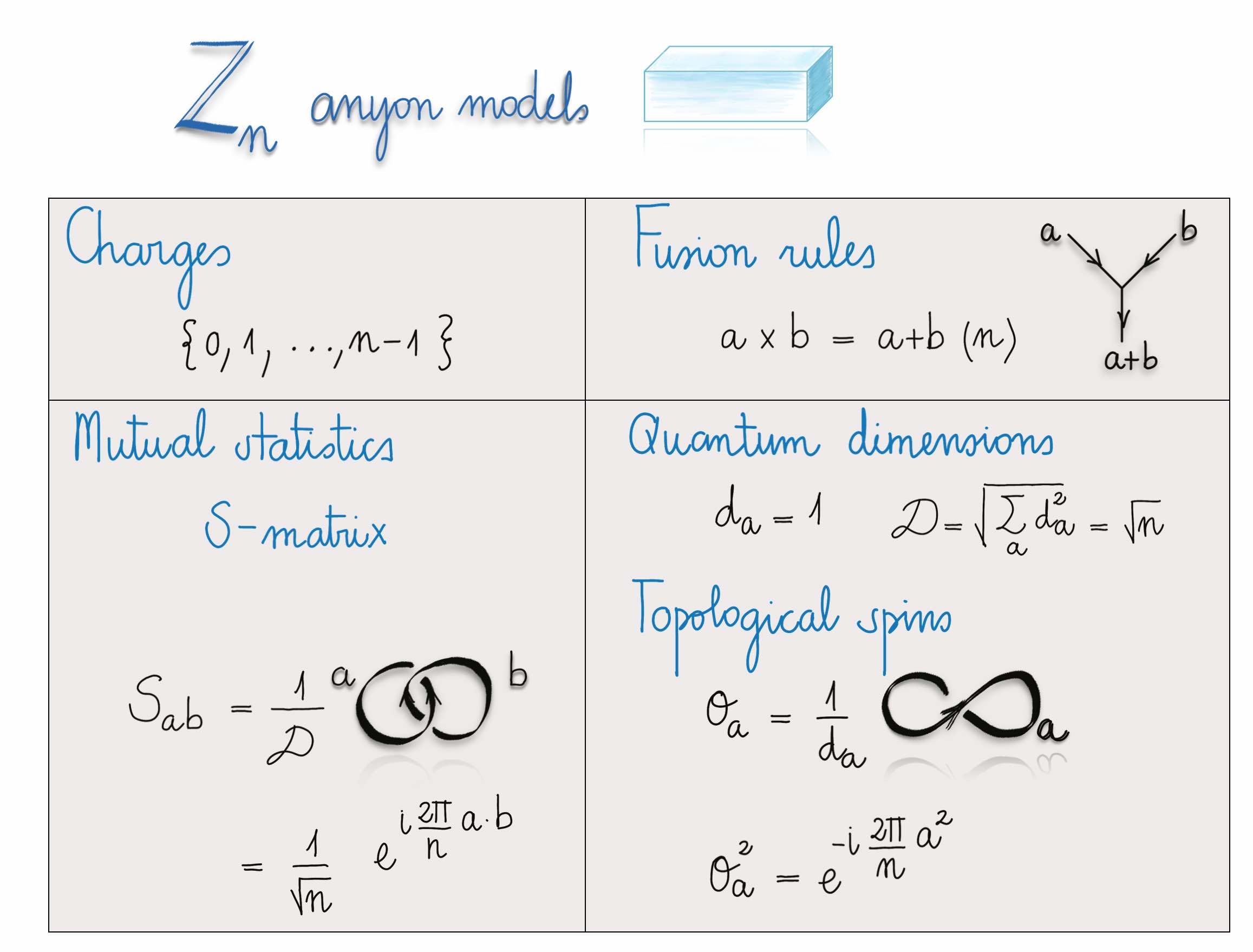}
\end{center}

\normalsize
\newpage
\pagestyle{fancy}
\fancyhf{}
\rhead{$\mathbb{Z}_n$ models}
\rfoot{\thepage}
\vspace*{0cm}
\section*{$\mathbb{Z}_n$ models}
\addcontentsline{toc}{section}{$\mathbb{Z}_n$ models}
A $\mathbb{Z}_n$ anyon model \footnote{The $\mathbb{Z}_n$ anyon models I define here are Abelian modular anyon models corresponding to $\mathbf{SU}(n)_1$ conformal field theories.}
is characterized by a set of $n$ charges:
\begin{eqnarray}
\{0,1,\cdots,n-1\},
\end{eqnarray}
which fulfill the fusion rules:
\begin{eqnarray}
a\times b=a+b\,(\text{mod}\,\,n).
\end{eqnarray}

The mutual braiding statistics of charges $a$ and $b$ is given by the element $S_{ab}$ of the $S$-matrix:
\begin{eqnarray}
S_{ab}=\frac{1}{\sqrt{n}}e^{i\frac{2\pi}{n}a\cdot b}.
\end{eqnarray}

The self statistics of charge $a$ is given by the topological spin $\theta_a$:
\begin{eqnarray}
\theta_{a}^2=e^{-i\frac{2\pi}{n}a^2}.
\end{eqnarray}

\begin{wrapfigure}{r}{0.5\textwidth}
\vspace*{-0.9cm}
\includegraphics[width=0.48\textwidth]{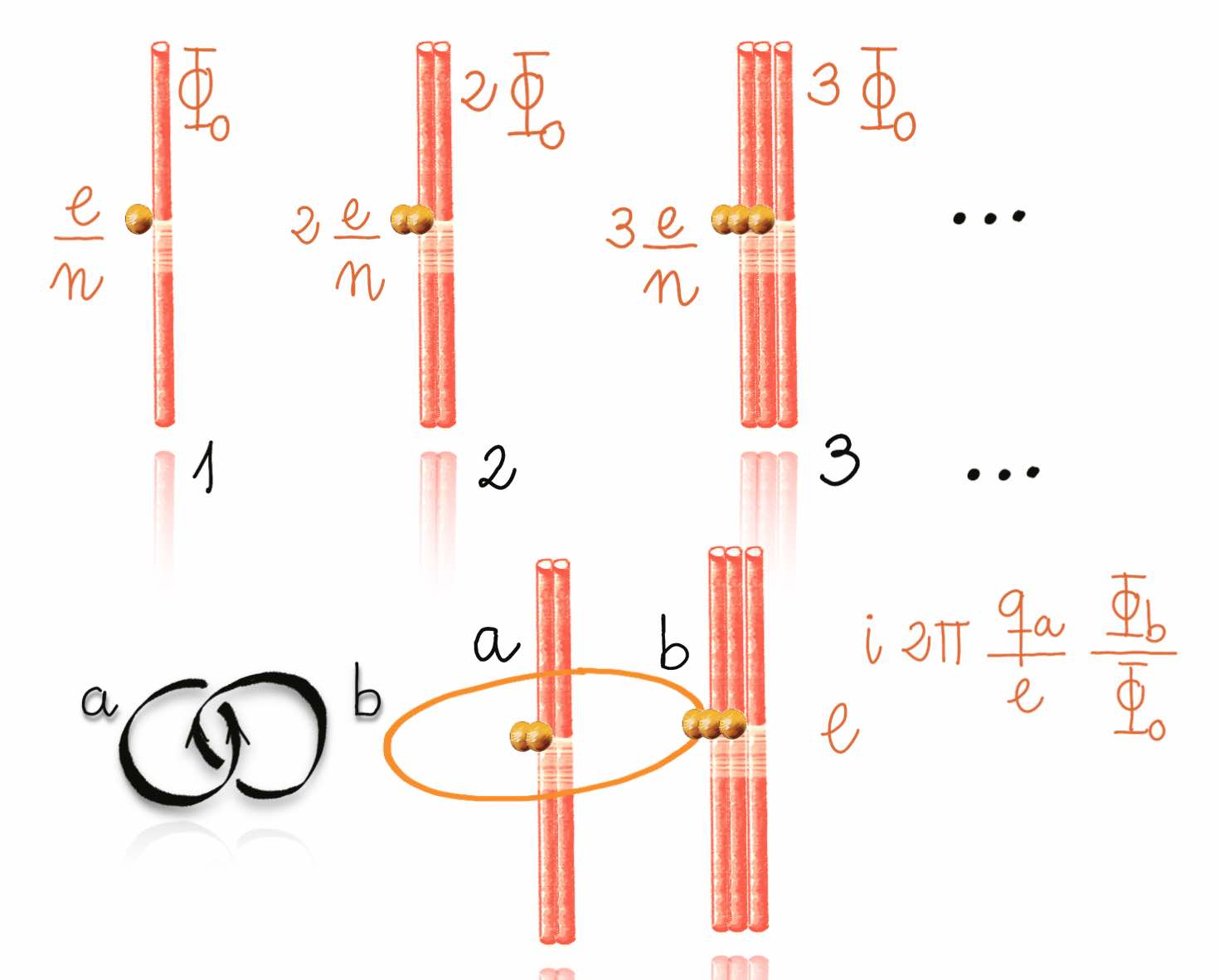}
\end{wrapfigure}
\vspace*{0.5cm}
A physical realization of a $\mathbb{Z}_n$ model can be obtained by representing the charges of the model with fractional electric charges attached to magnetic fluxes \cite{A1,A2,Preskill}.
A topological charge $a$ is represented by a fractional charge $q_a=a\frac{e}{n}$ attached to a flux $\phi_a=a\phi_0$, where $e$ is the electron charge and $\phi_0$ is the quantum of flux.

The mutual statistics $S_{ab}$ is obtained as the Aharonov-Bohm phase that charge-flux composites $q_a$ and $q_b$ acquire when going around each other.

\newpage
\pagestyle{fancy}
\fancyhf{}
\lhead{The building blocks}
\lfoot{\thepage}
\vspace*{0cm}
\begin{center}
\includegraphics[width=\textwidth]{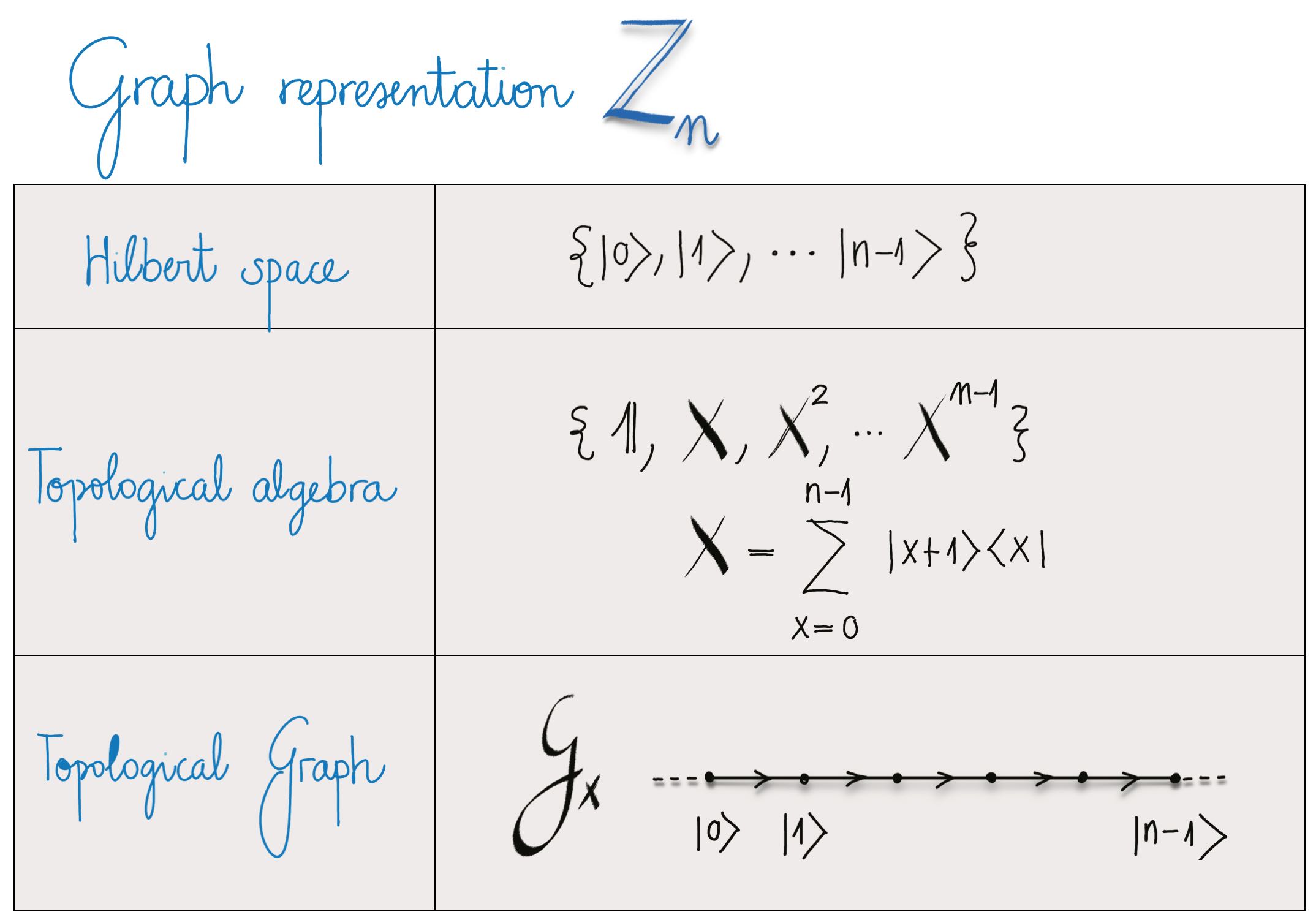}
\end{center}
\begin{center}
\includegraphics[width=0.85\textwidth]{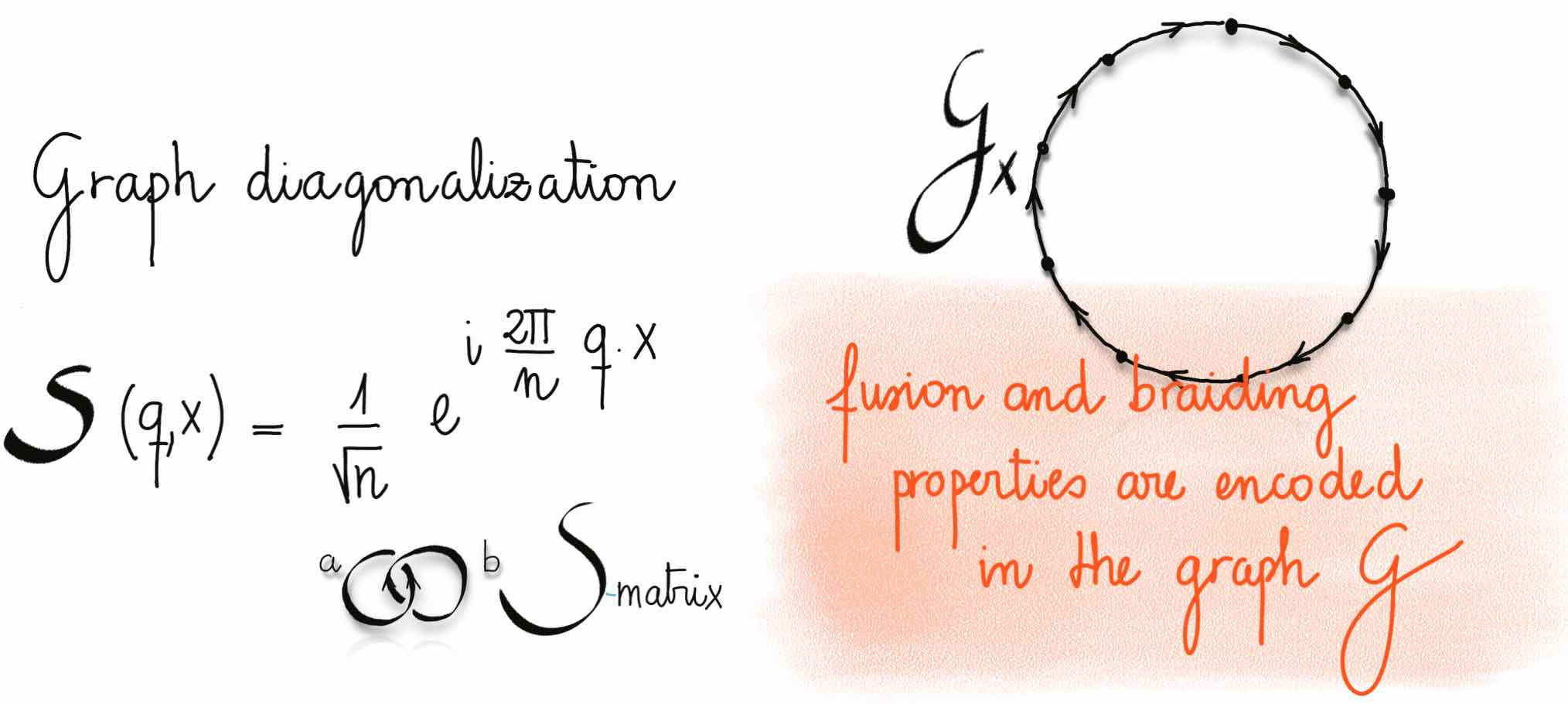}
\end{center}

\newpage
\pagestyle{fancy}
\fancyhf{}
\rhead{$\mathbb{Z}_n$ models}
\rfoot{\thepage}
\vspace*{0cm}
\section*{Graph representation of $\mathbb{Z}_n$ models}
\addcontentsline{toc}{section}{Graph representation of $\mathbb{Z}_n$ models}
{\em Hilbert space}. The Hilbert space corresponding to a $\mathbb{Z}_n$ model has dimension $n$. I denote the canonical basis by:
\begin{eqnarray}
\{\ket{0},\ket{1},\cdots,\ket{n-1}\}.
\end{eqnarray}
{\bf Topological algebra}. Following the definition introduced in the previous section, the topological algebra of the model is given by the set of operators:
\begin{eqnarray}
\mathcal{A}=\{\mathds{1},X,X^2,\cdots,X^{n-1}\},
\end{eqnarray}
with 
\begin{eqnarray}
X=\sum_{x=0}^{n-1}\ket{x+1}\bra{x},
\end{eqnarray}
where $\ket{x}=\ket{x \,(\text{mod}\,n)}$. The operator $X$ fulfills $X^n=\mathds{1}$.

{\bf Generating graph}. 
The charge $1$ represented by the operator $X$ is a generating or fundamental charge. 

The graph associated with it, 
$\mathcal{G}_X$, is the one in which each vertex is linked to its next (to the right) neighbour.
That is, it is an {\bf oriented lattice with periodic boundary conditions.}
Fusion rules and braiding rules of the model are encoded in this graph.

{\bf S-Matrix}. The $S$-matrix of the model is directly obtained by diagonalizing the operator $X$. The eigenstates of $X$ are Fourier transformed states 
of the form:
\begin{eqnarray}
\ket{q}=\frac{1}{\sqrt{n}}\sum_{x=0}^{n-1} e^{i\frac{2\pi}{n}q\cdot x}\ket{x}.
\end{eqnarray}
The unitary and symmetric matrix diagonalizing the algebra is thus:
\begin{eqnarray}
S_{qx}=\frac{1}{\sqrt{n}}e^{i\frac{2\pi}{n}q\cdot x},
\end{eqnarray}
which corresponds to the $S$-matrix of the $\mathbb{Z}_n$ anyon model.


\newpage
\pagestyle{fancy}
\fancyhf{}
\lhead{The building blocks}
\lfoot{\thepage}
\vspace*{3.5cm}
\begin{center}
\includegraphics[width=\textwidth]{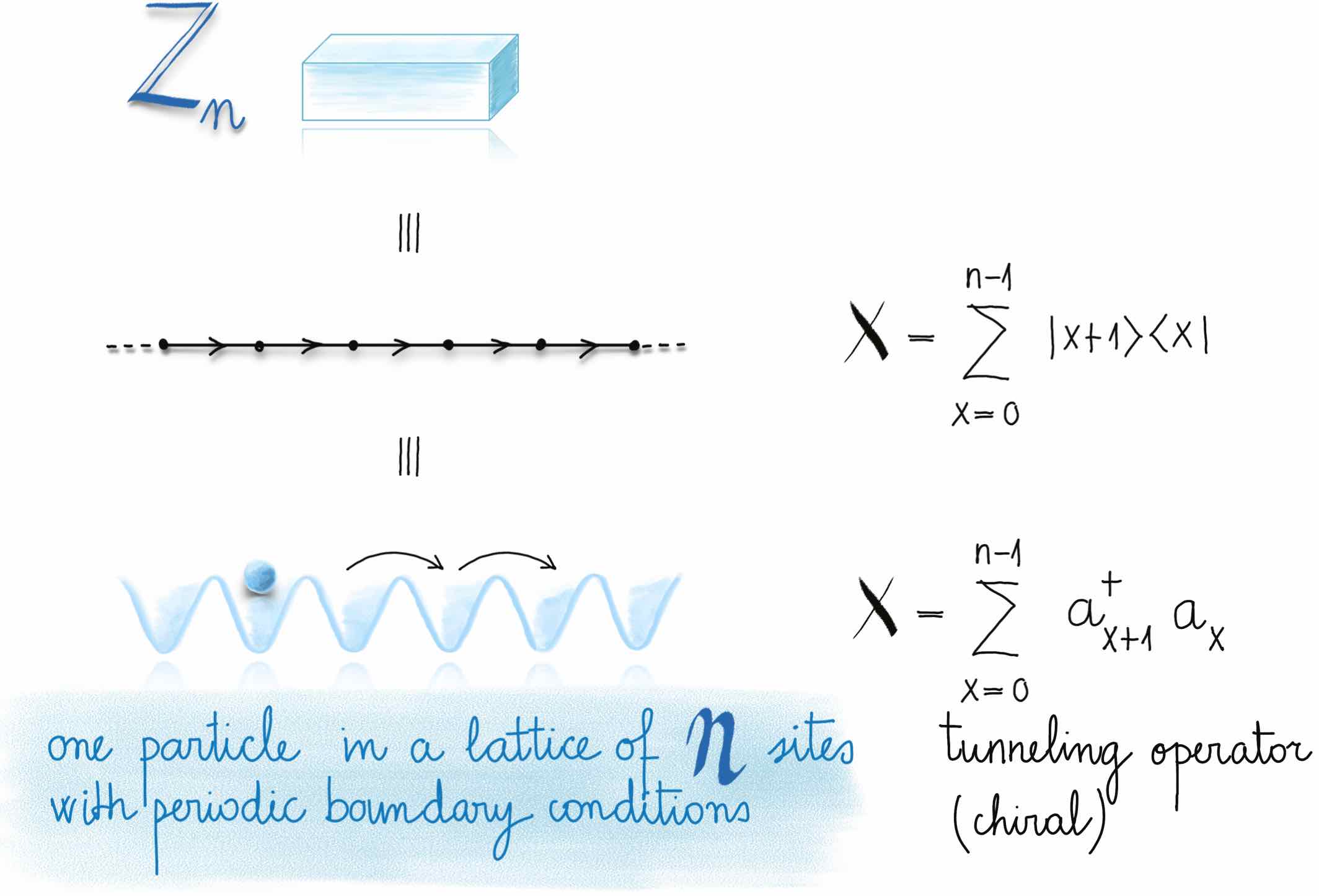}
\end{center}

\newpage
\pagestyle{fancy}
\fancyhf{}
\rhead{$\mathbb{Z}_n$ models}
\rfoot{\thepage}
\vspace*{1.5cm}
\section*{The leap to a particle in a lattice}
\vspace*{0.3cm}
\addcontentsline{toc}{section}{The leap to a particle in a lattice}
The generating graph contains complete information of the anyon model.
Fusion and braiding rules can be read from the graph.

The operator $X$ is the (chiral) translation operator in a one-dimensional lattice with periodic boundary conditions.

{\em Abstraction}. I identify a $\mathbb{Z}_n$ anyon model with a particle in a lattice of $n$ sites with periodic boundary conditions and chiral tunneling operator:

\begin{eqnarray}
X=\sum_{x=0}^{n-1}a^\dagger_{x+1}a_{x}^{},
\end{eqnarray}

where $a_{x}$($a^\dagger_{x}$) is the annihilation (creation) operator of a particle in the lattice site $x$.

This identification establishes the essence of the elementary pieces of the construction.

{\bf The building blocks are particles in lattices.}

{\bf The construction will assemble particles in lattices.}

\newpage
\thispagestyle{empty}
\vspace*{5.5cm}
\begin{center}
\includegraphics[width=\textwidth]{ConstructionTitleFigure.jpg}
\end{center}

\newpage
\thispagestyle{empty}
\pagestyle{fancy}
\fancyhf{}
\rhead{The construction}
\rfoot{\thepage}
\section*{}
\addcontentsline{toc}{section}{The Construction}
\hspace{1cm}
\includegraphics[width=0.8\textwidth]{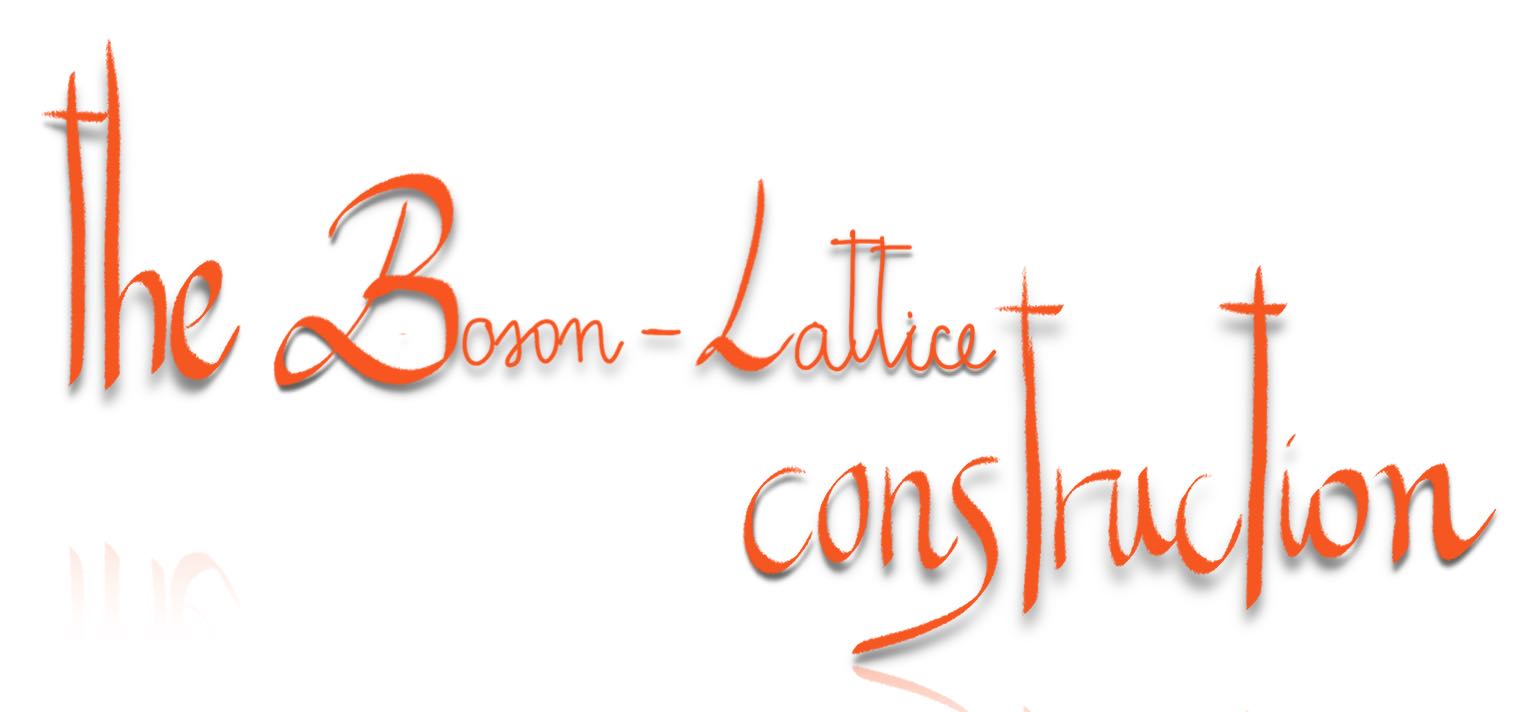}

\hspace*{2cm}
\parbox{12cm}{Using the language and the building blocks described in the previous sections, I present a formalism to systematically construct anyon models.
\parskip=8pt

An anyon model is built up by assembling $k$ identical building blocks of length $n$. The Hilbert space of the model is obtained by {\bf bosonization} of the Hilbert spaces of the building blocks.

Based on the graphs of the building blocks, I give a prescription to construct a graph in the bosonized Hilbert space. This graph is conceived such that it always corresponds to the generating graph of a modular anyon model. 

A one-to-one correspondence is established between the properties of the Boson-Lattice system (Fock states, tunneling connectivity patterns, eigenvalues and eigenstates) and the properties of the anyon model (topological charges, fusion rules, quantum dimensions, $S$ and $T$ matrices).

This Boson-Lattice construction systematically generates, by varying the number of bosons and the number of lattice sites, a series of well known tabulated anyon models. In particular, it generates anyon models corresponding to truncated Lie algebras such as $\mathbf{SU}(2)_k$, Fibonnaci, $\mathbf{SO}(3)_k$, or $\mathbf{SO}(5)_k$.
Interestingly, the construction also yields anyon models which are not tabulated.}

\newpage
\pagestyle{fancy}
\fancyhf{}
\lhead{The construction}
\lfoot{\thepage}
\vspace*{2cm}
\hspace*{1cm}
\includegraphics[width=0.85\textwidth]{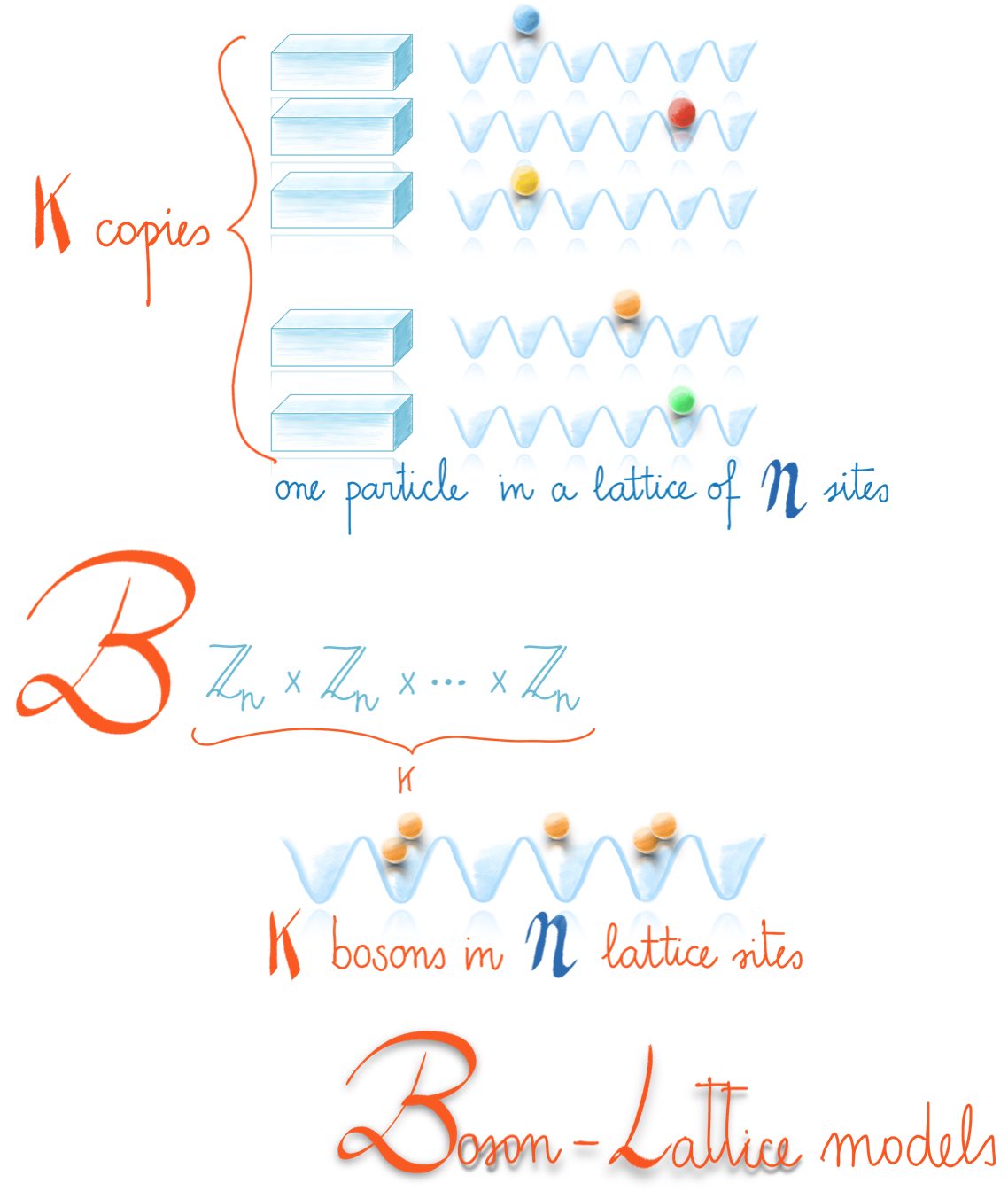}

\newpage
\pagestyle{fancy}
\fancyhf{}
\rhead{Bosonization}
\rfoot{\thepage}
\vspace*{0,5cm}
\section*{The bosonization idea}
\addcontentsline{toc}{section}{The bosonization idea}
In the last section I have shown that a building block anyon model (a $\mathbb{Z}_n$ model) is completely characterized by:

\begin{itemize}
\item {a Hilbert space $\mathcal{H}(1,n)$, corresponding to a single particle in a one-dimensional lattice of $n$ sites with periodic boundary conditions.}
\item {a generating graph $\mathcal{G}(1,n)$, corresponding to the chiral tunneling operator of the particle in such a lattice.}
\end{itemize}

I give now a prescription to assemble these building blocks in order to sequentially generate new anyon models.

To construct a new anyon model I consider $k$ identical building blocks of length $n$.

I define the Hilbert space $\mathcal{H}(k,n)$ associated with the new anyon model as the one resulting from bosonization (symmetrization) of the tensor product of the $k$ identical Hilbert spaces of the building blocks:

\begin{eqnarray}
\mathcal{H}(k,n)=\mathcal{S}\,\,\underbrace{\mathcal{H}(1,n)\otimes\cdots\otimes\mathcal{H}(1,n)}_{k\,\, \text{copies}}.
\end{eqnarray}

The Hilbert space $\mathcal{H}(k,n)$ is the one of $k$ bosons in a  one-dimensional lattice of $n$ lattice sites with periodic boundary conditions.

The new Hilbert space is constructed by making the $k$ particles become indistinguishable.
It is important to emphasize that in this bosonization strategy the particles that are made indistinguishable are not physical objects, but mathematical constructions used to encode an anyon model.



\newpage
\pagestyle{fancy}
\fancyhf{}
\lhead{The construction}
\lfoot{\thepage}
\vspace*{0.7cm}
\begin{center}
\includegraphics[width=0.85\textwidth]{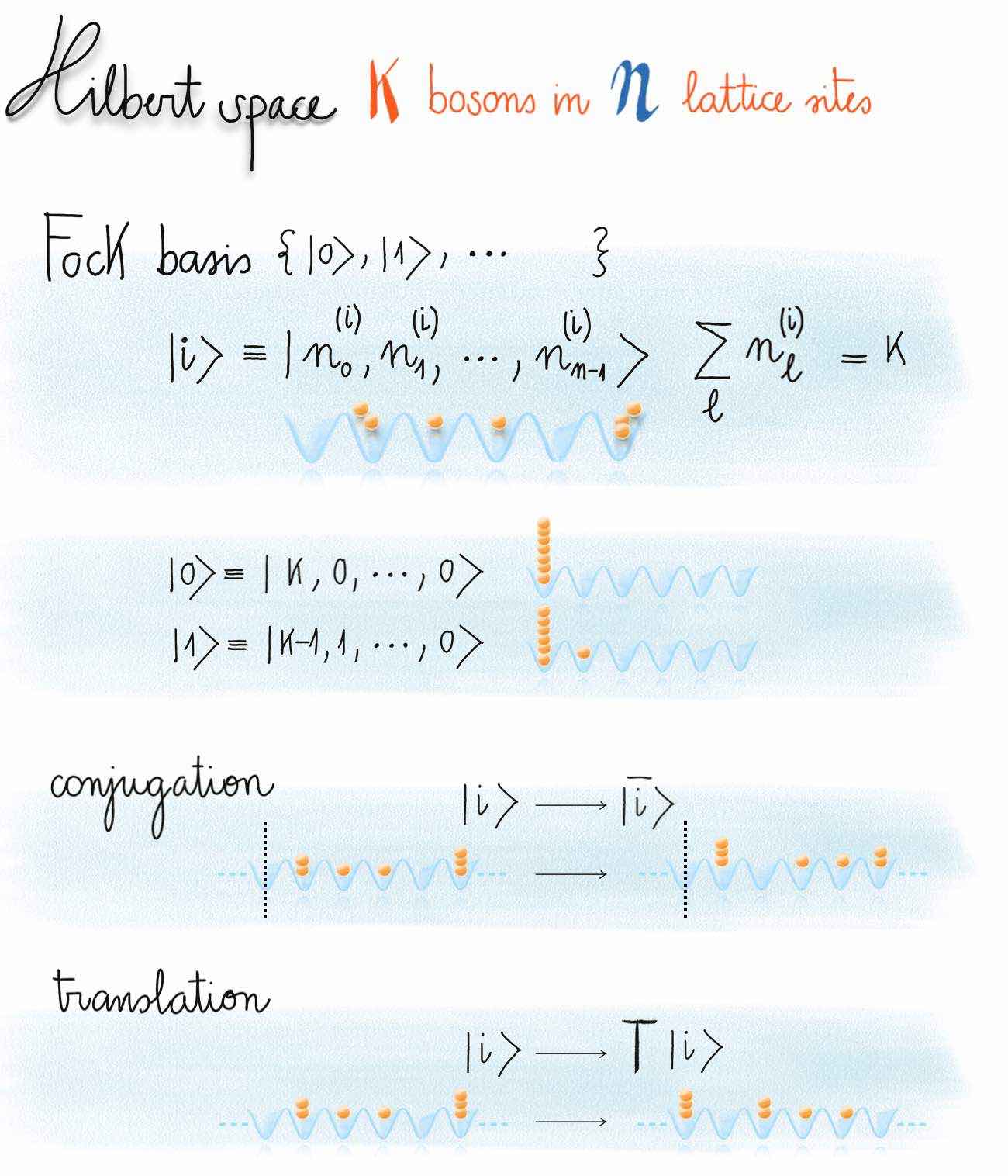}
\end{center}

\newpage
\pagestyle{fancy}
\fancyhf{}
\rhead{Bosonization}
\rfoot{\thepage}
\section*{The Boson-Lattice Hilbert space}
\addcontentsline{toc}{section}{The Boson-Lattice Hilbert space}

I give here some definitions in the Hilbert space $\mathcal{H}(k,n)$ which will be useful to describe the anyon model associated with it.

{\bf The Fock basis}. I consider the basis of Fock states. Each Fock state is characterized by the corresponding occupation numbers of the lattice sites:
\begin{eqnarray}
\ket{i}\equiv\ket{n_0^{(i)},n_1^{(i)},\cdots,n_{n-1}^{(i)}},
\end{eqnarray}
with $n_\ell^{(i)}$ being the occupation number of site $\ell$, and $\ell=0,\cdots,n-1$. The total number of bosons is equal to $k$, $\sum_\ell n_\ell^{(i)}=k$.

{\bf The trivial state}. I choose a reference state, $\ket{0}$, as the Fock state with all bosons occupying the same lattice site (for example, the site $\ell=0$.)
\begin{eqnarray}
\ket{0}\equiv\ket{k,0,\cdots,0}.
\end{eqnarray}
I call this state the trivial state.

{\bf The generating state}. I denote as $\ket{1}$ the Fock state obtained from the state $\ket{0}$ by transferring one boson to site $1$:
\begin{eqnarray}
\ket{1}\equiv\ket{k-1,1,\cdots,0}.
\end{eqnarray}
I call this state the generating state.

{\bf Conjugation}. I define the unitary operation $C$ as the one mapping each Fock state to its mirror image with respect to the site $0$:
\begin{eqnarray}
\ket{i}&\overset{C}{\longrightarrow} &\ket{\bar i}=C\ket{i} \nonumber\\
n_\ell^{(i)}&\overset{C}{\longrightarrow} &n_{n-\ell}^{(i)}.
\end{eqnarray}
As a reflection, the unitary $C$ fulfills $C^\dagger=C$ and thus $C^2=\mathds{1}$.

{\bf Global translation}. I define the unitary operation $T$ as the one mapping each Fock state to the one in which each boson has been moved one site to the right:
\begin{eqnarray}
\ket{i}&\overset{T}{\longrightarrow} &T\ket{i}\nonumber\\
n_\ell^{(i)}&\overset{T}{\longrightarrow} &n_{\ell-1}^{(i)}.
\end{eqnarray}
The unitary $T$ fulfills $T^\dagger=T^{n-1}$.


\newpage
\pagestyle{fancy}
\fancyhf{}
\lhead{The construction}
\lfoot{\thepage}
\vspace*{2cm}
\begin{center}
\includegraphics[width=\textwidth]{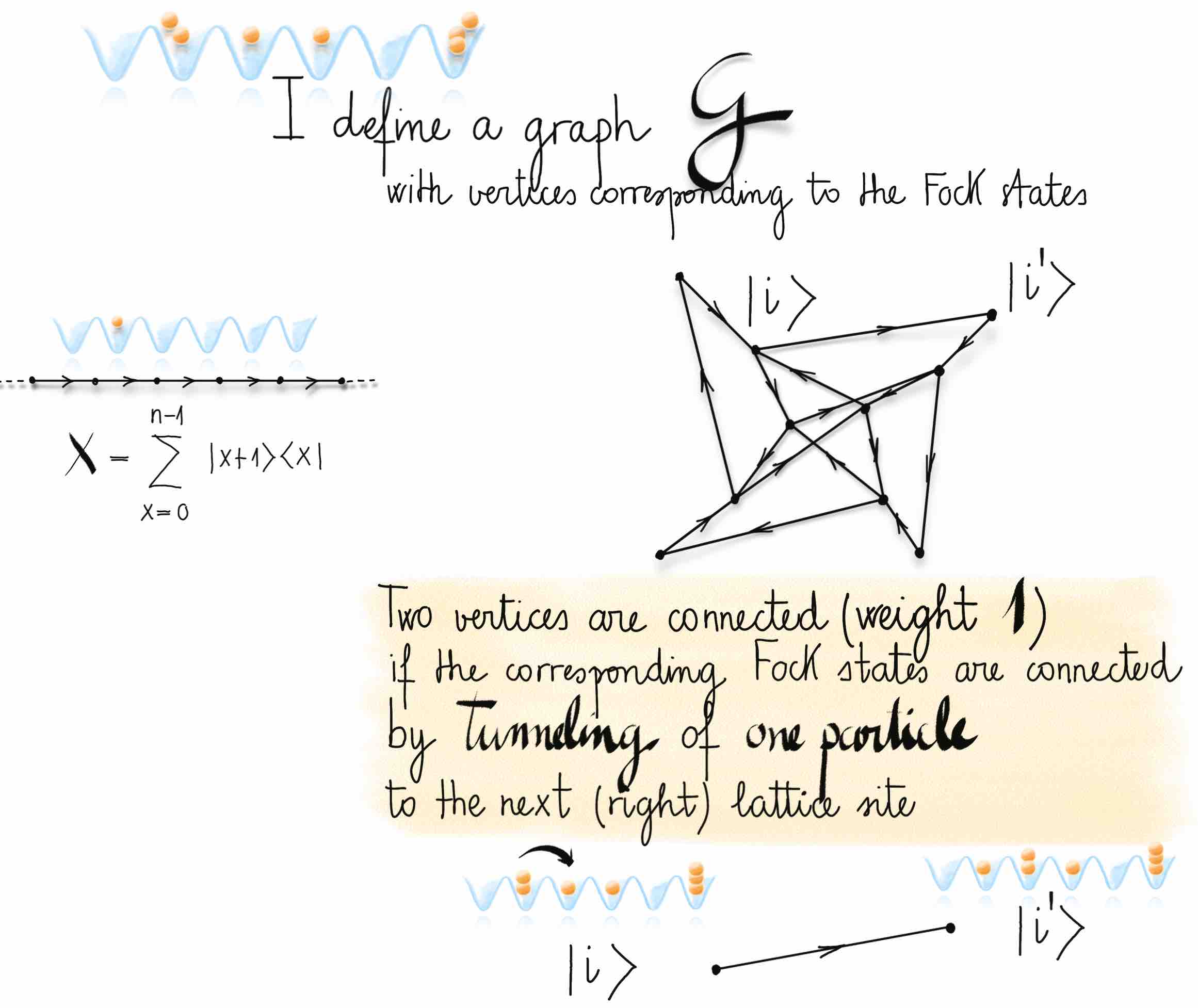}
\end{center}

\newpage
\pagestyle{fancy}
\fancyhf{}
\rhead{The Graph}
\rfoot{\thepage}
\vspace*{0cm}
\section*{The Boson-Lattice graph}
\addcontentsline{toc}{section}{The Boson-Lattice graph}

associated with the Hilbert space $\mathcal{H}(k,n)$ I define a graph $\mathcal{G}(k,n)$ which will be the generating graph of the corresponding anyon model. The conception of this graph is an essential step to develop the Boson-Lattice construction. 

\hspace*{0,5cm}\begin{mdframed}

{\bf Boson-Lattice Graph: Definition.}

The Boson-Lattice Graph $\mathcal{G}(k,n)$ associated with the Hilbert space $\mathcal{H}(k,n)$ of $k$ bosons in a periodic lattice of $n$ sites is defined as follows:

{\bf Vertices}. The vertices of the graph are in one to one correspondence with the Fock states in the Hilbert space $\mathcal{H}(k,n)$.

{\bf Connectivity and links-weight}. Two vertices are connected if the corresponding Fock states are connected by tunneling of one boson to the next (to the right) lattice site. 
The link  is given a weight $1$.

\end{mdframed}

The connectivity pattern of the Boson-Lattice graph is inspired by the connectivity of the building block graph, the generating graph of the anyon model $\mathbb{Z}_n$. There, two vertices are connected if the corresponding one-particle states are connected by tunneling of the particle to the next (to the right) lattice site.

It is illuminating to write down the expression for the operator $X$ corresponding to the graph $\mathcal{G}$. It can be written as:
\begin{eqnarray}
X=\sum_{i^{\vphantom{\prime}} \leadsto i^\prime}\ket{i^\prime} \bra{i^{\vphantom{\prime}}}&\longleftrightarrow &\mathcal{G},
\end{eqnarray}
where $i \leadsto i^\prime$ indicates that the sum runs over pairs of Fock states $\ket{i}$, $\ket{i^\prime}$, such that $\ket{i^\prime}$ can be obtained from $\ket{i}$ through tunneling of one particle to the next (to the right) lattice site.

Using creation and annihilation operators we can write $X$ as:
\begin{eqnarray}
X=\sum_{\ell=0}^{n-1}A^\dagger_{\ell+1}A^{\vphantom{\dagger}}_\ell.
\end{eqnarray}


\newpage
\pagestyle{fancy}
\fancyhf{}
\lhead{The construction}
\lfoot{\thepage}

Here,
\begin{eqnarray}
&&A^\dagger_{\ell}\ket{\,\cdots, n_\ell, \cdots\,}=\ket{\,\cdots, n_\ell+1, \cdots\,}\nonumber\\
&&A^{}_{\ell}\ket{\,\cdots, n_\ell, \cdots\,}=
\begin{cases}
\hspace*{0.1cm} 0 \hspace*{3cm} \text{if} \,\,n_\ell=0\\
\hspace*{0.1cm}\ket{\,\cdots, n_\ell-1, \cdots\,} \hspace*{1cm} \text{otherwise}.
\end{cases}
\label{CreationOperators}
\end{eqnarray}
The operators $A_\ell$ satisfy the commutation relations:
\begin{eqnarray}
\left[A^{\vphantom{\dagger}}_{\ell^{\vphantom{\prime}}},A^\dagger_{\ell^\prime}\right]=\delta_{\ell^{\vphantom{\prime}}\ell^\prime}P_\ell,
\label{CommutationRelationAoperators}
\end{eqnarray}
where $P_\ell$ is the projector onto the subspace of Fock states with $n_\ell=0$.
It is crucial to note that the operators $A_\ell$ are not one-particle bosonic operators. The operator $X$ is therefore a many-body operator different from the one-particle tunneling operator:
\begin{eqnarray}
X\ne\sum_{\ell=0}^{n-1}a^\dagger_{\ell+1}a_{\ell}^{\vphantom{\dagger}}.
\end{eqnarray}
Here, $a^\dagger_{\ell}$ $(a^{\vphantom{\dagger}}_{\ell})$ is the creation (annihilation) operator of one boson at site $\ell$.

\vspace*{1cm}
\begin{center}
\includegraphics[width=\textwidth]{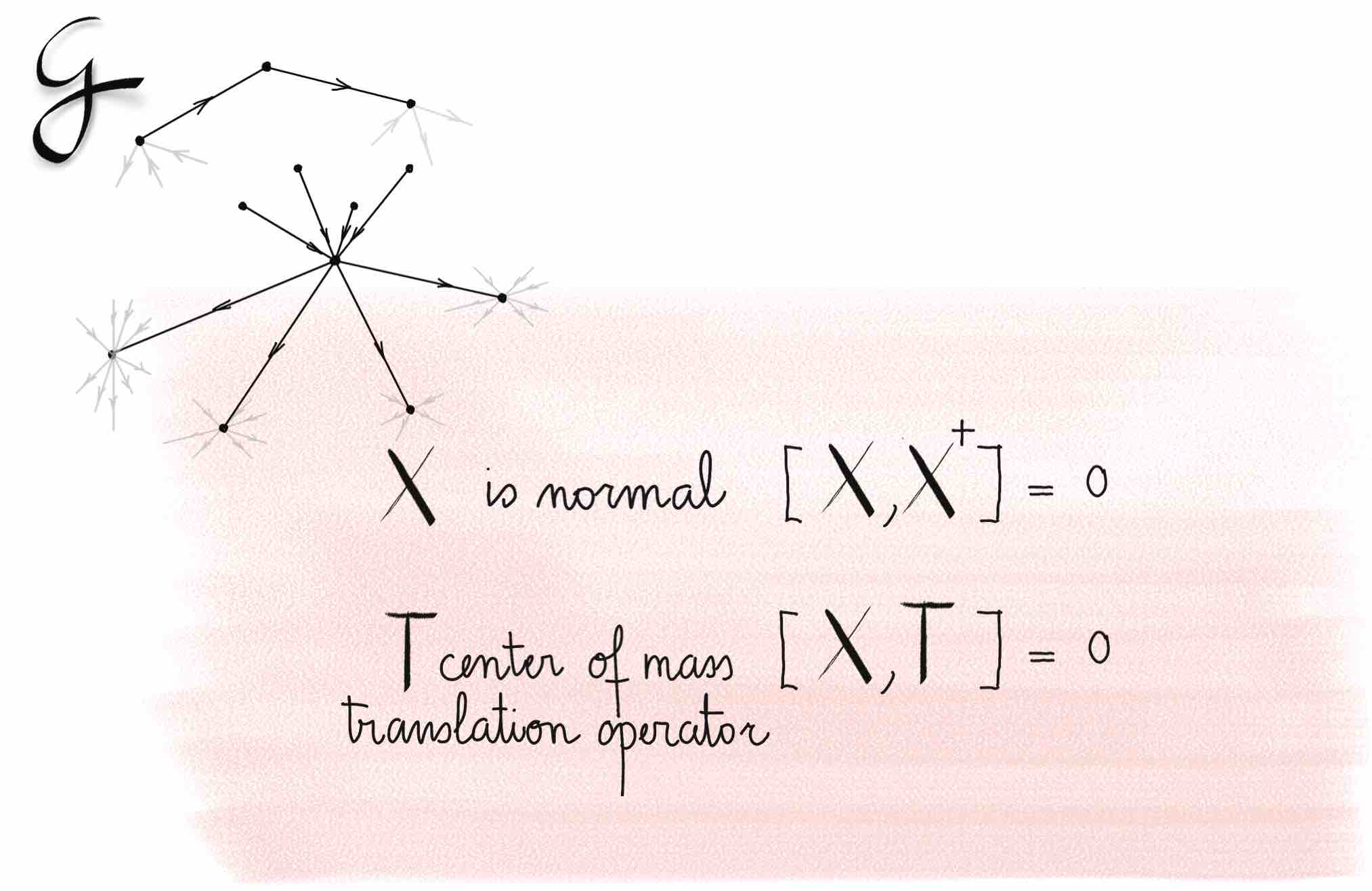}
\end{center}

\newpage
\pagestyle{fancy}
\fancyhf{}
\rhead{The Graph}
\rfoot{\thepage}

\vspace*{0.5cm}
\subsection*{Properties of the operator $X$}
\addcontentsline{toc}{section}{Properties of the operator $X$}

The operator $X$ corresponding to the Boson-Lattice graph fulfils the following properties. 

\begin{itemize}

\item {\bf T-invariance}: $T^\dagger XT=X$

Taking into account that $T^\dagger A_\ell T=A_{\ell-1}$, we have:
\begin{eqnarray}
T^\dagger XT=\sum_\ell A^\dagger_{\ell}A^{\vphantom{\dagger}}_{\ell-1}=X.
\end{eqnarray}

\item  {\bf Conjugation}: $CXC=X^\dagger$

Given that $CA_\ell C=A_{n-\ell}$, we have:
\begin{eqnarray}
CXC=\sum_\ell A^\dagger_{n-\ell-1}A^{\vphantom{\dagger}}_{n-\ell}=\sum_\ell A^\dagger_{\ell-1}A^{\vphantom{\dagger}}_{\ell}=X^\dagger.
\end{eqnarray}

\item {\bf $X$ is normal}: $\left[X,X^\dagger\right]=0$

From the commutation relations of the operators $A_\ell$ (\ref{CommutationRelationAoperators}), we have:
\begin{eqnarray}
\left[X,X^\dagger\right]&=&\sum_{\ell^{\vphantom{\prime}},\ell^\prime}
\left[
A^\dagger_{\ell^{\vphantom{\prime}}+1}
A^{\vphantom{\dagger}}_{\ell^{\vphantom{\prime}}},\,
A^\dagger_{\ell^\prime+1}
A^{\vphantom{\dagger}}_{\ell^\prime}
\right]=
\sum_{\ell^{\vphantom{\prime}}}
\left[
A^\dagger_{\ell^{\vphantom{\prime}}+1}
A^{\vphantom{\dagger}}_{\ell^{\vphantom{\prime}}},\,
A^\dagger_{\ell^{\vphantom{\prime}}}
A^{\vphantom{\dagger}}_{\ell^{\vphantom{\prime}}+1}\right]=\nonumber\\
&=&
\sum_{\ell^{\vphantom{\prime}}}
P_\ell-P_{\ell+1}+P_\ell P_{\ell+1}-P_\ell P_{\ell+1}=0.
\end{eqnarray}

\item $X\ket{0}=\ket{1}$
\item $\braket{a|X|b}=0,1$
\end{itemize}

These properties imply properties for the Boson-Lattice graph which will entitle it to be the generating graph of an anyon model.
\newpage
\pagestyle{fancy}
\fancyhf{}
\lhead{The construction}
\lfoot{\thepage}
\vspace*{3cm}
\begin{center}
\includegraphics[width=\textwidth]{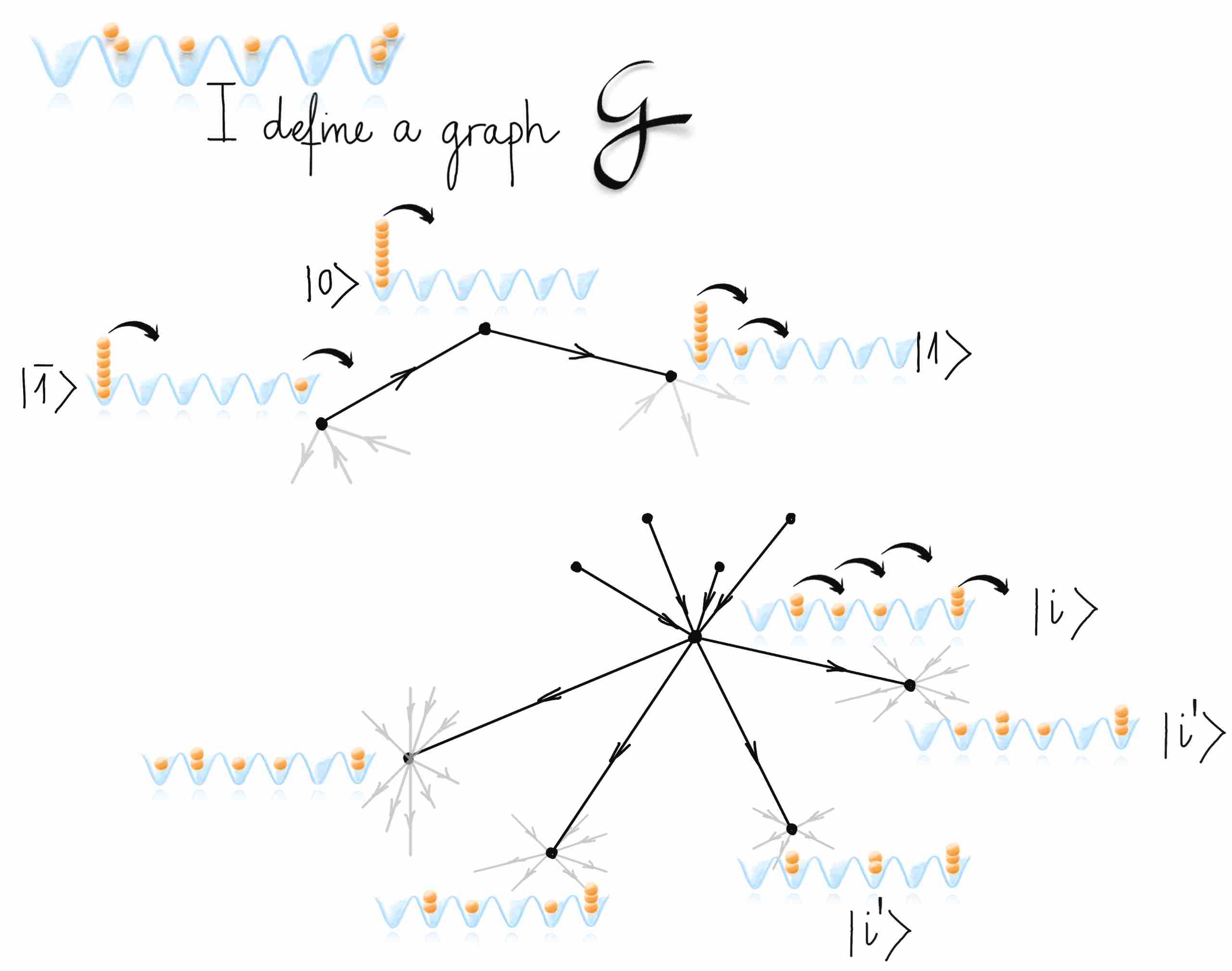}
\end{center}

\newpage
\pagestyle{fancy}
\fancyhf{}
\rhead{The Graph}
\rfoot{\thepage}
\subsection*{Boson-Lattice graph properties}
\addcontentsline{toc}{section}{Boson-Lattice graph properties}

As defined above the Boson-Lattice graph exhibits the following properties:

\begin{itemize}
\item {\bf $T$ invariance.} The graph is invariant under the unitary $T$, the global translation by one site. Given a vertex $\ket{a}$ the graph looks the same from the translated vertex $\ket{a^\prime}=T\ket{a}$:
\small
\begin{eqnarray}
\braket{c|X|a}=\braket{c|T^\dagger X T|a}=\braket{c^\prime|X|a^\prime}.
\end{eqnarray}
\normalsize
\item {\bf Conjugate graph.} Under the conjugation operation $C$, the arrows of the links are reversed. Given a vertex $\ket{a}$ the graph looks the same from the conjugate vertex $\ket{\bar a}=C\ket{a}$, but arrows are reversed:
\small
\begin{eqnarray}
\braket{c|X|a}=\braket{c|CCXCC|a}=\braket{\bar c|X^\dagger| \bar a}=\braket{\bar a|X| \bar c}.
\end{eqnarray}
\normalsize
\item {\bf From the vertex $\ket{0}$}  there is only one outgoing link (to the vertex $\ket{1}$) and one incoming link (from the vertex $\ket{\bar 1}=C\ket{1}$):
\small
\begin{eqnarray}
\delta_{1 b}=\braket{b|X|0}=\braket{0|X|\bar b}=\delta_{\bar1 \bar b}.
\end{eqnarray}
\normalsize
\item {\bf Links have weight 1}. There are no links with multiple lines.
\item {\bf Current conservation law}. At each vertex the number of incoming links is equal to the number of outgoing links. This number is equal to the number of occupied sites in the corresponding Fock state. 

This can be easily seen by noticing that a Fock state with $r$ occupied states can lead (by chiral tunneling of one particle) to $r$ different Fock states. Reversely, such Fock state can be obtained (by chiral tunneling of one particle) from $r$ different Fock states. 

\small
Formally, equality of number of incoming links $\#_i$ and outcoming links $\#_o$ can be shown as a consequence of $X$ being normal:
\begin{eqnarray}
&&\braket{a|XX^\dagger |a}=\braket{a|X^\dagger X |a} \\
&\Longleftrightarrow& \sum_b\braket{a|X|b}\braket{a|X|b}=\sum_b\braket{b|X|a}\braket{b|X |a} \label{CurrentSquare}\\
&\Longleftrightarrow& \#_i=\sum_b\braket{a|X|b}=\sum_b\braket{b|X |a}=\#_o\label{Current},
\end{eqnarray}
where (\ref{Current}) follows from (\ref{CurrentSquare}) since $\braket{a|X|b}=0,1$.
\normalsize
\item {\bf The graph is connected.} Given two arbitrary Fock states, there exists a sequence of consecutive tunneling moves of one particle to the next (to the right) lattice site that connects one Fock state with the other.

\end{itemize}


\newpage

\pagestyle{fancy}
\fancyhf{}
\lhead{The construction}
\lfoot{\thepage}
\vspace*{2cm}
\begin{center}
\includegraphics[width=\textwidth]{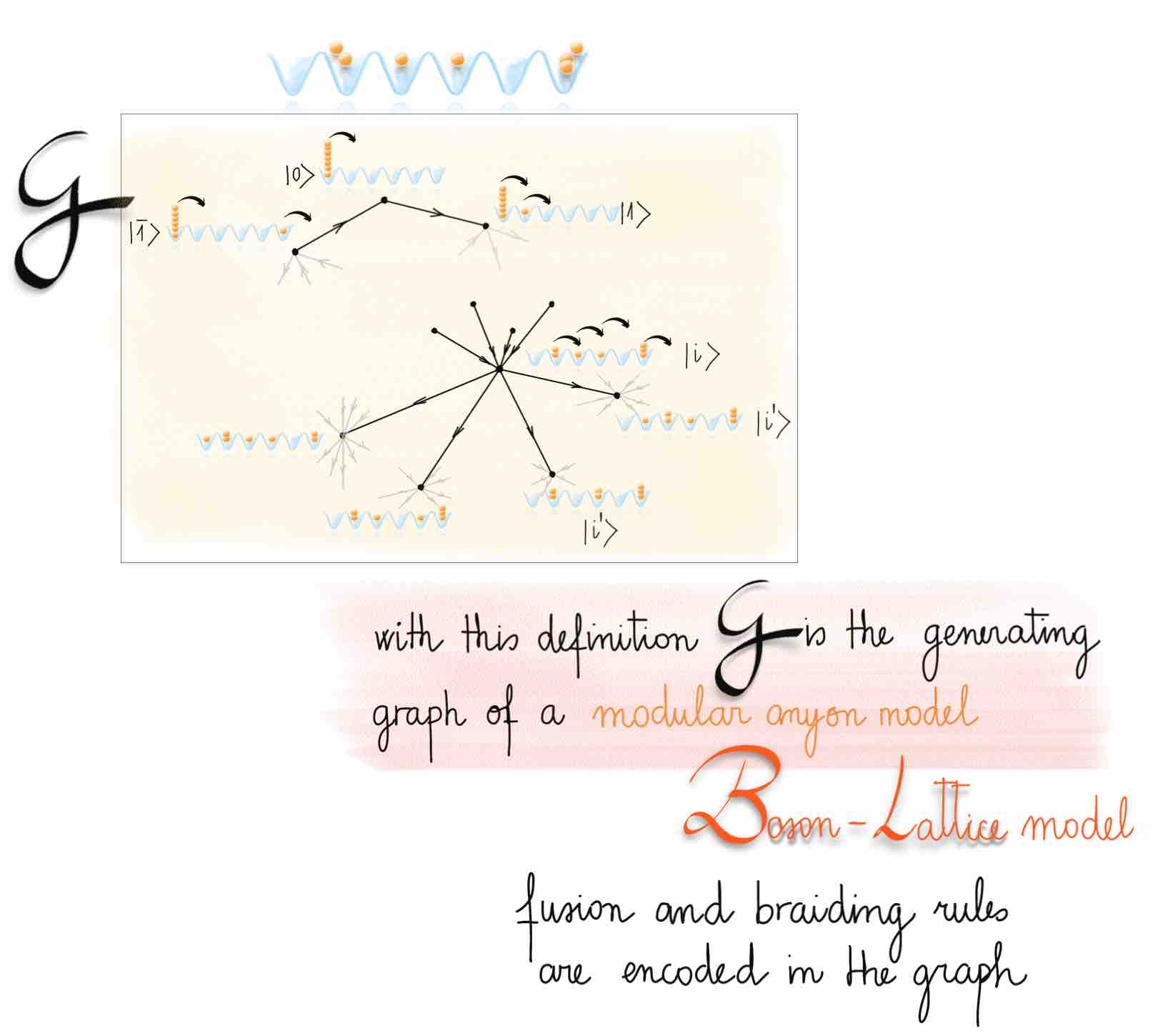}
\end{center}

\newpage
\pagestyle{fancy}
\fancyhf{}
\rhead{Central result}
\rfoot{\thepage}
\vspace*{0cm}
\hspace*{3cm}
\section*{Boson-Lattice central result}
\addcontentsline{toc}{section}{Boson-Lattice central result}

The central result of the Boson-Lattice construction I present here, states that the Boson-Lattice graph I have defined above is the generating graph of a modular anyon model.
The result is formulated as follows:

\vspace*{1cm}
\hspace*{1,5cm}
\parbox{12cm}{ {\bf The Boson-Lattice graph $\mathcal{G}(k,n)$ associated with the Hilbert space $\mathcal{H}(k,n)$ of $k$ bosons in a one-dimensional lattice of $n$ sites is the generating graph of a modular anyon model for any number of bosons $k$ and any number of lattice sites $n$.}
\parskip=8pt

The {\em topological charges} of the Boson-Lattice anyon model are in one to one correspondence with the {\em Fock states} of the Boson-Lattice system.

The {\em fusion rules and braiding rules} of the anyon model are encoded in the graph $\mathcal{G}$.

The graph $\mathcal{G}$ can be completed to a set of graphs which define the {\em topological algebra} of the anyon model.

The {\em $S$-matrix} of the anyon model is obtained from {\em diagonalization} of the topological algebra.}

\vspace{1,2cm}
It is remarkable that a graph defined through connectivity rules between Fock states of a boson lattice system is able to encode an anyon model. Furthermore,  series of tabulated modular anyon models can all be encoded in Boson-Lattice graphs.

In the following I describe a blueprint to obtain the properties of a Boson-Lattice anyon model from the Boson-Lattice graph. A correspondence is established between the elements characterizing the graph and the properties of the anyon model. The graph features guarantee that the obtained fusion and braiding rules are well defined and satisfy the required consistency equations.


\newpage
\pagestyle{fancy}
\fancyhf{}
\lhead{The construction}
\lfoot{\thepage}
\vspace*{2cm}
\begin{center}
\includegraphics[width=0.93\textwidth]{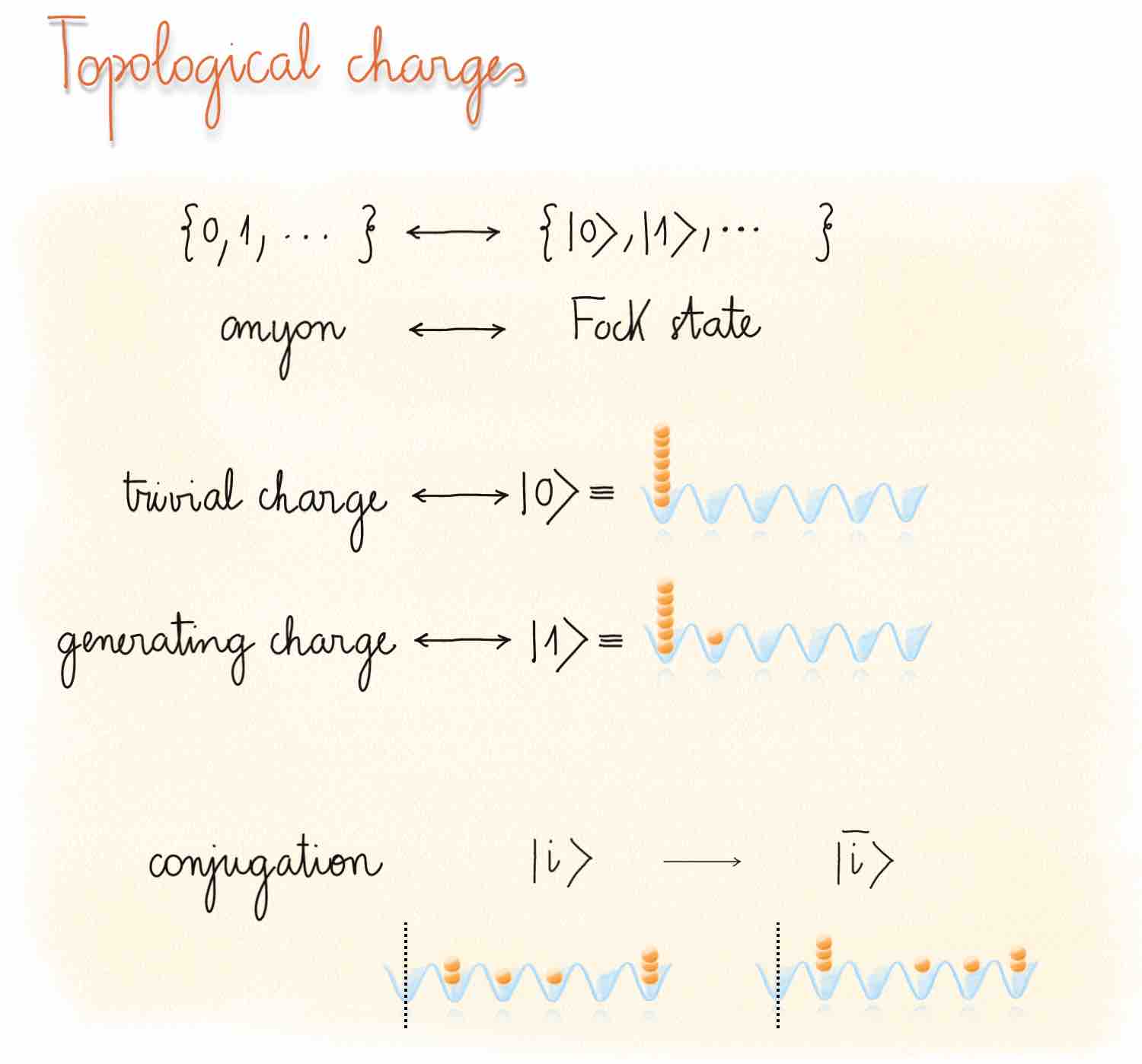}
\end{center}

\newpage
\pagestyle{fancy}
\fancyhf{}
\rhead{Topological charges}
\rfoot{\thepage}
\vspace*{0cm}
\section*{Topological charges}
\addcontentsline{toc}{section}{Topological charges}

\normalsize
The topological charges of the  anyon model associated with the Boson-Lattice system $\mathcal{H}(k,n)$ are in one to one correspondence with the Fock states of the system:
\begin{eqnarray}
\{a,b,\cdots,c\} &\longleftrightarrow &\{\ket{a},\ket{b},\cdots,\ket{c}\}.
\end{eqnarray}

The number of topological charges equals the dimension of $\mathcal{H}(k,n)$.

The trivial charge $0$ is represented by the Fock state $\ket{0}$, with all bosons in the same lattice site.

The charge $1$, represented by the Fock state $\ket{1}$, will be the generating charge of the model.

{\bf Conjugation}. Given a charge $a$, the conjugate charge $\bar a$ corresponds to the conjugate Fock state:
\begin{eqnarray}
\bar a \longleftrightarrow C\ket{a}=\ket{\bar a}.
\end{eqnarray}
Since $C^2=1$, we have that $\bar {\bar a}=a$.

{\bf Translation equivalence relation}. The global translation $T$ defines an {\em equivalence relation} in the set of charges. Two charges are equivalent if the corresponding Fock states are obtained from each other by applying a power of the operator $T$:
\begin{eqnarray}
a \sim b \Longleftrightarrow \ket{b}=T^r \ket{a} \hspace*{0.5cm} \text{for some} \hspace*{0.2cm} r=0,1,\cdots,n-1.
\end{eqnarray}
For each charge $a$ the class of {\em translated charges} is denoted by
\begin{eqnarray}
\{a, ta, \cdots, t^{n-1} a\},
\end{eqnarray}
in one to one correspondence with the set of translated Fock states
\begin{eqnarray}
\{\ket{a}, T\ket{a},  \cdots, T^{n-1}\ket{a}\}.
\end{eqnarray}
The class of the trivial charge will be denoted by:
\begin{eqnarray}
\{0, t, \cdots, t^{n-1} \}.
\end{eqnarray}


\newpage
\pagestyle{fancy}
\fancyhf{}
\lhead{The construction}
\lfoot{\thepage}
\vspace*{0cm}
\subsection*{Fusion rules of charge $1$}
\addcontentsline{toc}{section}{Fusion rules of charge $1$}

The Boson-Lattice graph $\mathcal{G}(k,n)$ is the topological graph corresponding to charge $1$:
\begin{eqnarray}
1\leftrightarrow\mathcal{G}\leftrightarrow X.
\end{eqnarray}
The fusion rules of charge $1$ can be therefore read from the connectivity pattern of $\mathcal{G}$, or, equivalently, from the matrix elements of the operator $X$:
\begin{eqnarray}
N_{1a}^b=\braket{b\,|X|\,a}.
\end{eqnarray}

The conjugate charge $\bar 1$ is assigned the topological graph $\mathcal{G}^*$, which corresponds to the operator $X^\dagger$:
\begin{eqnarray}
\bar1\leftrightarrow\mathcal{G}^*\leftrightarrow X^\dagger.
\end{eqnarray}
Its fusion rules are:
\begin{eqnarray}
N_{\bar 1 a}^b=\braket{b\,|X^\dagger|\,a}=\braket{a\,|X|\,b}.
\end{eqnarray}

The special properties of the graph assure that these fusion rules are well defined, since they fulfill:
\begin{itemize}
\item $N_{1 0}^a=\delta_{1a}$\\
since we have $X\ket{0}=\ket{1}$ and therefore $N_{1 0}^a=\braket{a\,|X|\,0}=\delta_{1a}$.
\item $N_{1 a}^0=\delta_{a \bar 1}$\\
since we have $N_{1 a}^0=\braket{0\,|X|\,a}=\braket{\bar a\,|X|\,0}=\braket{\bar 1|a}=\delta_{a \bar 1}$.
\item $N_{1 a}^b=N_{\bar 1 \bar a}^{\bar b}=N_{1 \bar b}^{\bar a}$\\
since we have $\braket{b\,|X|\,a}=\braket{\bar b\,|X^\dagger|\,\bar a}=\braket{\bar a\,|X|\,\bar b}$.
\end{itemize}

\newpage
\pagestyle{fancy}
\fancyhf{}
\rhead{Fusion rules}
\rfoot{\thepage}
\vspace*{0cm}
\subsection*{Fusion rules of charge $t$}
\addcontentsline{toc}{section}{Fusion rules of charge $t$}

The charge $t^r$ is assigned the operator $T^r$. 
Since we have that 
$N_{t^r a}^b=\braket{b\,|T^r|\,a}=\delta_{b,t^ra}$,  the charge $t^r$ is thus an Abelian charge with fusion rules:
\begin{eqnarray} 
t^r\times t^s&=&t^{r+s\,(n)}\nonumber\\
a\times t^r&=&at^r.
\end{eqnarray}

\vspace*{1cm}
\subsection*{The nucleus of the topological algebra}
\addcontentsline{toc}{section}{The nucleus of the topological algebra}

The set of operators 
\begin{eqnarray} 
\{X,X^\dagger,T\}
\end{eqnarray}
constitute the nucleus of the topological algebra of the anyon model.
They are normal, they commute with each other, and their fusion rules are well defined. As we see below, they can be completed to a topological algebra.


\newpage
\pagestyle{fancy}
\fancyhf{}
\lhead{The construction}
\lfoot{\thepage}
\vspace*{0cm}
\section*{X can be completed to a topological algebra}
\addcontentsline{toc}{section}{X can be completed to a topological algebra}

The operator $X$ as defined above can be completed to a topological algebra of operators:
\begin{eqnarray}
\mathcal{A}=\{\mathds{1},X_1,X_2,\cdots\},
\end{eqnarray}
where the operator $X_a$ is associated with the charge $a$, and $X_1\equiv X$.
The fusion rules of the model are then given by
\begin{eqnarray}
N_{a b}^c=\braket{c|X_a|b}.
\end{eqnarray}

The following result shows how to complete the topological algebra:
\hspace*{0,5cm}\begin{mdframed}
{\bf The algebra of polynomials}. 
For each charge $a$ there exists a unique operator $X_a$ of the form:
\begin{eqnarray}
X_a=\text{\large p}_a(X,X^\dagger,T),
\end{eqnarray}
where $\text{\large p}_a$ is a polynomial of integer coefficients of the operators $X,X^\dagger$ and $T$ satisfying:
\begin{eqnarray}
&&X_a\ket{0}=\ket{a},\\
&&\braket{c|X_a|b}=0,1,2,\cdots.
\end{eqnarray}
This set of polynomials defines the topological algebra of the anyon model.
\end{mdframed}

The existence of the algebra of polynomials follows from the connectivity properties of the Boson-Lattice graph. Since the graph is connected, every state $\ket{a}$ can be reached from the state $\ket{0}$ by consecutive application of the operator $X$. Equivalently, a combination of consecutive applications of the operators $X$, $X^\dagger$ and $T$ connects any state $\ket{a}$ with the state $\ket{0}$. Thus there is always a polynomial $X_a$ of integer coefficients of these operators that fulfills $X_a\ket{0}=\ket{a}$.

In the following section I will explicitly find these polynomials for a series of examples of Boson-Lattice models. 

\newpage
\pagestyle{fancy}
\fancyhf{}
\rhead{Fusion rules}
\rfoot{\thepage}
\thispagestyle{empty}

{\em The polynomials are unique.} Once we have found an algebra of polynomials satisfying the conditions above, we can be sure that no other exists. To prove that, we consider two different sets of polynomials $\{X_a\}$ and $\{X_a^\prime\}$. Since the  operators $X$,$X^\dagger$, and $T$ commute with each other, we have that $[X^{\phantom{\prime}}_a,X^{\phantom{\prime}}_b]=[X^\prime_a,X_b]=[X^\prime_a,X^\prime_b]=0$. Therefore:
\begin{eqnarray}
X_a^{\phantom{\prime}}\ket{b}=X_a^{\phantom{\prime}}X_b^{\phantom{\prime}}\ket{0}=X_b^{\phantom{\prime}}X_a^{\phantom{\prime}}\ket{0}=X_b^{\phantom{\prime}}X_a^\prime\ket{0}=X_a^\prime\ket{b},
\end{eqnarray}
and thus $X_a=X_a^\prime$.

{\em The polynomials define a topological algebra.}  By definition, the polynomials fulfill the necessary and sufficient conditions given in the first section for a topological algebra. The only non-trivial property we need to prove is that: 
\begin{itemize}
\item For each $X_a$  there exists $X_{\bar a}$ such that $X_{\bar a}=X_a^\dagger$.
\end{itemize}
This is shown by noticing that:
$X_a^\dagger=p_a^\dagger=Cp_a^{\phantom{\dagger}}C$,
and therefore
\begin{eqnarray}
X_a^\dagger\ket{0}=CX_a^{\phantom{\dagger}}C\ket{0}=\ket{\bar a},
\end{eqnarray}
so that $X_{\bar a}^{\phantom{\dagger}}=X_a^\dagger$.

\includegraphics[width=\textwidth]{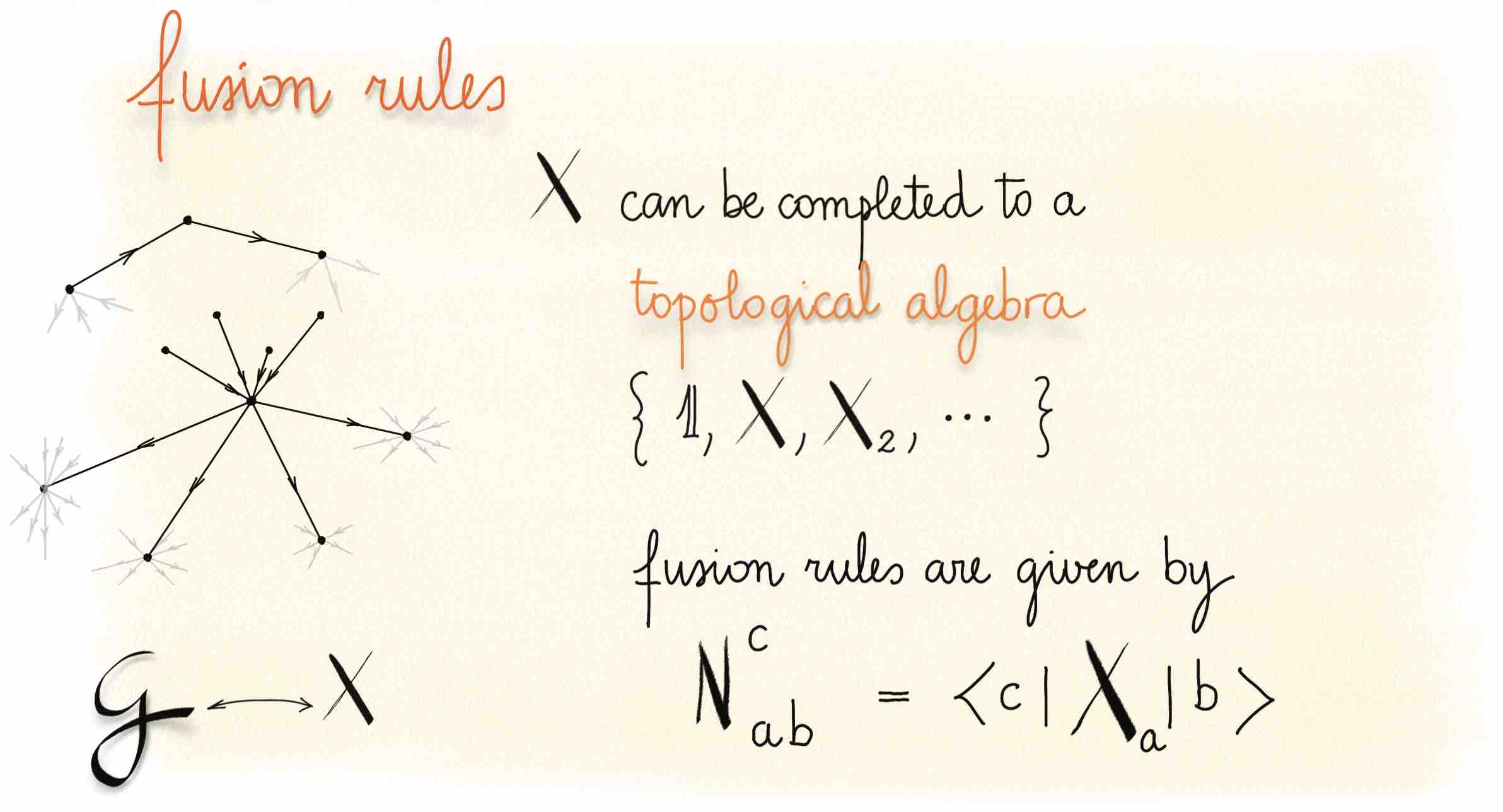}

\newpage
\pagestyle{fancy}
\fancyhf{}
\lhead{The construction}
\lfoot{\thepage}

\vspace*{2cm}
\begin{center}
\includegraphics[width=0.9\textwidth]{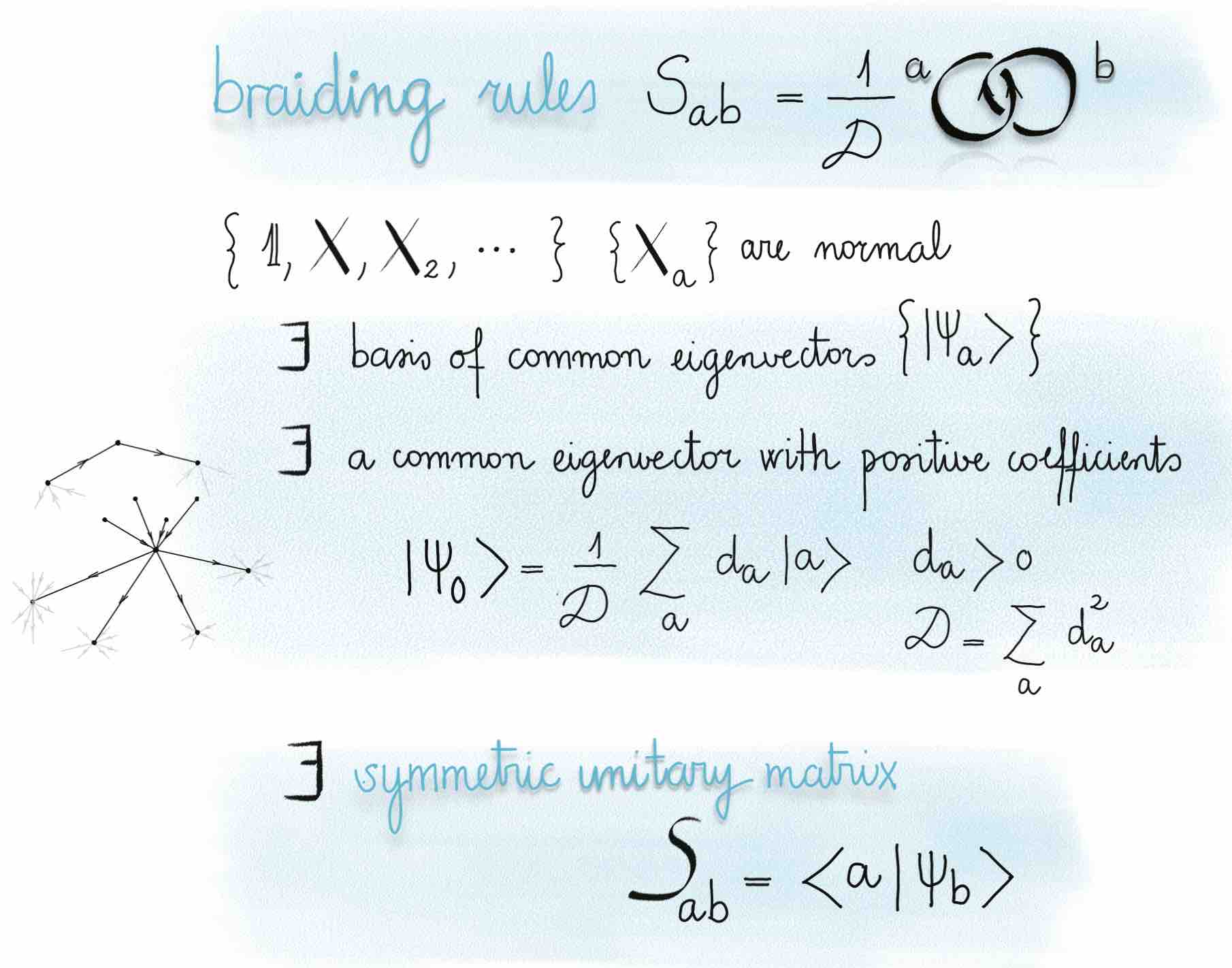}
\end{center}

\newpage
\pagestyle{fancy}
\fancyhf{}
\rhead{Braiding}
\rfoot{\thepage}
\vspace*{0.5cm}
\section*{S and T matrices of the Boson-Lattice model}
\addcontentsline{toc}{section}{S and T matrices of the Boson-Lattice model}
The braiding rules of the Boson-Lattice model are encoded in the Boson-Lattice graph $\mathcal{G}$. The special properties of this graph  guarantee that the braiding rules are well defined and correspond to those of a modular anyon model.

{\bf Quantum dimensions}.
The Boson-Lattice graph $\mathcal{G}$ is a strongly connected graph. Therefore the corresponding operator $X$ is a non-negative irreducible matrix. The Perron-Frobenius theorem states that the operator $X$ has a real eigenvalue $\lambda_0$ (largest in absolute value) with a corresponding eigenvector $\ket{\psi_0}$ whose components are all positive. Without loss of generality we can write this state as:
\begin{eqnarray}
\ket{\psi_0}=\frac{1}{\mathcal{D}}\sum_ad_a\ket{a},
\end{eqnarray}
where $d_a>0$, $d_0=1$ and $\mathcal{D}=\sqrt{\sum_a d_a^2}$. The components of this vector define the quantum dimensions of the anyon model. The topological charge $a$ has quantum dimension $d_a$ and the anyon model has quantum dimension $\mathcal{D}$.

{\bf The S-matrix}. The operators in the Boson-Lattice topological algebra are normal operators. Since the algebra is Abelian, it follows that there exists an orthonormal basis of common eigenvectors $\{\ket{\psi_a}\}$. The characteristic features of the Boson-Lattice algebra assure that these eigenvectors can be chosen such that the unitary matrix $S$:
\begin{eqnarray}
S_{ab}=\braket{a|\psi_b}
\end{eqnarray}
is a {\em symmetric} matrix. Moreover, a Boson-Lattice algebra is such that there is essentially a unique way of choosing this unitary matrix as a symmetric 
one\footnote{Different symmetric choices of a unitary matrix correspond to anyon models that are trivially related, for example, they can be mirror image models under parity ($S^*=CSC$).}. This matrix defines the S-matrix of the anyon model.

\newpage
\pagestyle{fancy}
\fancyhf{}
\lhead{The construction}
\lfoot{\thepage}

{\bf The T-matrix}. The special properties of the Boson-Lattice algebra also guarantee that the $S$-matrix defined above satisfies the following property. It can be written as:
\begin{eqnarray}
(S\mathcal{T})^3=\Theta S^2,
\label{ModularSTEquation}
\end{eqnarray}
where
\begin{eqnarray}
\mathcal{T}_{ab}=\theta_{a}\delta_{ab}
\end{eqnarray}
and
\begin{eqnarray}
\Theta=\frac{1}{\mathcal{D}}\sum_a d_a^2 \theta_a=e^{i2\pi c/8}.
\end{eqnarray}

The diagonal matrix $\mathcal{T}$ defines the $T$-matrix of the Boson-Lattice model 
\footnote{It can be shown that for a modular anyon model equation (\ref{ModularSTEquation}) is equivalent to equation (\ref{S-T-Relation}). Details of proof will be given elsewhere.}. 
The diagonal elements $\theta_a$ define the topological spins of the charges, and the constant $c$ is the central charge of the modular anyon model.

\vspace{1cm}
\begin{center}
\includegraphics[width=0.9\textwidth]{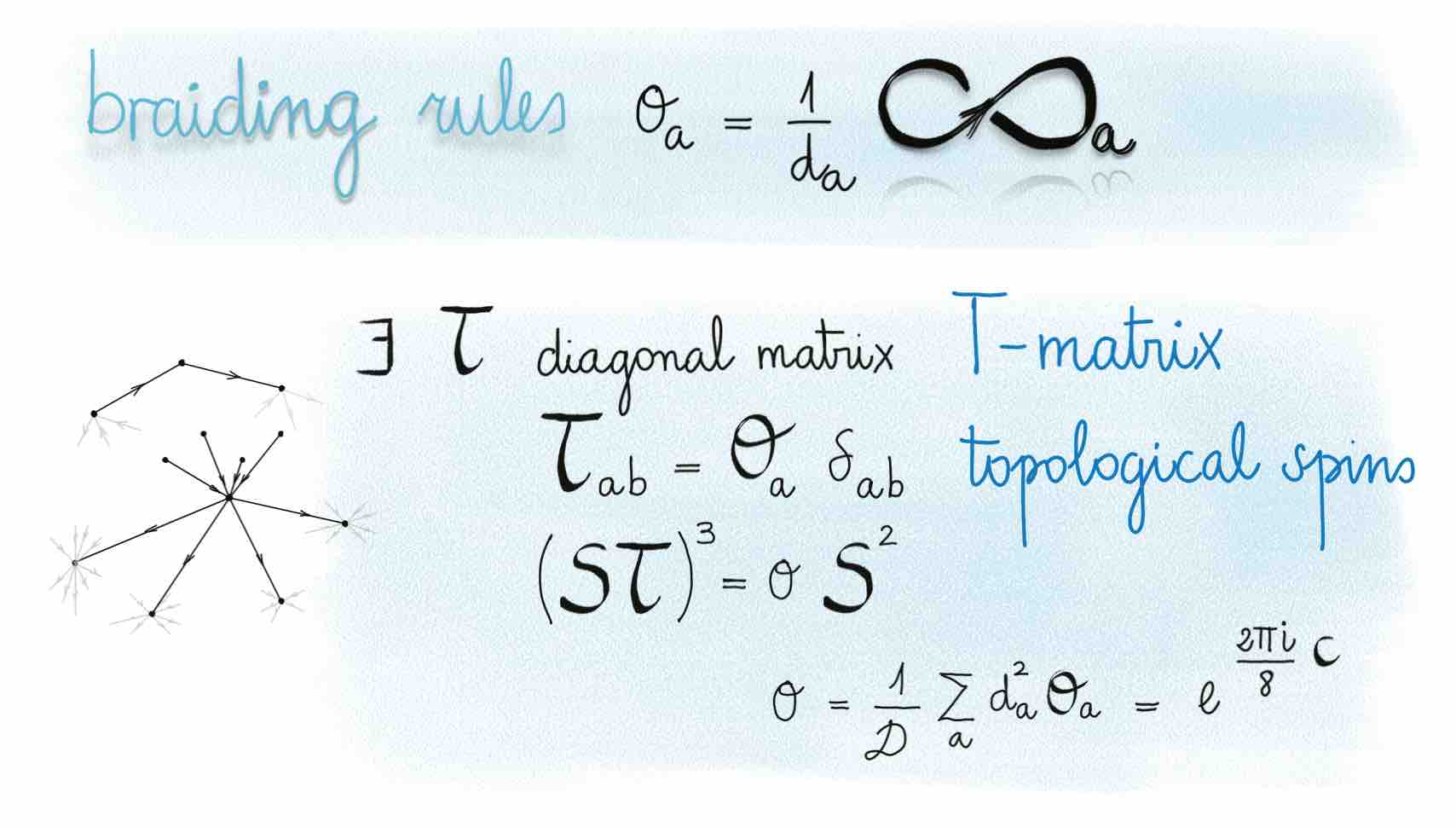}
\end{center}

\newpage
\pagestyle{fancy}
\fancyhf{}
\rhead{Braiding}
\rfoot{\thepage}
\section*{The special properties of Boson-Lattice graphs}
\addcontentsline{toc}{section}{The set of Boson-Lattice models}

Given an arbitrary graph, the set of conditions it has to fulfill to be the generating graph of a modular anyon model is highly demanding.
First, the graph has to be the generating graph of a topological algebra. Second, it has to admit a symmetric eigenbasis of eigenvectors. Finally, such eigenbasis has to fulfill the non-trivial condition given in equation (\ref{S-T-Relation}).

This array of conditions is so restrictive that it seems clear that a randomly chosen graph for a Boson-Lattice system has low chances to represent a modular anyon model. Moreover, there is in principle no reason to think that such fortunate graph could even exist for a Boson-Lattice system.

The Boson-Lattice graph I have defined succeeds in generating well defined modular anyon models for any number of bosons and lattice sites. The special connectivity properties of the graph make it possible to fulfill the non-trivial constellation of conditions that guarantee the existence of a modular anyon model.

Not less surprising is the fact that, as I show in the next section, series of known tabulated anyon models can all be encoded into Boson-Lattice graphs.

\begin{center}
\includegraphics[width=0.8\textwidth]{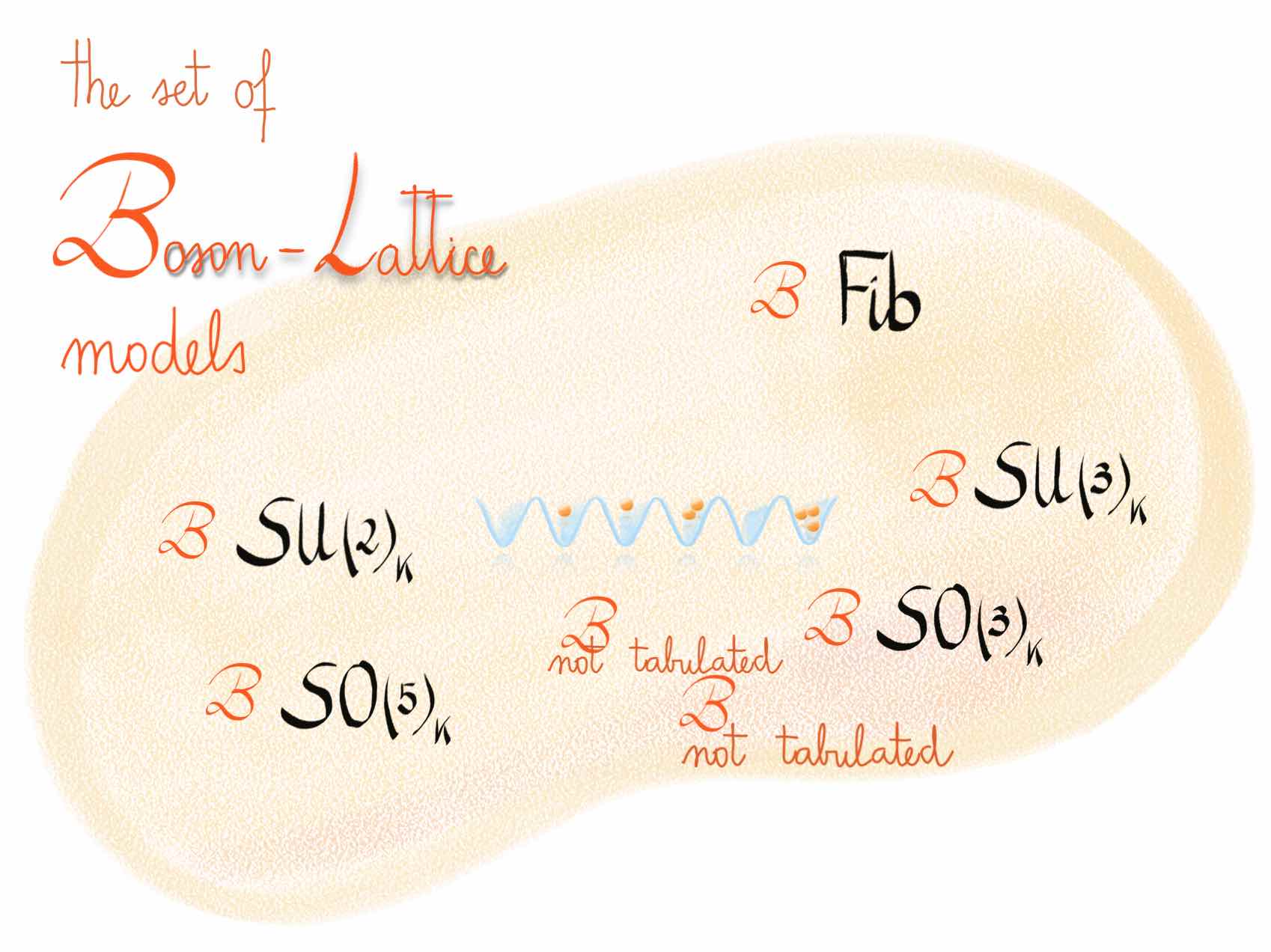}
\end{center}

\newpage
\thispagestyle{empty}
\vspace*{2cm}
\begin{center}
\includegraphics[width=\textwidth]{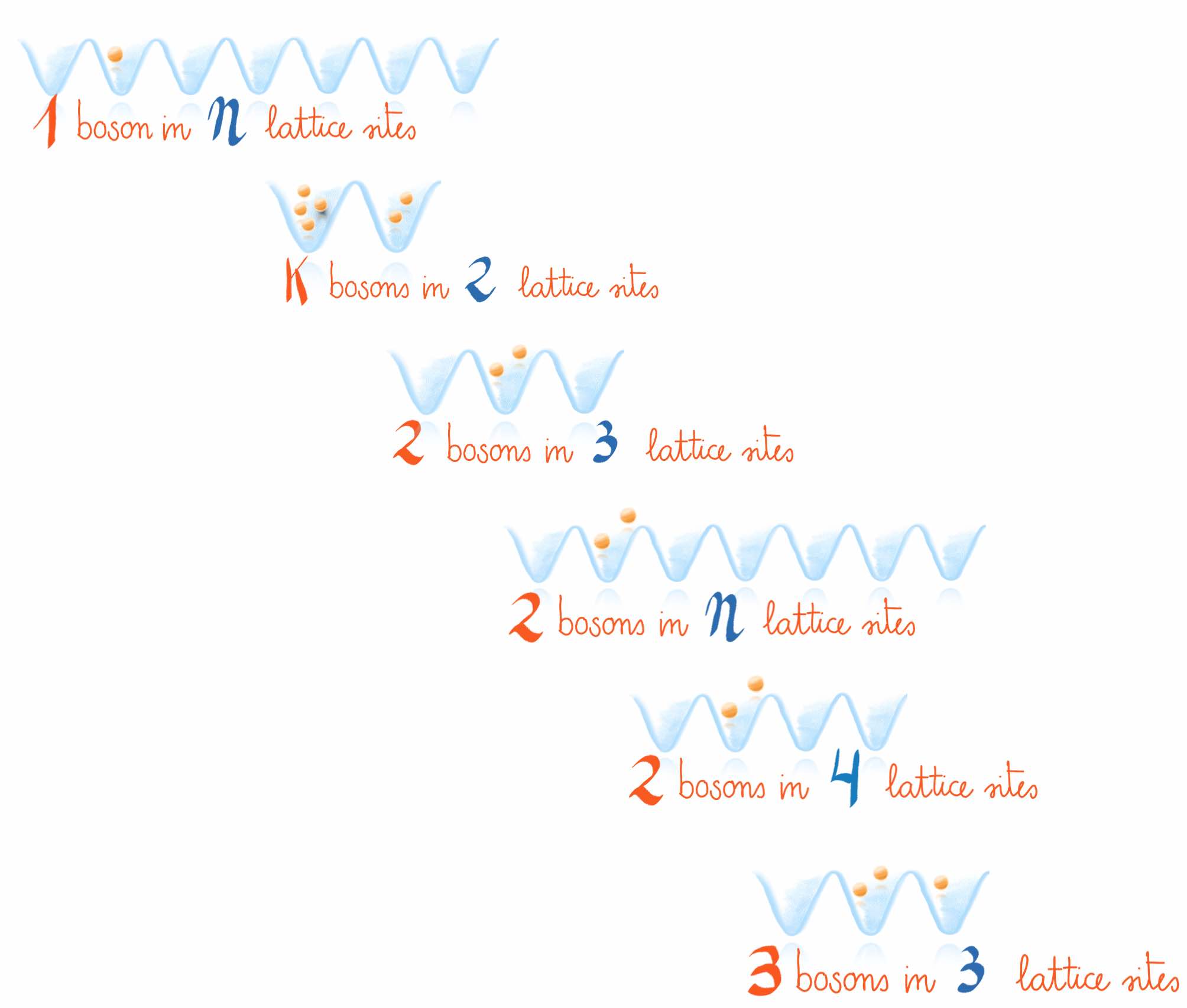}
\end{center}

\newpage
\thispagestyle{empty}
\vspace*{0.5cm}
\section*{}
\addcontentsline{toc}{section}{Examples}
\hspace*{0.5cm}
\includegraphics[width=0.75\textwidth]{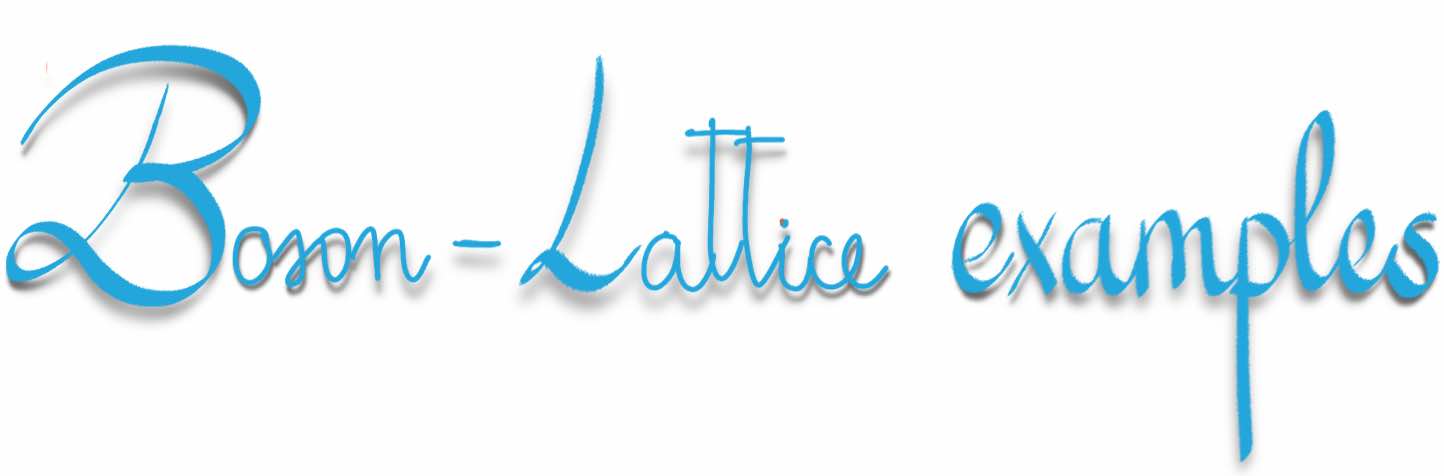}

\large
\vspace*{0cm}
\hspace*{2cm}
\parbox{11cm}{To see the Boson-Lattice construction at work I consider several examples of Boson-Lattice anyon models constructed with the formalism introduced above. 
\parskip=8pt

Given a system with $k$ bosons in a lattice with $n$ sites, I analyze the corresponding bosonic Hilbert space.
I identify the Boson-lattice graph and show that the corresponding operator $X$ can be completed to a topological algebra. This algebra encodes the fusion rules of the model.
{\em Diagonalization} of the topological algebra allows us to derive the braiding properties of the Boson-Lattice anyon model.

The set of fusion and brading rules obtained with the Boson-Lattice formalism define a well-defined anyon model. In some cases, the constructed Boson-Lattice models correspond to tabulated models, such as $\mathbf{SU}(2)_k$ or $\mathbf{SO}(3)_k$. Interestingly, we will see how the construction also yields other well-defined anyon models that are not tabulated.}


\newpage
\thispagestyle{empty}
\vspace*{6.5cm}
\hspace*{1cm}
\includegraphics[width=0.85\textwidth]{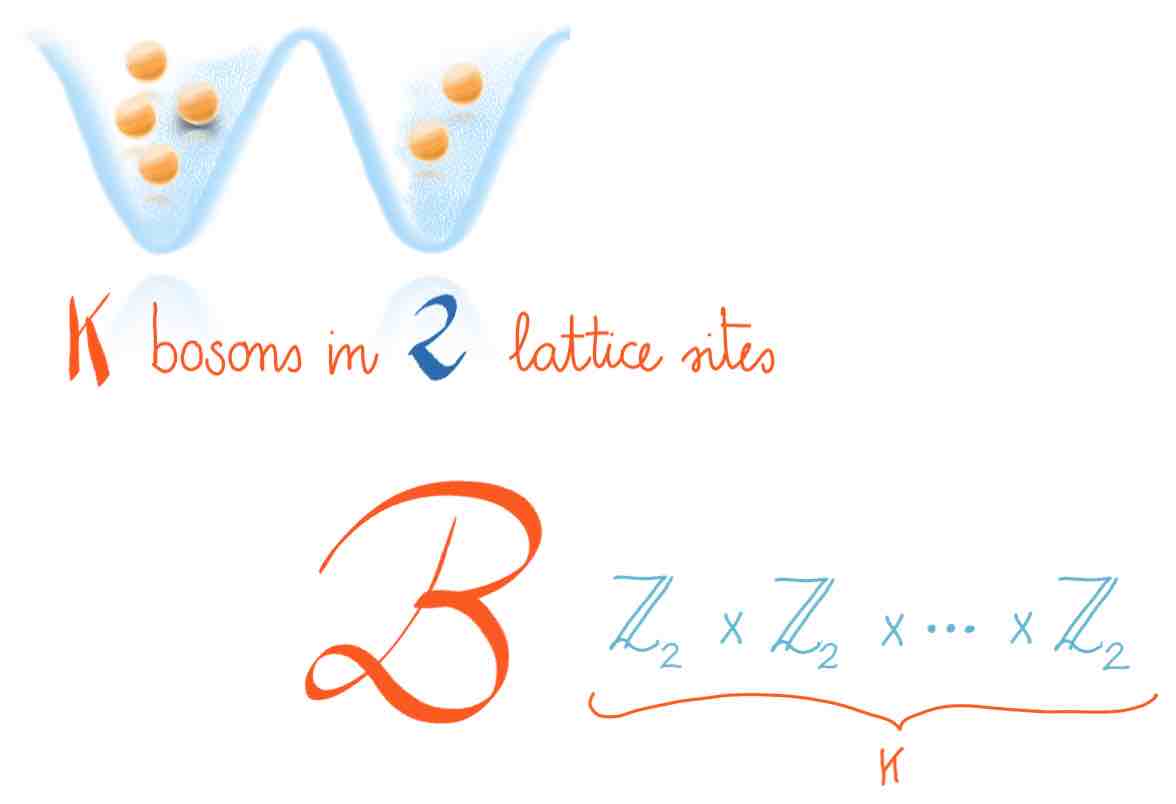}

\newpage
\pagestyle{fancy}
\fancyhf{}
\rhead{Boson-Lattice examples}
\rfoot{\thepage}
\section*{}
\addcontentsline{toc}{section}{$k$ bosons in $2$ lattice sites}
\vspace*{3cm}
\hspace*{1cm}
\includegraphics[width=0.7\textwidth]{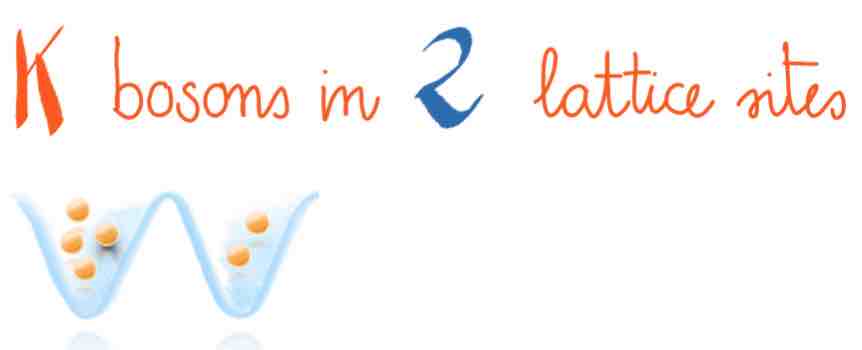}

\vspace{1cm}
\hspace*{2cm}
\parbox{11cm}{ 
\parskip=8pt
I analyze the anyon model corresponding to $k$ bosons in a lattice of $2$ sites.

I identify the Boson-Lattice generating graph, construct the topological algebra and characterize the fusion and braiding rules of the anyon model.
I show that it corresponds to the modular anyon model $\mathbf{SU}(2)_k$.

It is illuminating to see how in the language of the Boson-Lattice construction, the $\mathbf{SU}(2)_k$ anyon model corresponds to a particle in a lattice of $k+1$ sites with open boundary conditions.
The elements of the $S$-matrix of the anyon model acquire a physical interpretation as the eigenfunctions of such a particle.
}


\newpage
\pagestyle{fancy}
\fancyhf{}
\lhead{Boson-Lattice examples}
\lfoot{\thepage}
\vspace*{2.5cm}
\begin{center}
\includegraphics[width=\textwidth]{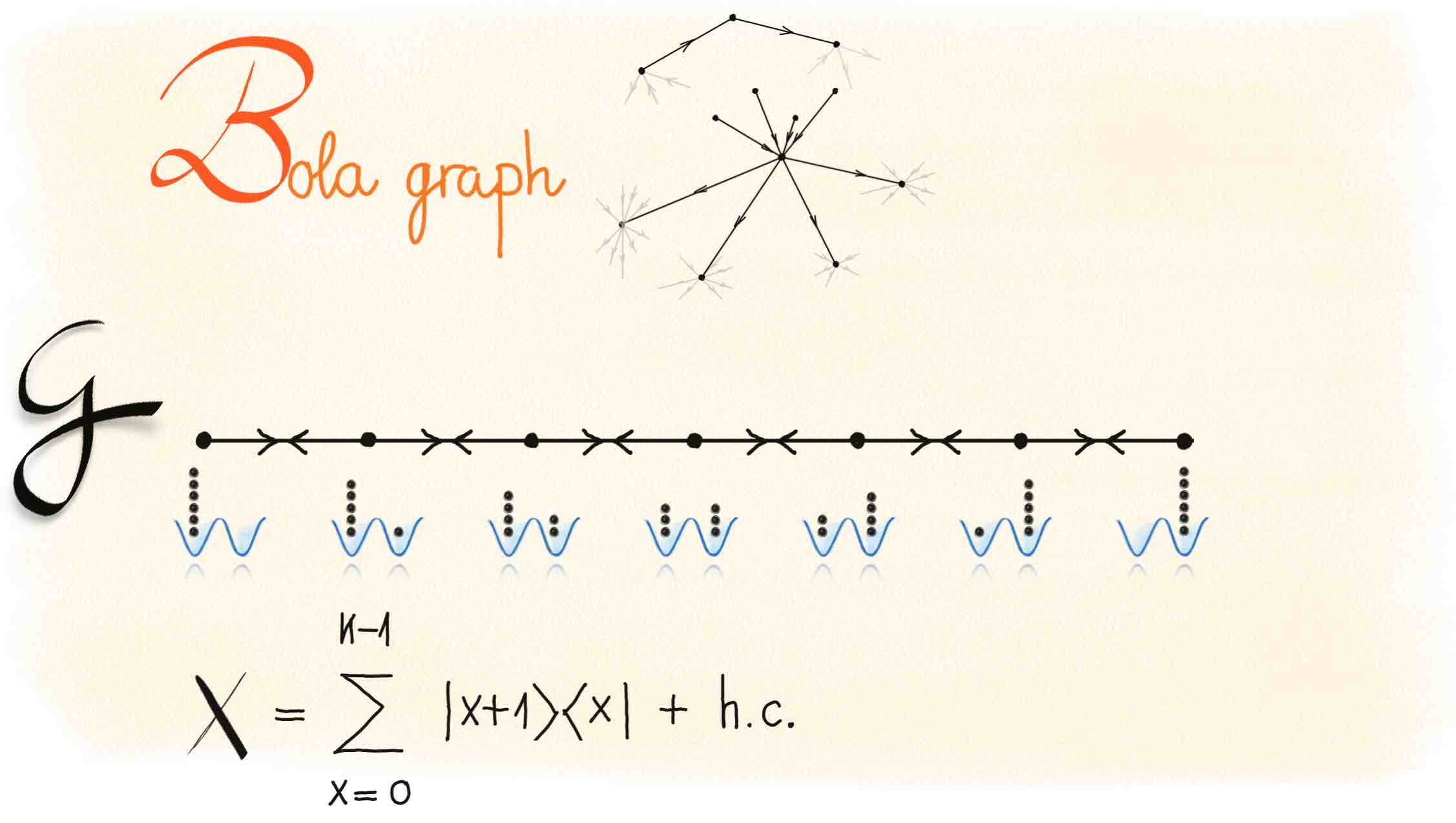}
\end{center}

\normalsize
\newpage
\pagestyle{fancy}
\fancyhf{}
\rhead{$k$ bosons in $2$ sites}
\rfoot{\thepage}
\vspace*{1cm}
\section*{Boson-Lattice graph}
\addcontentsline{toc}{section}{Boson-Lattice graph}

The Hilbert space of $k$ bosons in $2$ lattice sites has dimension $k+1$. The Fock states can be labelled by:
\begin{eqnarray}
\ket{x}=\frac{(a^\dagger_0)^{k-x}}{\sqrt{(k-x)!}}\frac{(a^\dagger_1)^{x}}{\sqrt{x!}}\ket{\text{vac}}, \quad x=0,\cdots,k,
\end{eqnarray}
where $a^\dagger_0$ $(a^\dagger_1)$ creates a particle in the $0$ ($1$) lattice site, and $\ket{x}$ denotes the Fock state with $k-x$ bosons in the $0$ site and $x$ bosons in the $1$ site.

The corresponding anyon model has $k+1$ topological charges, in one to one correspondence with the Fock states. We label them by:
\begin{eqnarray}
\{0,1,\cdots, k\} \quad \longleftrightarrow \quad \{\ket{0},\ket{1},\cdots, \ket{k}\}.
\end{eqnarray}

Following the prescription of the construction, the Boson-Lattice graph corresponds to a one-dimensional lattice of $k+1$ sites in which each vertex is connected to its two next neighbors. The two ending vertices are not connected to each other.

The operator $X$ corresponding to the Boson-Lattice generating graph can be written as:
\begin{eqnarray}
X=\sum_{x=0}^{k-1}\ket{x+1}\bra{x} + \text{h.c.}
\end{eqnarray}
This is a hermitian operator that corresponds to the (real) {\bf tunneling operator of one particle in a one-dimensional  lattice of $k+1$ sites with open boundary conditions}.

\newpage
\pagestyle{fancy}
\fancyhf{}
\lhead{Boson-Lattice examples}
\lfoot{\thepage}

\vspace*{1cm}
\begin{center}
\includegraphics[width=0.7\textwidth]{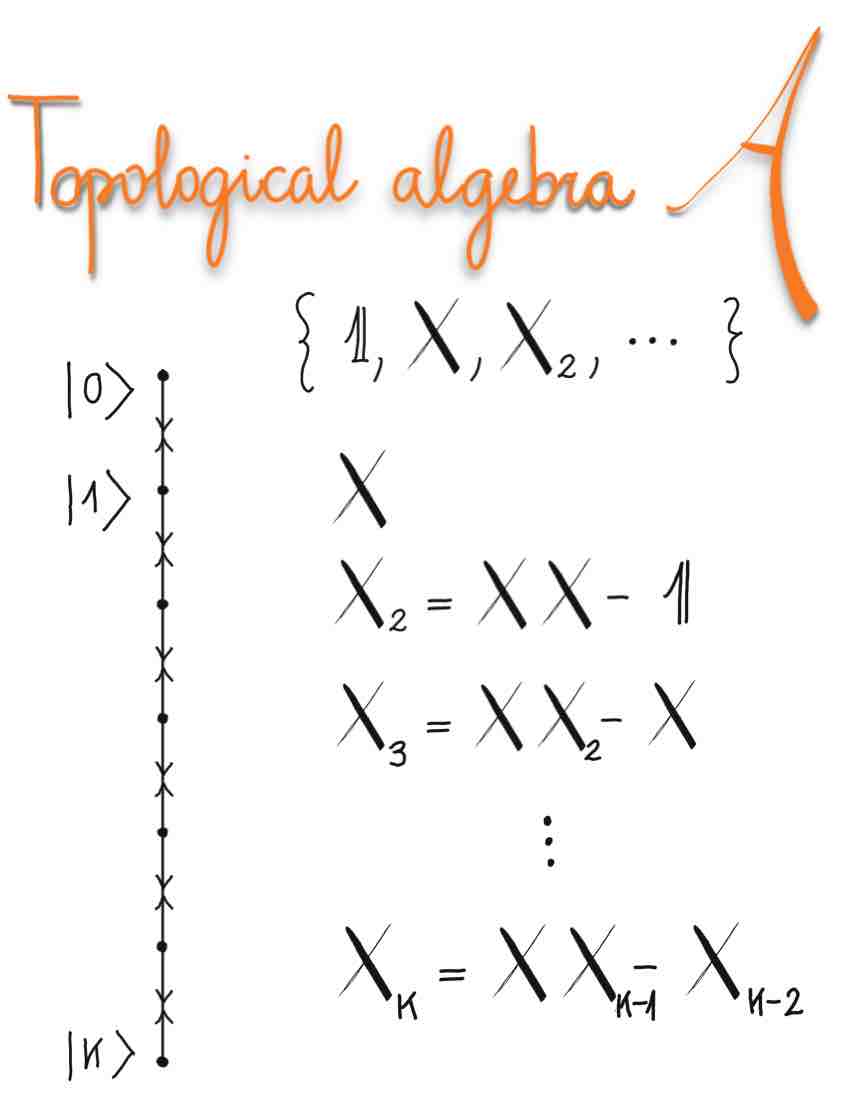}
\end{center}

\newpage
\pagestyle{fancy}
\fancyhf{}
\rhead{$k$ bosons in $2$ sites}
\rfoot{\thepage}

\vspace*{1.5cm}
\section*{Topological algebra}
\addcontentsline{toc}{section}{Topological algebra}

The generating operator $X$ can be completed to a topological algebra. To show this we search for the algebra of polynomials of $X$,
\begin{eqnarray}
\mathcal{A}=\{\mathds{1},\text{p}_1[X],\cdots,\text{p}_k[X]\},
\end{eqnarray} 
that satisfy 
\begin{eqnarray}
\text{p}_\ell[X]\ket{0}=\ket{\ell}, \quad \ell=0,\cdots,k.
\end{eqnarray}

By inspection of the generating graph $\mathcal{G}$ it is straightforward to see that the polynomials are obtained by the recursive relation:
\begin{eqnarray}
\text{p}_{\ell+1}[X]=X\text{p}_\ell[X]-\text{p}_{\ell-1}[X],
\end{eqnarray}
with $\text{p}_0[X]=\mathds{1}$ and $\text{p}_1[X]=X$. 

Explicitly, we obtain:
\begin{eqnarray}
\text{p}_0=\mathds{1},\,\,\,\text{p}_1=X, \,\,\,\text{p}_2=X^2-1, \,\,\,\text{p}_3=X^3-2X,\,\,\,\cdots
\end{eqnarray} 

Additionally, the following {\em boundary} equation is fulfilled:
\begin{eqnarray}
X\text{p}_k[X] -\text{p}_{k-1}[X]=0.
\end{eqnarray}

The polynomials above define by construction a topological algebra. To explicitly show that they correspond to non-negative matrices, we draw the corresponding topological graphs.

\newpage
\pagestyle{fancy}
\fancyhf{}
\lhead{Boson-Lattice examples}
\lfoot{\thepage}
\vspace*{1cm}
\begin{center}
\includegraphics[width=\textwidth]{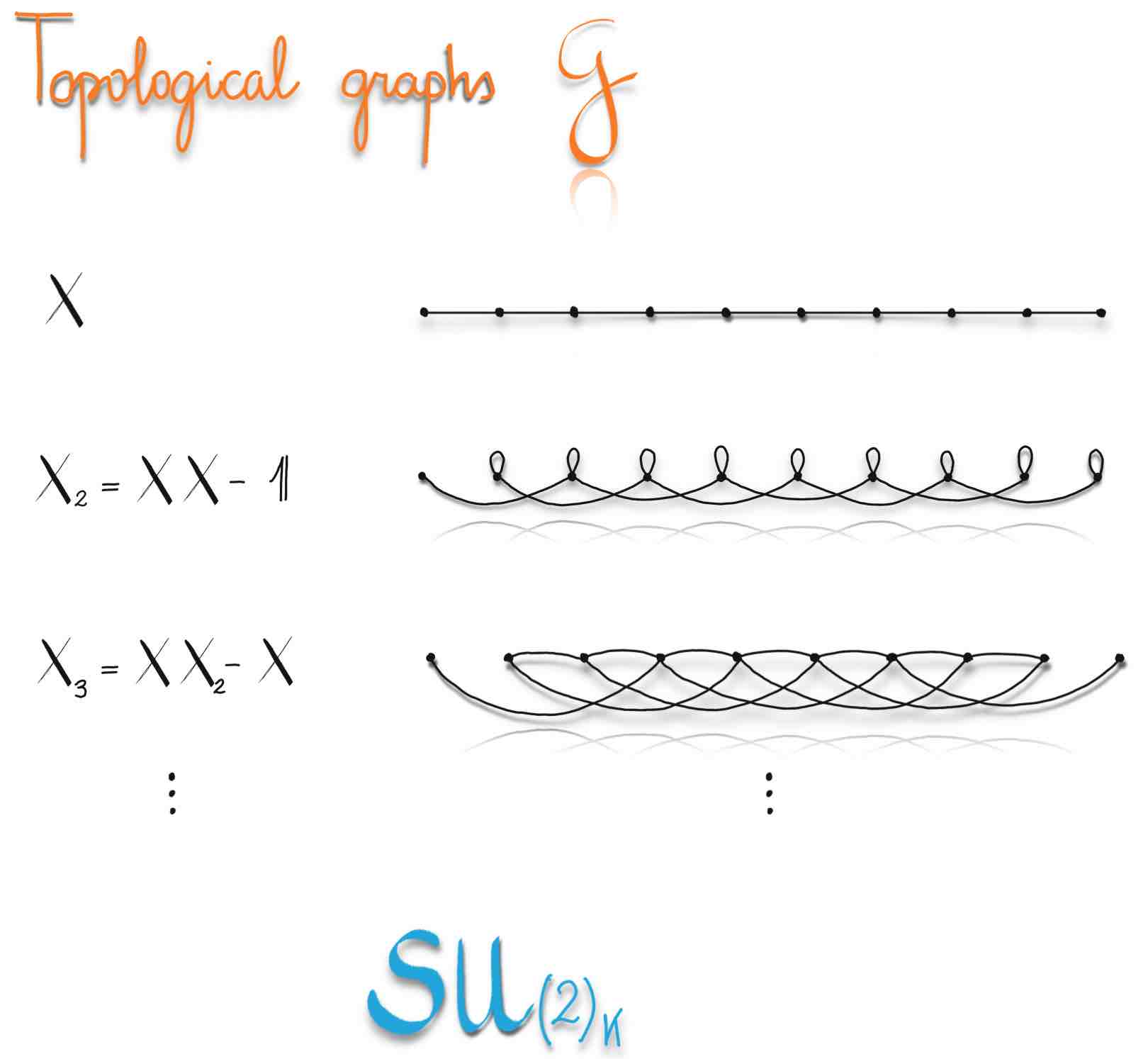}
\end{center}

\newpage
\pagestyle{fancy}
\fancyhf{}
\rhead{$k$ bosons in $2$ sites}
\rfoot{\thepage}
\vspace*{1cm}
\section*{Topological graphs and Fusion rules}
\addcontentsline{toc}{section}{Topological graphs and Fusion rules}

From the Boson-Lattice generating graph we can directly read the fusion rules of the generating charge $1$. These are:

\begin{center}
\begin{tabular}{llll}
& & Fusion rules &\\
& & &\\
$X\ket{0}=\ket{1}$ & &$1 \times 0=1$ &\\
$X\ket{x}=\ket{x-1}+\ket{x+1}$ &$\longleftrightarrow$& $1 \times x=(x+1)+(x-1)$& $1<x<k$\\
$X\ket{k}=\ket{k-1}$ & &$1 \times k=k-1$&
\end{tabular}
\end{center}

Interestingly, they exactly correspond to those of the topological charge $\frac{1}{2}$ in the anyon model 
$\mathbf{SU}(2)_k$ \footnote{Note that the charges of the anyon model $\mathbf{SU}(2)_k$ are usually labelled by $\{0,\frac{1}{2},1,\cdots,\frac{k}{2}\}$.
Here, they are labelled by $\{0,1,2,\cdots,k\}$.}.

By composing the generating graph with itself we can generate the topological graphs depicted in the figure.  
The fusion rules of the anyon model are directly obtained from these graphs. They are:

\begin{eqnarray}
x_1\times x_2=\sum_{x=|x_1-x_2|}^m x,
\end{eqnarray} 
where $m=\text{min}\{x_1+x_2,2k-x_1-x_2\}$.

They exactly correspond to the fusion rules of the anyon model $\mathbf{SU}(2)_k$.

\newpage
\pagestyle{fancy}
\fancyhf{}
\lhead{Boson-Lattice examples}
\lfoot{\thepage}
\vspace*{1cm}
\begin{center}
\includegraphics[width=0.85\textwidth]{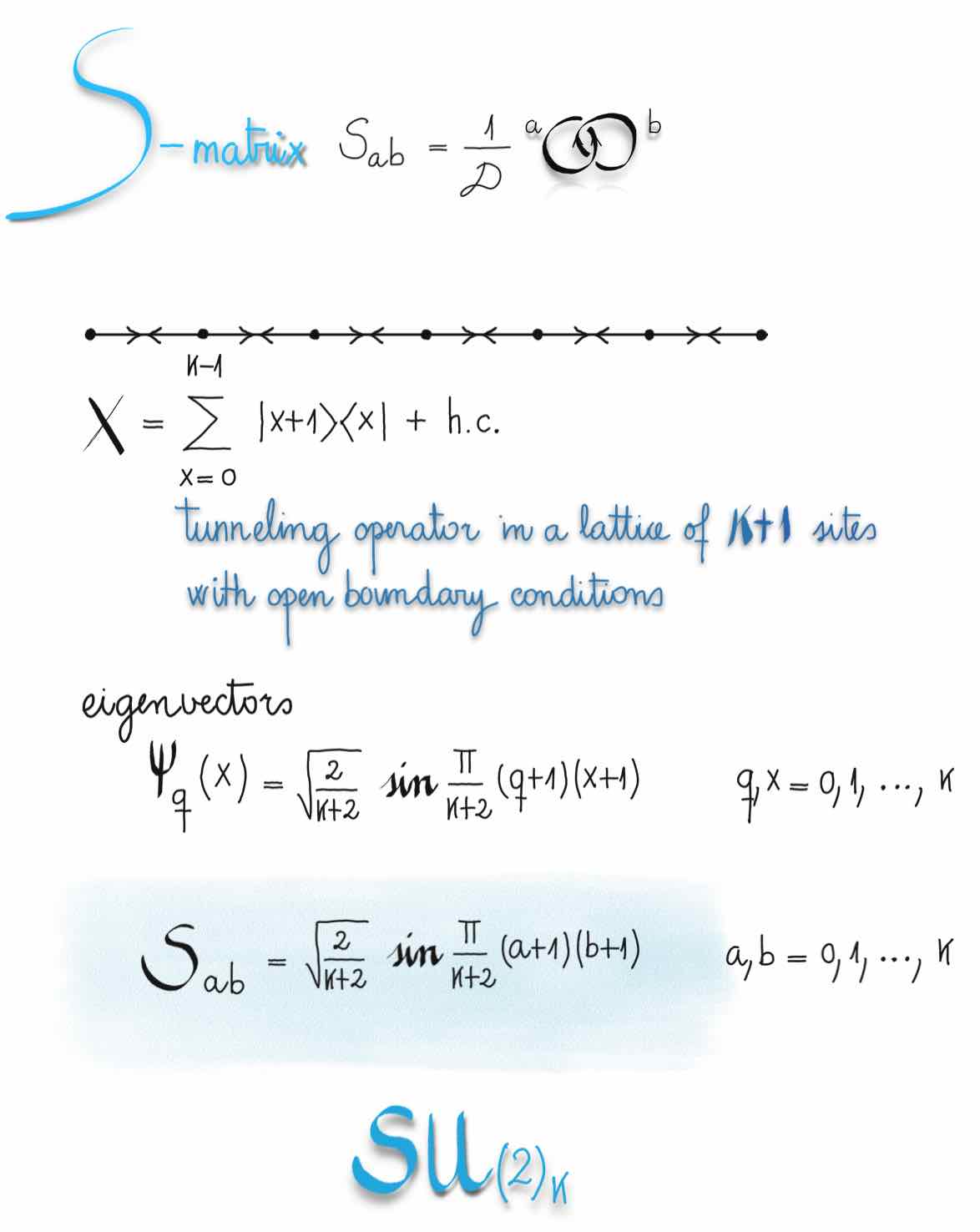}
\end{center}

\newpage
\pagestyle{fancy}
\fancyhf{}
\rhead{$k$ bosons in $2$ sites}
\rfoot{\thepage}
\vspace*{1cm}
\section*{S-matrix}
\addcontentsline{toc}{section}{S-matrix}

Diagonalization of the operator $X$ yields eigenstates of the form:
\begin{eqnarray}
\ket{\psi_q}=\sum_x S_{qx}\ket{x},
\end{eqnarray} 
where
\begin{eqnarray}
S_{qx}=\frac{\sqrt{2}}{\sqrt{k+2}}\sin\frac{\pi}{k+2}(q+1)(x+1).
\end{eqnarray} 
They are sine functions that vanish at the boundaries of the one-dimensional lattice. They are the eigenmodes of a particle in a lattice of $k$ sites with open boundary conditions. 

Pleasingly, they define a unitary symmetric matrix that exactly corresponds to the $S$-matrix of the $\mathbf{SU}(2)_k$ anyon model.

{\bf It is remarkable that something as physical (and simple) as the tunneling Hamiltonian of a particle in a one-dimensional lattice with open boundaries, can encode the apparently complex mathematical properties of the anyon model $\mathbf{SU}(2)_k$.}


\newpage
\thispagestyle{empty}
\vspace*{4.5cm}
\begin{center}
\includegraphics[width=0.78\textwidth]{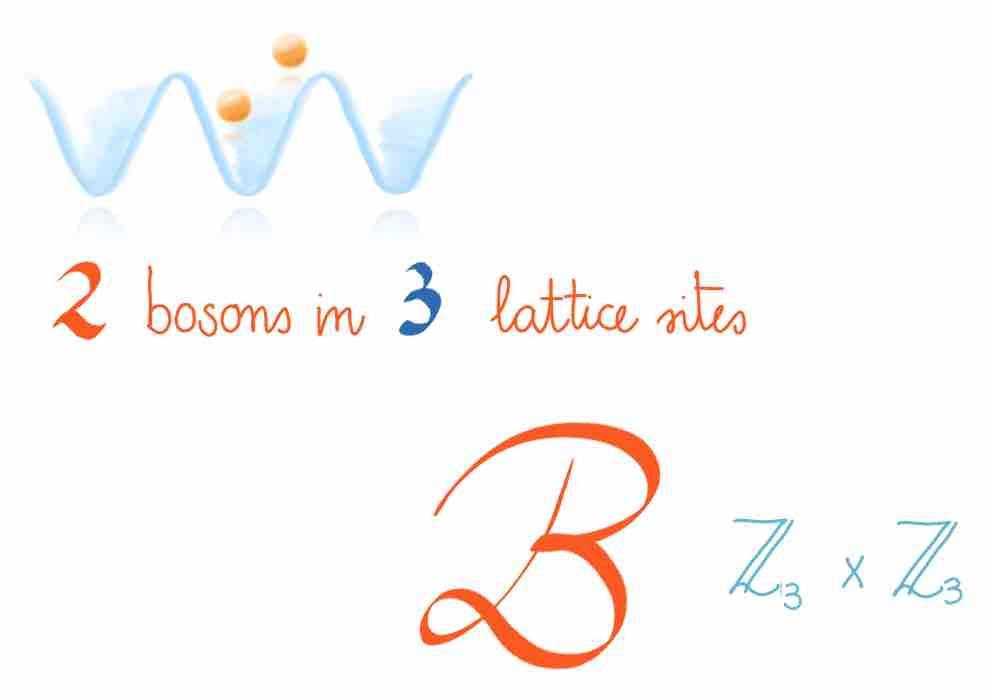}
\end{center}

\newpage
\pagestyle{fancy}
\fancyhf{}
\rhead{Boson-Lattice examples}
\rfoot{\thepage}
\vspace*{3cm}
\section*{}
\addcontentsline{toc}{section}{$2$ bosons in $3$ lattice sites}
\hspace*{1cm}
\includegraphics[width=0.75\textwidth]{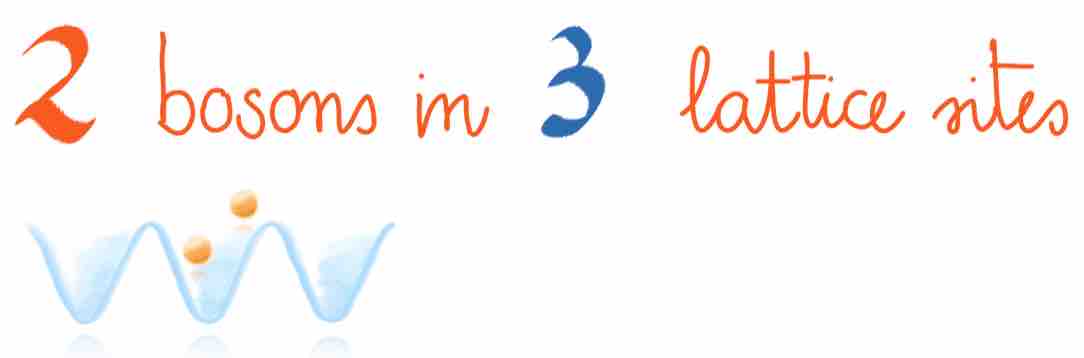}

\large
\hspace*{3cm}
\parbox{9.5cm}{\parskip=8pt

I analyze the anyon model corresponding to $2$ bosons in a lattice of $3$ sites.

I identify the Boson-Lattice generating graph, construct the topological algebra and characterize the fusion and braiding rules of the anyon model.

I show that it corresponds to the modular anyon model \bf{Fib} $\times \,\mathbb{Z}_3$.}

\normalsize
\newpage
\pagestyle{fancy}
\fancyhf{}
\lhead{Boson-Lattice examples}
\lfoot{\thepage}
\section*{Boson-Lattice graph}
\addcontentsline{toc}{section}{Boson-Lattice graph}

The Hilbert space of $2$ bosons in $3$ lattice sites has dimension $6$. I denote the states in the Fock basis by
\begin{equation}
\{\ket{0},\ket{1},\ket{2},\ket{3},\ket{4},\ket{5}\}.
\end{equation}
The corresponding anyon model has thus $6$ topological charges.

Following the prescription given in the previous section, the Boson-Lattice generating graph $\mathcal{G}$ is the one depicted below.

\vspace{0.5cm}
\begin{center}
\includegraphics[width=0.8\textwidth]{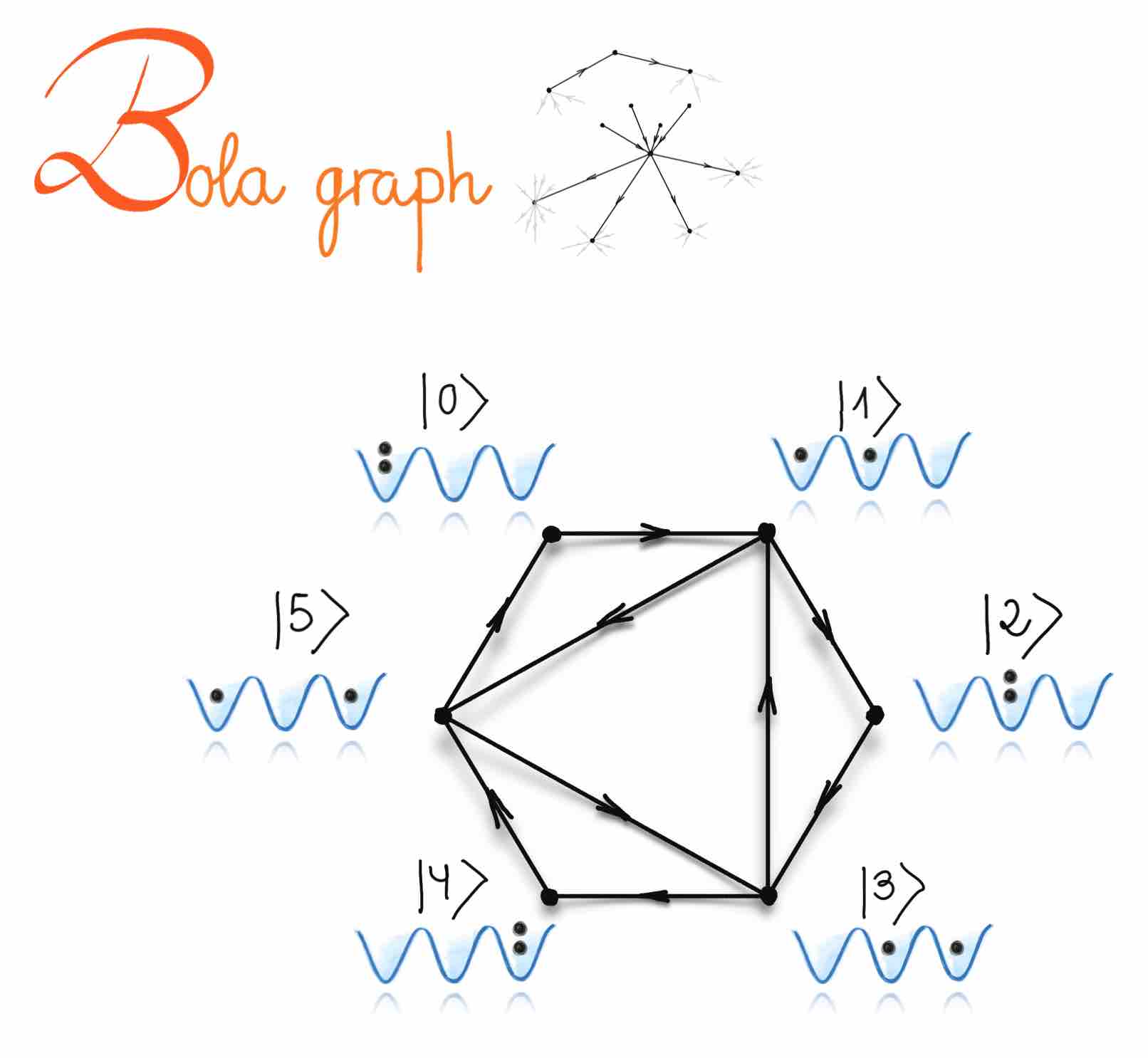}
\end{center}

\newpage
\pagestyle{fancy}
\fancyhf{}
\rhead{$2$ bosons in $3$ sites}
\rfoot{\thepage}
\section*{Topological Algebra}
\addcontentsline{toc}{section}{Topological Algebra}

For this model, the completion of the generating operator $X$ to a topological algebra is straightforward. 
We have that
\begin{equation}
\begin{array}{lll}
\ket{0}=\mathds{1}\ket{0} & \ket{2}=T\ket{0} & \ket{4}=T^2\ket{0}  \\

\ket{1}=X\ket{0} & \ket{3}=XT\ket{0} & \ket{5}=XT^2\ket{0}.
\end{array}
\end{equation}

Therefore the set of operators
\begin{eqnarray}
\mathcal{A}=\{\mathds{1},X,T,XT,T^2,XT^2\}
\end{eqnarray}
is the algebra of polynomials (of the operators $X$ and $T$) we are looking for. They commute with each other, they have positive entries (since both $X$ and $T$ have positive entries), and we have $X^\dagger=XT^2$, $T^\dagger=T^2$ and $(XT)^\dagger=XT$.
They define the topological algebra.

\begin{center}
\includegraphics[width=1.03\textwidth]{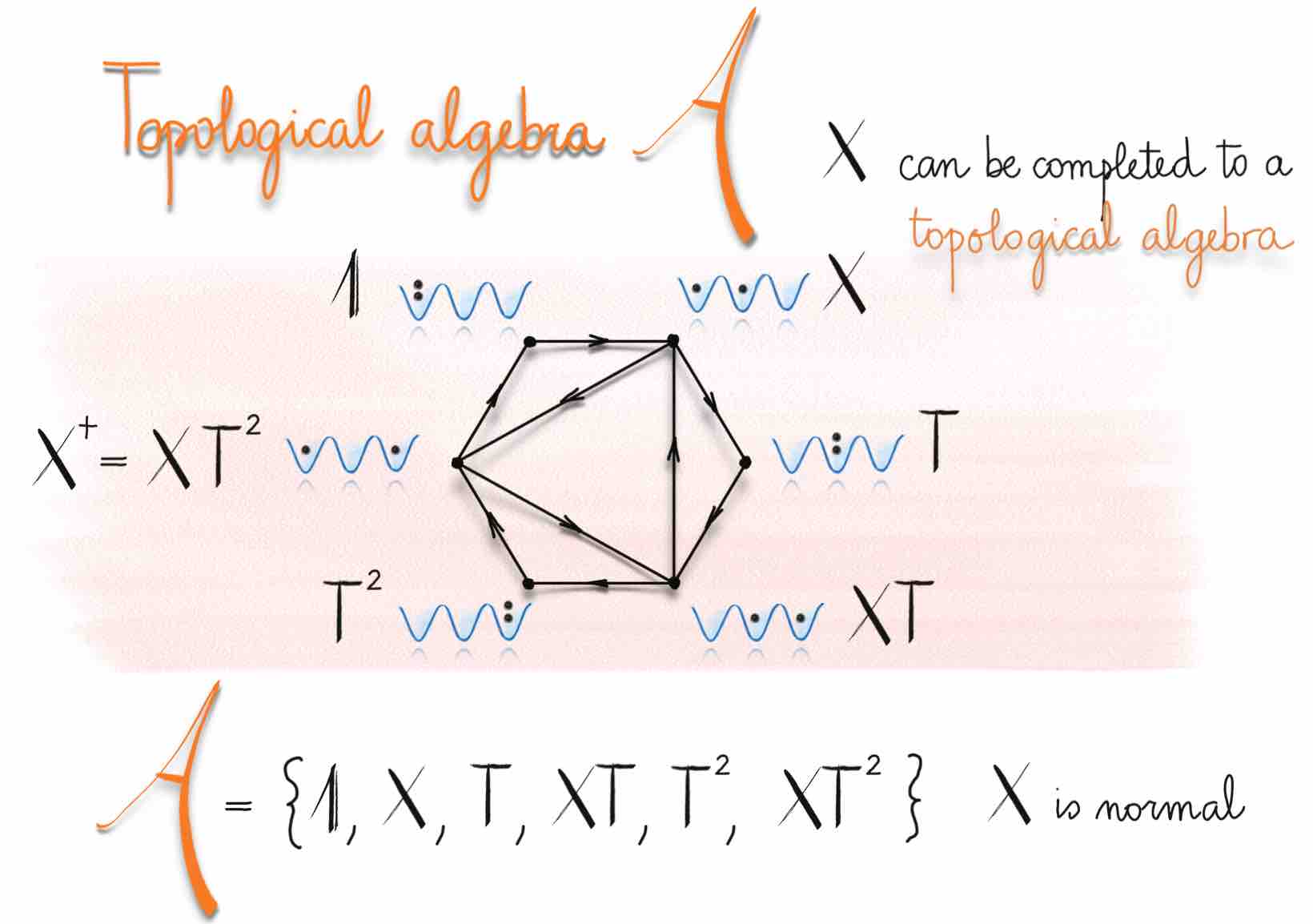}
\end{center}

\newpage
\pagestyle{fancy}
\fancyhf{}
\lhead{Boson-Lattice examples}
\lfoot{\thepage}
\vspace*{0cm}
\section*{Topological graphs}
\addcontentsline{toc}{section}{Topological graphs}
It is illuminating to draw the set of graphs corresponding to the topological algebra. This can be done just by composition of the graphs corresponding to the operators $X$ and $T$. 

Nicely, the graph corresponding to the operator $XT$ decomposes into three identical copies of the generating graph of the Fibonacci model.

\vspace*{1cm}
\begin{center}
\includegraphics[width=0.97\textwidth]{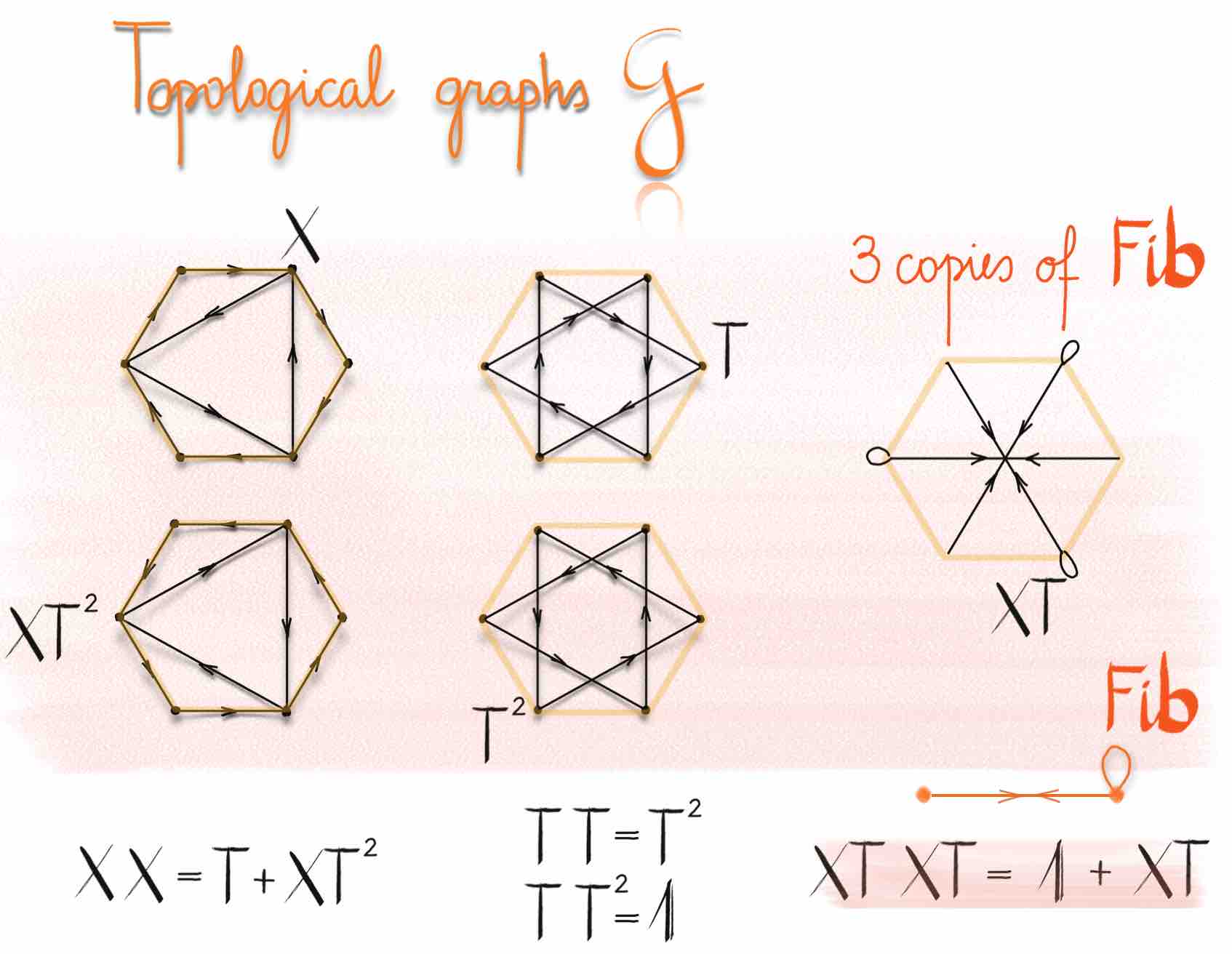}
\end{center}

\newpage
\pagestyle{fancy}
\fancyhf{}
\rhead{$2$ bosons in $3$ sites}
\rfoot{\thepage}
\vspace*{0cm}
\section*{Equivalence to  $\text{Fib}\times\mathbb{Z}_3$}
\addcontentsline{toc}{section}{The anyon model $\text{Fib}\otimes\mathbb{Z}_3$}

The topological algebra
\begin{eqnarray}
\mathcal{A}=\{\mathds{1},X,T,XT,T^2,XT^2\}
\end{eqnarray}
has two subalgebras:
\begin{eqnarray}
\begin{array}{lll}
\text{{\bf Fibonacci} subalgebra} &\{\mathds{1},XT\} &XT \cdot XT=\mathds{1}+XT\\
&&\\
\mathbb{Z}_3 \,\,\text{subalgebra}&\{\mathds{1},T,T^2\}& T \cdot T=T^2\\
&&T\cdot T^2=\mathds{1}.
\end{array}
\end{eqnarray}
It is useful to write the operators $XT$ and $T$ as:
\begin{eqnarray}
\begin{array}{rlrll}
XT&=&X_{\text{Fib}}&\otimes &\mathds{1}_3\\
T&= &\mathds{1}_2 &\otimes &X_{\mathbb{Z}_3},
\end{array}
\end{eqnarray}
where 
\begin{eqnarray}
X_{\text{Fib}}=\left[
\begin{array}{cc}
0&1\\
1&1
\end{array}
\right],
\quad
\quad
X_{\mathbb{Z}_3}=\left[
\begin{array}{ccc}
0&0&1\\
1&0&0\\
0&1&0
\end{array}
\right],
\end{eqnarray}
and $\mathds{1}_2$ ($\mathds{1}_3$) is the $2\times 2$ ($3\times 3$) identity matrix.
Defining 
\begin{eqnarray}
&&\mathcal{A}_{\text{Fib}}=\{\mathds{1}_2,X_{\text{Fib}}\}\\
&&\mathcal{A}_{\mathbb{Z}_3}=\{\mathds{1}_3,X_{\mathbb{Z}_3}^{\phantom{2}},X_{\mathbb{Z}_3}^2\},
\end{eqnarray}
the topological algebra $\mathcal{A}$ can be written as the tensor product of the topological algebra of the Fibonacci model and the topological algebra of the $\mathbb{Z}_3$ model:
\begin{eqnarray}
\mathcal{A}=\mathcal{A}_{\text{Fib}}\otimes \mathcal{A}_{\mathbb{Z}_3}.
\end{eqnarray}

The fusion rules of the Boson-lattice model are therefore those of 
the anyon model 
\begin{eqnarray}
\text{Fib}\times\mathbb{Z}_3.
\end{eqnarray}

\newpage
\pagestyle{fancy}
\fancyhf{}
\lhead{Boson-Lattice examples}
\lfoot{\thepage}
\section*{S-Matrix}
\addcontentsline{toc}{section}{S-Matrix}
Diagonalization of the graphs yields a unique (up to conjugation) symmetric and unitary matrix, which defines de $S$-matrix of the anyon model.
This matrix can be written as the tensor product:
\begin{eqnarray}
S=S_{\text{Fib}}\otimes S_{\mathbb{Z}_3}, 
\end{eqnarray}
where $S_{\text{Fib}}$ and $S_{\mathbb{Z}_3}$ are, respectively, the $S$-matrix of the Fibonnaci anyon and the $\mathbb{Z}_3$ anyon models.

\vspace*{1cm}
\begin{center}
\includegraphics[width=0.9\textwidth]{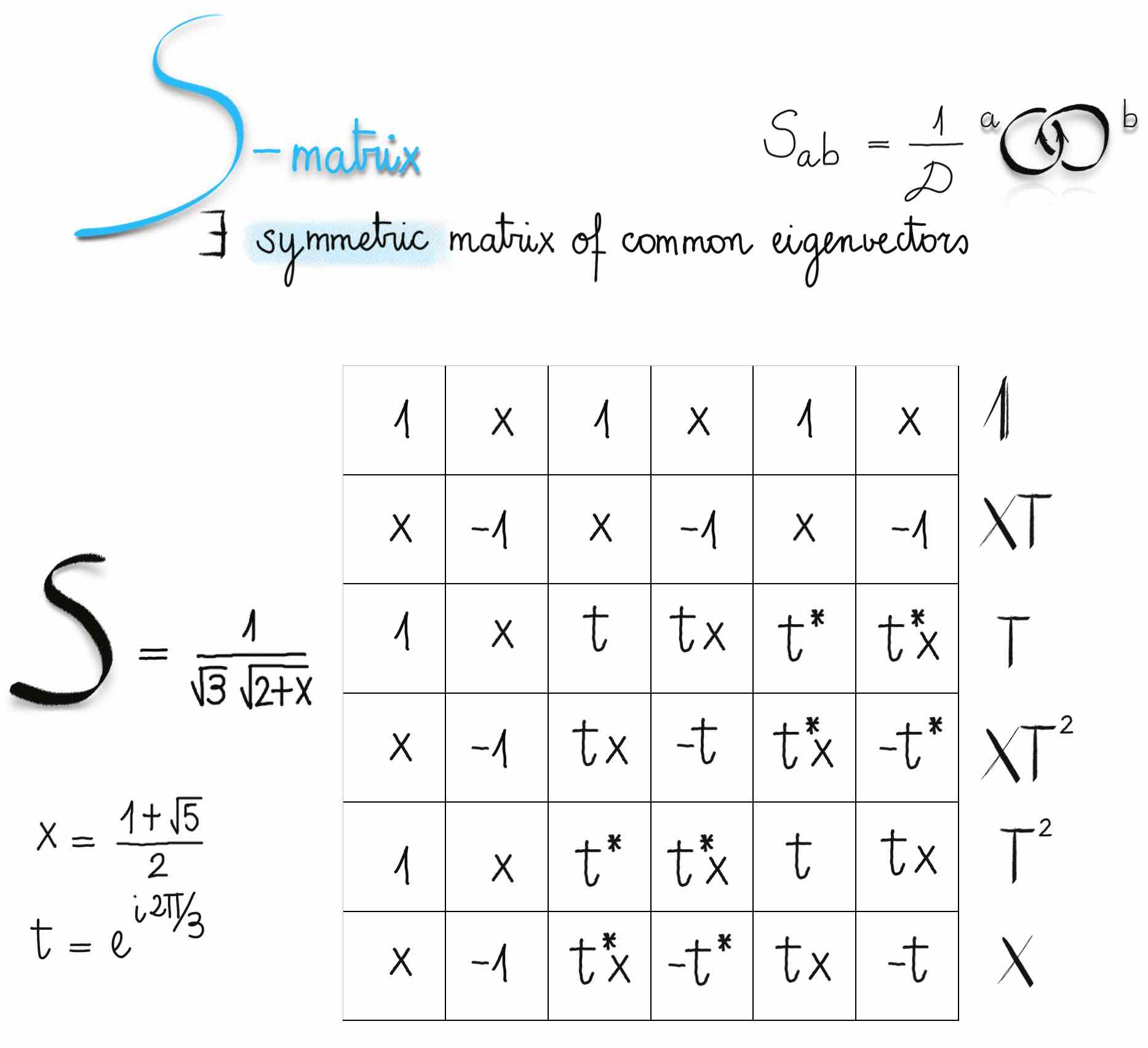}
\end{center}

\newpage
\pagestyle{fancy}
\fancyhf{}
\rhead{$2$ bosons in $3$ sites}
\rfoot{\thepage}
\normalsize
\section*{T-Matrix}
\addcontentsline{toc}{section}{T-Matrix}
The topological $T$-matrix is determined by equation (\ref{S-T-Relation}), relating the $S$-matrix to the $T$-matrix. It can be written as:
\begin{eqnarray}
T=T_{\text{Fib}}\otimes T_{\mathbb{Z}_3}, 
\end{eqnarray}
where $T_{\text{Fib}}$ and $T_{\mathbb{Z}_3}$ are, respectively, the $T$-matrix of the Fibonnaci anyon and the $\mathbb{Z}_3$ anyon models.
\vspace*{1cm}
\begin{center}
\includegraphics[width=0.85\textwidth]{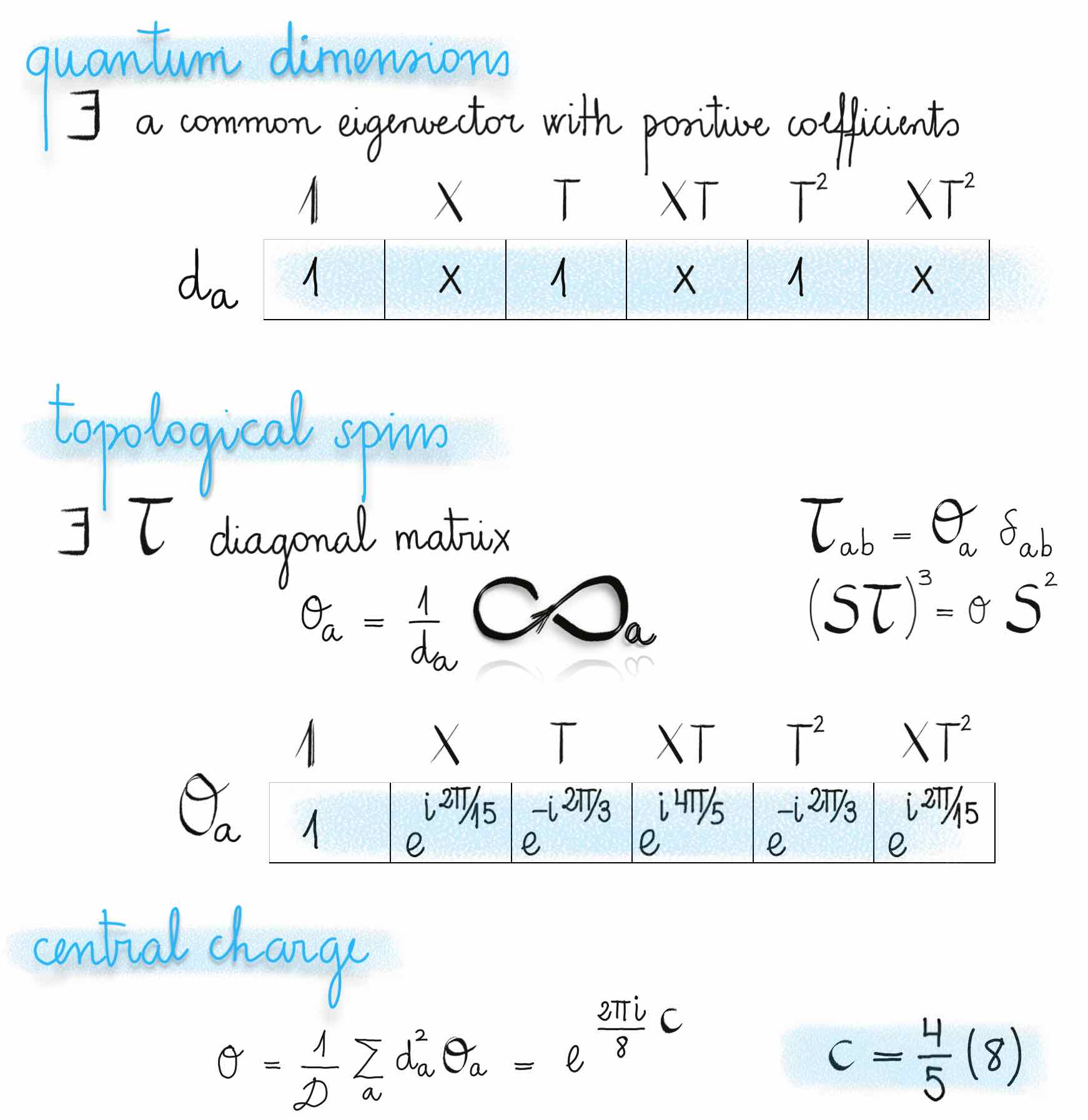}
\end{center}


\newpage
\thispagestyle{empty}
\vspace*{5cm}
\hspace*{1cm}
\includegraphics[width=0.8\textwidth]{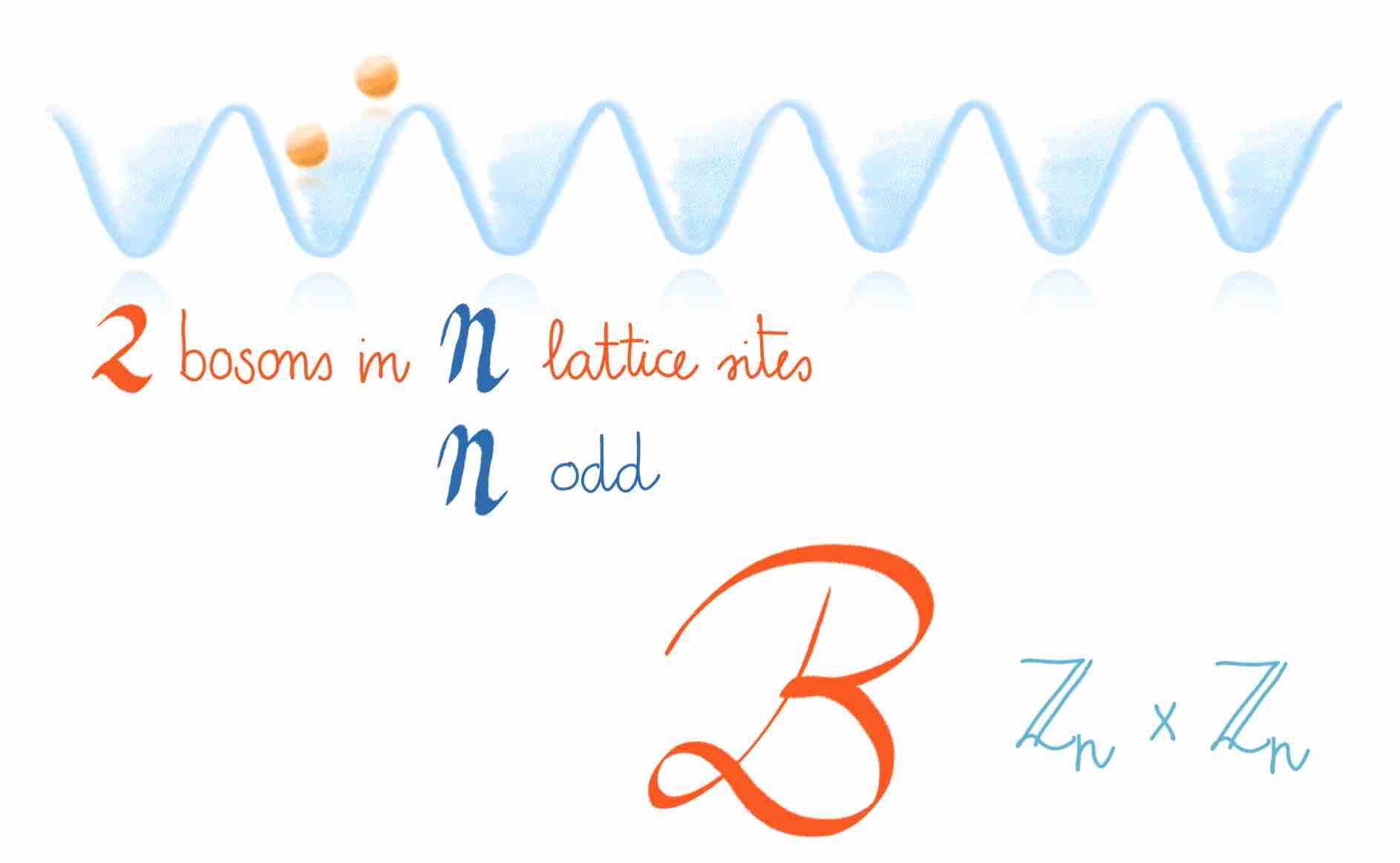}

\newpage
\pagestyle{fancy}
\fancyhf{}
\rhead{Boson-Lattice examples}
\rfoot{\thepage}
\vspace*{3cm}
\section*{}
\addcontentsline{toc}{section}{$2$ bosons in $n$ lattice sites}
\includegraphics[width=0.7\textwidth]{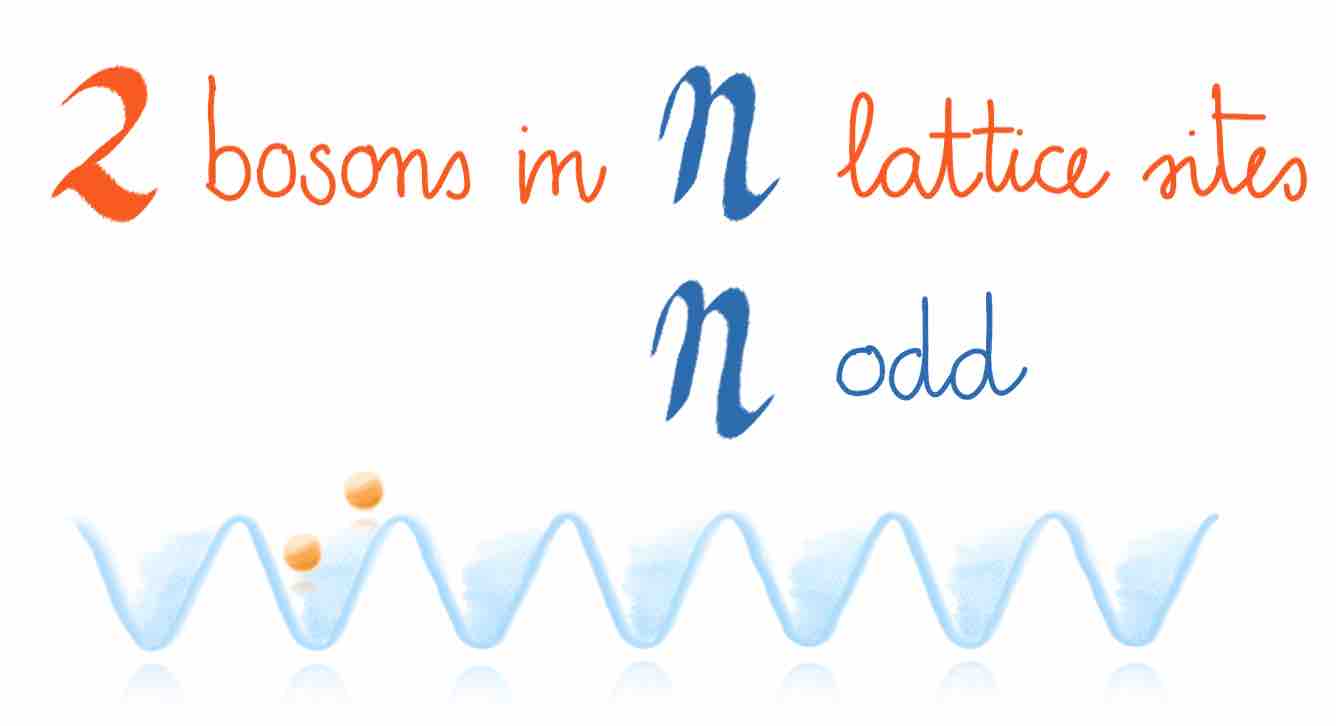}

\large
\hspace*{3cm}
\parbox{9cm}{\parskip=8pt

I analyze the  anyon model corresponding to $2$ bosons in a lattice of $n$ sites, with $n$ odd.

I identify the Boson-Lattice generating graph, construct the topological algebra, and characterize the fusion and braiding rules of the anyon model.

I show that it corresponds to the modular anyon model  {\bf SO}$(3)_n\times \,\mathbb{Z}_n$.

Notice that for $n=3$ we have {\bf SO}$(3)_3 \equiv $ {\bf Fib}, recovering the result of the previous section.}

\newpage
\pagestyle{fancy}
\fancyhf{}
\lhead{Boson-Lattice examples}
\lfoot{\thepage}
\normalsize
\section*{Hilbert space}
\addcontentsline{toc}{section}{Fock basis}
The Hilbert space of $2$ bosons in $n=2\ell+1$ lattice sites has dimension $(\ell+1)\times n$. The Fock states can be labelled by the relative distance $r$ between the two particles and the position $x$ of one of them (for example, the one with the smallest value of $x$):
\begin{eqnarray}
\ket{r,x}=T^x\ket{r,0}, \quad &&r=0,\cdots,\ell \nonumber\\
 &&x=0,\cdots,n-1.
\end{eqnarray}
Since $n$ is odd, for a fixed relative position $r$ there are always $n$ possible Fock states. 

\vspace*{0.5cm}
\begin{center}
\includegraphics[width=0.85\textwidth]{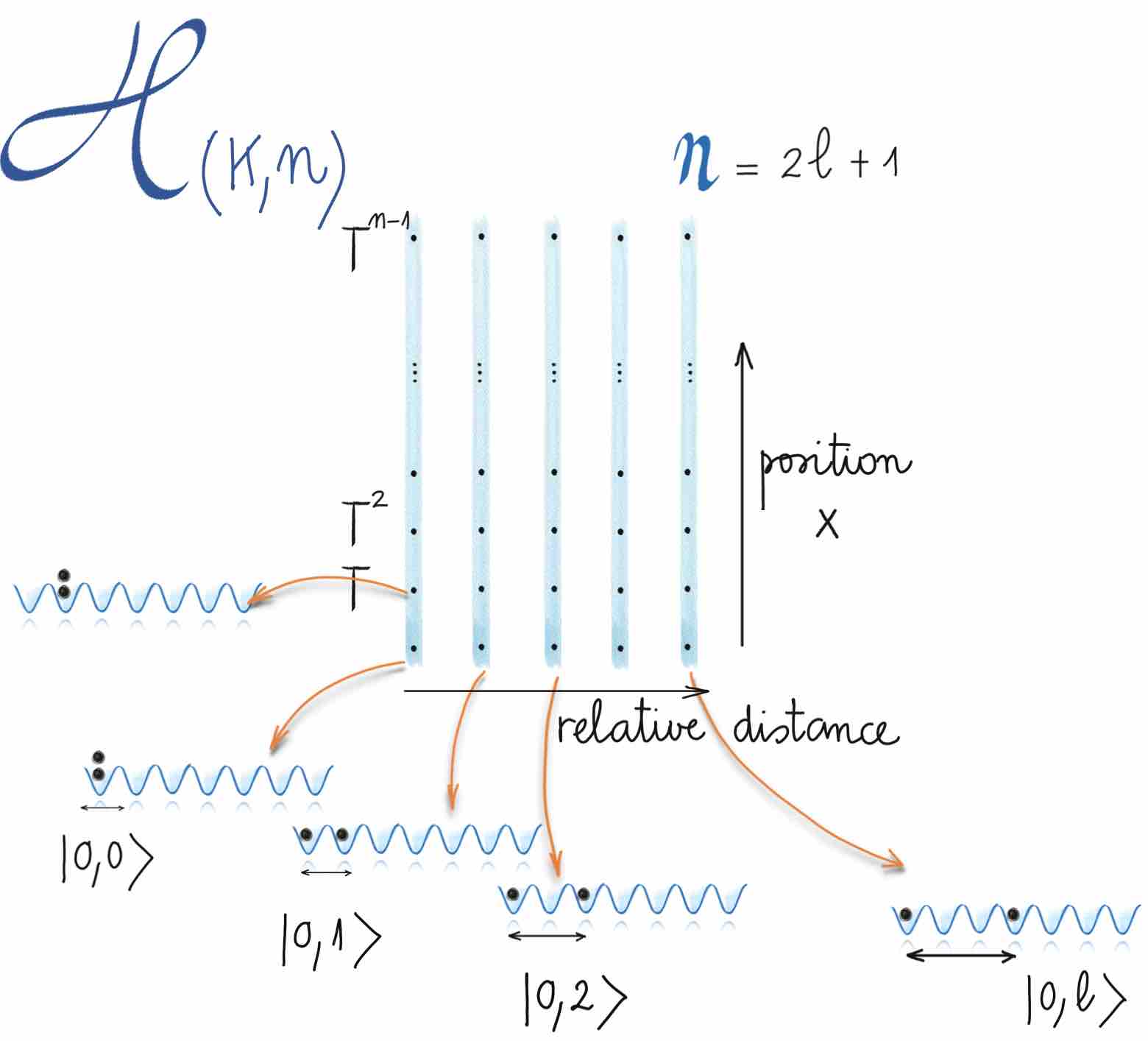}
\end{center}


\newpage
\pagestyle{fancy}
\fancyhf{}
\rhead{$2$ bosons in $n$ sites}
\rfoot{\thepage}
\section*{Boson-Lattice graph}
\addcontentsline{toc}{section}{Boson-Lattice graph}
The corresponding anyon model has $(\ell+1)\times n$ topological charges.
Following the prescription of the construction,  the Boson-Lattice generating graph $\mathcal{G}$ is depicted below for
$n=5$, $n=7$ and $n=9$.

\vspace*{1cm}
\includegraphics[width=1.02\textwidth]{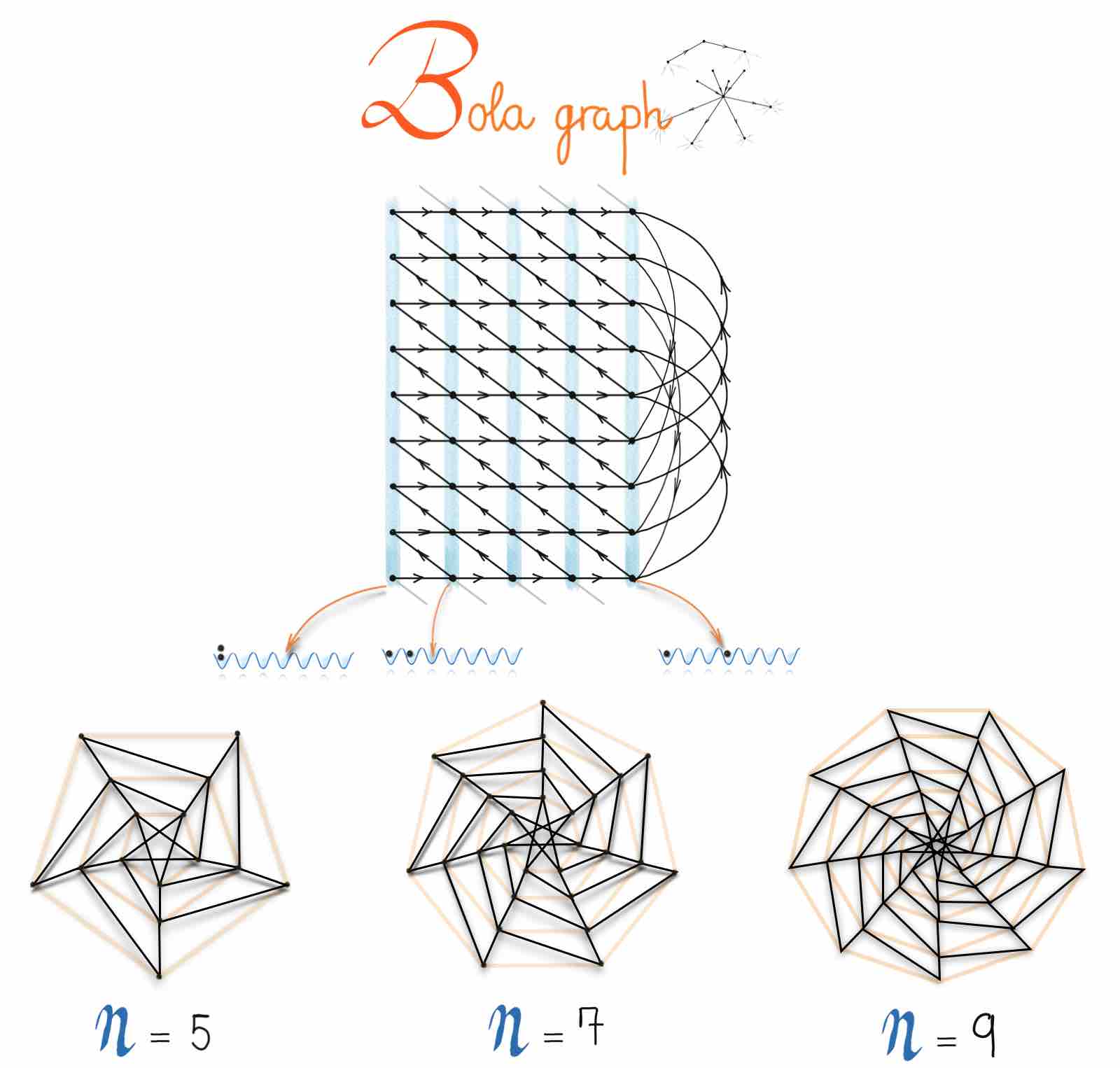}


\newpage
\pagestyle{fancy}
\fancyhf{}
\lhead{Boson-Lattice examples}
\lfoot{\thepage}

\section*{Topological Algebra}
\addcontentsline{toc}{section}{Topological Algebra}

By inspection of the generating graph $\mathcal{G}$ we can directly see that the algebra of polynomials we are looking for is:
\begin{eqnarray}
\mathcal{A}=\{X_rT^x \quad r=0,\cdots,\ell;\,x=0,\cdots,n-1\},
\end{eqnarray}
where $X_0=\mathds{1}$, $X_1=X$ and $X_r,$ (for $r=2,\cdots,\ell$)  are polynomials of the operators $X$ and $T$, obtained through the recursive relation:
\begin{eqnarray}
X_r=XX_{r-1}-X_{r-2}\,T.
\end{eqnarray}

\vspace*{1cm}
\begin{center}
\includegraphics[width=0.98\textwidth]{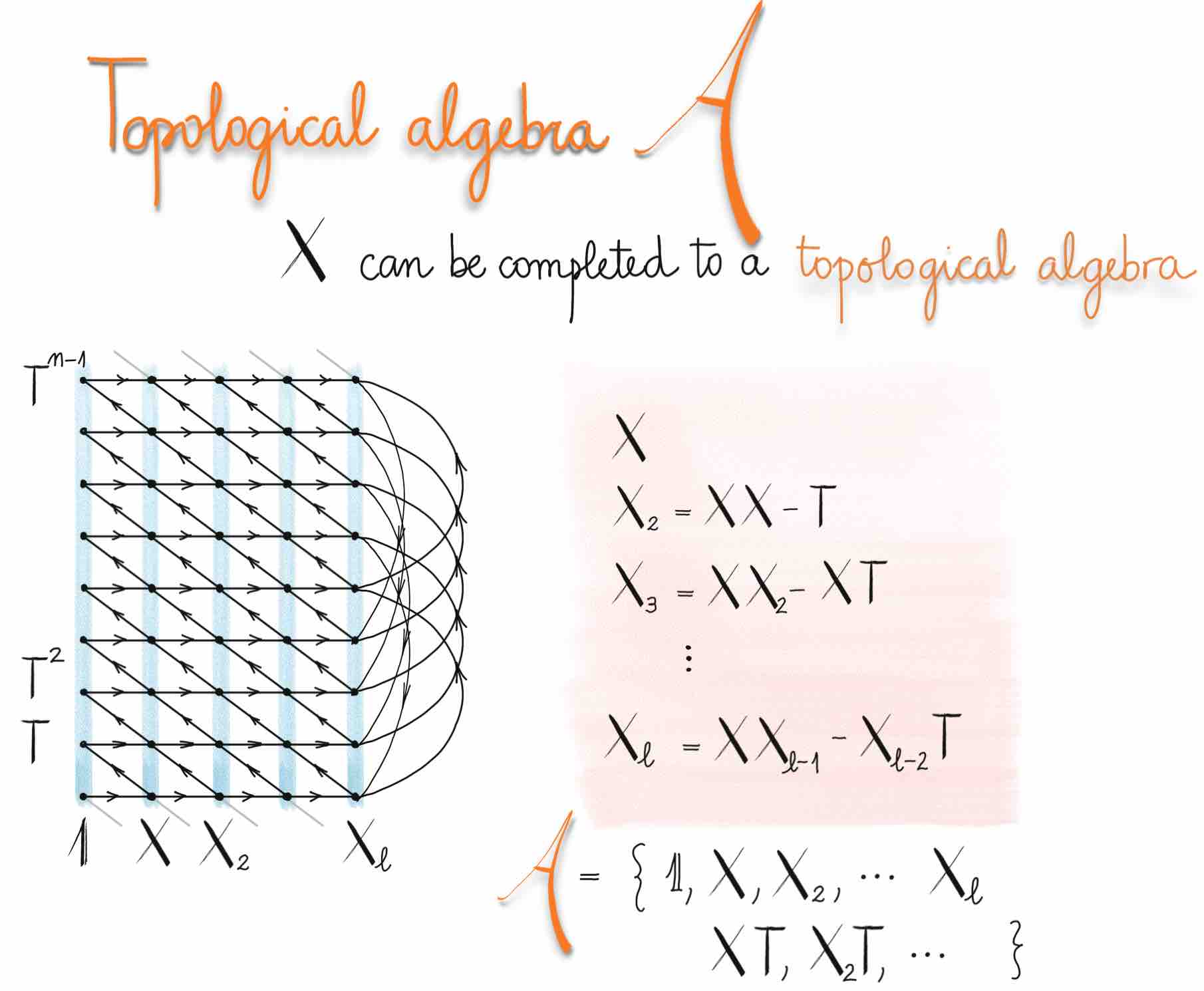}
\end{center}

\newpage
\pagestyle{fancy}
\fancyhf{}
\rhead{$2$ bosons in $n$ sites}
\rfoot{\thepage}

\section*{The $\mathbf{SO}(3)_k$ subalgebra}
\addcontentsline{toc}{section}{The $\mathbf{SO}(3)_k$ subalgebra}

To show that $\mathcal{A}$ is a topological algebra, let me consider the operator $Q=XT^\ell\in \mathcal{A}$. This is the generating operator of the subalgebra
\begin{eqnarray}
\{\mathds{1},Q=Q_1,Q_2,\cdots, Q_\ell\}, \quad Q_r=X_rT^{r\ell}.
\end{eqnarray}
It fulfills:
\begin{eqnarray}
\mathds{1} \times Q &=& Q \nonumber\\
\,Q \times Q_r&=&Q_{r-1}+Q_{r+1} \nonumber\\
\,Q \times Q_\ell&=&Q_{\ell-1}+Q_\ell.
\end{eqnarray}
The corresponding graph is depicted below. It is very similar to the generating graph of the anyon model $\mathbf{SU}(2)_\ell$, except for the important fact that the graph has now a loop at the last vertex. Is this graph a topological graph?

\vspace*{1cm}
\begin{center}
\includegraphics[width=\textwidth]{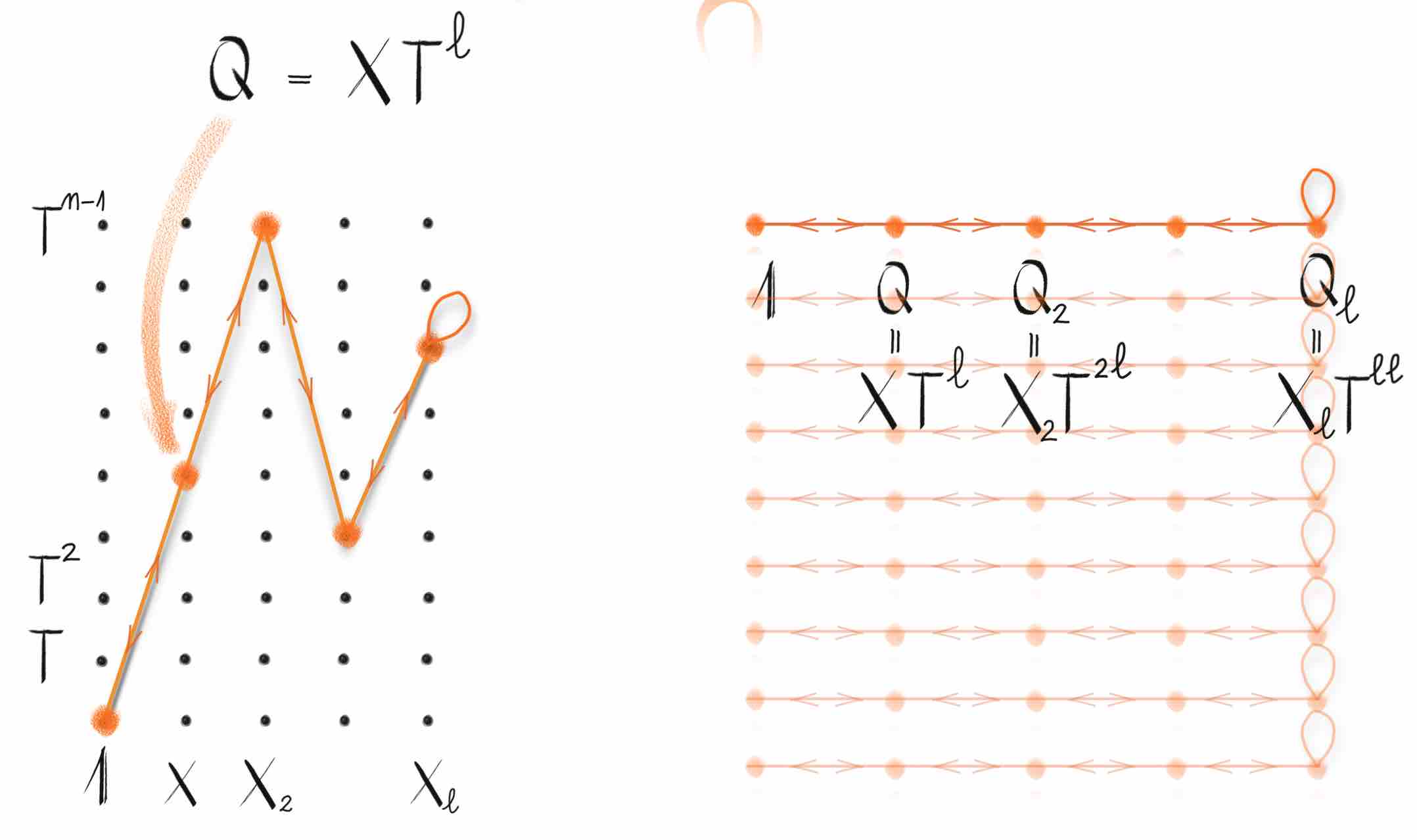}
\end{center}


\newpage
\pagestyle{fancy}
\fancyhf{}
\lhead{Boson-Lattice examples}
\lfoot{\thepage}

\vspace*{0cm}
Let me consider the operator $Q_2$ in the subalgebra above. Reordering the elements of the subalgebra
\begin{eqnarray}
Q_i \longrightarrow \widetilde{Q}_i=Q_{p(i)},
\end{eqnarray}
with the permutation $p$ such that
\begin{eqnarray}
\{0,1,\cdots,\ell\} &\longrightarrow &\{0,2,4,\cdots, \ell, \ell-1,\cdots,3,1\} \hspace{1cm}\text{$\ell$ even}\nonumber\\
\{0,1,\cdots,\ell\} &\longrightarrow &\{0,2,4,\cdots, \ell-1, \ell,\cdots,3,1\} \hspace{1cm}\text{$\ell$ odd},
\end{eqnarray}
we have that $\widetilde{Q}\equiv Q_2$ fulfills:
\begin{eqnarray}
\mathds{1} \times \widetilde{Q} &=& \widetilde{Q} \nonumber\\
\,\widetilde{Q} \times \widetilde{Q}_r&=&\widetilde{Q}_{r-1}+\widetilde{Q}_r+\widetilde{Q}_{r+1} \nonumber\\
\,\widetilde{Q} \times \widetilde{Q}_\ell&=&\widetilde{Q}_{\ell-1}+\widetilde{Q}_\ell.
\end{eqnarray}
These fusion rules exactly correspond to the ones of the anyon model $\mathbf{SO}(3)_n$.

\vspace*{1cm}

\begin{center}
\includegraphics[width=0.8\textwidth]{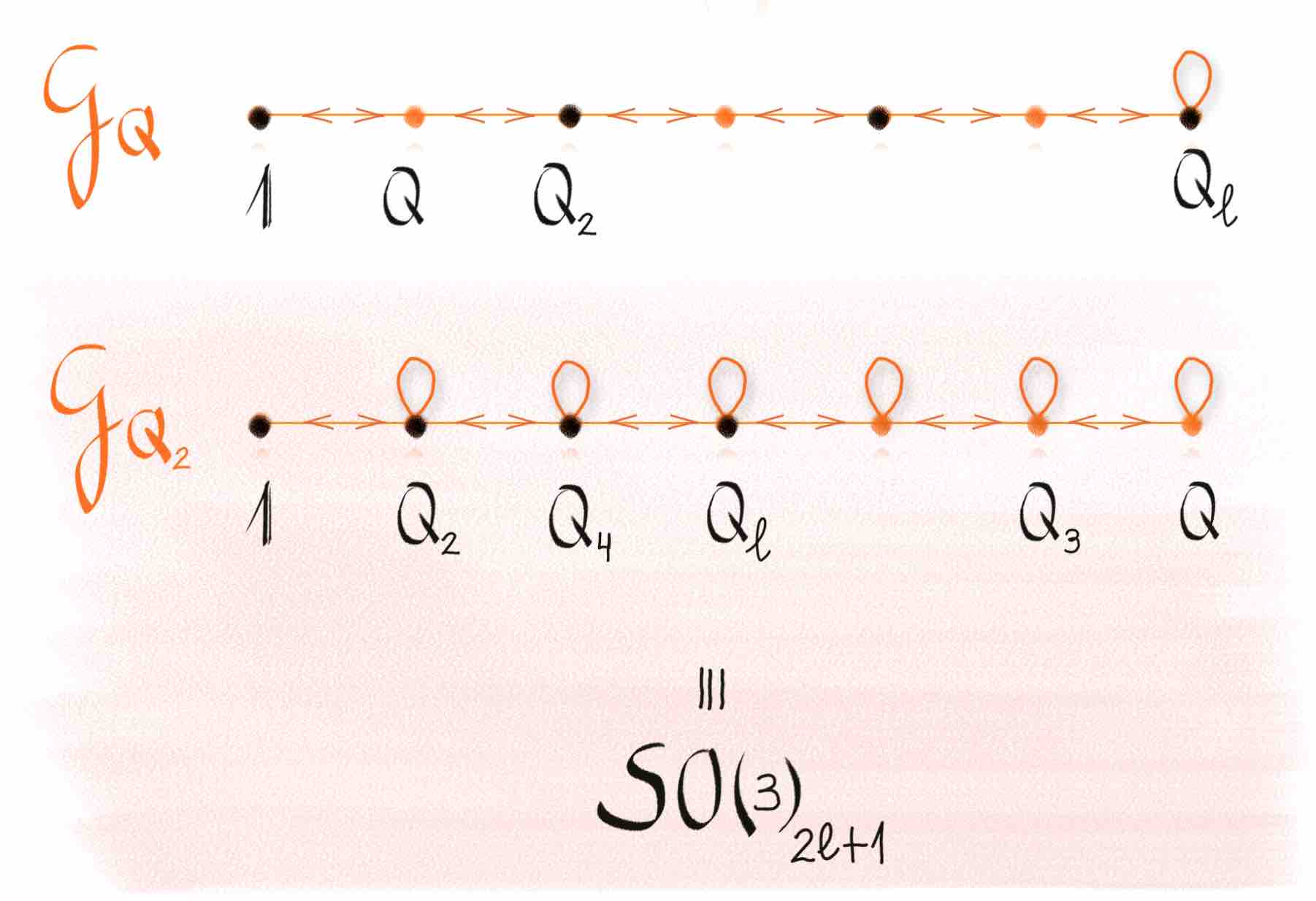}
\end{center}


\newpage
\pagestyle{fancy}
\fancyhf{}
\rhead{$2$ bosons in $n$ sites}
\rfoot{\thepage}
\vspace*{1cm}
The generating operator $Q$ corresponds to the tunneling Hamiltonian of one particle in a one-dimensional lattice of $\ell+1$ sites with open boundaries, and in the presence of an impurity potential at the last site, $\ell$. 
The eigenmodes of this Hamiltonian define a unitary symmetric matrix, which exactly coincides with the topological $S$-matrix of the modular anyon model  $\mathbf{SO}(3)_n$.

It is again remarkable that the Hamiltonian of a particle in a one-dimensional lattice with open boundaries (and an impurity potential at the last site) can encode the mathematical properties of the anyon model $\mathbf{SO}(3)_n$.

From the results above, we can conclude that the anyon model corresponding to the Boson-Lattice system of $2$ particles in $n$ sites, with $n$ odd, is $\mathbf{SO}(3)_n \times \mathbb{Z}_n$.

\begin{center}
\includegraphics[width=0.8\textwidth]{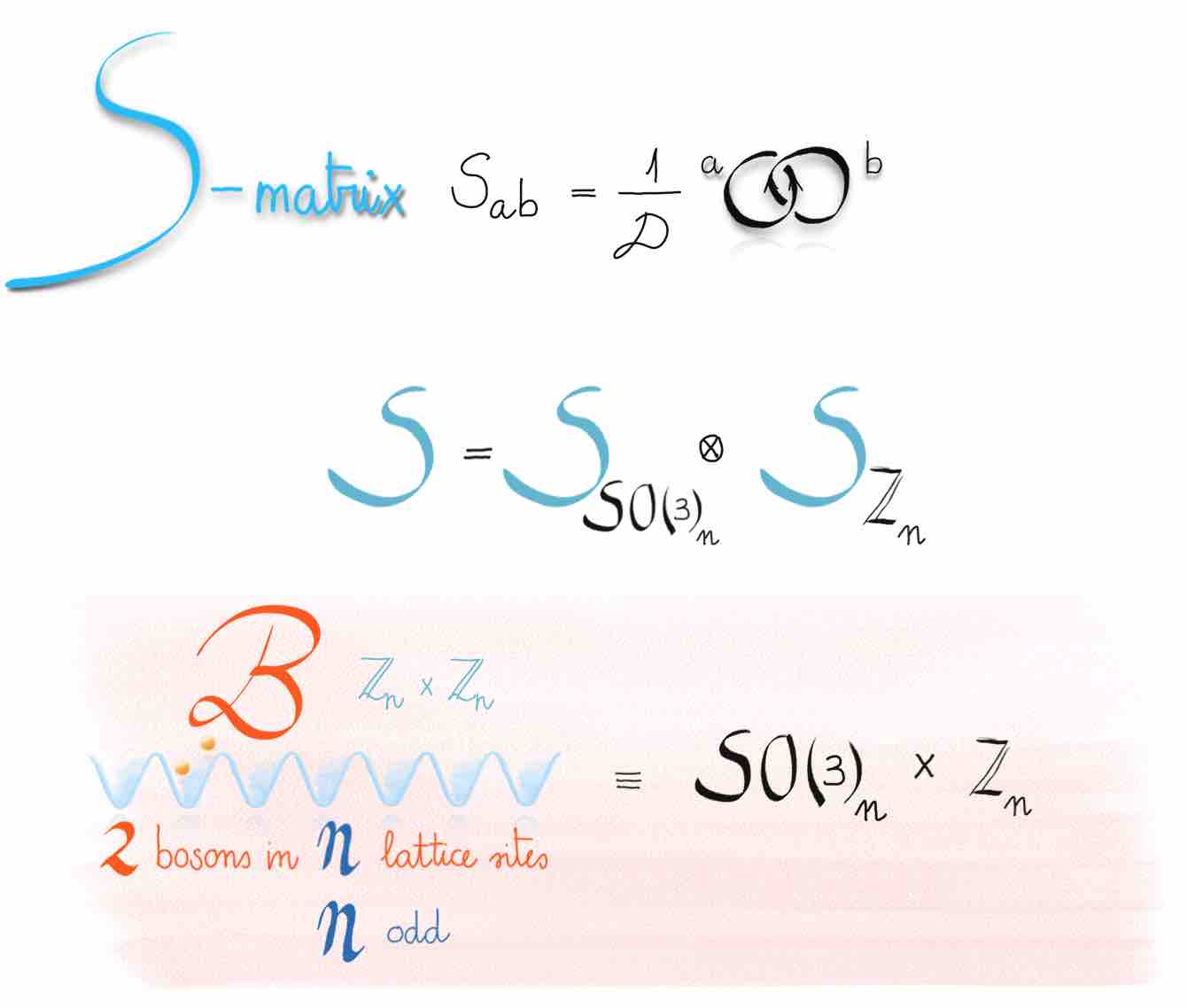}
\end{center}


\newpage
\thispagestyle{empty}
\vspace*{5cm}
\hspace*{1cm}
\includegraphics[width=0.8\textwidth]{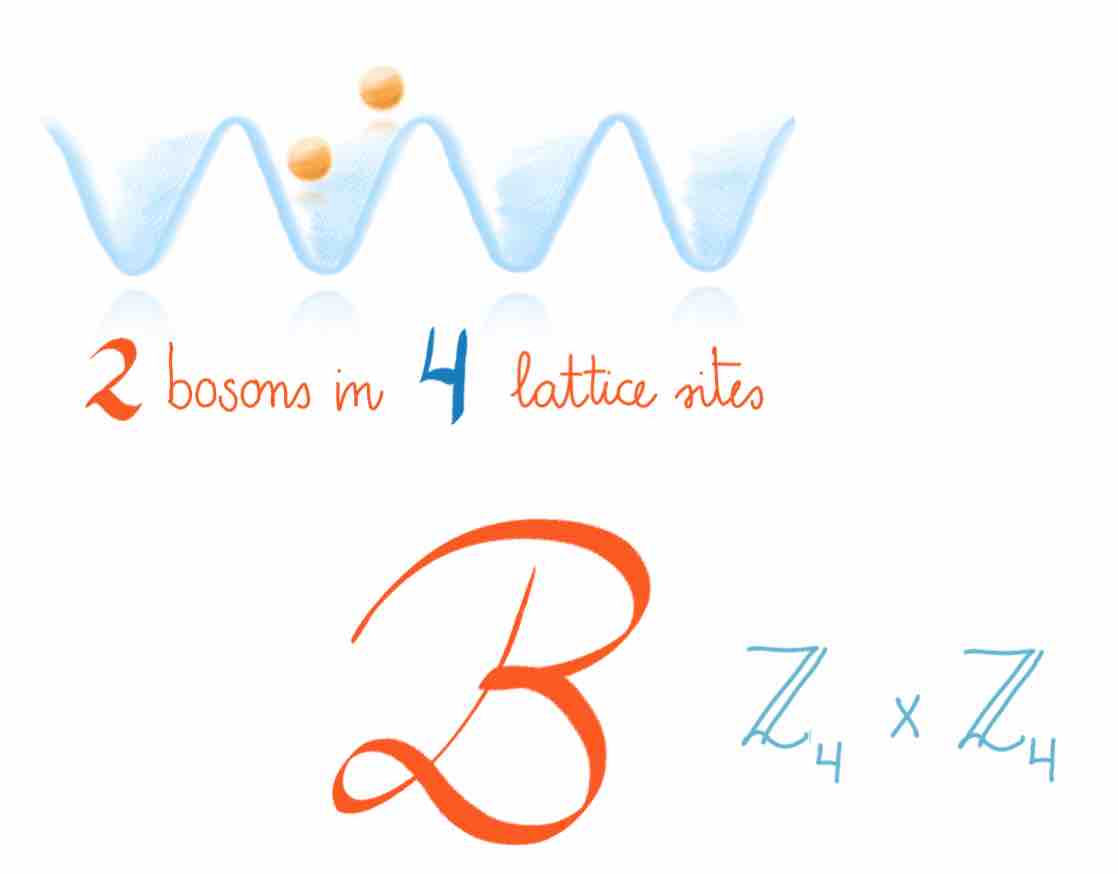}

\newpage
\pagestyle{fancy}
\fancyhf{}
\rhead{Boson-Lattice examples}
\rfoot{\thepage}
\vspace*{3cm}
\section*{}
\addcontentsline{toc}{section}{$2$ bosons in $4$ lattice sites}
\includegraphics[width=0.7\textwidth]{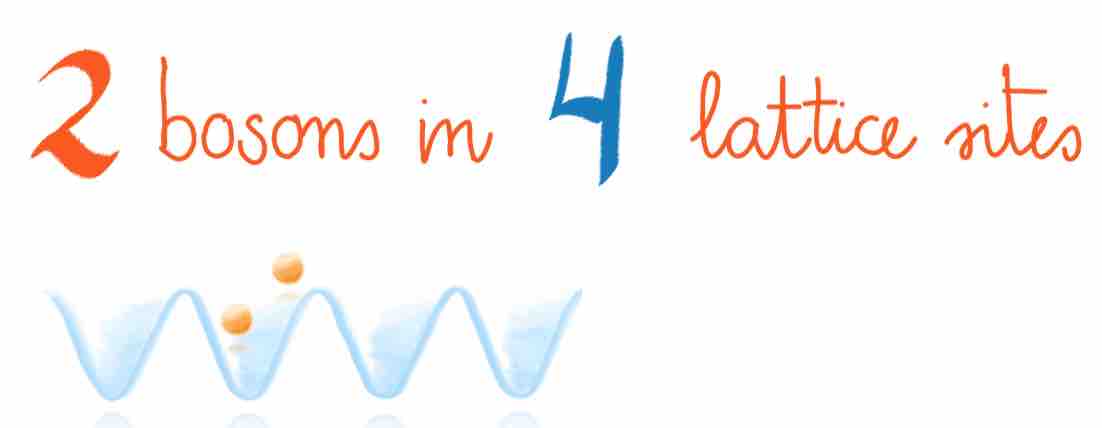}

\large
\hspace*{3cm}
\parbox{10cm}{\parskip=8pt

I analyze the anyon model corresponding to $2$ bosons in a lattice of $4$ sites, as an illustration of the interesting case of $2$ bosons in a lattice with an {\em even} number of sites.

I identify the Boson-Lattice  graph, construct the topological algebra, and characterize the fusion and braiding rules of the anyon model.

In contrast to the case of $2$ bosons in a lattice with an odd number of sites, the topological algebra of this Boson-Lattice model is not decomposable as the tensor product of two topological algebras.

The Boson-Lattice model of $2$ bosons in $4$ sites is a well defined anyon model, which is to my knowledge {\em not tabulated}.}

\normalsize
\newpage
\pagestyle{fancy}
\fancyhf{}
\lhead{Boson-Lattice examples}
\lfoot{\thepage}
\normalsize
\section*{Boson-Lattice graph}
\addcontentsline{toc}{section}{Boson-Lattice graph}
The Hilbert space of $2$ bosons in $4$ lattice sites has dimension $10$. The corresponding anyon model has therefore $10$ topological charges.
Following the prescription given in the previous section, we construct the Boson-Lattice generating graph $\mathcal{G}$ depicted below.

\vspace*{1cm}
\begin{center}
\includegraphics[width=0.9\textwidth]{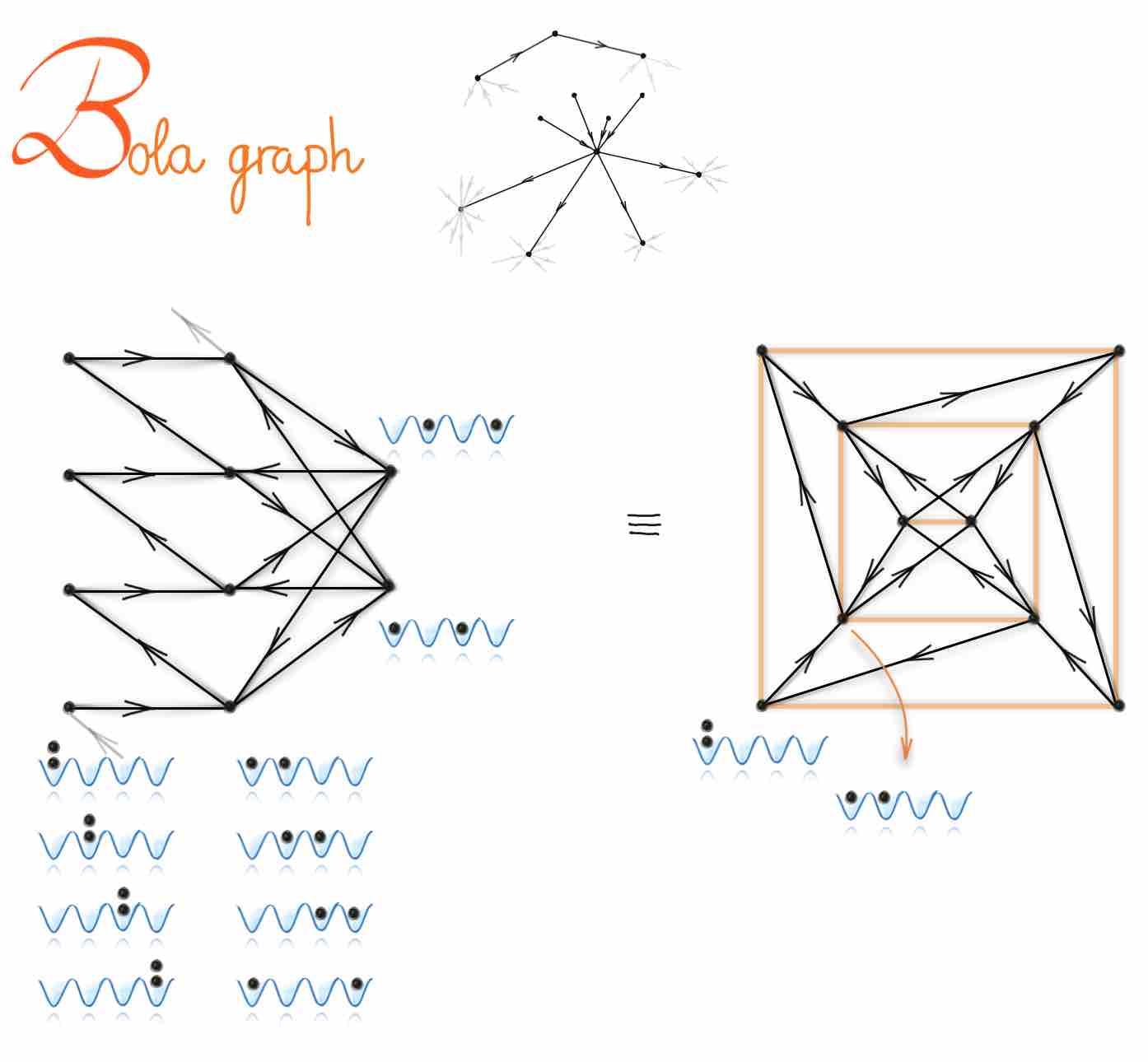}
\end{center}


\newpage
\pagestyle{fancy}
\fancyhf{}
\rhead{$2$ bosons in $4$ sites}
\rfoot{\thepage}
\section*{Topological Algebra}
\addcontentsline{toc}{section}{Topological Algebra}
By inspection of the generating graph $\mathcal{G}$ we can see that the set of polynomials  we are looking for is:
\begin{eqnarray}
\mathcal{A}=\{\mathds{1},T, T^2,T^3,X,XT,XT^2,XT^3,Q,TQ\},
\end{eqnarray}
where the operator $Q$ fulfills:
\begin{eqnarray}
&&Q=XX-T \\
&&QT^2=Q.
\end{eqnarray}

\vspace*{0.3cm}
\begin{center}
\includegraphics[width=0.95\textwidth]{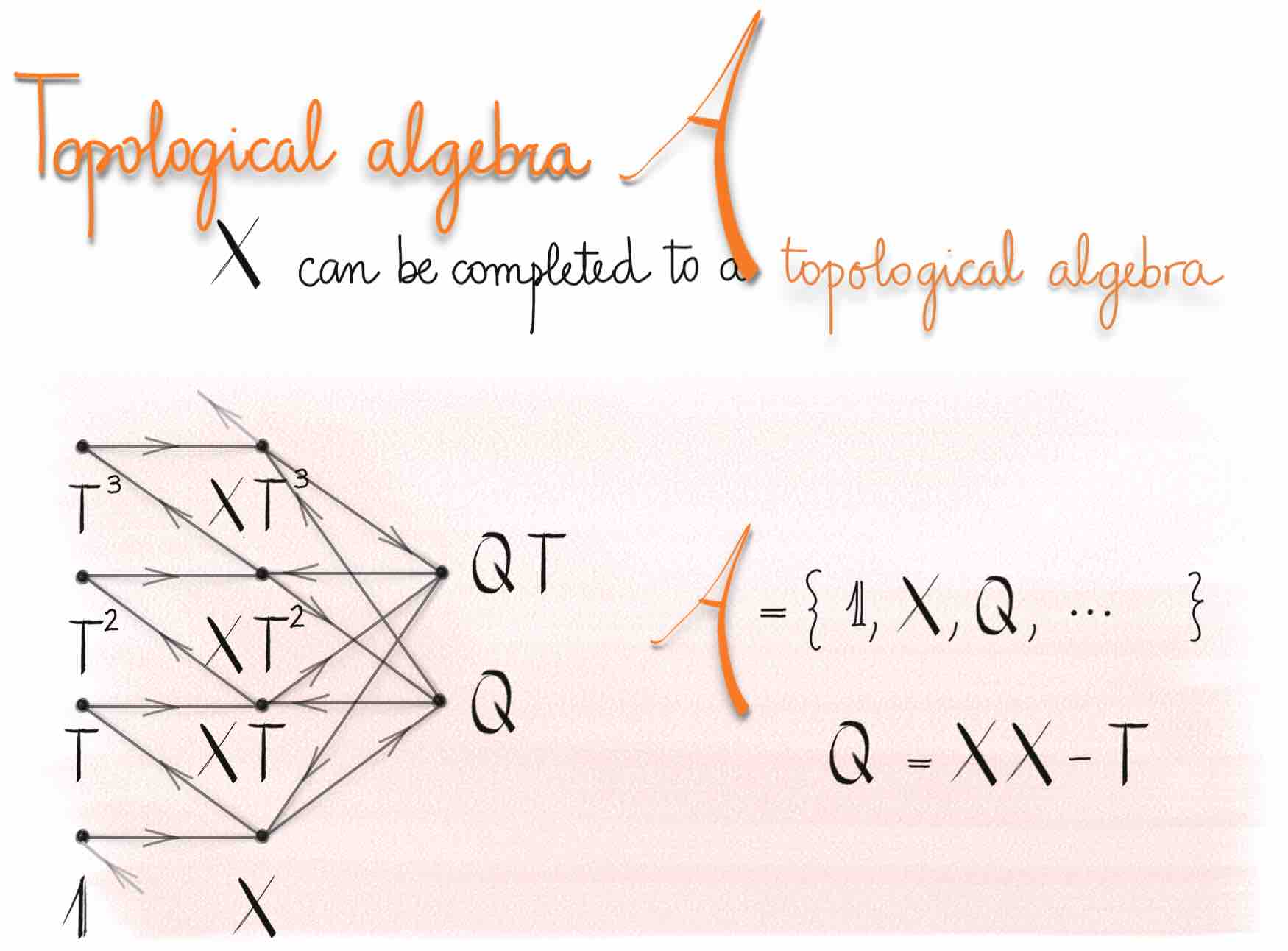}
\end{center}

\newpage
\pagestyle{fancy}
\fancyhf{}
\lhead{Boson-Lattice examples}
\lfoot{\thepage}
\vspace*{0cm}
\section*{Topological graphs}
\addcontentsline{toc}{section}{Topological graphs}
To show that $\mathcal{A}$ is a topological algebra, it suffices to show that $Q$ has non-negative entries. By composing the graph $X$ with itself and subtracting the graph corresponding to the operator $T$, we obtain the graph of the operator $Q$.
This is a non-negative graph (links have weight either $0$ or $1$) that decomposes into two independent graphs.

\vspace*{1cm}
\includegraphics[width=0.9\textwidth]{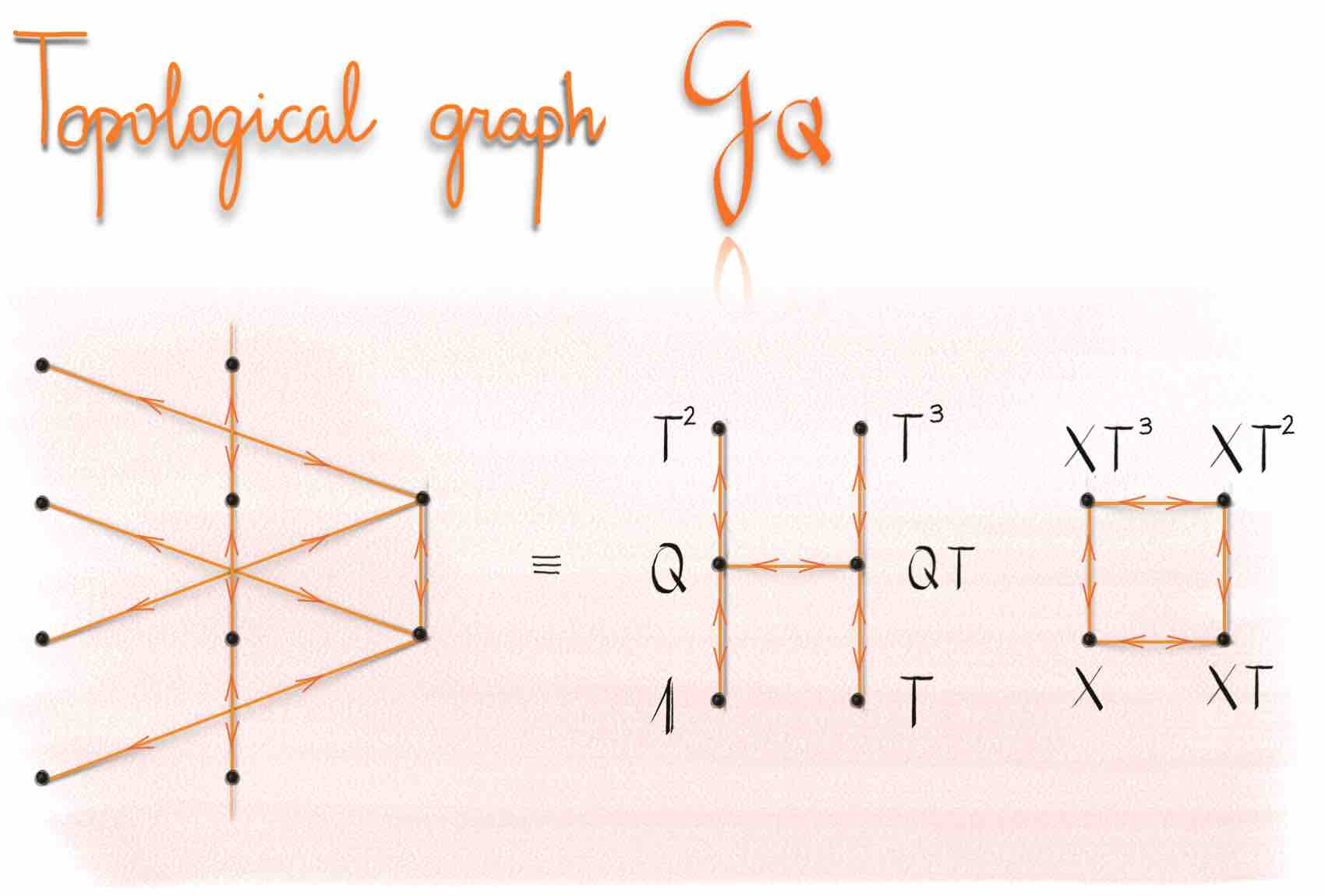}

\newpage
\pagestyle{fancy}
\fancyhf{}
\rhead{$2$ bosons in $4$ sites}
\rfoot{\thepage}
\vspace*{0cm}
\normalsize
\section*{Fusion rules}
\addcontentsline{toc}{section}{Fusion Rules}

The fusion rules of the model can be directly read from the topological graphs below.

\begin{center}
\includegraphics[width=0.9\textwidth]{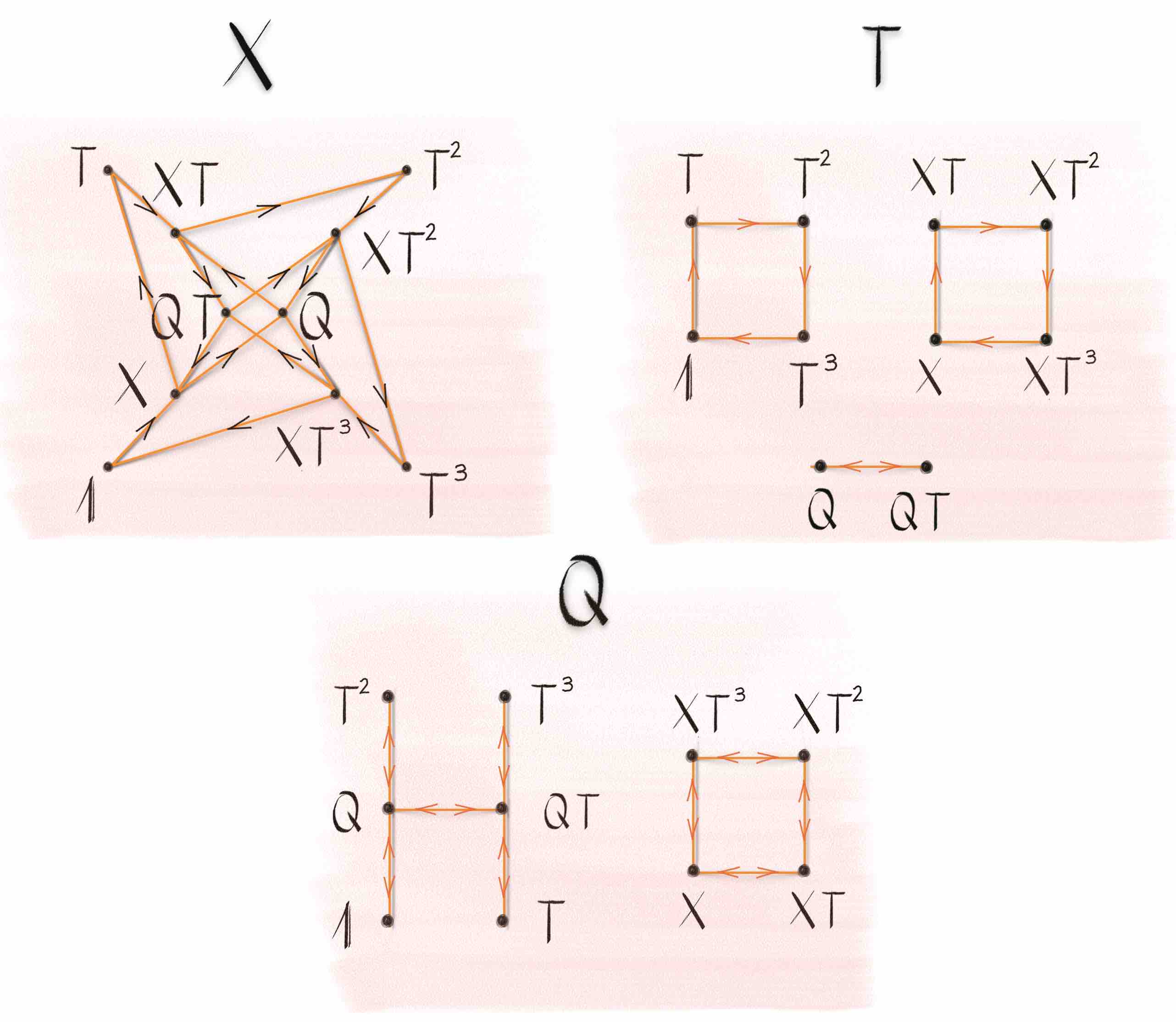}
\end{center}
\newpage
\pagestyle{fancy}
\fancyhf{}
\lhead{Boson-Lattice examples}
\lfoot{\thepage}
\normalsize
\section*{S-Matrix}
\addcontentsline{toc}{section}{S-Matrix}
Diagonalization of the graphs yields a unique (up to conjugation) symmetric and unitary matrix, which defines the $S$-matrix of the anyon model.

\vspace*{1cm}
\begin{center}
\includegraphics[width=0.9\textwidth]{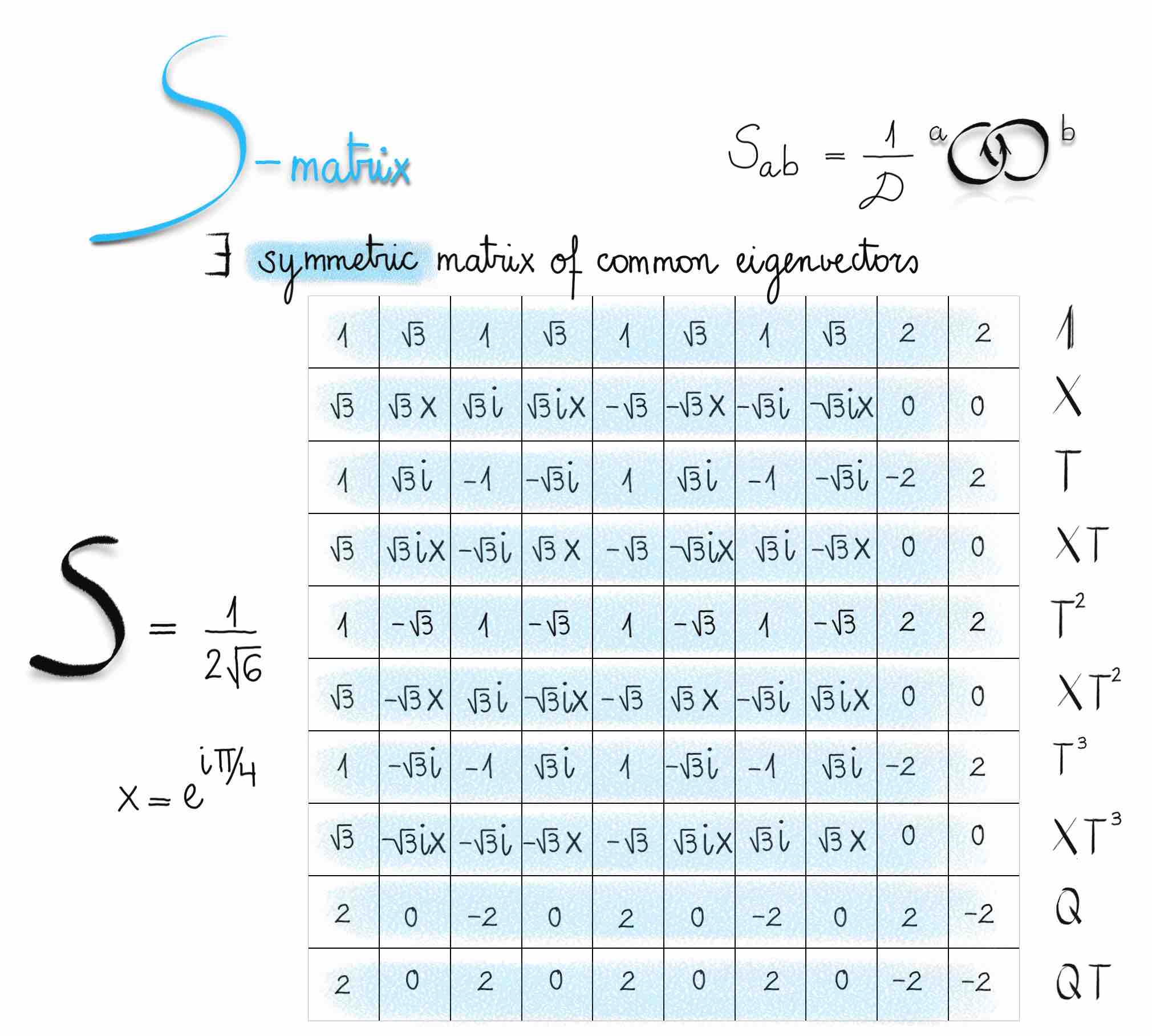}
\end{center}

\newpage
\pagestyle{fancy}
\fancyhf{}
\rhead{$2$ bosons in $4$ sites}
\rfoot{\thepage}
\normalsize
\section*{T-Matrix}
\addcontentsline{toc}{section}{T-Matrix}
The topological $T$-matrix is uniquely determined by equation (\ref{S-T-Relation}), relating the $S$-matrix to the $T$-matrix. 

\vspace*{1.5cm}
\begin{center}
\includegraphics[width=0.9\textwidth]{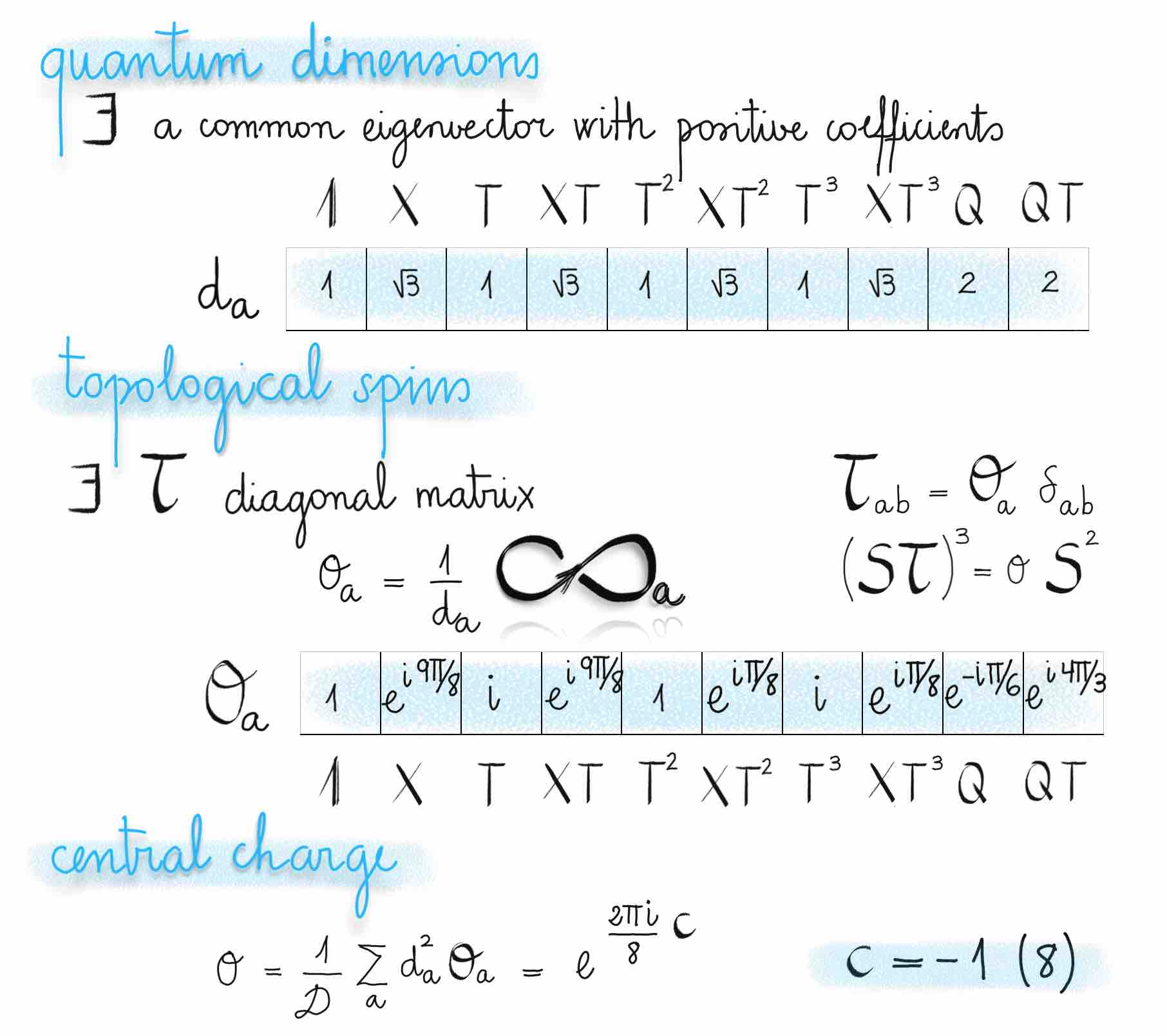}
\end{center}


\newpage
\thispagestyle{empty}
\vspace*{5cm}
\hspace*{1cm}
\includegraphics[width=0.8\textwidth]{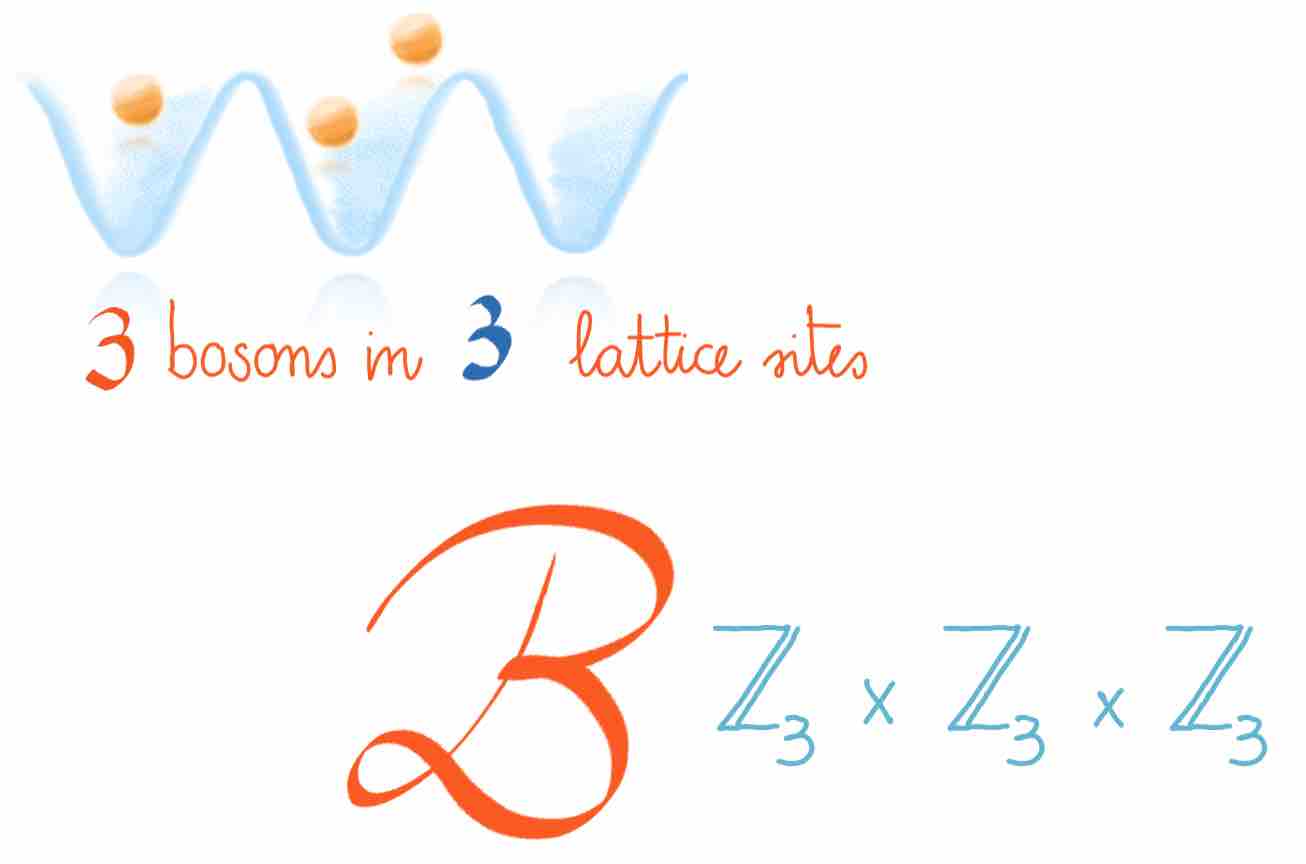}

\newpage
\pagestyle{fancy}
\fancyhf{}
\rhead{Boson-Lattice examples}
\rfoot{\thepage}
\vspace*{3cm}
\section*{}
\addcontentsline{toc}{section}{$3$ bosons in $3$ lattice sites}
\hspace*{1cm}
\includegraphics[width=0.7\textwidth]{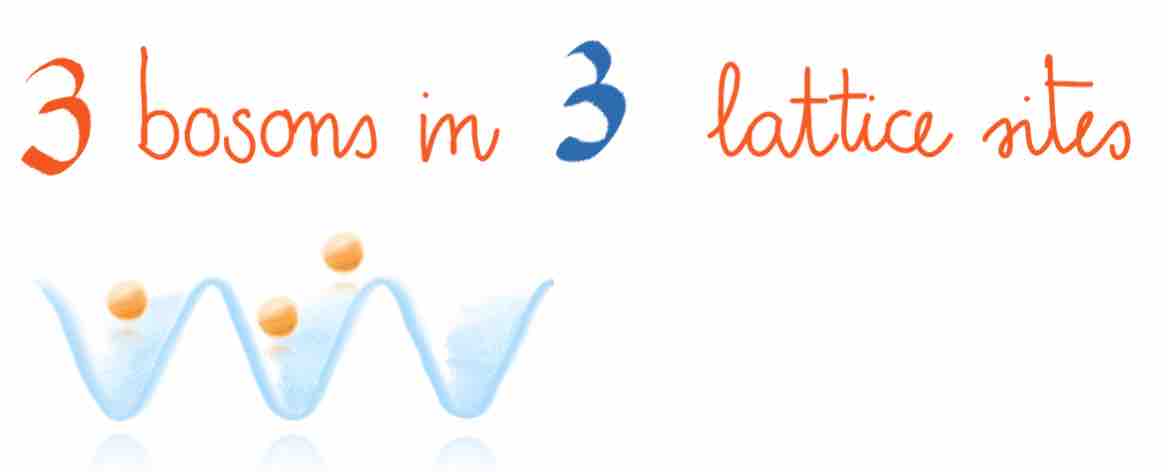}

\large
\hspace*{3cm}
\parbox{9cm}{\parskip=8pt

I analyze the  anyon model corresponding to $3$ bosons in a lattice of $3$ sites.

I identify the Boson-Lattice  graph, construct the topological algebra, and characterize the fusion and braiding rules of the anyon model.

The anyon model corresponding to $3$ bosons in a lattice of $3$ sites is a well defined anyon model, which is to my knowledge {\em not tabulated}.}


\newpage
\pagestyle{fancy}
\fancyhf{}
\lhead{Boson-Lattice examples}
\lfoot{\thepage}
\normalsize
\section*{Boson-Lattice graph}
\addcontentsline{toc}{section}{Boson-Lattice graph}
The Hilbert space of $3$ bosons in $3$ lattice sites has dimension $10$. The corresponding anyon model has therefore $10$ topological charges.
Following the prescription given in the previous section, we construct the Boson-Lattice generating graph $\mathcal{G}$ as depicted below.

\vspace*{1cm}
\begin{center}
\includegraphics[width=0.9\textwidth]{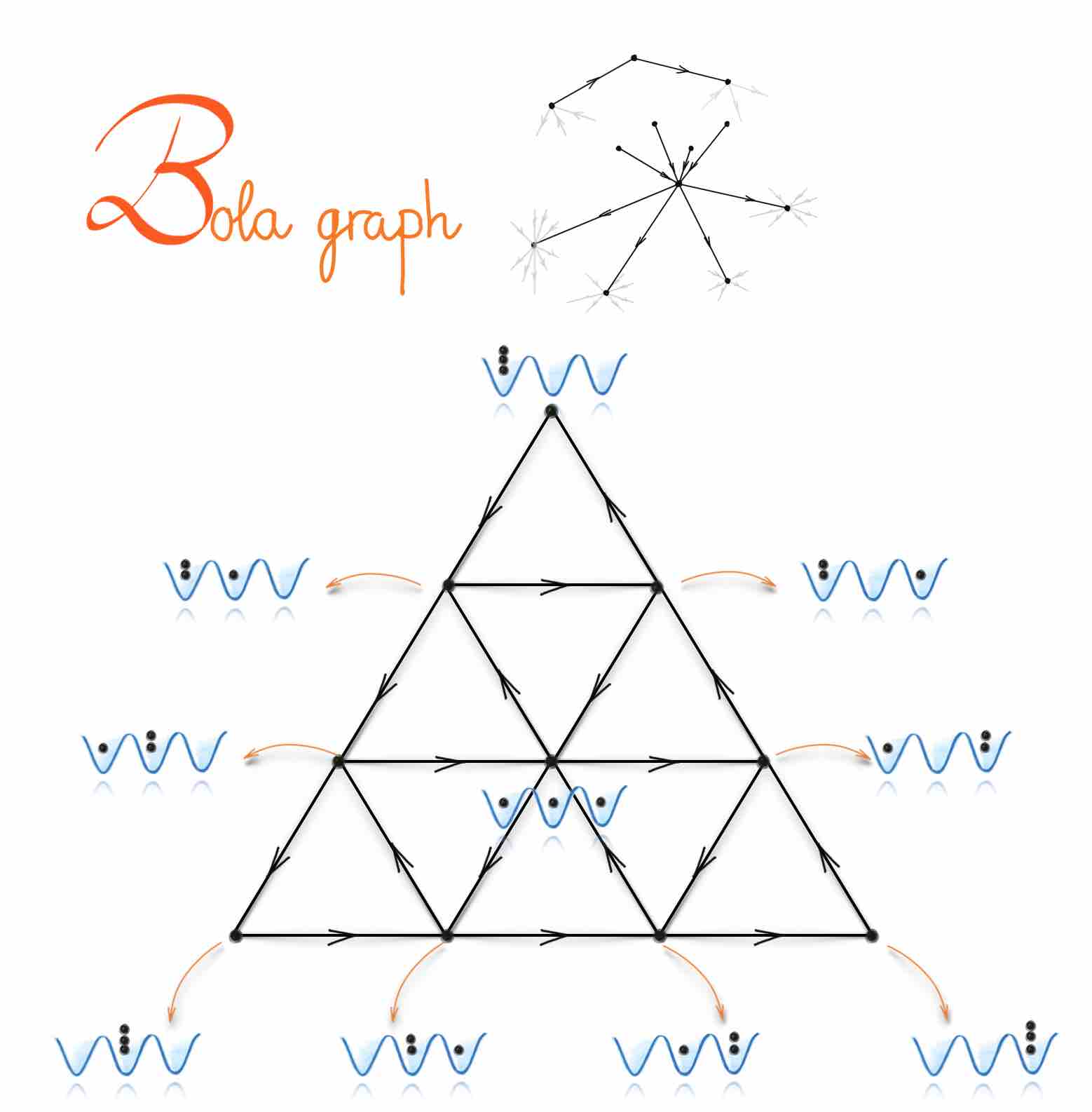}
\end{center}


\newpage
\pagestyle{fancy}
\fancyhf{}
\rhead{$3$ bosons in $3$ sites}
\rfoot{\thepage}
\section*{Topological Algebra}
\addcontentsline{toc}{section}{Topological Algebra}
By inspection of the generating graph $\mathcal{G}$ we can see that the set of polynomials  we are looking for is:
\begin{eqnarray}
\mathcal{A}=\{\mathds{1},T, T^2,X,XT,XT^2,X^\dagger,X^\dagger T,X^\dagger T^2,Q\},
\end{eqnarray}
where the operator $Q$ fulfills:
\begin{eqnarray}
&&Q=XX^\dagger-\mathds{1}\\
&&QT=Q.
\end{eqnarray}

\begin{center}
\includegraphics[width=1.05\textwidth]{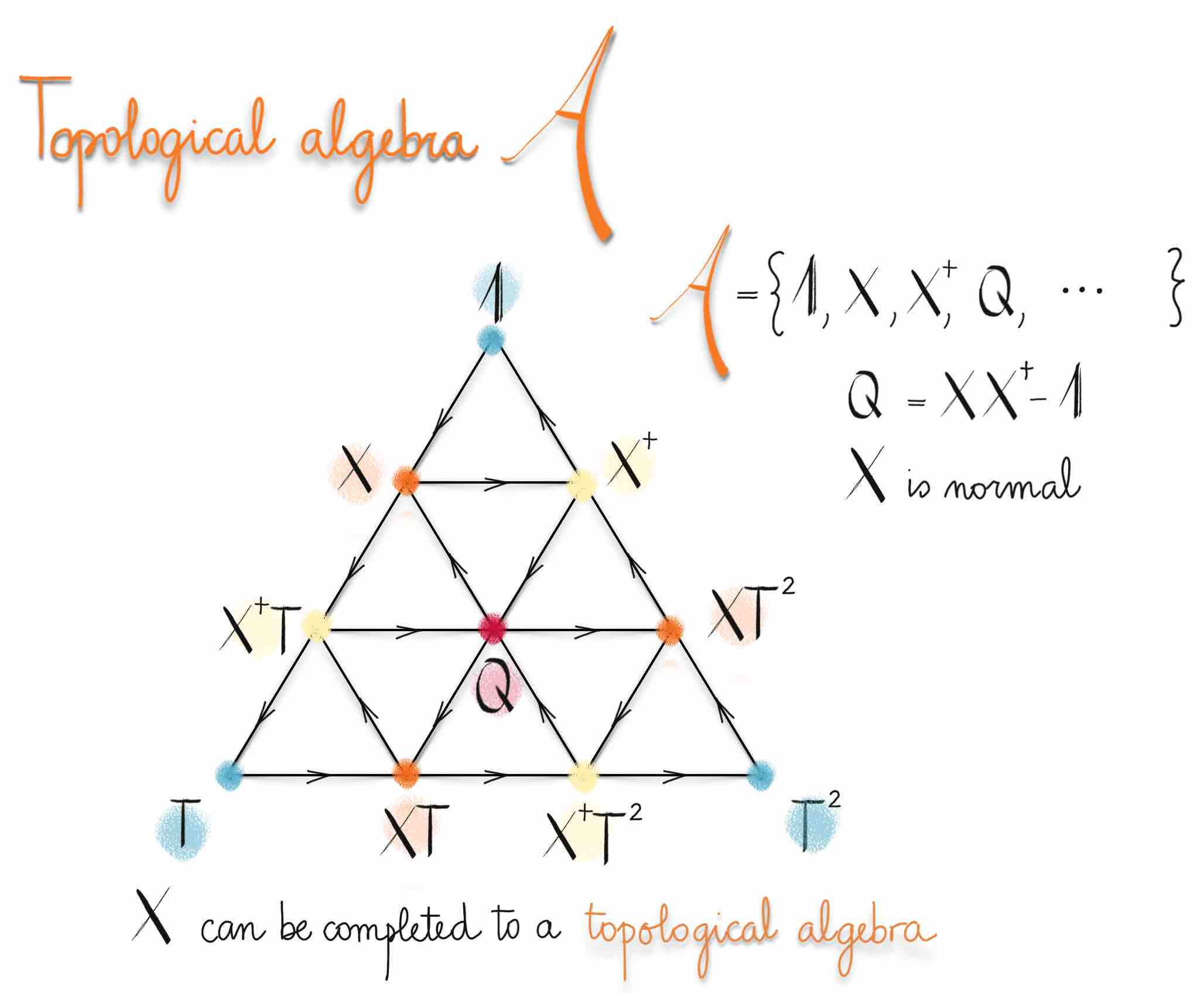}
\end{center}

\newpage
\pagestyle{fancy}
\fancyhf{}
\lhead{Boson-Lattice examples}
\lfoot{\thepage}
\normalsize
\section*{Topological graphs}
\addcontentsline{toc}{section}{Topological graphs}
By composing the graphs of  $X$ and $X^\dagger$ and subtracting the identity graph, we obtain the graph of the operator $Q$.
This is a non-negative graph that decomposes into three independent graphs.

\vspace*{1cm}
\begin{center}
\includegraphics[width=0.75\textwidth]{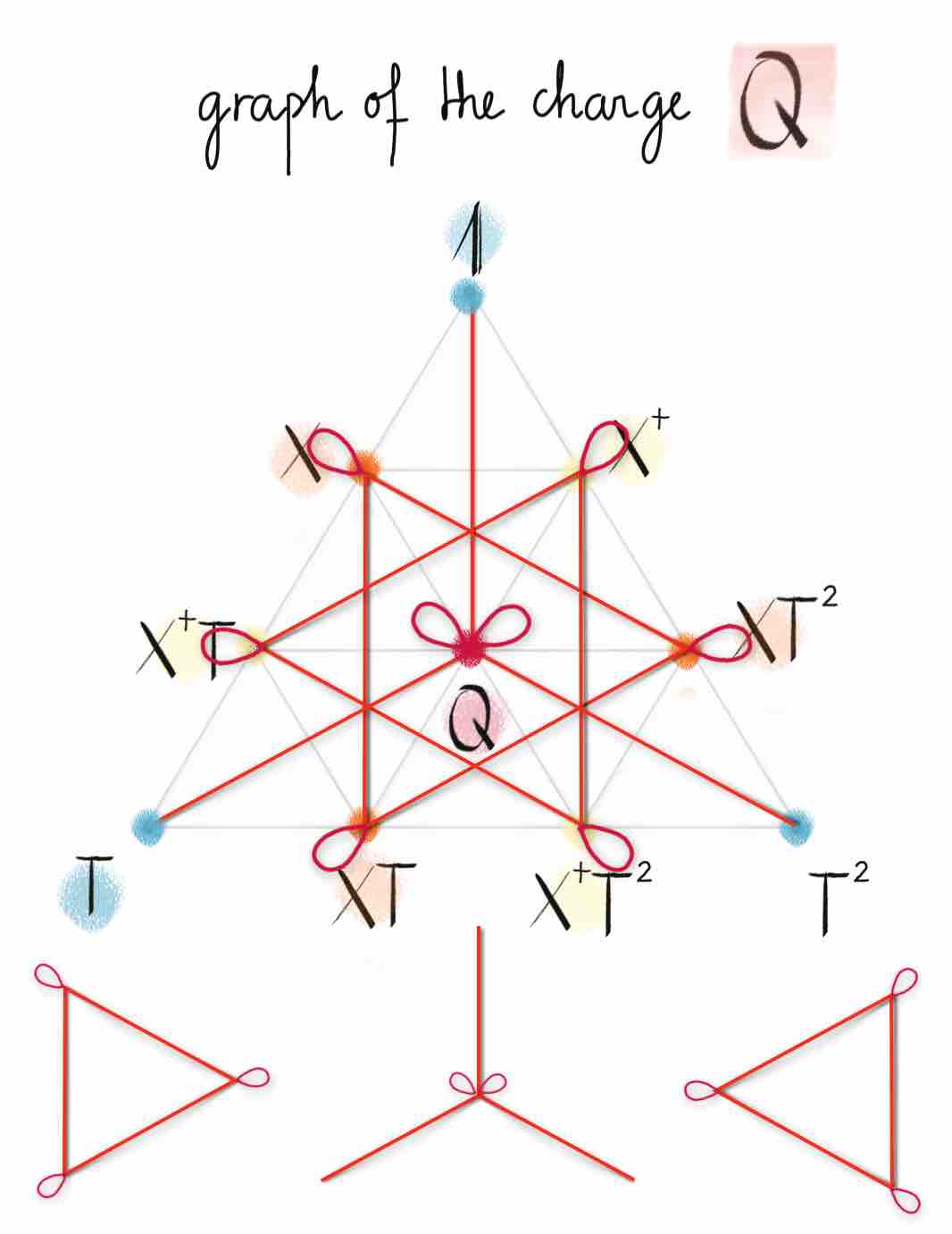}
\end{center}

\newpage
\pagestyle{fancy}
\fancyhf{}
\rhead{$3$ bosons in $3$ sites}
\rfoot{\thepage}
\normalsize
\section*{Fusion rules}
\addcontentsline{toc}{section}{Fusion Rules}

The fusion rules of the model can be directly read from the topological graphs below.
Interestingly, this model has multiplicities larger than $1$, as it can be seen from the double loop in the graph corresponding to the charge $Q$. We have:
\begin{eqnarray}
Q\times Q=1+T+T^2+2Q.
\end{eqnarray}
This model is not tabulated in the tables by Bonderson \cite{Bonderson}, which are restricted to multiplicity-free anyon models.

\vspace*{1cm}
\begin{center}
\includegraphics[width=0.45\textwidth]{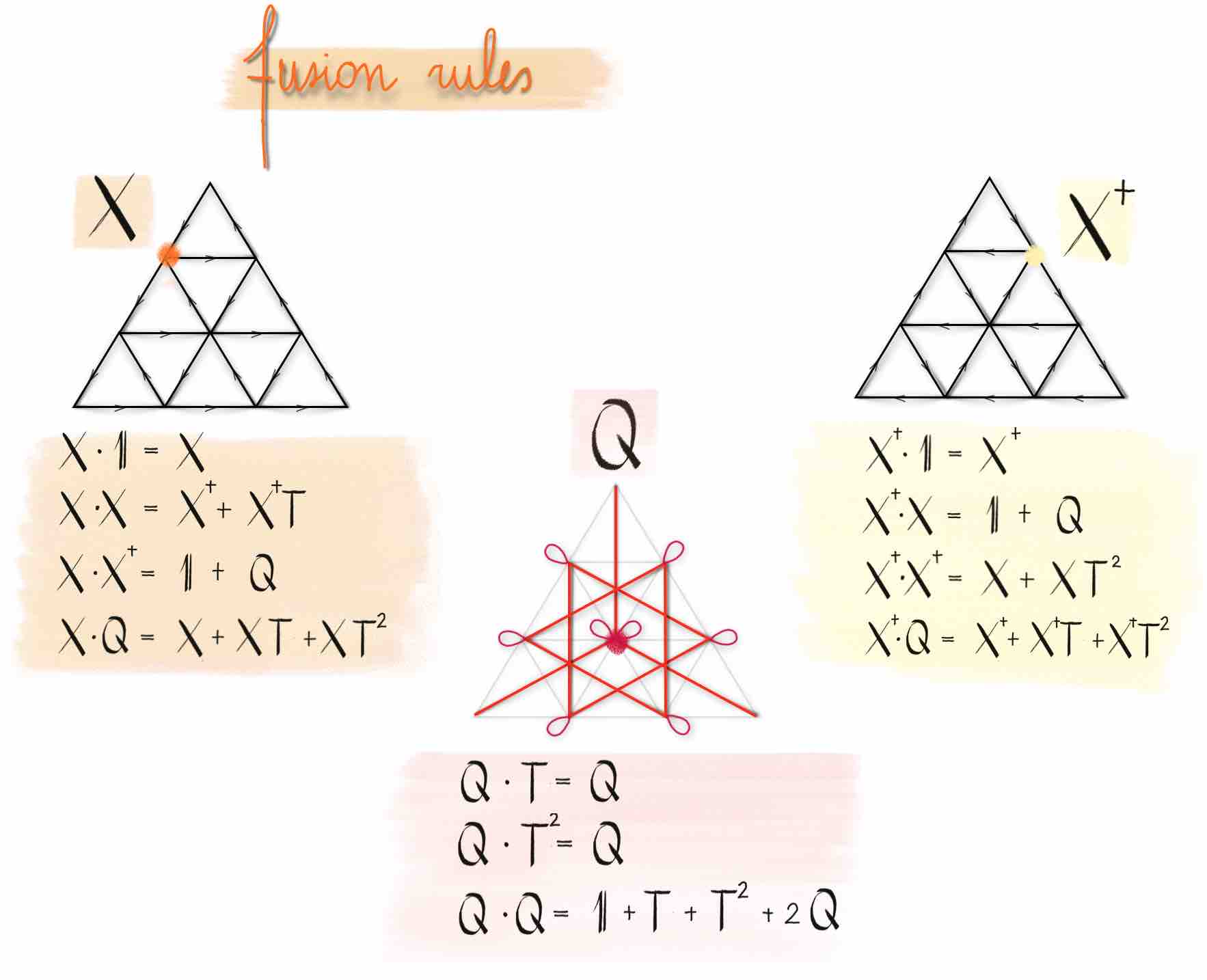}
\end{center}

\newpage
\pagestyle{fancy}
\fancyhf{}
\lhead{Boson-Lattice examples}
\lfoot{\thepage}
\normalsize
\section*{S-Matrix}
\addcontentsline{toc}{section}{S-Matrix}
Diagonalization of the graphs yields a unique (up to conjugation) symmetric and unitary matrix, which defines the $S$-matrix of the anyon model.

\vspace*{1cm}
\includegraphics[width=0.9\textwidth]{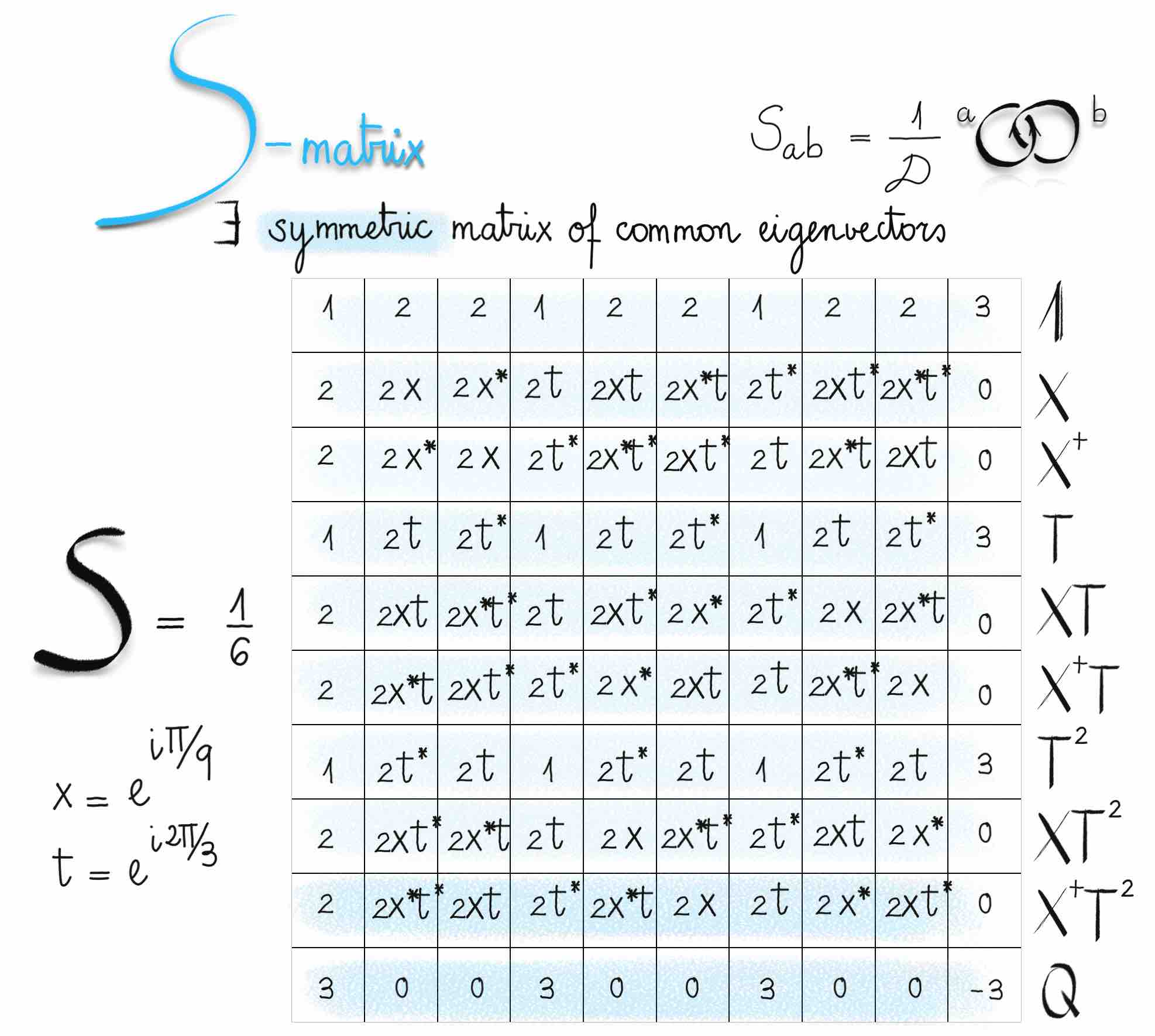}

\newpage
\pagestyle{fancy}
\fancyhf{}
\rhead{$3$ bosons in $3$ sites}
\rfoot{\thepage}
\normalsize
\section*{T-Matrix}
\addcontentsline{toc}{section}{T-Matrix}
The topological $T$-matrix is uniquely determined by equation (\ref{S-T-Relation}), relating the $S$-matrix to the $T$-matrix. 

\vspace*{1cm}
\includegraphics[width=0.9\textwidth]{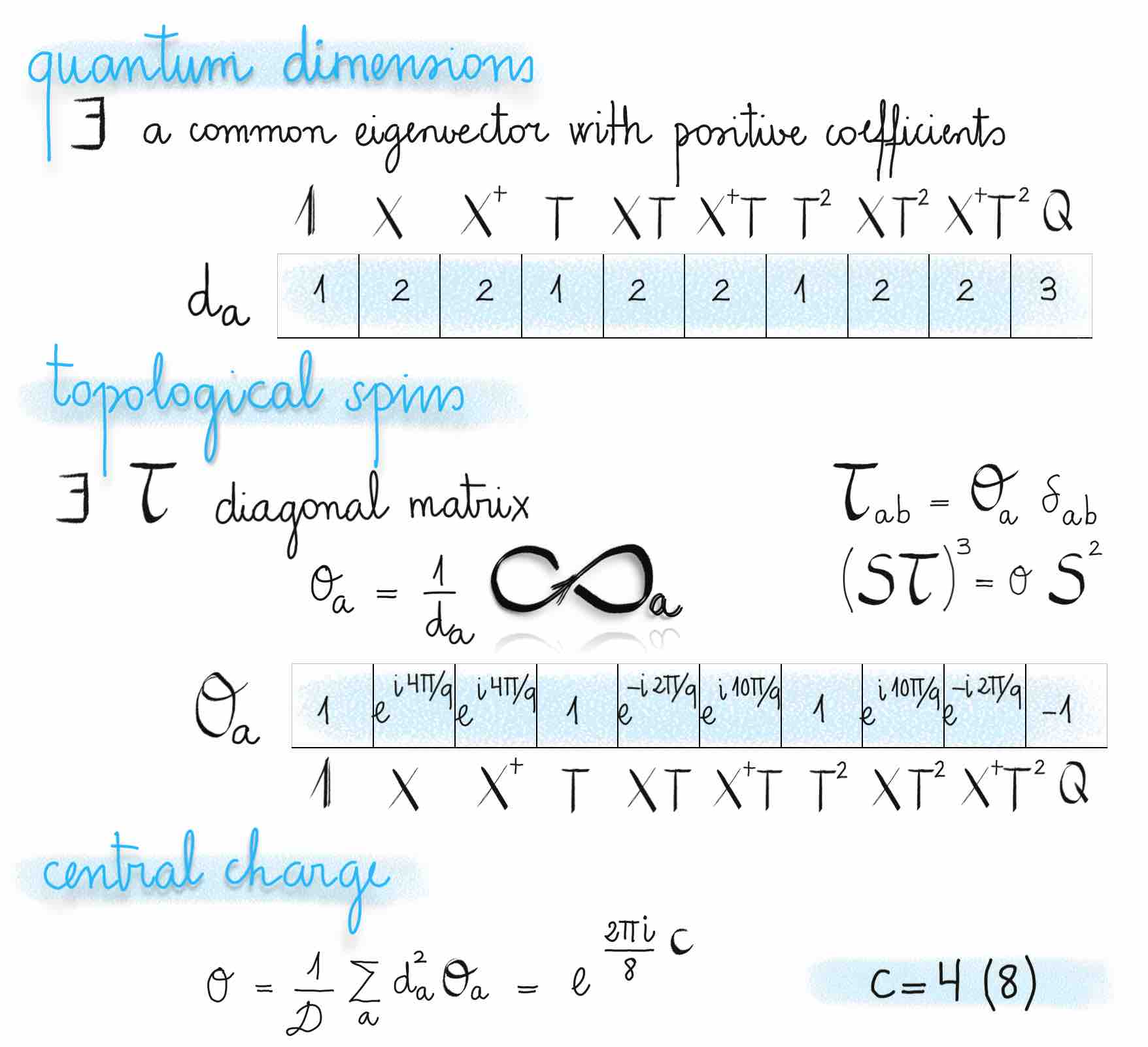}


\newpage
\thispagestyle{empty}
\vspace*{5cm}
\hspace*{1cm}
\includegraphics[width=0.8\textwidth]{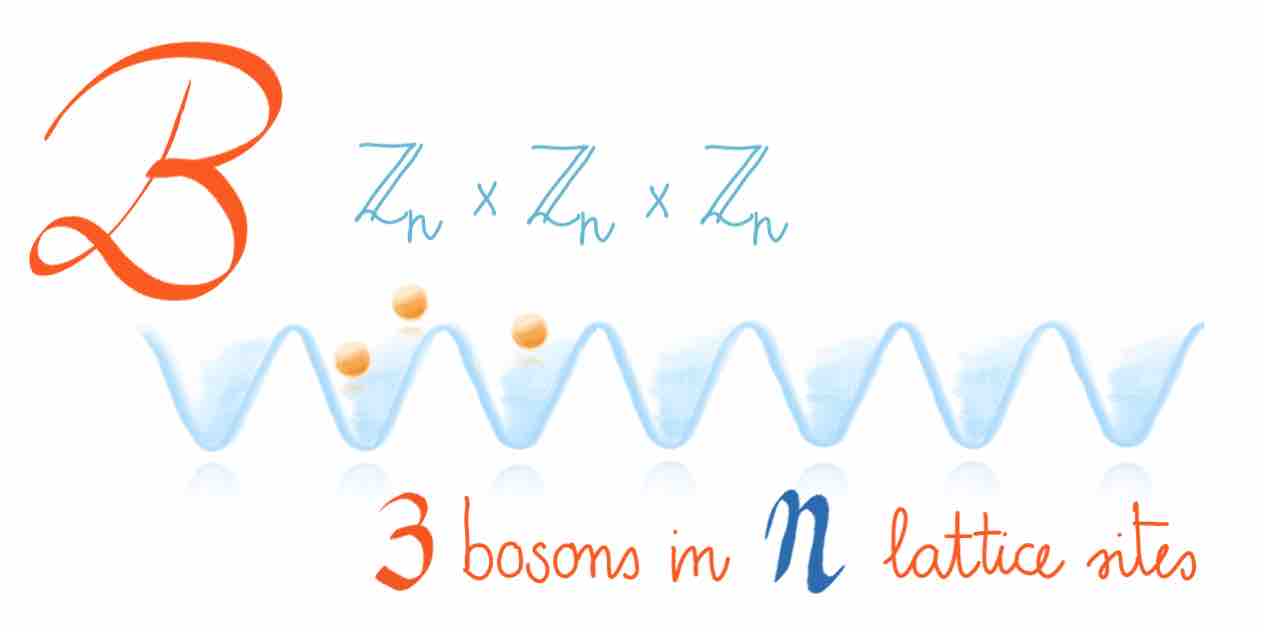}

\newpage
\pagestyle{fancy}
\fancyhf{}
\rhead{Boson-Lattice examples}
\rfoot{\thepage}
\vspace*{3cm}
\section*{}
\addcontentsline{toc}{section}{$3$ bosons in $N$ lattice sites}
\hspace*{1cm}
\includegraphics[width=0.65\textwidth]{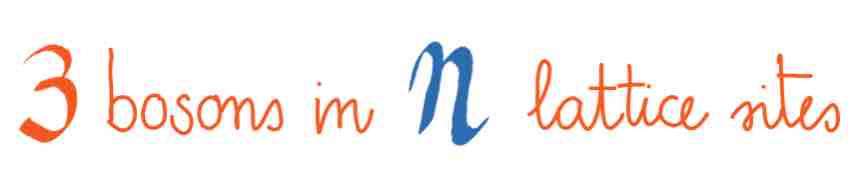}

\large
\hspace*{3cm}
\parbox{9cm}{\parskip=8pt

I analyze the  anyon model corresponding to $3$ bosons in a lattice of $N$ sites.}


\newpage
\pagestyle{fancy}
\fancyhf{}
\lhead{Boson-Lattice examples}
\lfoot{\thepage}
\normalsize
\section*{Boson-Lattice graph}
\addcontentsline{toc}{section}{Boson-Lattice graph}
Following the prescription given in the previous section, we construct the Boson-Lattice generating graph $\mathcal{G}$ as depicted below.
By considering the graph of equivalence classes (a vertex corresponds to the class of Fock states that can be obtained from each other by a global translation) the graph takes the form of a pyramid with a triangular tiling. 

\vspace*{2cm}
\begin{center}
\includegraphics[width=0.9\textwidth]{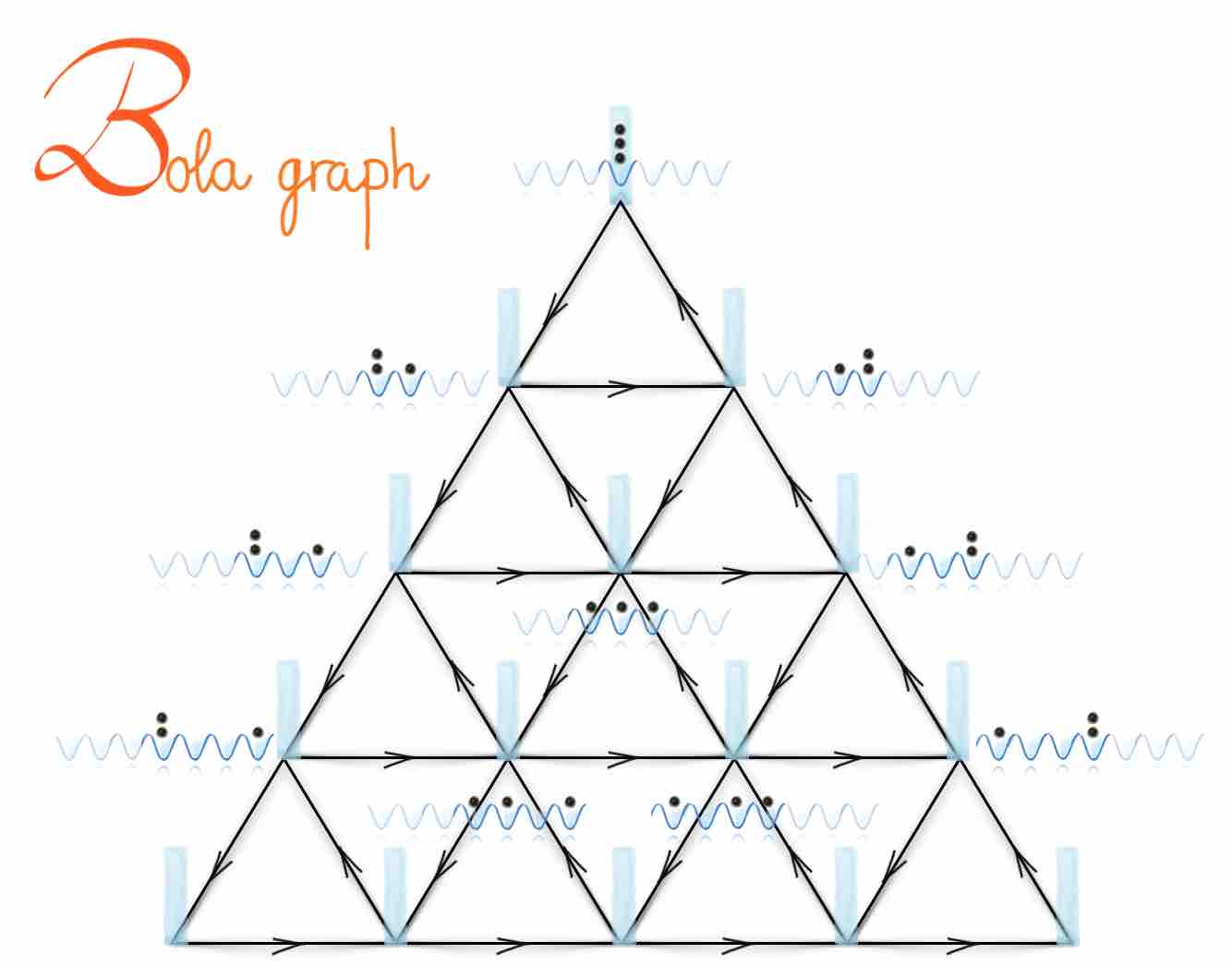}
\end{center}


\newpage
\pagestyle{fancy}
\fancyhf{}
\rhead{$3$ bosons in $n$ sites}
\rfoot{\thepage}
\normalsize
\section*{Topological Algebra}
\addcontentsline{toc}{section}{Topological Algebra}
The pyramidal structure of the graph allows to obtain the topological algebra in a recursive way.
Polynomials corresponding to charges at a certain level of the pyramid are obtained from polynomials at the previous levels as depicted below.

\vspace*{2cm}
\begin{center}
\includegraphics[width=0.92\textwidth]{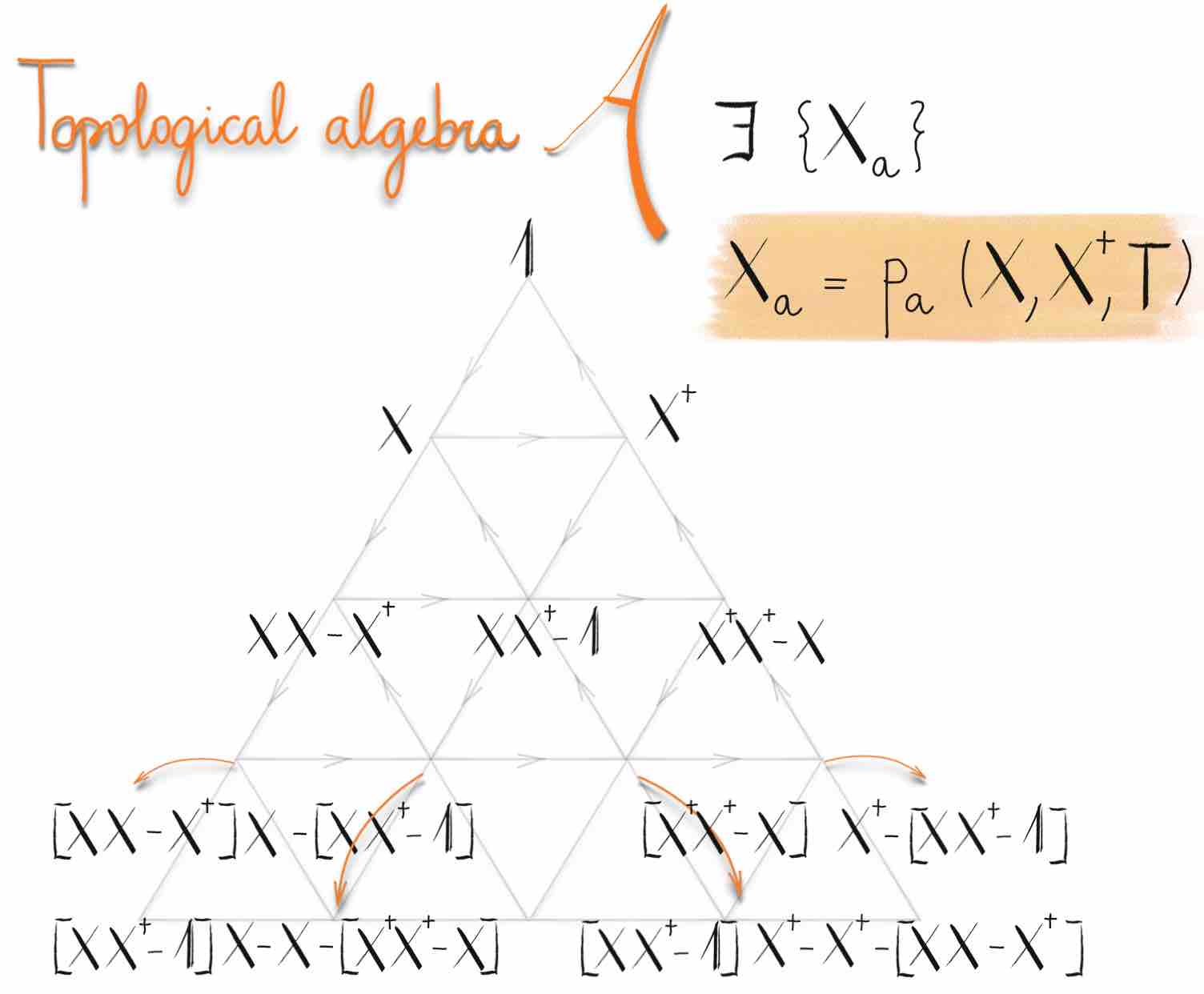}
\end{center}

\newpage
\pagestyle{fancy}
\fancyhf{}
\lhead{Boson-Lattice examples}
\lfoot{\thepage}
\normalsize
\section*{Topological graphs}
\addcontentsline{toc}{section}{Topological graphs}
The pyramidal structure assures that the polynomials correspond to non-negative matrices.
This can be graphically seen by drawing the corresponding graphs. Graphs at a certain level of the pyramid contain the graphs corresponding to the previous level, so that the operators remain non-negative after subtraction of graphs from the upper level. 

\vspace*{1cm}
\begin{center}
\includegraphics[width=0.95\textwidth]{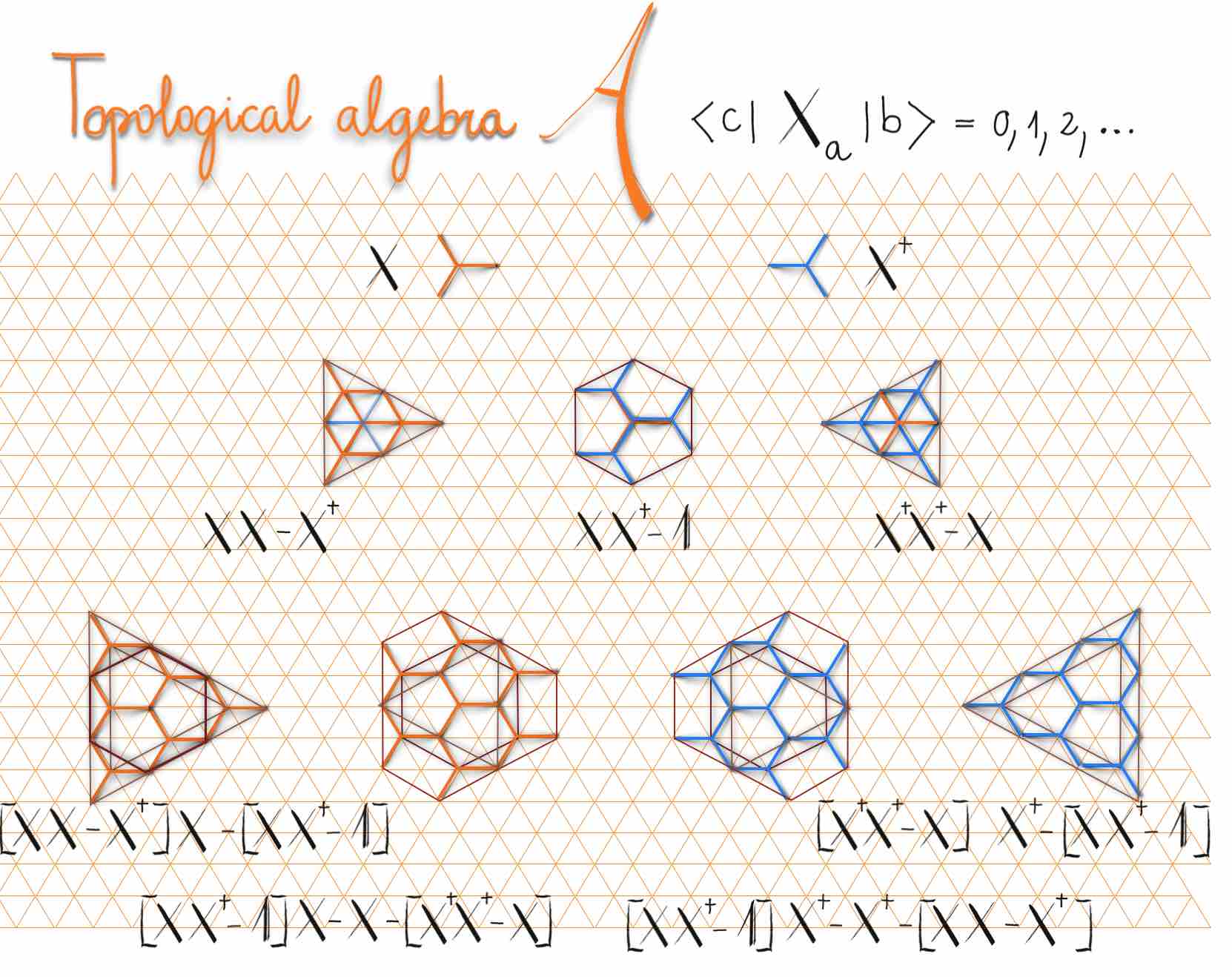}
\end{center}

\newpage
\pagestyle{fancy}
\fancyhf{}
\rhead{$4$ bosons in $n$ sites}
\rfoot{\thepage}
\normalsize
\section*{The general Boson-Lattice graph}
\addcontentsline{toc}{section}{4 Bosons in N Lattice sites}
The Boson-Lattice graphs for $k$ bosons in $n$ lattice sites correspond to multidimensional pyramidal structures. For $4$ bosons in $n$ lattice sites, the graph corresponds to a pyramid with a tetrahedral tiling. 
The pyramidal structure allows for the existence of a topological algebra of polynomials of the operators $X$, $X^\dagger$ and $T$. 

\vspace*{1cm}
\begin{center}
\includegraphics[width=0.9\textwidth]{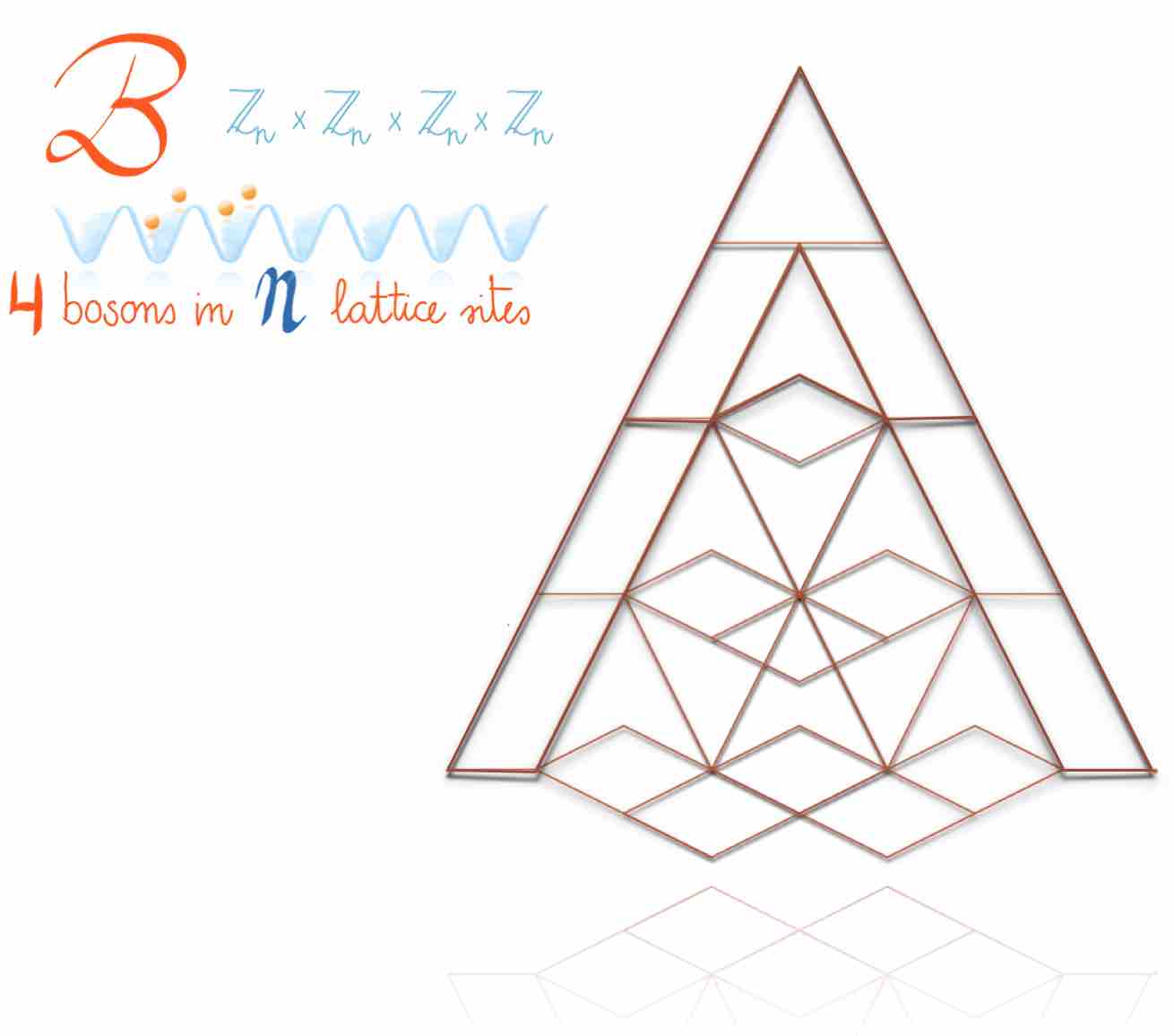}
\end{center}


\newpage
\thispagestyle{empty}
\vspace*{2,5cm}
\begin{center}
\includegraphics[width=0.9\textwidth]{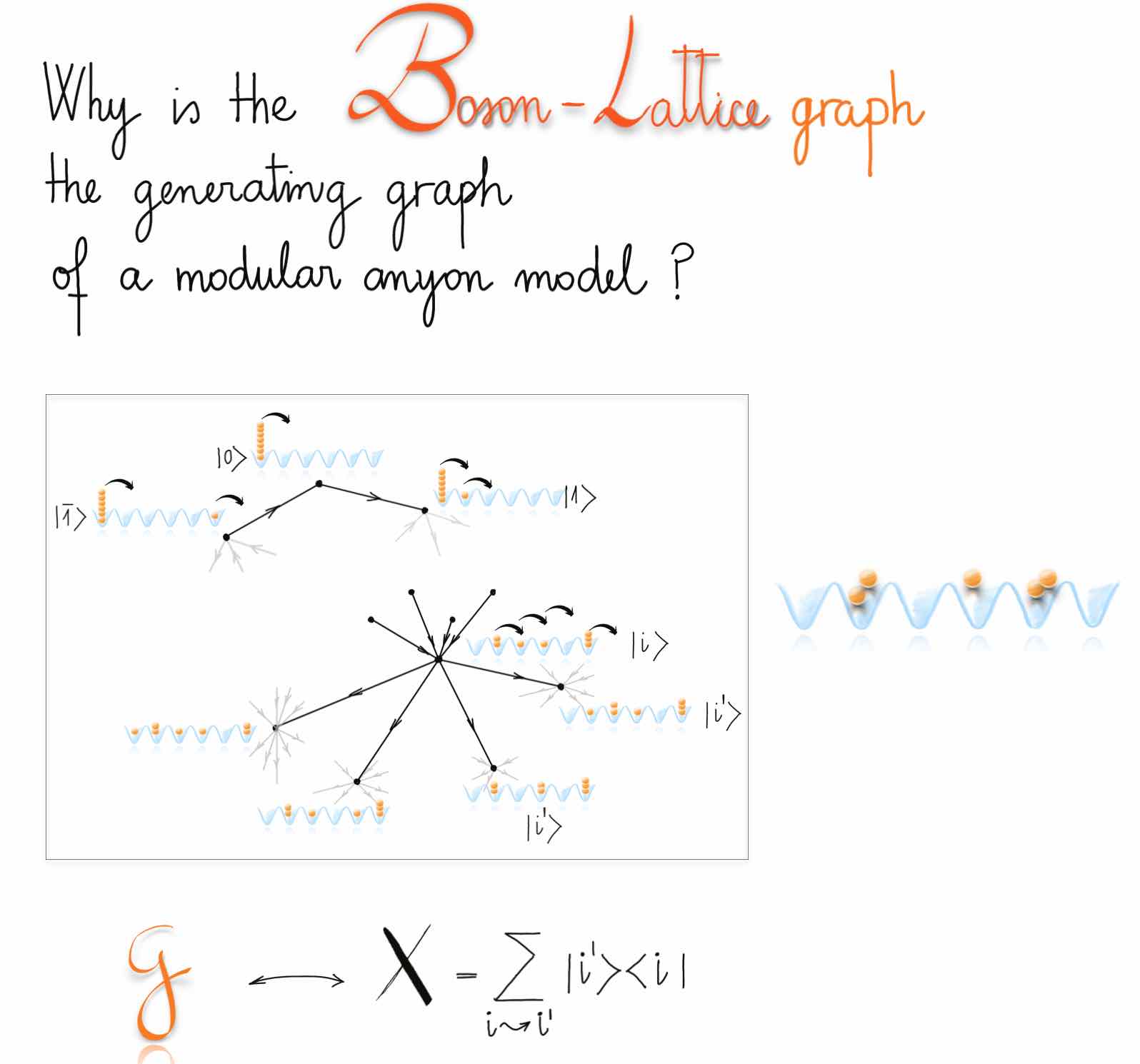}
\end{center}

\newpage
\thispagestyle{empty}
\pagestyle{fancy}
\fancyhf{}
\rhead{Why does the construction work?}
\cfoot{\thepage}
\section*{}
\addcontentsline{toc}{section}{Boson-Lattice construction and gravity}
\vspace*{1cm}
\begin{center}
\includegraphics[width=\textwidth]{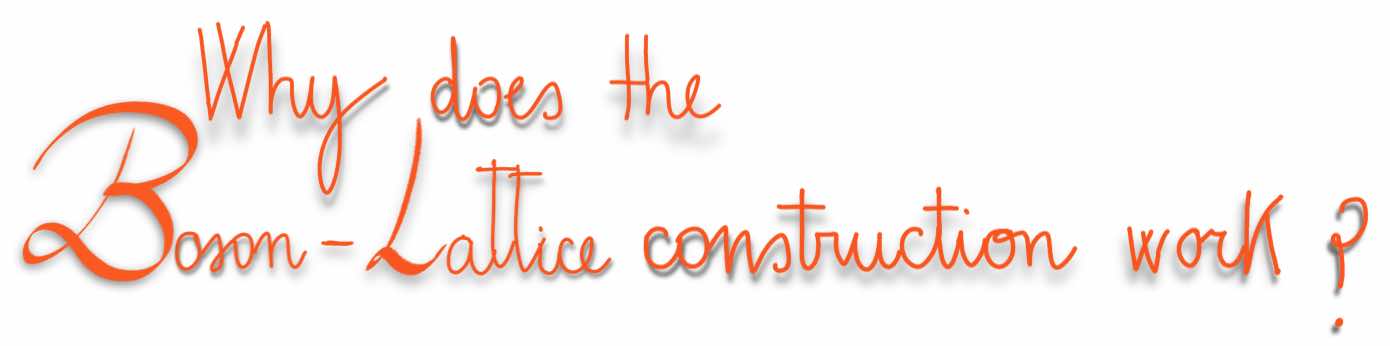}
\end{center}

\large
\hspace*{2cm}
\parbox{11.5cm}{\parskip=8pt

The Boson-Lattice construction succeeds in systematically generating well defined anyon models.
The Boson-Lattice graph is conceived such that 1) it defines a set of well defined fusion rules, and 2) there is always a symmetric choice of eigenvectors of the graph that can represent the $S$-matrix of a modular anyon model.

But why is this always like this? Why does the connectivity graph of Fock states of a bosonic lattice system always have a symmetric matrix of eigenvectors? Beyond possible mathematical general proofs that I can give, I am here interested in finding an intuitive argument able to explain this fact.

To this aim I take a closer look into the generating Boson-Lattice operator. I discover an illuminating connection between Boson-Lattice graphs and curved-space geometries.
This connection throws light into the success of the construction. 

But even more, this connection anticipates a beautiful duality between anyon models and curved space-time geometries, between anyon models and gravity.}


\newpage
\pagestyle{fancy}
\fancyhf{}
\lhead{Why does the construction work?}
\lfoot{\thepage}

\section*{Tunneling in a curved space}
\addcontentsline{toc}{section}{Tunneling in a curved space}

\normalsize
The generating operator corresponding to the Boson-Lattice graph has the form:
\begin{eqnarray}
X=\sum_{i^{\vphantom{\prime}} \leadsto i^\prime}\ket{i^\prime} \bra{i^{\vphantom{\prime}}}=\sum_{x}A^\dagger_{x+1}A^{\vphantom{\dagger}}_x,
\end{eqnarray}
where $i \leadsto i^\prime$ indicates that Fock state $\ket{i^\prime}$ can be obtained from $\ket{i}$ by tunneling of one particle, and the operators
$A^{\vphantom{\dagger}}_x$ are defined in Eq. \ref{CreationOperators}.

Though the operator $X$ connects Fock states that are connected by one-particle tunneling, $X$ is a many-body operator different from the one-particle tunneling operator:
\begin{eqnarray}
X\ne\sum_{x}a^\dagger_{x+1}a_{x}^{\vphantom{\dagger}}.
\end{eqnarray}
But how is then $X$ written in terms of bosonic creation and annihilation operators $a_x$? It is not difficult to see that the operator $A_x$ can be written as:
\begin{eqnarray}
A_x=\frac{1}{\sqrt{n_x+1}}\,a_x,
\end{eqnarray}
where $n_x=a^\dagger_{x}a_{x}^{\vphantom{\dagger}}$ is the density operator at site $x$.
\begin{center}
\includegraphics[width=\textwidth]{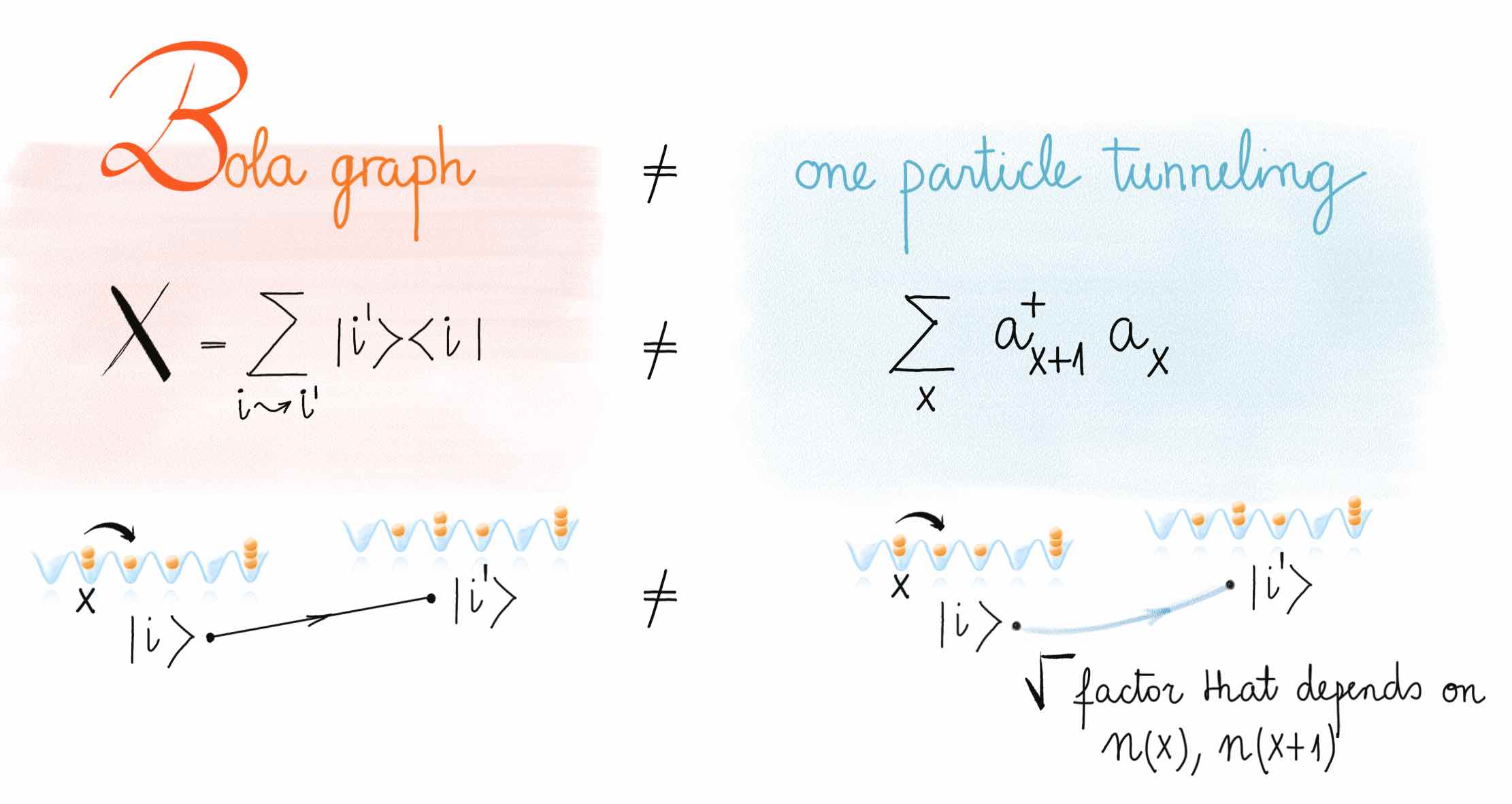}
\end{center}

\newpage
\pagestyle{fancy}
\fancyhf{}
\rhead{Tunneling in a curved space}
\rfoot{\thepage}
This means that the generating operator has the form:
\begin{eqnarray}
X=\sum_{x}a^\dagger_{x+1}\,g_x\,a_{x}^{\vphantom{\dagger}},
\end{eqnarray}
where
\begin{eqnarray}
g_x=\frac{1}{\sqrt{n_x+1}}\frac{1}{\sqrt{n_{x+1}+1}}.
\end{eqnarray}
This reveals that the Boson-Lattice graph is a {\bf correlated tunneling} operator: the tunneling amplitude for a particle to hope from site $x$ to site $x+1$ depends on the density of particles at sites $x$ and $x+1$. Effectively, {\em a particle tunnels in a space that has been curved by the presence (by the mass) of the other particles}. The metric in that curved space is given by $g_x$. 

\vspace*{1cm}
\begin{center}
\includegraphics[width=\textwidth]{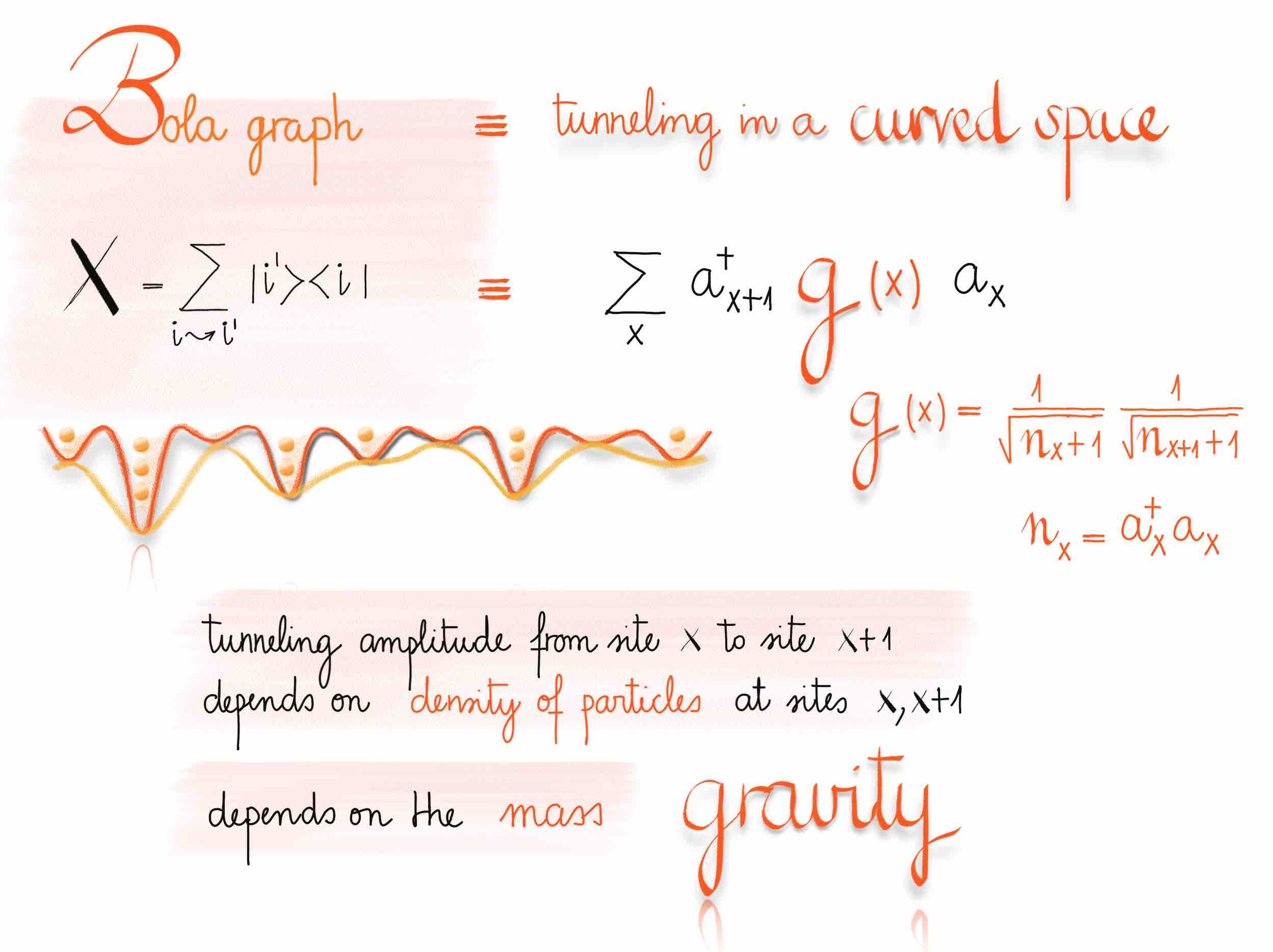}
\end{center}

\newpage
\pagestyle{fancy}
\fancyhf{}
\lhead{Why does the construction work?}
\lfoot{\thepage}
\vspace*{2,5cm}
\begin{center}
\includegraphics[width=\textwidth]{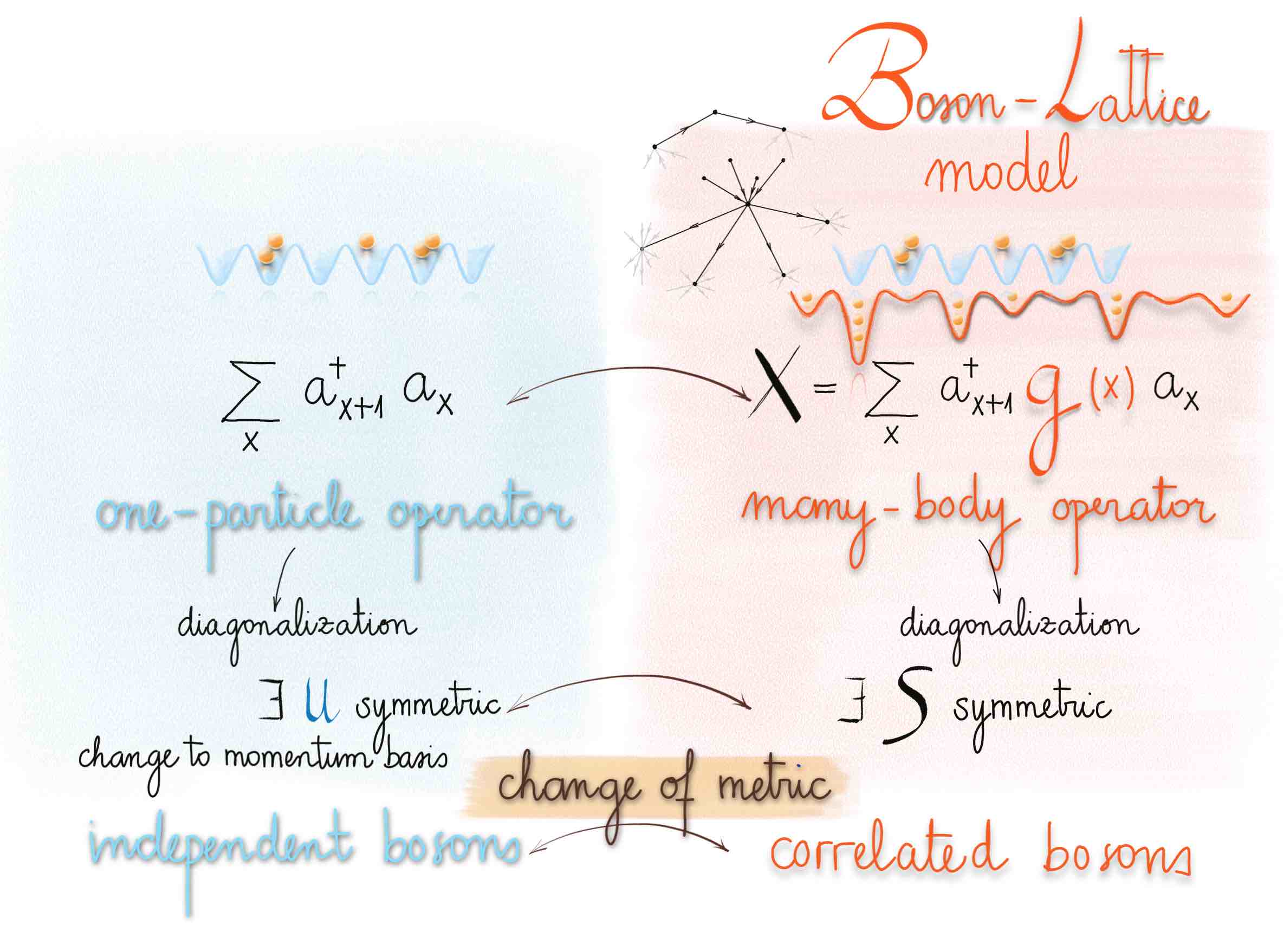}
\end{center}

\newpage
\pagestyle{fancy}
\fancyhf{}
\rhead{Flat vs. curved space}
\rfoot{\thepage}
\vspace*{1,6cm}
\section*{Flat space vs. curved space}
\addcontentsline{toc}{section}{Flat space versus curved space}

The connection above sheds light on the success of the Boson-Lattice construction.
A Boson-Lattice graph can be obtained from a one-particle tunneling operator by deforming the metric from flat to curved.
\begin{eqnarray}
\sum_{x}a^\dagger_{x+1}a_{x}^{\vphantom{\dagger}}\longleftrightarrow \sum_{x}a^\dagger_{x+1}g_xa_{x}^{\vphantom{\dagger}}.
\end{eqnarray}
It is clear that for the one-particle tunneling operator there exists a symmetric unitary matrix $U$ of eigenvectors. This corresponds to the Fock basis in momentum space:
\begin{eqnarray}
U_{ab}=\braket{a|\widetilde{b}}=\braket{\widetilde{a}| b},
\end{eqnarray}
where
\begin{eqnarray}
\ket{a}&\equiv&\ket{n_0,\cdots,n_{n-1}}
\propto[a^\dagger_0]^{n_0}\cdots [a^\dagger_{n-1}]^{n_{n-1}}\ket{\text{vac}}\\
\ket{\widetilde{b}}&\equiv&\ket{\widetilde{n}_0,\cdots,\widetilde{n}_{n-1}}\propto
[\widetilde{a}^\dagger_0]^{\widetilde{n}_0}\cdots [\widetilde{a}^\dagger_{n-1}]^{\widetilde{n}_{n-1}}\ket{\text{vac}},
\end{eqnarray}
with 
\begin{eqnarray}
\widetilde{a}_q=\frac{1}{\sqrt{n}}\sum_{x=0}^{n-1} e^{i\frac{2\pi}{n}q\cdot x}a_x.
\end{eqnarray}
This strongly suggests the existence of a unitary symmetric matrix of eigenvectors of the operator $X$, which is obtained after the metric transformation from flat to curved space:
\begin{eqnarray}
U_{ab}\longleftrightarrow S_{ab}.
\end{eqnarray}


\newpage
\thispagestyle{empty}
\vspace*{2,5cm}
\begin{center}
\includegraphics[width=1.05\textwidth]{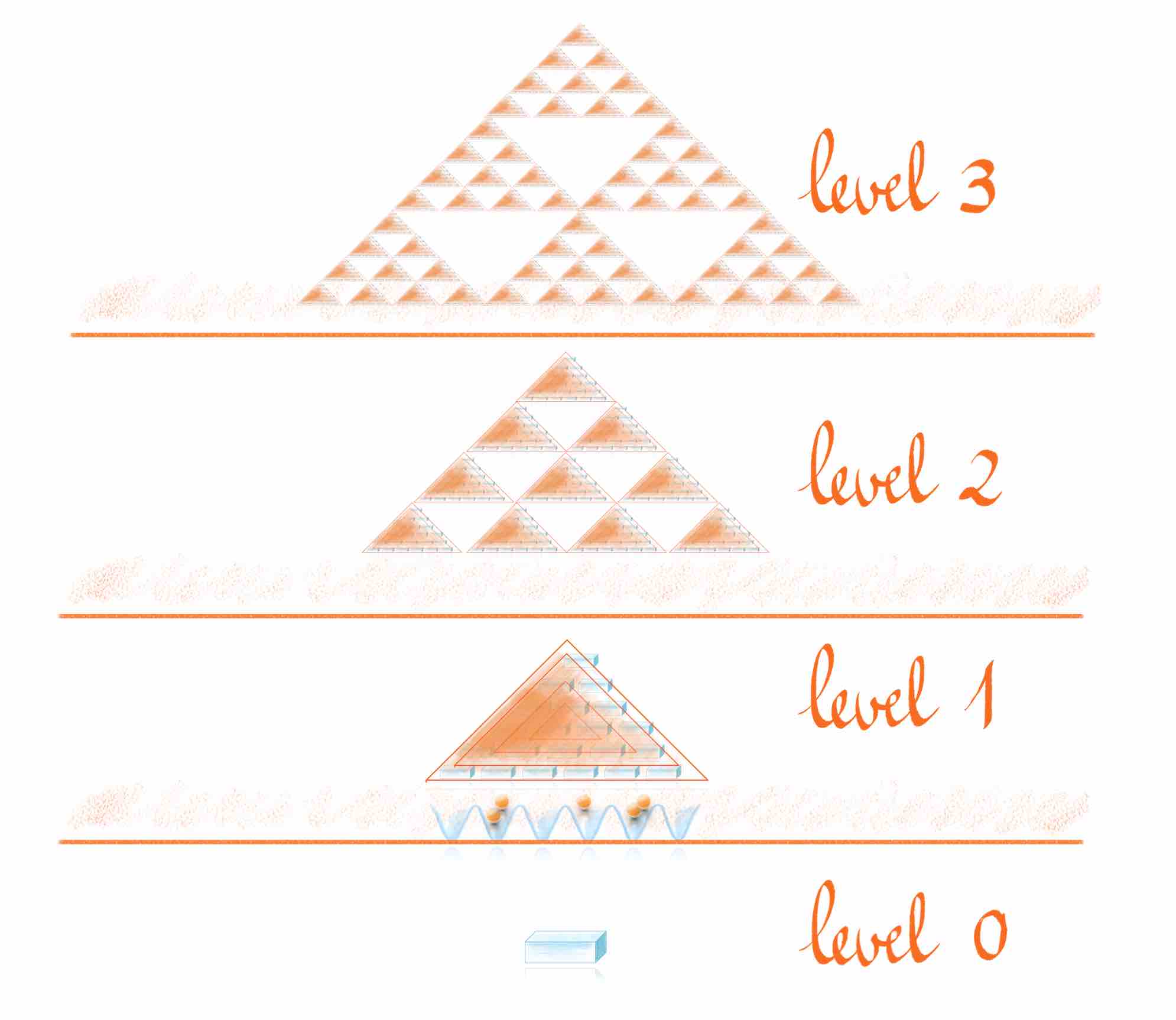}
\end{center}

\newpage
\thispagestyle{empty}
\pagestyle{fancy}
\fancyhf{}
\rhead{Higher levels of the construction}
\rfoot{\thepage}
\section*{}
\addcontentsline{toc}{section}{Higher levels of the construction}
\hspace*{1cm}
\includegraphics[width=0.9\textwidth]{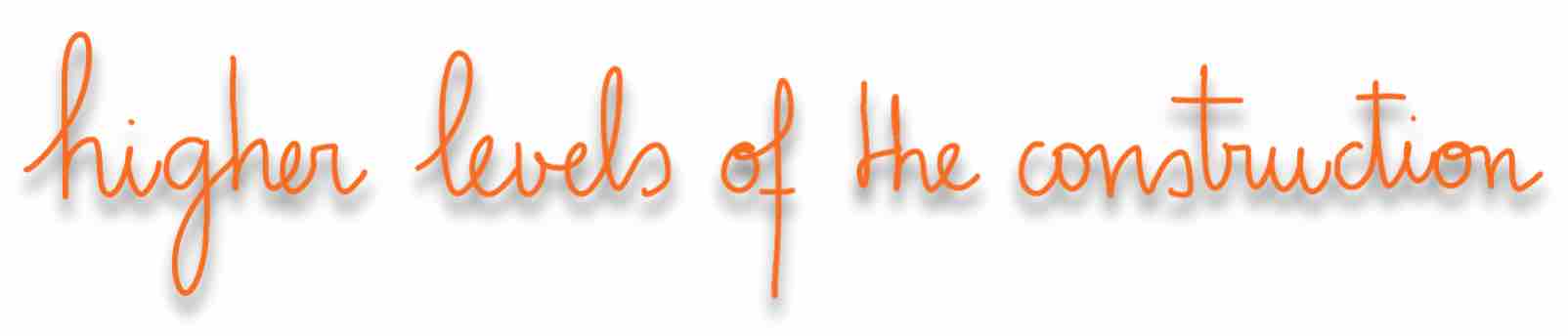}

\hspace*{2.5cm}
\parbox{11.5cm}{\parskip=8pt The Boson-Lattice construction is a fractal construction. Anyon models arising from assembling the building blocks $\mathbb{Z}_n$ can be used themselves as elementary pieces to generate more complex anyon models at a second level of the construction. 

Remarkably, the principle of assembly is the same at any level of the construction. An anyon model generated at a certain level is identified with a particle in a lattice given by the corresponding Boson-Lattice graph. Anyon models at the next level are then constructed by assembling bosons in that graph.
The bosonization principle of assembly generates well defined anyon models at any level of the construction.}

\begin{center}
\includegraphics[width=0.65\textwidth]{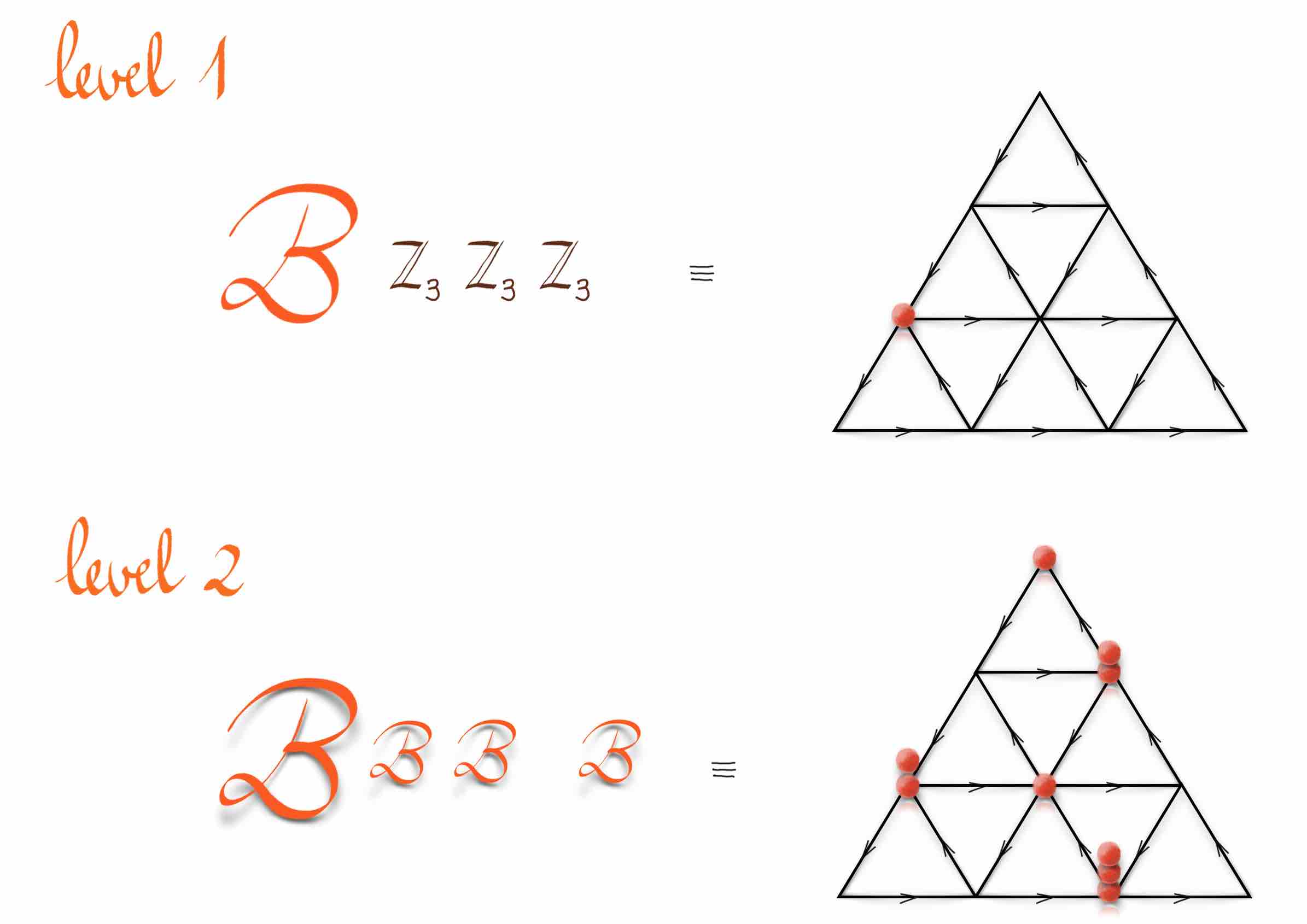}
\end{center}

\newpage
\thispagestyle{empty}
\pagestyle{fancy}
\fancyhf{}
\lhead{Higher levels of the construction}
\lfoot{\thepage}
\vspace*{1,5cm}
\begin{center}
\includegraphics[width=\textwidth]{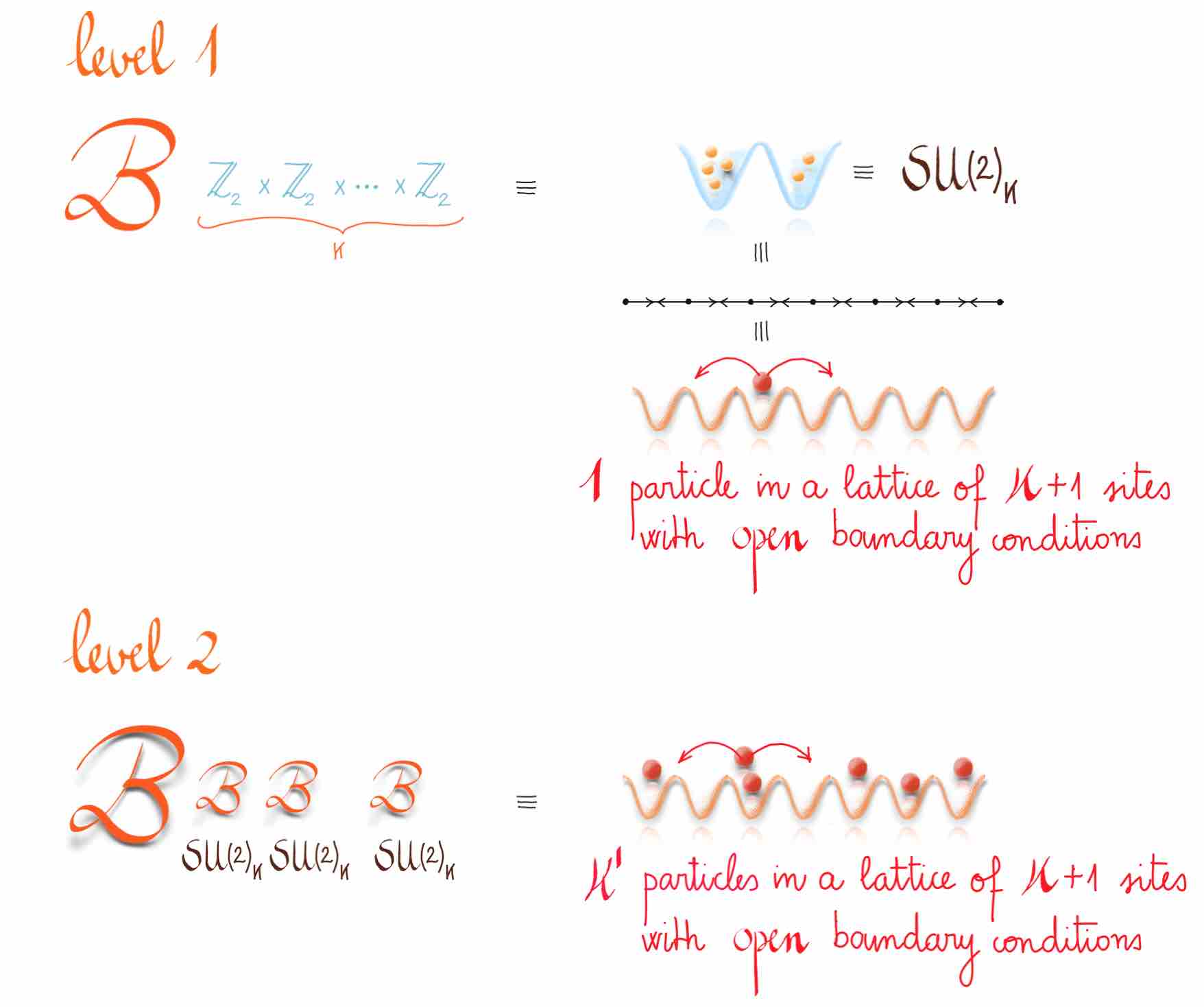}
\end{center}

\newpage
\pagestyle{fancy}
\fancyhf{}
\rhead{Higher levels of the construction}
\rfoot{\thepage}
\vspace*{1cm}
\section*{Second level of Boson-Lattice construction}
\addcontentsline{toc}{section}{Boson-Lattice construction at the second level}
To illustrate how the construction works at higher levels, let me consider the case in which the anyon models $\mathbf{SU}(2)_k$ are used as building blocks. 
The Boson-Lattice graph encoding the anyon model $\mathbf{SU}(2)_k$ is a one-dimensional lattice of $k+1$ sites with open boundaries. 
Following the bosonization principle, the Boson-Lattice graph of the model obtained by assembling $k^\prime$ copies of $\mathbf{SU}(2)_k$ corresponds to the connectivity graph of Fock states of $k^\prime$ bosons in a one-dimensional lattice of $k+1$ sites with open boundaries. 

For example, the anyon model resulting from assembling two identical copies of $\mathbf{SU}(2)_2$ is described by the Hilbert space of $2$ bosons in a lattice of $3$ sites with open boundary conditions.
This Hilbert space has dimension $6$. The corresponding anyon model has therefore $6$ topological charges.
The Boson-Lattice graph is constructed following the connectivity pattern prescription.
The fusion rules and braiding rules can be obtained. They exactly correspond to the anyon model $\mathbf{SO}(5)_2$.

By varying the number $k^\prime$ of $\mathbf{SU}(2)_k$ building blocks the whole tower of anyon models $\mathbf{SO}(5)_{k^\prime}$ can be constructed.

\begin{center}
\includegraphics[width=0.85\textwidth]{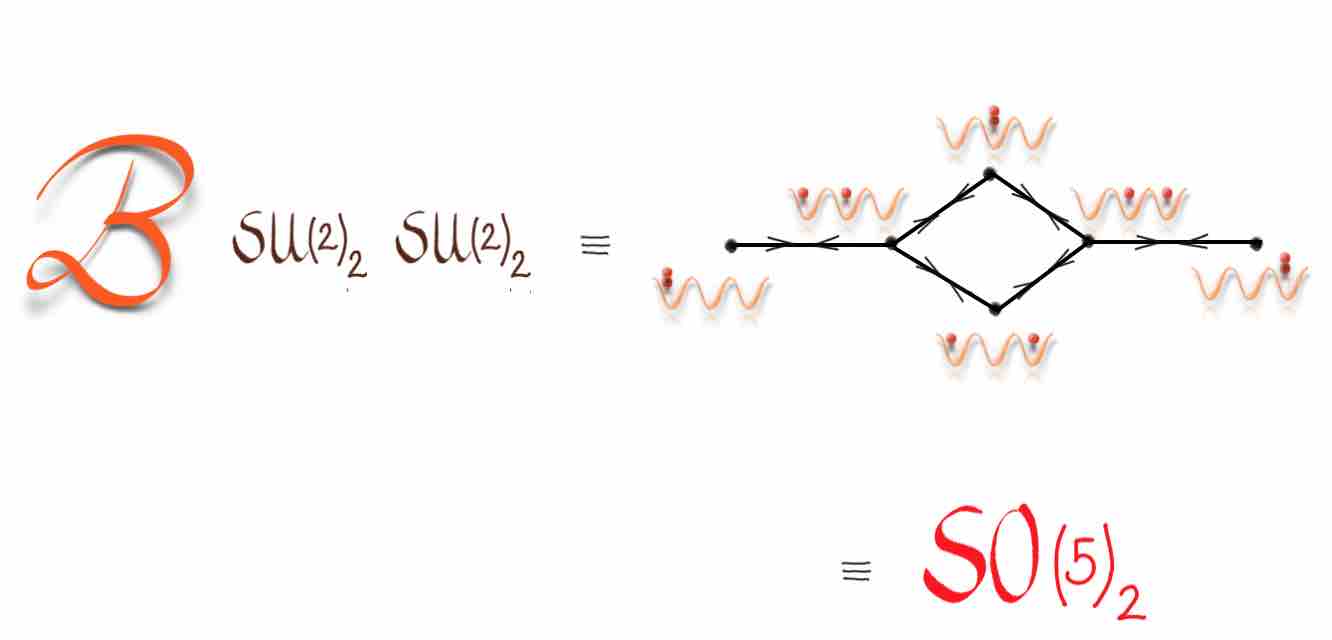}
\end{center}


\newpage
\thispagestyle{empty}
\vspace*{2,5cm}
\begin{center}
\includegraphics[width=\textwidth]{IntroGraph.jpg}
\end{center}

\newpage
\thispagestyle{empty}
\pagestyle{fancy}
\fancyhf{}
\rhead{Closures and openings}
\rfoot{\thepage}
\vspace*{1cm}
\section*{}
\begin{center}
\includegraphics[width=0.5\textwidth]{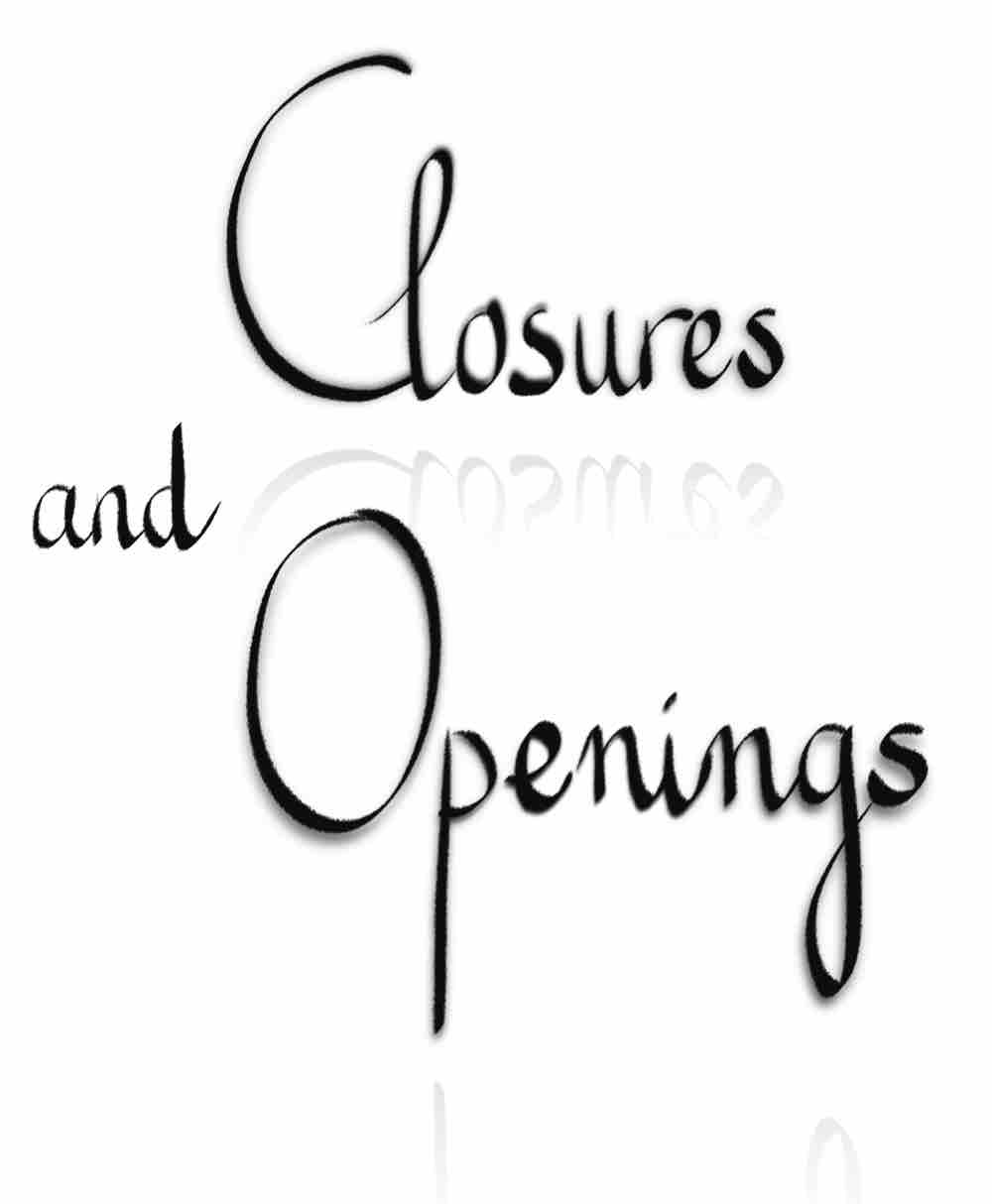}
\end{center}
\addcontentsline{toc}{section}{Closures and openings}


\newpage
\thispagestyle{empty}
\vspace*{2cm}
\begin{center}
\includegraphics[width=0.85\textwidth]{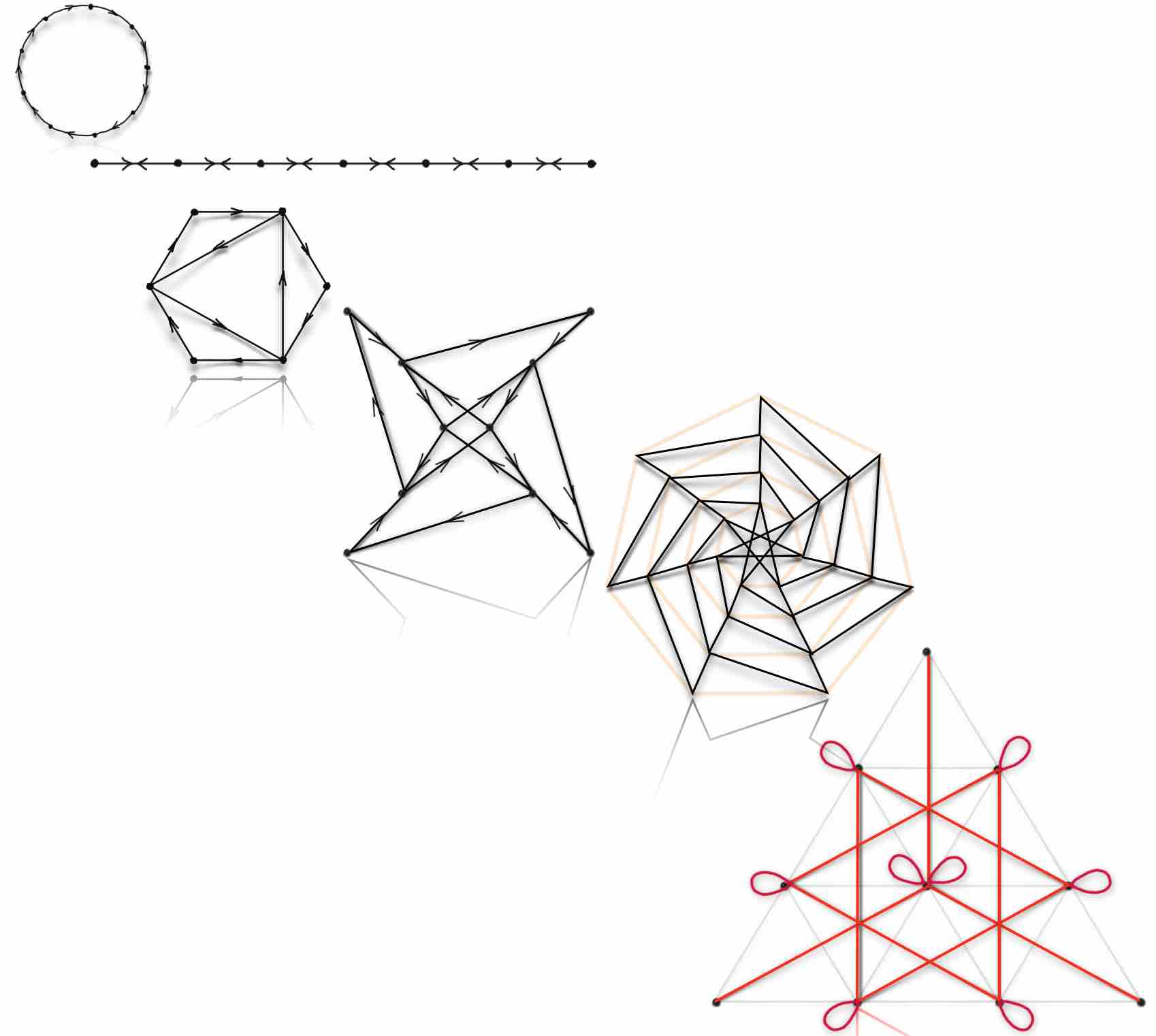}
\end{center}

\newpage
\pagestyle{fancy}
\fancyhf{}
\rhead{Closures and openings}
\rfoot{\thepage}
\vspace*{2cm}
\subsection*{The language of topological graphs}
\addcontentsline{toc}{section}{The language of graphs}

The language of topological graphs I have developed provides a visual and enlightening way to encode anyon models.

One single graph is able to encode complete information about the fusion rules of an anyon model. 
Unlike tables of fusion rules in which the essential information about an anyon model might be sometimes difficult to grasp, hidden inside a huge collection of numbers, a glance at a topological graph can rapidly tell us about the fundamental properties of an anyon model. Is the model Abelian or non-Abelian? Does it decompose into a product of simpler models? Which charges are obtained from which? They are all questions that can be immediately answered after a quick examination of the connectivity pattern of a topological graph. 

Braiding properties can also be nicely extracted from topological graphs. I have emphasized and discussed how diagonalization of a topological graph can yield important information, sometimes complete, about the topological $S$ and $T$ matrices of an anyon model. In some cases diagonalization of a topological graph might even turn obvious, as it is the case for anyon models such as $\mathbb{Z}_n$, $\mathbf{SU(2)}_k$, or $\mathbf{SO}(3)_k$, for which the corresponding graphs are familiar lattice patterns, whose eigenstates and eigenvalues are evident to us.

Graph representation of matrices is often used in physics and mathematics. Here, I have revealed the strength of a graph language to represent anyon models.


\newpage
\thispagestyle{empty}
\vspace*{2cm}
\begin{center}
\includegraphics[width=0.8\textwidth]{IntroBosonLattice.jpg}
\includegraphics[width=0.6\textwidth]{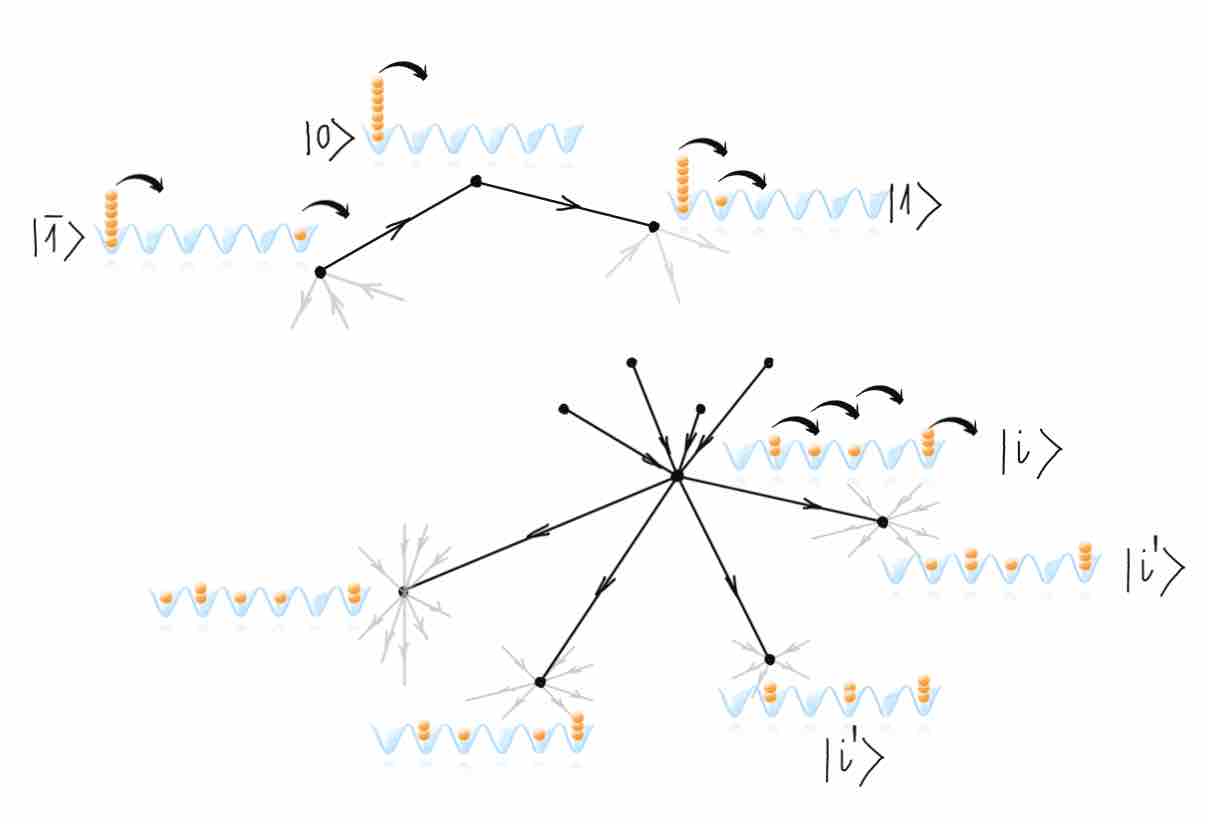}
\end{center}

\newpage
\pagestyle{fancy}
\fancyhf{}
\rhead{Closures and openings}
\rfoot{\thepage}
\vspace*{2cm}
\subsection*{The Boson-Lattice construction}
\addcontentsline{toc}{section}{The Boson-Lattice construction}

When trying to conceive a construction for anyon models, 
I had always clear that the Abelian models $\mathbb{Z}_n$ could serve as appropriate building blocks.
As for the principle of assembly, I had in mind some kind of procedure that would make copies of elementary anyon models indistinguishable, though I did not have a befitting language to articulate the idea of {\em indistinguishability} of anyon models.

The first key idea came through a visual abstraction. The graph language helped me to represent the Abelian model $\mathbb{Z}_n$ by a particle in a one-dimensional lattice. Triggered by this idea the principle of {\em bosonization} naturally arose as an appropriate way to implement the notion of indistinguishability.

There are exponentially many ways to define a graph in a boson lattice system. To define the Boson-lattice graph as the connectivity graph of Fock states was a guess. A guess that happened to be right.

The Boson-Lattice construction represents the collapse of two languages into one. It reveals that the mathematical language to describe anyon models can be the same as the one describing boson-lattice systems. It provides us with a physically meaningful language (the one of bosons, Fock states and tunneling Hamiltonians) to describe the mathematical properties (fusion rules and braiding rules) of anyon models.
I believe this physical language can help us  in our way to fill the explanatory gap between the mathematical and the physical sides of topological orders.


\newpage
\thispagestyle{empty}
\vspace*{2cm}
\begin{center}
\includegraphics[width=\textwidth]{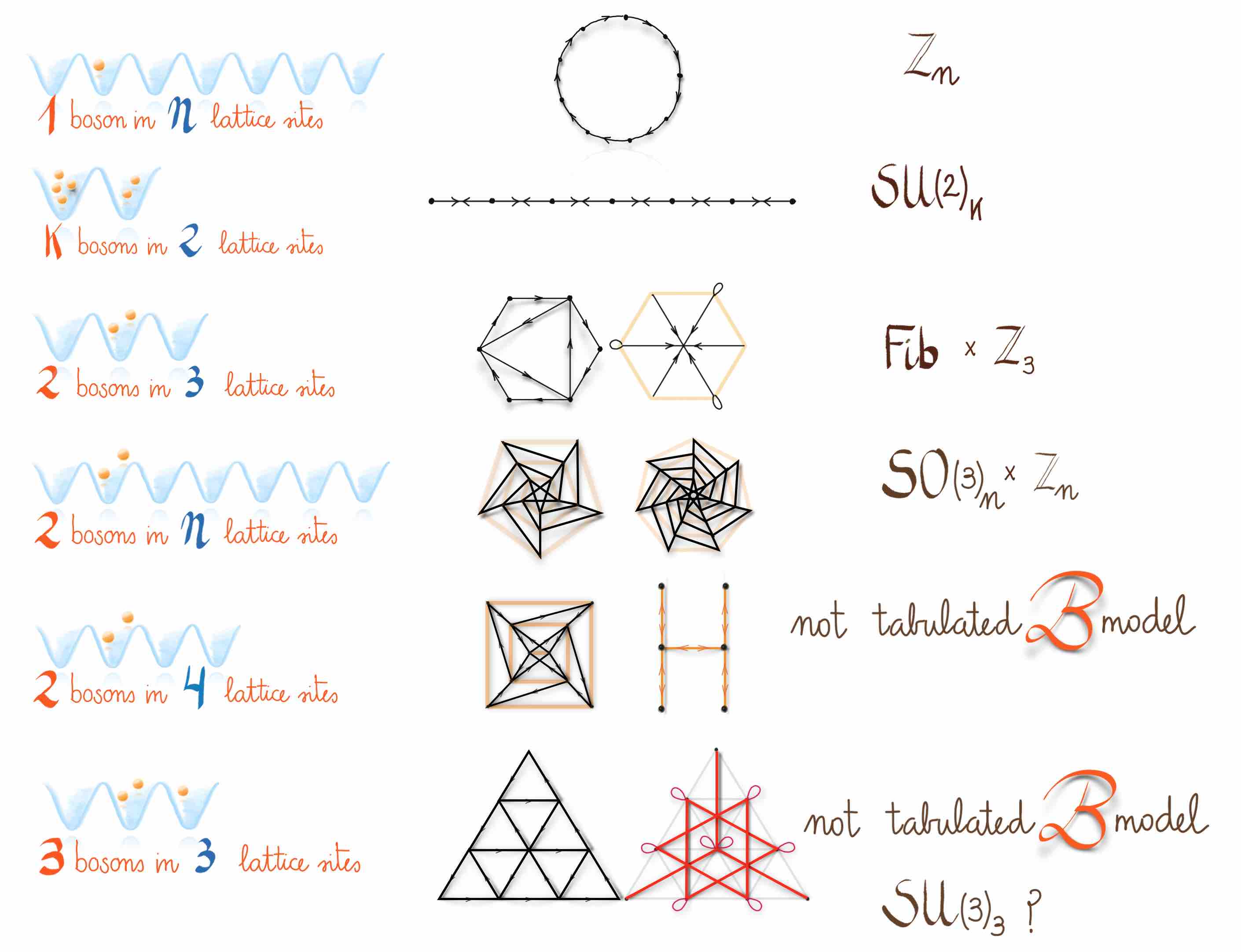}
\end{center}

\newpage
\pagestyle{fancy}
\fancyhf{}
\rhead{Closures and openings}
\rfoot{\thepage}
\vspace*{1cm}
\subsection*{The construction works}
\addcontentsline{toc}{section}{The construction works}
The Boson-Lattice construction provides an orderly systematic way to construct anyon models. By assembling different numbers of identical building blocks (by varying the number of bosons $k$) and by changing their size (varying the number of lattice sites $n$) the construction succeeds in generating towers of well known tabulated anyon models.

The construction reveals a skeleton for the space of anyon models. It tells us that towers of more and more complex anyon models can be constructed by assembling identical anyon models and making them indistinguishable.

\begin{center}
\includegraphics[width=0.8\textwidth]{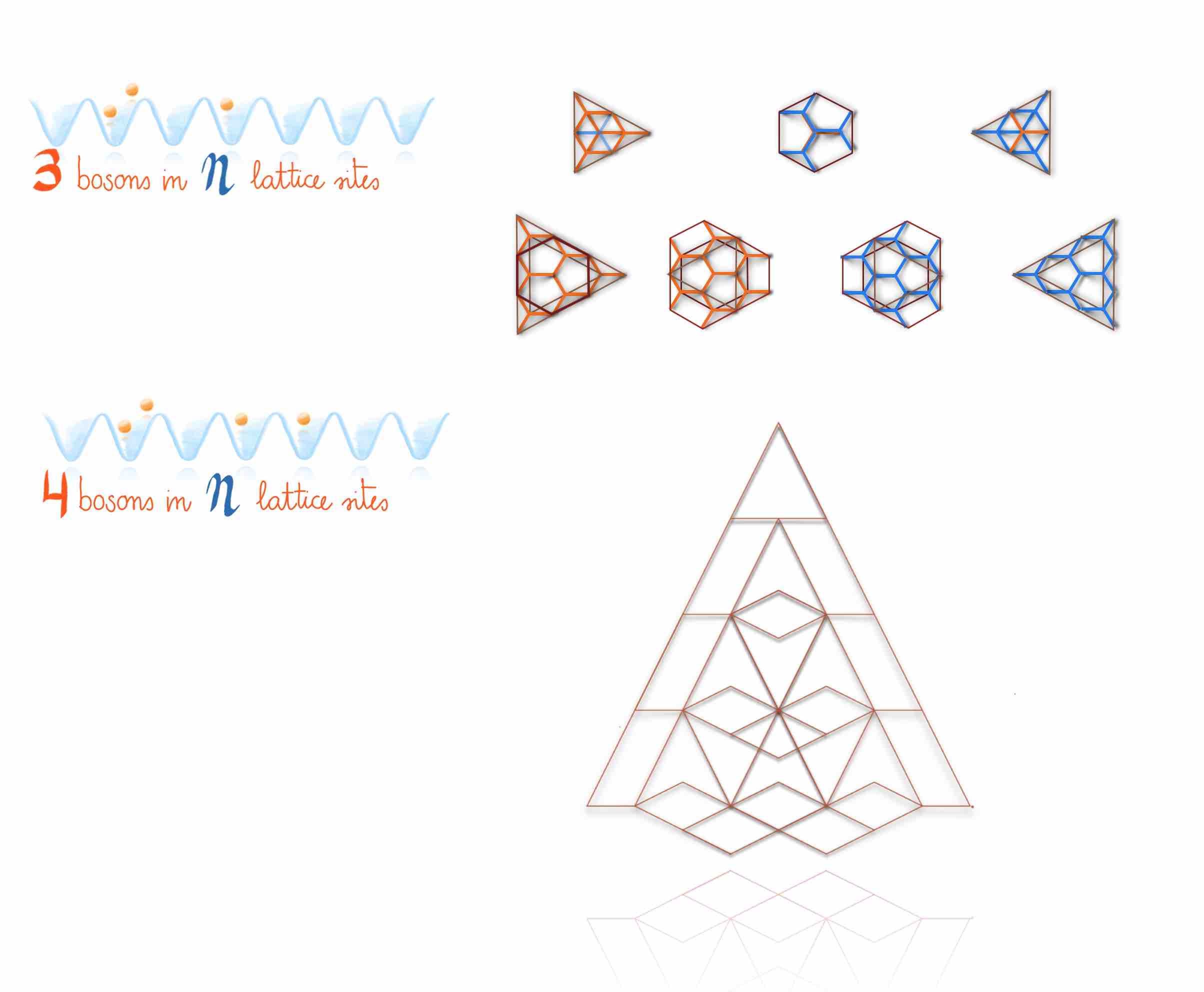}
\end{center}


\newpage
\thispagestyle{empty}
\vspace*{3cm}
\begin{center}
\includegraphics[width=0.85\textwidth]{IntroGraph.jpg}
\end{center}

\newpage
\pagestyle{fancy}
\fancyhf{}
\rhead{Closures and openings}
\rfoot{\thepage}
\vspace*{3cm}
\subsection*{Conformal and topological quantum field theories}
\addcontentsline{toc}{section}{Conformal and topological quantum field theories}
I have focused in this work on the construction of {\em modular} anyon models, for which corresponding conformal field theories and topological quantum field theories exist.
It would be very interesting to delineate a graph of correspondences between the Boson-Lattice construction and conformal and topological quantum field theories. 

How are known concepts in Chern-Simons theory and conformal field theory expressed in the Boson-Lattice language? 
Might the Boson-Lattice construction bring light into our understanding of the anatomy of conformal field theories and topological quantum field theories?

The Boson-Lattice construction can also serve to build up non-modular anyon models, for which no conformal field theories exist.


\newpage
\thispagestyle{empty}
\vspace*{3cm}
\begin{center}
\includegraphics[width=0.73\textwidth]{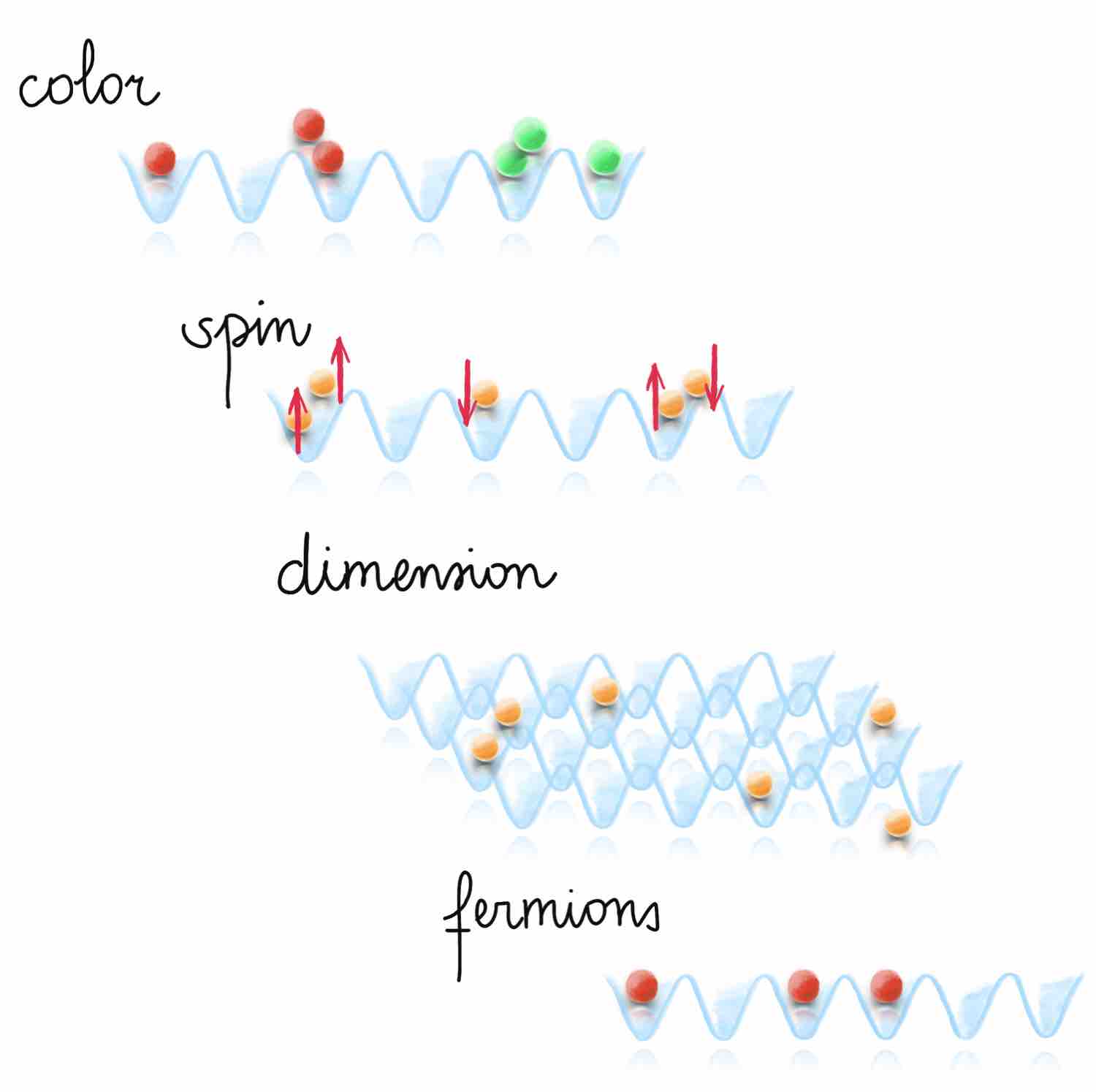}
\end{center}

\newpage
\pagestyle{fancy}
\fancyhf{}
\rhead{Closures and openings}
\rfoot{\thepage}
\vspace*{3cm}
\subsection*{Generalizations of the Boson-Lattice construction}
\addcontentsline{toc}{section}{Generalizations of the Boson-Lattice construction}
The Boson-Lattice construction can be enriched by adding internal degrees of freedom to the bosons participating. For example, we can consider bosons with color or spin, or also change the dimensionality of the lattice. In this way, anyon models corresponding to non-chiral topological orders, such as {\em quantum doubles}, can be generated.

Moreover, the construction can be extended to fermions. By considering Fermion-Lattice graphs other towers of anyon models are built up.

Additionally, other anyon models beyond $\mathbb{Z}_n$ can be used as initial building blocks at the first level of the construction.


\newpage
\thispagestyle{empty}
\vspace*{2.5cm}
\begin{center}
\includegraphics[width=0.95\textwidth]{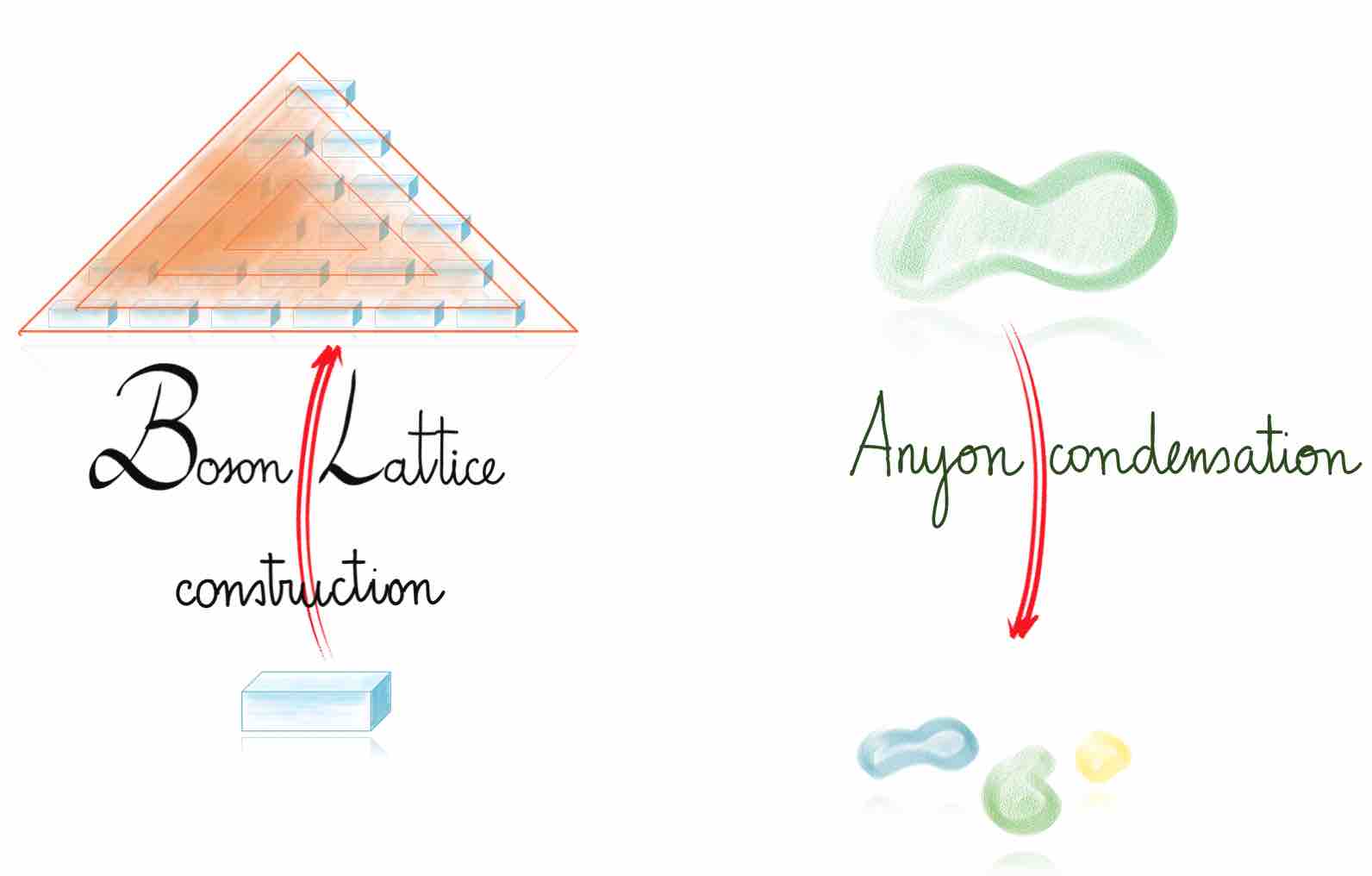}
\end{center}

\newpage
\pagestyle{fancy}
\fancyhf{}
\rhead{Closures and openings}
\rfoot{\thepage}
\vspace*{3cm}
\subsection*{Anyon condensation}
\addcontentsline{toc}{section}{Anyon condensation}

In light of the Boson-Lattice construction, the reverse process of disintegrating anyon models into simpler pieces acquires an enlightened perspective.

Within the Boson-Lattice picture, anyon condensation is a condensation of actual bosons. Moreover, the concept of topological symmetry breaking corresponds to actual symmetry breaking in the Boson-lattice system.

I find extremely interesting to investigate how anyon condensation is precisely described in the Boson-Lattice language, and, moreover, whether the Boson-Lattice construction can guide us to develop a systematic framework to describe anyon condensation, for which no fully general description is known.


\newpage
\thispagestyle{empty}
\vspace*{4.5cm}
\begin{center}
\includegraphics[width=0.9\textwidth]{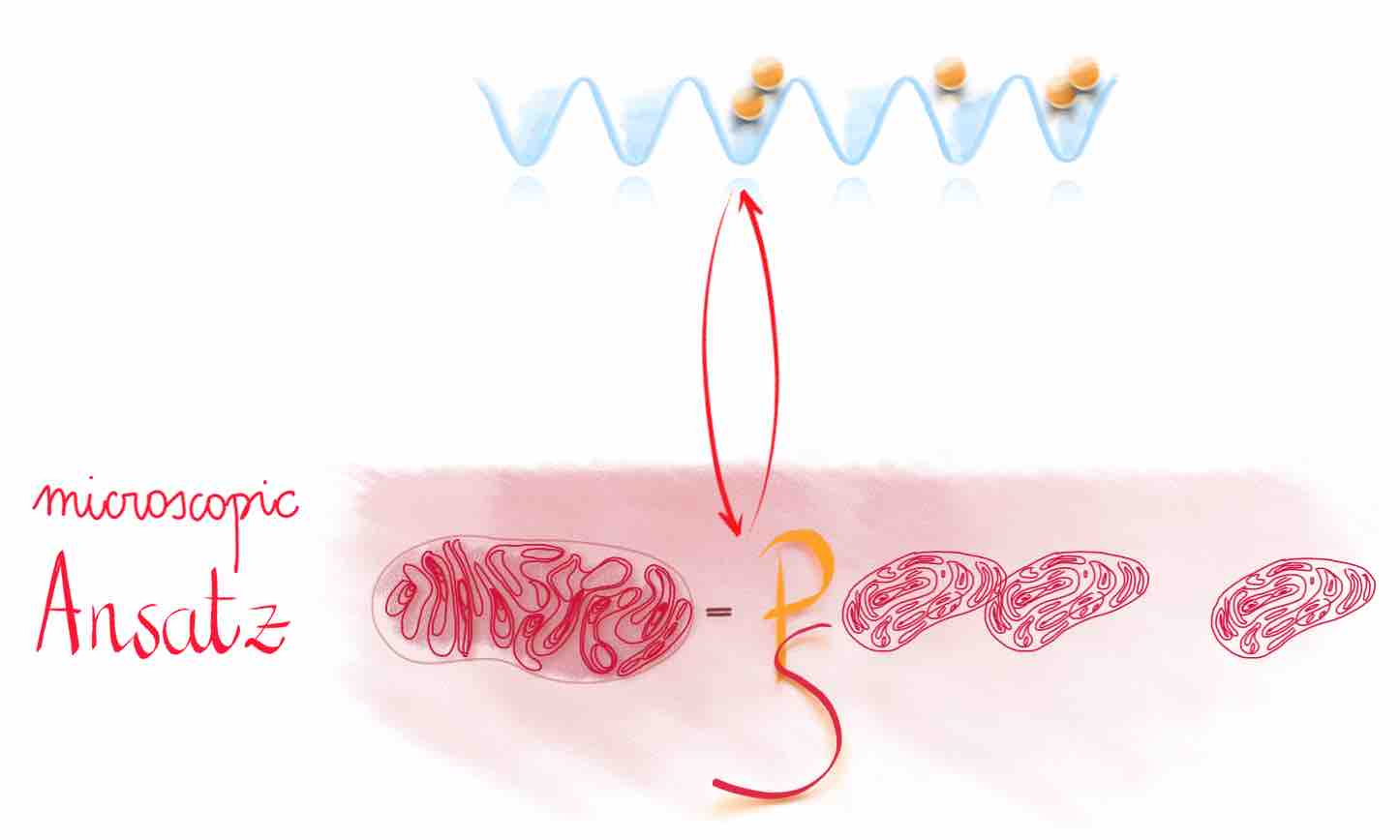}
\end{center}

\newpage
\pagestyle{fancy}
\fancyhf{}
\rhead{Closures and openings}
\rfoot{\thepage}
\vspace*{1.5cm}
\subsection*{A dual construction at the physical level}
\addcontentsline{toc}{section}{}

Especially interesting to me is the prospect of developing a dual construction at the microscopic physical level. I believe the Boson-Lattice construction for anyon models can inspire the blueprint of a construction of many-body wave functions and Hamiltonians for the corresponding topological orders.

A dual bosonization principle of assembly could be used to orderly build up complex topologically ordered many-body wave functions from elementary ones.
Such systematic framework would reveal a dual anatomy in the phase space of topologically ordered systems.

In previous work \cite{Paredes1,Paredes2,Paredes3}, I have proposed an Ansatz for the systematic construction of non-Abelian topologically ordered states by symmetrizing identical copies of simpler states like Abelian topological states. This type of construction is known to be the one behind non-Abelian quantum Hall states \cite{NA1,NA4,NA5,Barkeshli1,Barkeshli2,Barkeshli3}. I have revealed its potential to describe the structure of general non-Abelian topological states, describing, for example, other families of interesting non-Abelian states such as quantum doubles or string-net condensates.

How does the notion of indistinguishability of wave functions match the one of indistinguishability of anyon models?
In which way is the symmetrization of many-body wave functions connected to the Boson-Lattice construction?

I believe the Boson-lattice construction can also guide the systematic construction of parent Hamiltonians for topological states.
Boson-Lattice graphs can encode the physical ingredients (local degrees of freedom, interactions between them) of the corresponding topological order. 
They constitute dual entities able to simultaneously embody the mathematical and physical parts of topological orders.

It would be alluring to investigate the connections between the many-body wave functions and Hamiltonians emerging from the Boson-Lattice construction and those of seminal topological systems and models, such as {\em fractional quantum Hall systems} \cite{NA1,NA2,NA3,NA4,NA5,NA6,NA7}, {\em quantum loop models} \cite{Kitaev2, QLM1, QLM2, QLM3}, and {\em string-net models} \cite{SNC1,SNC2,SNC3,SNC4}.


\newpage
\thispagestyle{empty}
\vspace*{2cm}
\begin{center}
\includegraphics[width=0.9\textwidth]{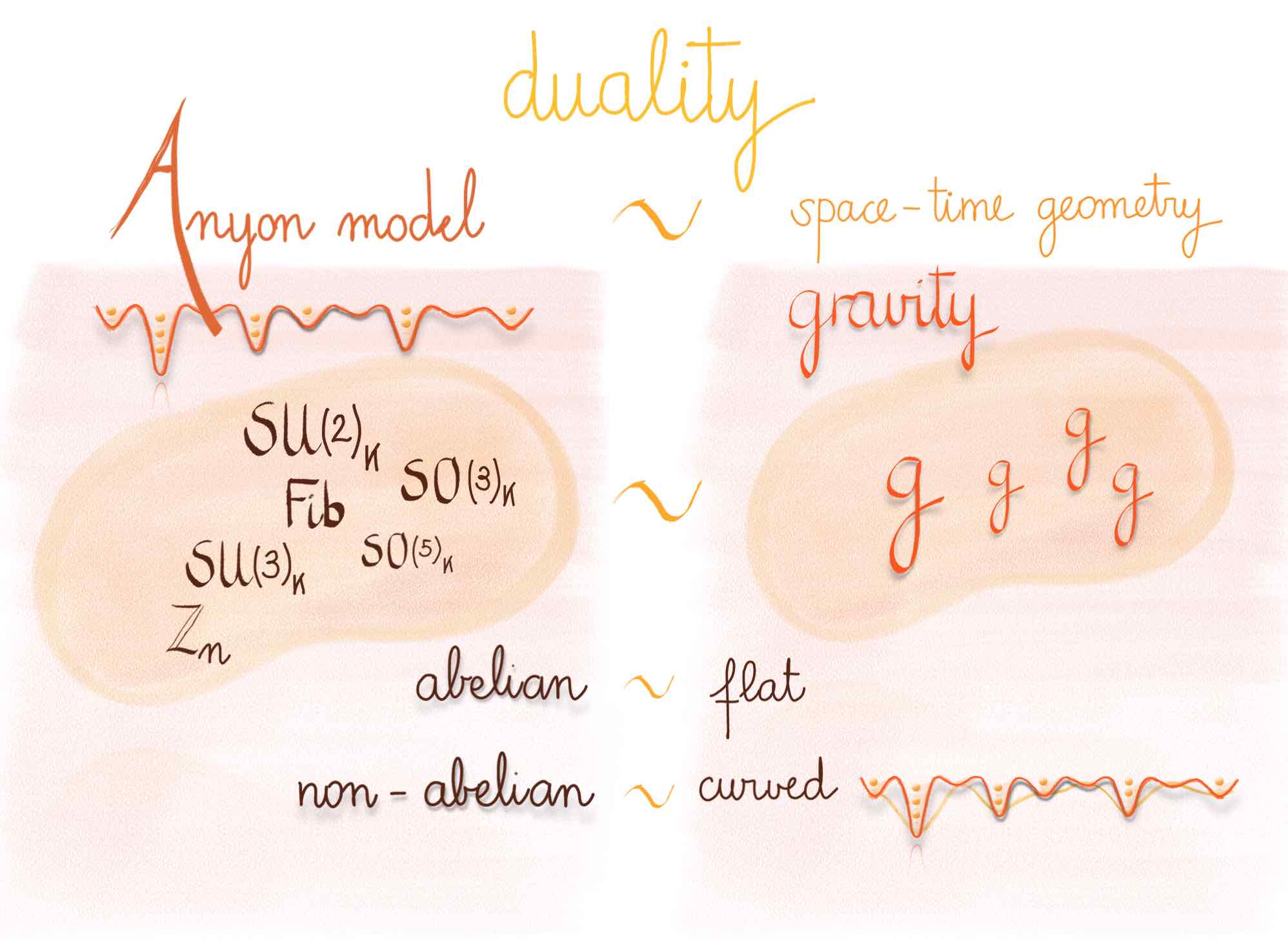}
\end{center}

\newpage
\pagestyle{fancy}
\fancyhf{}
\rhead{Closures and openings}
\rfoot{\thepage}
\vspace*{3.5cm}
\subsection*{Duality between anyon models and gravity}
\addcontentsline{toc}{section}{Duality between anyon models and gravity}

I have drawn a connection between anyon models and curved-space geometries.
Within the Boson-Lattice approach an anyon model corresponds to a system of bosons in a lattice whose metric has been curved.
Abelian models are associated with flat geometries, whereas non-Abelian models correspond to curved ones. 

This connection is extremely attractive to me. What type of space-geometries do anyon models correspond to? 
How can time be included into the theory?

I  believe that this connection anticipates a beautiful duality between anyon models and gravity, which I feel compelled to unveil.


\newpage
\thispagestyle{empty}

\vspace*{4cm}

Acknowledgment of DFG funding:

This work was supported by the Deutsche Forschungsgemeinschaft (DFG, German Research Foundation) under Project No. 277974659 via Research Unit FOR 2414.


\newpage
\thispagestyle{empty}
\pagestyle{fancy}
\fancyhf{}
\rhead{References}
\cfoot{\thepage}

\newpage
\thispagestyle{empty}
\vspace*{7cm}
\begin{center}
\includegraphics[width=0.4\textwidth]{ConstructionTitleFigure.jpg}
\end{center}

\end{document}